\documentclass{vldb}

\usepackage{booktabs} 
\usepackage{times} 
\usepackage{color} 
\usepackage{balance} 
\usepackage{url} 
\usepackage{tabularx}
\usepackage{siunitx} 
\usepackage{epsfig,tabularx,subfigure,multirow, graphicx}
\usepackage{enumitem}

\usepackage{hyperref} 

\usepackage{amsthm,amssymb,amsmath}  

\newcolumntype{L}[1]{>{\raggedright\arraybackslash}p{#1}}
\newcolumntype{C}[1]{>{\centering\arraybackslash}p{#1}}
\newcolumntype{R}[1]{>{\raggedleft\arraybackslash}p{#1}}

\usepackage[linesnumbered,ruled,vlined]{algorithm2e}
\SetKwRepeat{Do}{do}{while}
\SetCommentSty{mycommfont}

\long\def\comment#1{}

\newcommand{\nop}[1]{}

\newtheorem{example}{\bf Example}

\theoremstyle{remark}

\theoremstyle{definition}
\newtheorem{definition}{\bf Definition}

\vldbTitle{Task Assignment in Spatial Crowdsourcing [Experiments and Analyses]  (Technical Report)}
\vldbAuthors{Peng Cheng, Xun Jian and Lei Chen}
\vldbDOI{https://doi.org/TBD}

\author{
	{Peng Cheng, Xun Jian, Lei Chen}
	\vspace{1.6mm}\\
	\fontsize{10}{10}\selectfont\itshape
	The Hong Kong University of Science and Technology, Hong Kong, China\\
	\fontsize{9}{9}\selectfont\ttfamily\upshape
	\{pchengaa, xjian, leichen\}@cse.ust.hk
}

\begin{document}

\title{Task Assignment in Spatial Crowdsourcing [Experiments and Analyses] (Technical Report)}
\maketitle

\begin{abstract}
Recently, with the rapid development of mobile devices and the crowdsourcing
platforms, the spatial crowdsourcing has attracted much attention
from the database community. Specifically,  spatial crowdsourcing
refers to sending a location-based request to workers according to
their positions, and workers need to physically move to specified locations to conduct tasks.
Many works have studied task assignment problems in spatial 
crowdsourcing, however, their problem settings are different 
from each other. Thus, it is hard to compare the 
performances of existing algorithms on task assignment in spatial crowdsourcing.
In this paper, we present a comprehensive experimental comparison 
of most existing algorithms on task assignment in spatial crowdsourcing. 
Specifically, we first give general definitions about spatial workers 
and spatial tasks based on definitions in the existing works such that the existing 
algorithms can be applied on the same synthetic and real data sets. 
Then, we provide a uniform implementation for all the tested algorithms 
of task assignment problems in spatial crowdsourcing (open sourced). Finally, based on 
the results on both synthetic and real data sets, we discuss the strengths 
and weaknesses of tested algorithms, which can guide future research 
on the same area and practical implementations of spatial crowdsourcing systems.
\end{abstract}

\section{Introduction}
\label{sec:introduction}

With the ubiquity of smart devices equipped with various sensors (e.g., GPS) 
and the convenience of wireless mobile networks (e.g., 5G), nowadays people 
can easily participate in \textit{spatial tasks} requiring to be conducted at specified 
locations that are close to their current locations, such as taking photos/videos \cite{GoogleMapStreetView}, delivering packages \cite{taskrabbit}, and/or  reporting waiting times 
of hot restaurants \cite{fieldagent}. As a result, a new framework, namely \textit{spatial crowdsourcing} \cite{kazemi2012geocrowd}, which 
enables \textit{spatial workers} to conduct spatial tasks, has emerged in both academia 
(e.g., the database community) and industry (e.g., Uber\cite{uber}). In spatial crowdsourcing systems (e.g., gMission \cite{chen2014gmission, gmissionhkust} 
and MediaQ \cite{kim2014mediaq}), the active workers can only conduct 
 spatial tasks close enough to them such that they can 
physically move to the required locations before the deadlines of 
tasks. Therefore, 
studying and designing effective strategies for helping workers to
conduct spatial tasks to maximize the overall \textit{utility} (defined in Section \ref{sec:definition}) of systems is the major 
goal of the existing studies in spatial crowdsourcing 
\cite{kazemi2012geocrowd, kazemi2013geotrucrowd, to2015server, 
	deng2013maximizing, li2015oriented, cheng2016task, cheng2015reliable, 
	hassan2014multi, pournajaf2014spatial, hu2016crowdsourced}.

In a spatial crowdsourcing platforms, spatial tasks are keeping on arriving and being completed, 
and workers are free to join or leave. In addition, the platforms have no 
information about the future arrival tasks and workers. In general, there are two modes to assign workers to tasks: 1) batch-based mode \cite{kazemi2012geocrowd, to2015server, cheng2016task, hu2016crowdsourced}, where the platforms periodically assign the available workers to the opening tasks in the current timestamps; 2) online mode \cite{deng2013maximizing, li2015oriented}, where the platforms immediately assign suitable tasks to the worker when he/she joins in the platform (In platforms with more workers than tasks, when a new task is created, the system will assign the most suitable worker to it). Specifically, we illustrate the spatial crowdsourcing with the following examples:

\begin{example}
(Car-hailing Services) Car-hailing services allow riders to post their travel requests to the system, then suitable cabs will be dispatched to them based on the locations of riders and cabs. Many industrial applications (e.g., Uber \cite{uber} and DiDi Chuxing \cite{didi}) provide car-hailing services. In car-hailing systems, a cab and a travel request can be treated as a worker and a task, respectively. Car-hailing systems usually try to match a travel request with a closest cab such that the travelling distance for the cab to pick up the rider is minimized and the waiting time of the rider is also minimized. The existing car-hailing systems can either work on batch-based mode (i.e., assigning available cabs to travel requests every 2 seconds) or online mode (i.e., assigning the most suitable cab to the travel request immediately when it appears). Thus, task assignment in car-hailing services can be modeled as a spatial crowdsourcing problem.
\end{example}

\begin{example}
(Mobile Audit Services) Mobile audit services allow companies to create their location-specific in-store audit projects, which can be reporting the on-shelf status of commodities, checking the prices of goods inside stores, and surveying the thinking of shoppers towards particular products. Then, shoppers with Mobile audit Apps (e.g., Field Agent \cite{fieldagent}) installed (noted as agent) will be assigned to proper tasks (e.g., closest tasks) and contribute their efforts. Since the agents are not totally correct, some quality control mechanisms (.e.g, Majority Voting \cite{kazemi2013geotrucrowd}) are used to aggregate the answers from different agents for the same task such that the returned answers are credible. With the returned answers, the business companies can analyze the almost real-time results and react quickly. The mobile audit system also can process the tasks on batch-based mode or online mode, which can also be modeled as a task assignment problem in spatial crowdsourcing.
\end{example}

To handle the task assignment problems in spatial crowdsourcing, existing studies proposed various algorithms to overcome the dynamics of the spatial tasks and workers, and to address problems with different utility definitions. For example, in \cite{kazemi2012geocrowd, kazemi2013geotrucrowd, to2015server, deng2013maximizing, li2015oriented, tong2016online} the utility of the spatial crowdsourcing system is defined as the number of finished task while in \cite{cheng2015reliable} it is defined as the reliability and diversity of finished tasks.
However, no existing work has compared the algorithms 
tailored for different settings in spatial crowdsourcing, thus the 
results in different works cannot be compared directly and it is difficult for users to know which algorithm to apply in the real applications.

In this paper, we provide a fair comparison study over the existing algorithms under a general spatial crowdsourcing definition to show their pros and cons.
Currently, there is no highly customizable spatial crowdsourcing platforms 
to run comparison experiments for all the existing spatial crowdsourcing 
algorithms. We utilize an open source simulation tool to generate data 
sets either following given distributions (e.g., Normal distribution, 
Zipf distribution, Skewed distribution and Uniform distribution) or 
based on real spatial/temporal data sets (e.g., Gowalla and Twitter), 
which can help to compare algorithms with different parameters 
more accurately.
In addition, we set up a common experiment setting for all the existing notable
methods, and show the performances 
of the methods on important spatial crowdsourcing metrics, 
such as running time, numbers of finished 
tasks, and average moving distance. As a result, our uniform implementation \cite{datasetGenerator, algorithms} can 
avoid the ``noises'' from implementation skills (e.g., Java v.s. C++), settings and metrics, 
which enable us to report the true contributions of algorithms.

To summarize, we try to make the following contributions:

\begin{itemize}
	\item We propose a general definition for task assignment in spatial crowdsourcing in Section \ref{sec:definition}, which can be a footstone for the future studies in this area.
	
	\item We provide uniform baseline implementations for the most notable algorithms in both batch-based and online mode. These implementations adopt common basic operations and offer a benchmark for comparing with future studies in this area.
	
	\item We propose an objective and sufficient experimental evaluation and test the performances of  the most notable algorithms over extensive benchmarks in Section \ref{sec:exp}.
	
	\item We discuss the advantages and disadvantages of two task assignment modes (batch-based mode and online mode) based on the results of our experimental evaluation in Section \ref{sec:summary}.
\end{itemize}

Section \ref{sec:sat_algorithms} and Section \ref{sec:wst_algorithms} introduce existing algorithms in batch-based mode and online mode respectively. Section \ref{sec:conclusion} concludes this paper.

\section{Problem Definition}
\label{sec:definition}

In this section, we give a general definition of task assignment in spatial crowdsourcing, which is based on the definitions in existing studies \cite{kazemi2012geocrowd, kazemi2013geotrucrowd,
	deng2013maximizing, cheng2015reliable}.

\begin{definition} $($Dynamic Moving Workers$)$ Let
	$W$ $=\{w_1,$ $w_2,$ $...,$ $w_n\}$ be a set of $n$
	workers. Each worker $w_i$ ($1\leq i\leq n$) is located at position
	$l_i(p)$ at timestamp $p$, can move with velocity $v_i$, specifies a square working area with side length $a_i$, and has a reliability value $r_i\in[0,1]$ and a capacity value $c_i$.
	\qquad $\blacksquare$ \label{definition:worker}
\end{definition}

In Definition \ref{definition:worker},  worker
$w_i$ can move dynamically with speed $v_i$ in any direction, and at
each timestamp $p$, he/she is located at  location $l_i(p)$. He/she
prefers to conduct the tasks within his/her square working area
centering at the spatial place $l_i(p)$ with side length of  $a_i$. Based on 
the historical performance of each worker, we can estimate his/her reliability 
values $r_i\in[0,1]$, which indicates the probability that he/she can correctly finish the assigned 
task. Moreover, each worker may accept at most $c_i$ tasks at the same 
time and conduct them one by one. In spatial crowdsourcing systems, a worker $w_i$ can be either available or busy. Here being available 
means the worker can be assigned with more tasks while being busy indicates 
the number of assigned tasks to worker $w_i$ reaches the his/her capacity  $c_i$ and no more tasks can be assigned unless he/she finishes
or rejects some assigned tasks.

\begin{definition}
	(Spatial Tasks) Let $T=\{t_1, t_2,
	..., t_m\}$ be a set of time-constrained spatial tasks. Each task $t_j$ ($1\leq j\leq m$) is published at timestamp $s_j$, locate
	at a specific location $l_j$, and is associated with a deadline $e_j$. To guarantee the 
	quality, task $t_j$ may require $b_j$ answers and specify a required 
	quality level $q_j$.\qquad
	$\blacksquare$ \label{definition:task}
\end{definition}

Usually, a task
requester creates a time-constrained spatial task $t_j$ at timestamp $s_j$, which
requires workers to physically reach a specific location $l_j$ before its deadline $e_j$. In order to tackle the intrinsic error rate (unreliability) of workers, different accuracy control techniques are used in existing studies \cite{cheng2015reliable, kazemi2013geotrucrowd, dawid1979maximum, ipeirotis2010quality}. Without loss of generality and for the ease of presentation, in this paper, we consider the spatial tasks with binary (Yes/No) choices and use Majority Voting \cite{kazemi2013geotrucrowd} to aggregate the answers from different workers such that the expected quality scores of tasks are satisfied. (To avoid draws, we can require $b_j$ to be an odd number.) For example, to check the stock status of a particular product (e.g., Coke Cola) in a store, the question of a spatial crowdsourcing task can be ``Whether the coke cola in the store has enough stock?'' and the answer could be ``Yes'' or ``No''.
Specifically, for a task $t_j$ with $b_j$ answers, we report the majority answer choice (selected by no less than $\frac{b_j+1}{2}$ workers) as the final result for task $t_j$. Let 
the set $W_j$ be the workers that answer task $t_j$.
We can compute the expected accuracy of a task as follows:

\begin{equation}
\Pr(W_j) = \sum_{x = \frac{b_j+1}{2}}^{b_j}\sum_{W_{j}^x}
\Big(\prod_{w_i \in W_{j}^x}r_{ij}\prod_{w_i \in W_j - W_{j}^x}(1-r_{ij})\Big),
\label{eq:task_accuracy}
\end{equation}

\noindent where $W_{j}^x$ indicates the subsets with exact $x$ workers out of the worker set $W_j$ who answered task $t_j$. Particularly, $\Pr(W_j)$ can represent the probability that the final answer of task $t_j$ is correct.

\begin{definition}
(Assignment Instance Set) At timestamp $p$, an \textit{assignment instance
		set}, denoted by $I_p$, is a set of worker-and-task assignment pairs
	in the form $\langle w_i, t_j\rangle$, where a spatial task $t_j$ is assigned to a worker $w_i$ while satisfying the constraints of workers and tasks. The utility of worker-and-task pair $\langle w_i, t_j\rangle$ is noted as $U(w_i, t_j)$.\qquad
	$\blacksquare$ \label{definition:instance}
\end{definition}

Here, each worker-and-task pair $\langle w_i, t_j\rangle$ in $I_p$ 
indicates the required location $l_j$ of 
task $t_j$ is in the working area of worker $w_i$ and he/she can reach $l_j$ 
before its arrival deadline $e_j$. Moreover, the capacity constraint of worker 
$w_i$ is satisfied, which means the number of assigned tasks for worker $w_i$ 
is not larger than his/her capacity $c_i$.  Assigning worker $w_i$ to task $t_j$ has utility $U(w_i, t_j)$, which can be defined in different forms. For example, in \cite{kazemi2012geocrowd}, it is simply defined as $U(w_i, t_j)=1$, which means only the number of assigned tasks is concerned.

Now we give the  formal definition of the task assignment in general spatial crowdsourcing (TA-GSC) problem as follows:
\begin{definition}
(TA-GSC Problem) Given a set of dynamic moving workers $W$ and a set of spatial tasks $T$, the TA-GSC problem is to find a task assignment instance set $I$ to maximize the total utility $\sum_{\langle w_i, t_j\rangle \in I} U(w_i, t_j)$ such that the following constraints are satisfied:
\begin{itemize}
	\item Working Area Constraint: worker $w_i$ can only be assigned to tasks located within his/her working area;
	
	\item Deadline Constraint: worker $w_i$ can only be assigned to tasks that he/she can arrive at before their deadlines;
	
	\item Capacity Constraint: at any time, worker $w_i$ can be assigned with at most $c_i$ tasks;
\end{itemize}
\end{definition}

\begin{figure}[ht]
	\centering
	\scalebox{0.25}[0.25]{\includegraphics{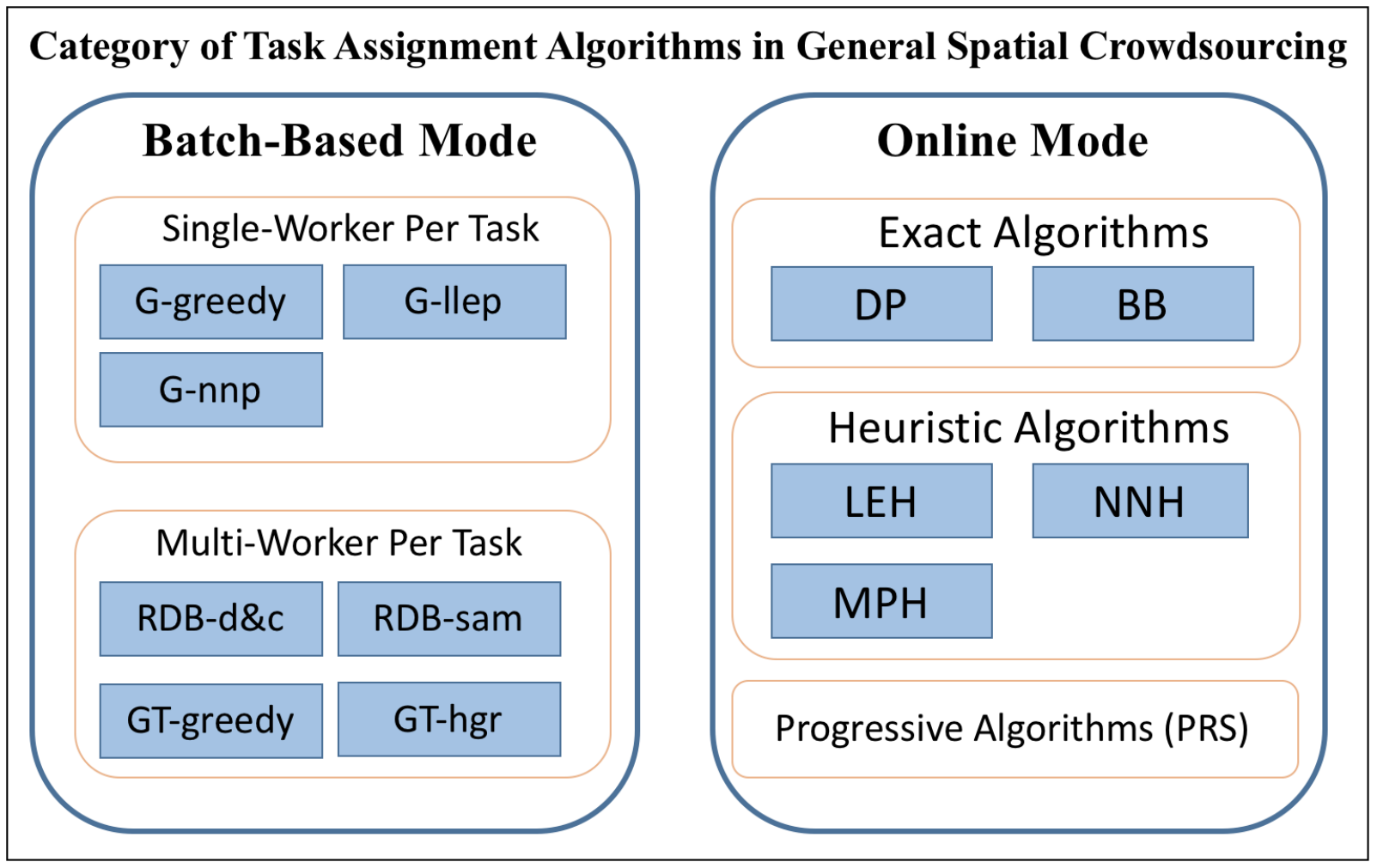}}\vspace{-2ex}
	\caption{\small Taxonomy of Task Assignment Algorithms for General Spatial Crowdsourcing.}
	\label{fig:taxonomy}\vspace{-3ex}
\end{figure}

Under this definition, we tested the notable existing  algorithms to solve the task assignment problems in general spatial crowdsourcing. Figure \ref{fig:taxonomy} shows a taxonomy of the tested algorithms. We present them one-by-one in the following two sections.

Table \ref{table0} summarizes the commonly used symbols.

\begin{table}
	\begin{center}
		\caption{\small Symbols and Descriptions.} \label{table0}
		{\small\scriptsize
			\begin{tabular}{l|l}
				{\bf Symbol} & {\bf \qquad \qquad \qquad\qquad\qquad Description} \\ \hline \hline
				$W$   & a set of dynamically moving workers\\
				$T$   & a set of time-constrained spatial tasks\\
				$l_i(p)$  & the position of worker $w_i$ at timestamp $p$\\
				$a_i$   & the side length of the working area of worker $w_i$\\
				$v_i$   & the moving velocity of worker $w_i$\\
				$r_i$   & the reliability value of worker $w_i$\\
				$c_i$   & the capacity of worker $w_i$\\
				$s_j$   & the timestamp of creating task $t_j$\\
				$e_j$   & the deadline of arriving at the location of task $t_j$\\
				$l_j$   & the position of task $t_j$\\
				$b_j$   & the number of required answers of task $t_j$\\
				$q_j$   & the required quality level of task $t_j$\\
				$\langle w_i, t_j\rangle$ & the worker-and-task assignment pair\\
				$U(w_i, t_j)$ & the utility value of the worker-and-task assignment pair $\langle w_i, t_j\rangle$\\
				\hline
				\hline
			\end{tabular}
		}
	\end{center}\vspace{-4ex}
\end{table}

\section{Algorithms in Batch-based Mode}
\label{sec:sat_algorithms}

In this section, we introduce the typical batch-based algorithms for 
\textsf{TA-GSC} problems, which periodically assign the ``current'' available workers to unfinished spatial tasks. The general framework of batch-based algorithm for \textsf{TA-GSC} problems is shown in Algorithm \ref{alg:framework}. In each iteration of the framework, it uses the batch-based algorithms to match the available workers to unfinished tasks, then notifies the workers to conduct their assigned tasks.

From the 
perspective of the number of required answers of each task, the batch-based algorithms can be categorized into two groups: 1) Single-worker per task algorithms, where each task needs one worker to answer; 2) Multi-worker per task algorithms, where each task needs more than one worker to answer. The batch problem in each iteration of Algorithm \ref{alg:framework} (lines 2 - 8) of the first group algorithms can be reduced to the maximum flow problem while that is NP-hard for the second group of algorithms. We introduce them one-by-one in the rest of this section.

\subsection{Single-Worker Per Task algorithms}
For single-worker per task algorithms introduced in this section, the utility function of assigning worker $w_i$ to task $t_j$ is defined as $U(w_i, t_j)=1$, which indicates the algorithms wants to maximize the assigned number of tasks. Then, the \textsf{TA-GSC} problem in each batch/iteration can be reduced to the maximum flow problem.
We first represent 
the reduction of the maximum flow problem when each task only needs one worker to answer, then introduce the single-worker per task algorithms.

\begin{algorithm}[t]
	\DontPrintSemicolon
	\KwIn{\small A time interval $\varPhi$}
	\KwOut{\small A set of worker-and-task assignment pairs within the time interval $\varPhi$}
	
	\While{current time $\varphi$ is in $\varPhi$}{
		retrieve all the available spatial tasks to $T$\;
		retrieve all the available workers to $W$\;
		\ForEach{$w_i \in W$}{
			obtain a set, $T_i$, of valid tasks for worker $w_i$\;
		}
		
		use batch-based task assignment algorithms to obtain a good assignment set $I$\;
		
		\ForEach{$\langle w_i, t_j \rangle \in I$}{
			inform worker $w_i$ to conduct task $t_j$\;
		}
	}
	
	\caption{The Framework of Batch-based Algorithms}
	\label{alg:framework}
\end{algorithm}

\subsubsection{Reduction to Maximum Flow Problem}
When each task needs only one worker, the problem 
to maximize the number of assigned tasks in each batch/iteration can be reduced to the maximum flow problem. 
For a set of available workers $W$ 
and a set of available tasks $T$, we can create a flow network 
graph $G=(V, E)$ with $V$ as the set of vertices, and $E$ as
the set of edges. The set $V$ contains 
$|W| +|T| + 2$ vertices. Each worker $w_i$ maps to 
a vertex $w_i$ and each task $t_j$ maps to a vertex $t_j$ in graph $G$. 
In addition, we create a $src$ vertex and a $dest$ vertex. We first 
connect $src$ vertex and every worker vertex $w_i$ and set the capacity 
for each of these edges as the capacity $c_i$ of worker $w_i$ since each 
worker can buffer at most $c_i$ tasks. Each task vertex $t_j$ is linked to 
the $dest$ vertex and the capacity is set to 1, as each task only needs 
one worker to perform. What is more, as each worker $w_i$ can only 
accept the tasks located inside their working areas $a_i$, for every 
worker vertex $w_i$ we add edges to all the tasks vertices 
that the corresponding tasks are inside the spatial working area $a_i$, and set the capacity of each edge to 1.

\begin{figure}[h]\centering
	\subfigure[][{\scriptsize An Example of $W$ and $T$}]{
		\scalebox{0.4}[0.4]{\includegraphics{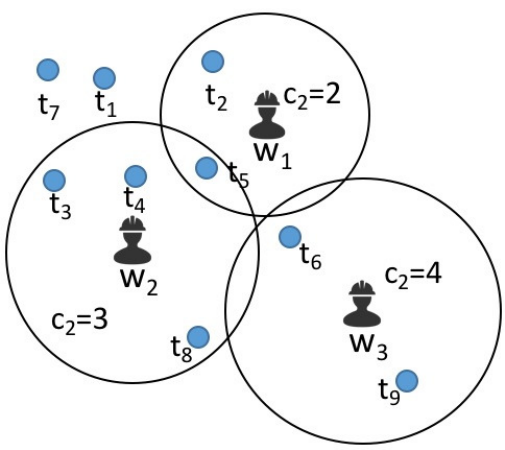}}
		\label{subfig:flow_layout}}
	\subfigure[][{\scriptsize Flow Network graph $G= (V,E)$}]{
		\scalebox{0.12}[0.11]{\includegraphics{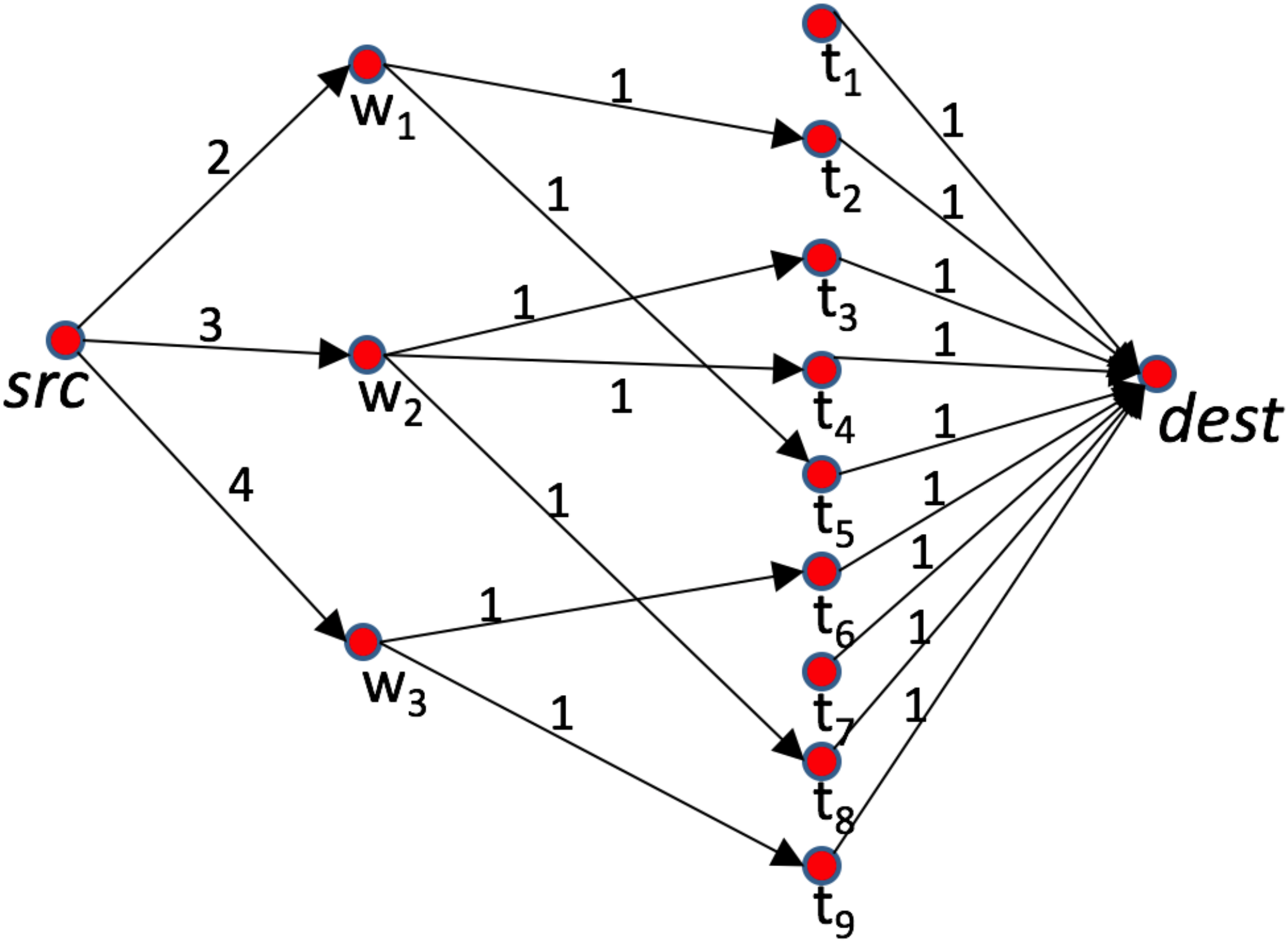}}
		\label{subfig:flow_network}}\vspace{-2ex}
	\caption{\small An Example of the Reduction of Maximum Flow Problem.}
	\label{fig:maximum_flow_example}
\end{figure}

Figure \ref{fig:maximum_flow_example} illustrates an example of this 
reduction. In Figure \ref{subfig:flow_layout}, each 
worker $w_i$ has a capacity value $c_i$ and a round working area around him/her. At the same time, Figure 
\ref{subfig:flow_network} shows the reduced maximum flow network graph. 
One link from worker vertex $w_i$ to task vertex $t_j$ exists only when the 
task $t_j$ is located inside the working area $a_i$ of worker $w_i$. For example, 
worker vertex $w_3$ is connected to task vertex $t_6$ as task $t_6$  locates
inside the working area of worker $w_3$. 

With the reduction of the maximum flow problem, existing maximum flow 
algorithms can be used to solve these task assignment problem for each batch/iteration. The Ford-Fulkerson algorithm \cite{kleinberg2006algorithm} is one 
well-known algorithm to compute the maximum flow. The idea behind Ford-Fulkerson algorithm is that it starts sending flow from the source vertex to the destination vertex, as long as there is a path between the two with available capacity. 
Note that, greedily applying the Ford-Fulkerson algorithm for each batch in Algorithm \ref{alg:framework} (denoted as {\sf G-greedy}) does not necessarily result in a globally optimal 
answer for the entire time span $\Phi$ \cite{kazemi2012geocrowd}. Two heuristic algorithms  
are designed to improve the results obtained by  {\sf G-greedy}.

\subsubsection{Least Location Entropy Priority Algorithm}
\label{sec:llep}

Least 
location entropy priority algorithm ({\sf G-llep}) \cite{kazemi2012geocrowd} 
gives higher priority to the tasks located in worker-sparse areas 
(areas with low workers densities). The intuition of this algorithm 
is that for a task located in worker-sparse areas, it is less likely 
that the task can have a potential worker to select in future timestamps. 
In other words, if a task located in worker-dense area is not assigned 
to any workers at the current timestamp, it has a higher possibility to 
be assigned to some other worker in the future timestamps compared 
with tasks in worker-sparse areas. 

The algorithm utilizes \textit{location entropy}  \cite{cranshaw2010bridging} 
to measure the total number of workers in a location as well as the relative 
proportion of their future visits to that location. A location with high location 
entropy indicates many workers visit that location with equal proportions. In 
other words, for a given location, if only a small number of workers often 
visit it, its location entropy is low. 

For a given location $l$, let $O_l$ be the set of visits to it, $W_l$ 
be the set of distinct workers that visited $l$, and $O_{w, l}$ be the set of 
visits belonging to worker $w$. Note that, here one visit of worker $w_i$ to a location $l$ means worker $w_i$ appears around location $l$ with distance $dis(w_i, l)\leq a_i$. Then, the location entropy for $l$ is calculated as follows:

\vspace{-3ex}
\begin{equation}
Entropy(l) =-\sum_{w \in W_l} P_l(w) \cdot \log P_l(w),
\end{equation}
\vspace{-2ex}

\noindent where $P_l(w)=\frac{|O_{w,l}|}{|O_l|}$ is the fraction of total 
visits to $l$ made by worker $w$. The location entropies will be updated every batch and one visit of worker $w_i$ to location $l$ here means the worker's working area covers location $l$ at the moment when the batch process starts. According to the suggestions in \cite{kazemi2012geocrowd}, we can discretize the whole spatial space into a grid with small cells (e.g., 30 meters $\times$ 30 meters), then just update the location entropy of each cell (when the working area  of worker $w_i$ overlaps with a cell in one batch, we count that as a visit of worker $w_i$ to the cell) and use each cell's location entropy as that of the tasks located in the cell.

For each location $l$, the entropy of 
it can be treated as its cost value, then the optimization goal of this algorithm 
is to assign as many tasks as possible with minimum total cost associated to the assigned tasks in each timestamp, which can be reduced to the minimum-cost 
maximum flow problem \cite{ahuja1988network}. 
To solve the minimum-cost maximum flow problem, one of the 
well-known techniques \cite{ahuja1988network} is to first find the maximum flow in the 
network, then use linear programming method to minimize the 
total cost of the flow. Let $G_p=(V,E)$ be the flow network graph for 
timestamp $p$. For each edge $(u,v)\in E$, the capacity is $c(u,v) > 0$, 
the flow $f(u,v) \geq 0$, and the cost is $a(u,v) \geq 0$. 
The cost of sending the flow $f(u,v)$ is $f(u,v)\cdot a(u,v)$. 
Denote the maximum flow sent from $src$ vertex to $dest$ vertex as $f_{max}$, 
then the linear programming to minimize the total cost can be represented as below:

\begin{alignat*}{2}
& \text{minimize}&  & \sum_{(u,v)\in E} f(u,v)\cdot a(u,v) \\
& \text{s.t.}&    \quad & 
\begin{aligned}[t]
&f(u,v) \leq c(u,v),\\
&f(u,v) = -f(v,u),\\
&\sum_{w \in V}f(u,w) = 0 \text{ for all $u \neq src, dest$}\\
&\sum_{w \in V}f(src, w) = f_{max} \text{ and } \sum_{w \in V}f(w, dest) = f_{max} 
\end{aligned}
\end{alignat*}

{\sf G-llep} maximizes the number of assigned tasks first, then minimizes 
the total cost guaranteeing that the total number of assigned tasks is maximized.

\subsubsection{Nearest Neighbor Priority Algorithm}

Nearest neighbor priority 
algorithm ({\sf G-nnp}) \cite{kazemi2012geocrowd}  first
maximizes the number of assigned tasks first, then minimizes the total moving 
distance of workers. The intuition of {\sf G-nnp} is that if the moving distances can be 
reduced, workers can finish their assigned tasks faster as the moving distances are shorter, then the overall number of finished tasks can be potentially improved.

In {\sf G-nnp}, the \textit{travel cost} $d(w,t)$ of worker $w$ to task $t$ is defined as
the Euclidean distance between them. In the network flow 
graph, each edge between a worker vertex and a task vertex is associated with a 
weight equaling the travel cost of the worker to the task. Then the problem turns into 
the minimum-cost maximum flow problem and the technique in Section \ref{sec:llep} 
with a different cost function can be applied to it.

\subsection{Multi-Worker Per Task Algorithms}
In real systems, workers may make mistakes or submit wrong answers 
deliberately such that the received answers are not totally reliable. 
To guarantee the reliability of tasks, existing works assign more than one worker to the same task (Multi-worker per task), then aggregate the answers from workers to obtain a reliable final answer for each task. 
In the rest part of this section, we introduce three multi-worker per task algorithms.

\subsubsection{Sampling-Based Algorithm}
\label{sec:sampling}

The sampling algorithm ({\sf RDB-sam}) is proposed to solve reliable diversity 
based spatial crowdsourcing problem (\textsf{RDB-SC}) \cite{cheng2015reliable}, 
which 
tries to maximize the minimum reliability score of tasks. \textsf{RDB-SC} is proved to be 
NP-hard, thus not tractable. {\sf RDB-sam}, as an approximation algorithm, 
can achieve a worker-and-task assignment strategy with high reliable-and-diversity 
score on the fly. We generally introduce the algorithm as follows. The algorithm first 
estimates the number of sample size $k$, where each sample is a possible 
assignment instance set (Definition \ref{definition:instance}). Then, it randomly generates $k$ samples
and reports the one with the highest reliability score as the final result.

{\sf RDB-sam} provides a method to estimate a 
sample size $K$ such that the ``best'' sample among the $K$ samples 
can achieve a ($\epsilon, \delta$)-bound, which means the ``best'' 
sample is within top $\epsilon$ of the entire population with probability $\delta$. 
For a given batch \textsf{TA-GSC} problem, {\sf RDB-sam} conducts a binary search within $\left(\frac{p\cdot M\cdot e
	-1+p}{1-p+e\cdot p}, M\right]$, such that $\widehat{K}$ is the
smallest $K$ value such that $Pr\{X \leq (1-\epsilon) \cdot N\} \leq 1-\delta$ 
(variable X be the rank of the largest sample, $S_K$, in the entire 
population and $N$ is the size of the entire population), where 
$p = \prod_{j=1}^n \frac{1}{deg(w_j)}$, $M = (1-\epsilon)\cdot N$, and 
$e$ is the base of the natural logarithm.

\subsubsection{Divide-and-Conquer Based Algorithm}

When each task needs more than one worker to conduct, the 
complexity of algorithms for \textsf{TA-GSC} problems will increase dramatically with the increase of the number of tasks and workers. To improve the efficiency, the divide-and-conquer based (\textsf{RDB-d\&c}) algorithm \cite{cheng2015reliable}  keeps dividing the whole problem instance into several subproblem 
instances, solves the subproblems instances, then 
merge the results of subproblem instances, which creates a trade-off between efficiency and effectiveness.

Since a worker may exist in more than one subproblems and \textsf{RDB-d\&c} solves each subproblem without coordinating
with other subproblems, the total assigned 
tasks of a worker may exceed his/her capacity of tasks. 
To satisfy the capacity constraint, \textsf{RDB-d\&c} first estimates the cost 
of replacing the worker in each subproblem, then it greedily substitutes 
the worker having lower replacing cost with the ``best'' available worker 
in the current situation. Here, the ``best'' available worker is the worker 
who can most improve the overall utility  and is not fully assigned with tasks.  
If conflicts between subproblems happens frequently, 
the time cost of reconciling conflicts will be enlarged and the running time will increase.

\subsubsection{Heuristic-Enhanced Greedy Algorithm}

Heuristic-Enhanced Greedy Algorithm ({\sf GT-hgr}) \cite{kazemi2013geotrucrowd}  assumes only when the aggregate reputation score
$ARS(t_i)$ of a task $t_i$ is higher than its required quality level $q_j$,  task $t_j$ is treated as 
a finished task. For a given task $t_j$ and its assigned workers $W_j$, 
its aggregate reputation score ($ARS(t_i)$)  is the probability that at least $\frac{|W_j|+1}{2}$ workers perform the task t correctly, which can be calculated with Equation (\ref{eq:task_accuracy}).

The utility function $U(w_i, t_j)$ of \textsf{GT-hgr} is defined as follows:

\begin{equation}
U(w_i, t_j)=\left\{
\begin{array}{ll}
\frac{1}{|W_j|}, & ARS(t_j) \geq q_j \\
0, & ARS(t_j) < q_j
\end{array}
\right. \label{eq:gt_utility}
\end{equation}
\noindent where $|W_j|$ is the number of workers assigned to task $t_j$. The idea of this definition is that only when the required quality level $q_j$ of task $t_j$ is satisfied, the system utility can increase 1 (i.e., $\sum_{w_i \in W_j}U(w_i, t_j)=1$, when $ARS(t_j) \geq q_j$). In addition, \textsf{TA-GSC} problem is proved NP-hard with the utility defined as Equation \ref{eq:gt_utility} by reducing from 
\textit{maximum 3-dimensional matching problem} (M3M) \cite{michael1979computers}.

{\sf GT-hgr} utilizes three heuristics to improve the result of a basic greedy 
algorithm ({\sf GT-greedy}), which greedily assign a task to one correct match until no 
further tasks can be assigned. Here one correct match is a task-and-workers 
pair $\langle t_j, W_j\rangle$ whose aggregate 
reputation score $ARS(W_j)$ is not less than the required quality level $q_j$ of task $t_j$. 
The first heuristic is filtering heuristic, which can reduce the size of correct 
matches by pruning the dominated correct matches. For two correct 
matches $\langle t_j, W_j\rangle$ and $\langle t_j, W'_j\rangle$, if 
$W_j \subseteq W'_j$, match $\langle t_j, W_j\rangle$ dominates 
$\langle t_j, W'_j \rangle$. The second heuristic is least worker assigned 
heuristic, which associates a higher priority for matches with fewer workers. 
The last heuristic is least aggregate distance  heuristic, which prefers 
the match with smaller summation of moving distances of the workers in that match.

\section{Algorithms in Online Mode}
\label{sec:wst_algorithms}

\begin{algorithm}[t]
	\DontPrintSemicolon
	\KwIn{\small An available worker $w_i$}
	\KwOut{\small A set of suitable tasks for worker $w_i$ to conduct}
		Obtain a set of valid tasks for worker $w_i$\;
		Use online task assignment algorithms to obtain a set, $T_i$, with the most number of suitable tasks for worker $w_i$\;
		
		Notify worker $w_i$ to conduct tasks in $T_j$\;
		
	\caption{The Framework of Online Algorithms}
	\label{alg:online_framework}
\end{algorithm}

In the online mode, 
the servers do not trace the locations of workers and just recommend a task plan 
for each worker when he/she is querying the suitable tasks, which indicates a 
route for the worker to go and conduct as many tasks as possible by the way \cite{deng2013maximizing, Deng2016TaskSI}. The utility function for the online mode algorithms discussed in this section is simply defined as $U(w_i, t_j) = 1$.

\begin{figure}[ht]
	\centering
	\scalebox{0.2}[0.2]{\includegraphics{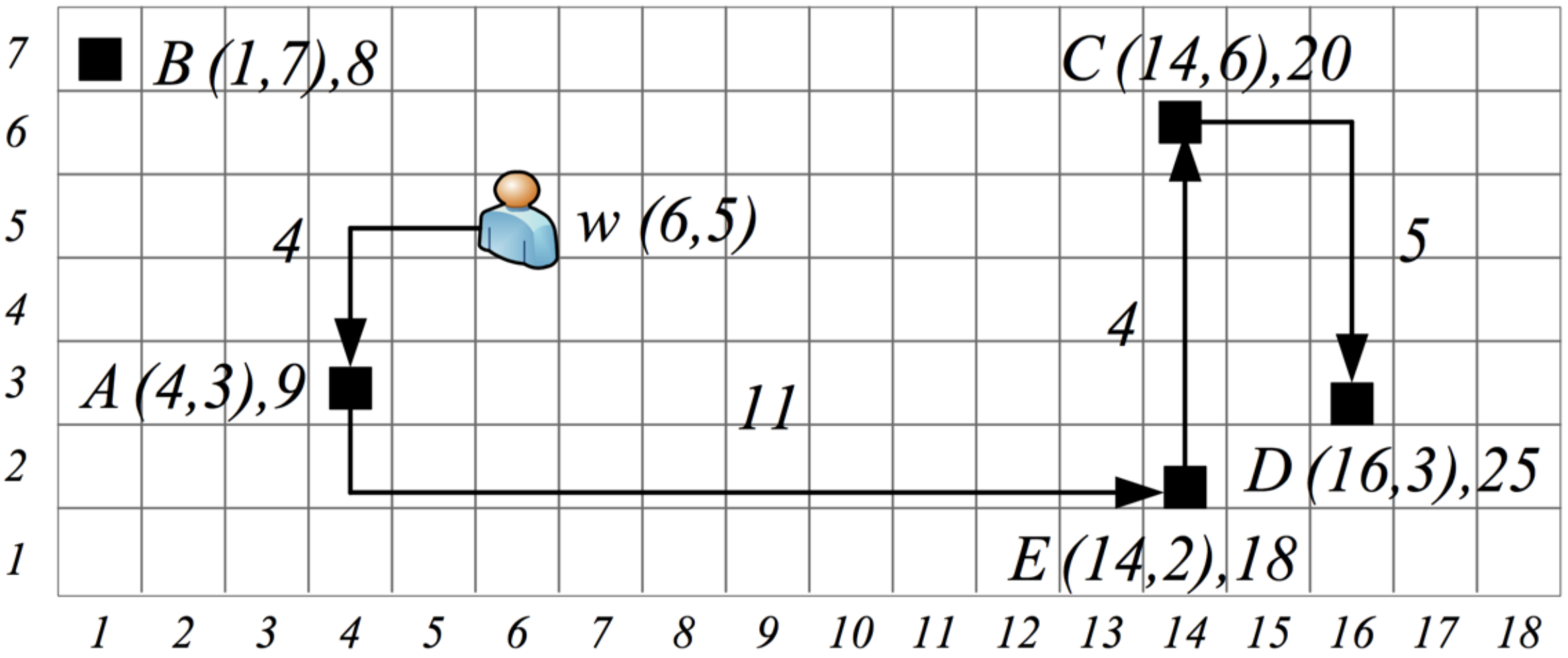}}\vspace{-2ex}
	\caption{\small Running example of MTS.}
	\label{fig:mts_running_example}
\end{figure}

The framework of the \textsf{TA-GSC} algorithms in online mode is shown in Algorithm \ref{alg:online_framework}.
One example is shown in Figure \ref{fig:mts_running_example} \cite{deng2013maximizing}, where the
worker is located at $(6,5)$ and five tasks $A$ to $E$ are located at five different locations 
with their deadlines. The result of this example is that the worker can finish 
at most four tasks following the order $A\rightarrow E \rightarrow C \rightarrow D$.

The \textsf{TA-GSC} problem in online mode can be reduced from a specialized version of Traveling Salesman 
Problem (TSP) called sTSP, which is a NP-hard problem \cite{deng2013maximizing}. 
Exact algorithms, such as \textit{dynamic programming algorithm} and 
\textit{branch-and-bound algorithm} \cite{deng2013maximizing}, can solve the problem for each single worker exactly. However, for the entire time period, exact algorithms still 
achieve only approximated results. In addition, to improve the efficiency, 
some heuristic algorithms and progressive algorithms are proposed \cite{deng2013maximizing, Deng2016TaskSI}. 
In the rest of this section, we will briefly introduce them.

\subsection{Exact Algorithms}
As the server in online mode tries to provide the longest 
tasks sequence for each worker such that he/she can conduct as many 
tasks as possible. Although this problem is
proved NP-hard,  the dynamic programming algorithm and 
the branch-and-bound algorithm \cite{deng2013maximizing} can solve small scale problems.

\subsubsection{Dynamic Programming Algorithm}
The dynamic programming algorithm (\textsf{DP}) \cite{deng2013maximizing} iteratively expands the sets of tasks in the ascending order of set sizes, and ignores the order of task sequence but examines the sets of tasks.
Given a worker $w$, and a set of tasks $T$, let $opt(T, j)$ be the maximum number 
of tasks that worker $w$ can complete under the constraints of tasks 
and starts from the current locations of $w$ and ends at the tasks $t_j$, 
and $R$ be the corresponding task sequence to achieve the optimum value. 
In addition, they denote the second-to-last task in $R$ as task $t_x$. Then, the 
recurrent formula is given as below

{\scriptsize
\begin{equation}
opt(T, t_j)=\left\{
\begin{array}{ll}
1, & \text{if } |T| = 1 \\
\underset{t_i\in T, t_x\neq t_j}{\max}\{opt(T-\{t_j\}, t_x)+\delta_{xj}\}, & \text{otherwise}
\end{array}
\right. \label{eq:recurrent_formula}
\end{equation}

\begin{equation}
\delta_{xj}=\left\{
\begin{array}{ll}
1, & \text{if } t_x \text{ can be finished after connecting } t_j \text{in the end of } R'\\
0, & \text{otherwise}
\end{array}
\right. \notag
\end{equation}
}
\noindent where $R'$ is a task sequence without task $t_j$.
With the recurrent formula in Equation \ref{eq:recurrent_formula}, the algorithm can be implemented based on existing dynamic programming framework.

To further reduce the running time of the dynamic programming algorithm, 
the Apriori principle \cite{agrawal1994fast} can be utilized to remove the invalid sets such 
that the problem space can be smaller. The observation is that if a task set is 
invalid, then all of its supersets must be invalid. When exploring the task sets, 
if one invalid task set is founded, all its supersets can be safely removed.
However, when most of the task sets are valid, the optimization strategy may not 
be effective as the cost of generating candidate sets may surpass the benefits from removing invalid task sets.

\subsubsection{Branch-and-Bound Algorithm}
\label{sec:bb_mts}
The branch-and-bound algorithm (\textsf{BB}) \cite{deng2013maximizing} searches the whole problem space with pruning and directing.
The search space of branch-and-bound algorithm can be represented as 
a tree, then the algorithm conducts a depth-first search with effective 
directing and pruning. Specifically, for each node, the algorithm expands it 
to a set of candidate task nodes. One observation is that a node's candidate 
task set in the search tree is the subset of its parent's candidate task set, 
which can improve the speed of expanding nodes. With the candidate task set 
of node $r$, the algorithm can estimate the upper bound $ub\_r$ of the 
maximum task sequence along the node $r$ in the search tree in the equation below
\begin{equation}
ub\_r = level(r) + |cand\_r|
\end{equation}

\noindent where $level(r)$ indicates the level of node $r$ in the search tree, 
and $|cand\_r|$ represents the size of the candidate task set of node $r$. Then 
the algorithm can safely prune the branch of node $r$ when its upper bound 
$ub\_r$ is smaller than the best current known solution $curMax$. To determine 
the best searching order, the algorithm sorts the current searching branches 
by their upper bounds (ub) or lower bounds (lb), which can be estimated with approximation 
algorithms in Section \ref{sec:mts_heuristic}. In addition, if the upper bound of a node $r$ is 
less than the lower bound of any other node, the node $r$ can be safely pruned.

\begin{figure}[ht]
	\centering
	\scalebox{0.18}[0.18]{\includegraphics{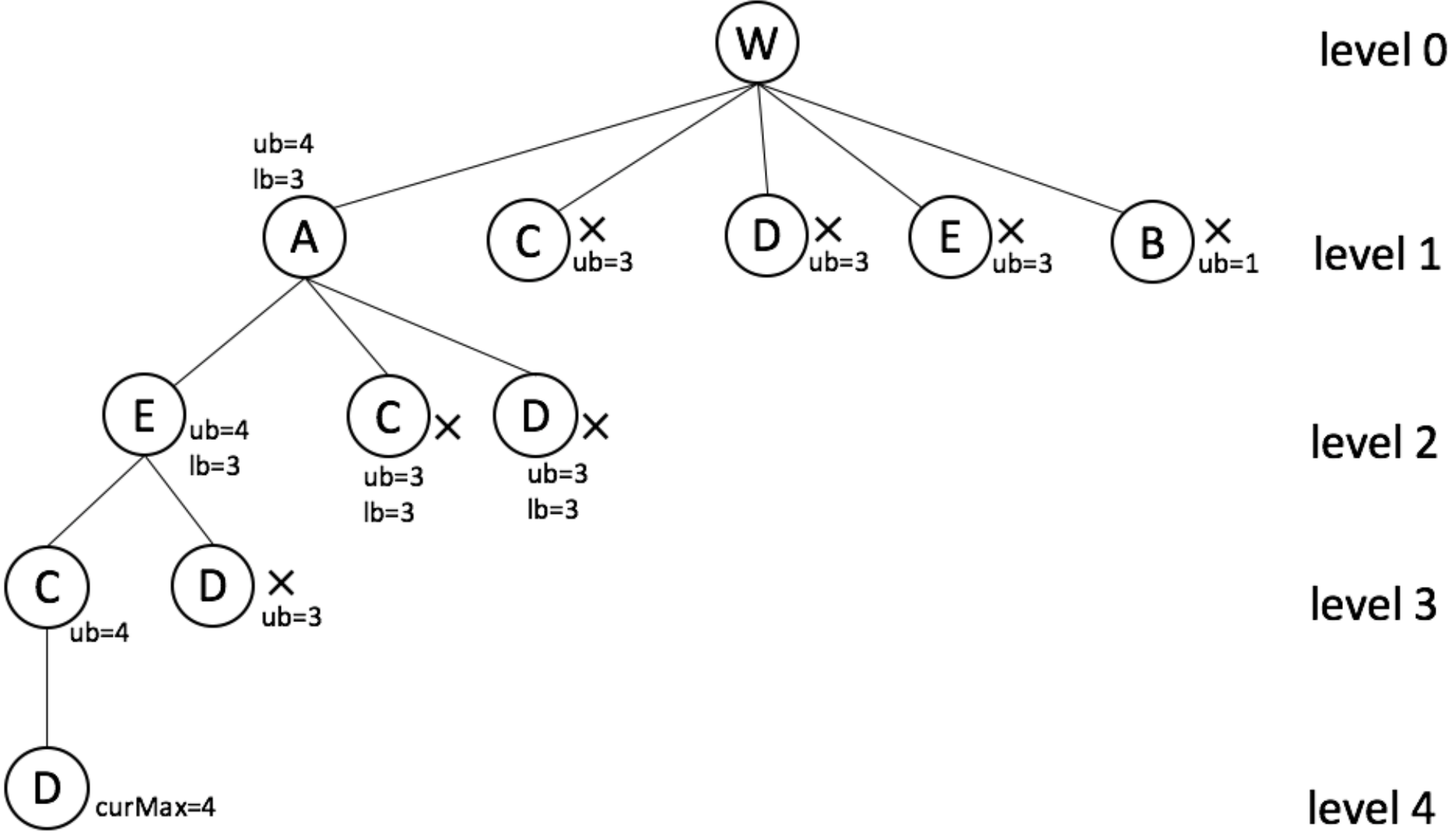}}
	\caption{\small An overview of branch-and-bound algorithm.}
	\label{fig:mts_bb}
\end{figure}

Figure \ref{fig:mts_bb} displays an overview of the branch-and-bound algorithm 
of solving the example shown in Figure \ref{fig:mts_running_example}. On level 1, 
the five nodes are ordered by their upper bounds and node B can be pruned as its 
upper bound is less than the lower bound of node A. Then after visiting node D on 
level 4, the algorithm finds the current best known result $curMax=4$. 
When $curMax=4$, the algorithm prunes all other nodes as their upper 
bounds are all less than 4.

\begin{table*}[t!]\centering
	\caption{\small Algorithms Comparison}\label{tab:algorithms_comp}
	{\small 
		\begin{tabular}{l|l|c|l|l}
			{\bf Algorithms} & {\bf  Time Complexity} & \textbf{Assignment Mode} & \textbf{Maximizing Goal} &\textbf{Randomization}\\
			\hline \hline
			MaxFlow Greedy (\textsf{G-greedy}) \cite{kazemi2012geocrowd}& $O(E\max|f|)$& Batch-based &  the number of assigned tasks&Deterministic\\
			MaxFlow with least location entropy priority (\textsf{G-llep}) \cite{kazemi2012geocrowd}& $O(E\max|f|)$ & Batch-based & the number of assigned tasks&Heuristic\\
			MaxFlow with nearest neighbor priority (\textsf{G-nnp}) \cite{kazemi2012geocrowd}& $O(E\max|f|)$ & Batch-based & the number of assigned tasks&Heuristic\\
			Trustworthy greedy (\textsf{GT-greedy}) \cite{kazemi2013geotrucrowd}&- & Batch-based & the number of correct matches&Randomized\\
			Heuristic-enhanced greedy (\textsf{GT-hgr}) \cite{kazemi2013geotrucrowd}& -& Batch-based& the number of correct matches&Heuristic\\
			Divide and conquer (\textsf{RDB-d\&c}) \cite{cheng2015reliable}& $O(m\cdot n^2)$ & Batch-based &  the minimum reliability&Heuristic\\
			Sampling (\textsf{RDB-sam}) \cite{cheng2015reliable}&- & Batch-based&  the minimum reliability&Randomized\\
			Dynamic programming (\textsf{DP}) \cite{deng2013maximizing}& $O(n \cdot m^2\cdot2^m)$& Online& the number of scheduled tasks&Deterministic\\
			Brach and bound (\textsf{BB})\cite{deng2013maximizing} & $O(n\cdot m!)$& Online& the number of scheduled tasks&Deterministic\\
			Heuristic ensemble algorithm (\textsf{HA}) \cite{deng2013maximizing}& $O(n \cdot \log (m))$& Online& the number of scheduled tasks&Heuristic\\
			Progressive algorithm (\textsf{PRS}) \cite{deng2013maximizing}& - & Online& the number of scheduled tasks&Heuristic\\
			\hline
		\end{tabular}}
	\end{table*}

\subsection{Heuristic Algorithms}
\label{sec:mts_heuristic}

Exact algorithms  give the exact result for each  single worker's request but 
not the entire time period, and their time complexities and memory consumption
increase exponentially as the number of tasks grows such that they are 
not efficient enough for real-world applications. In this section, we 
 briefly introduce three heuristics to return results to the workers quickly \cite{deng2013maximizing}.

\noindent\textbf{Least expiration time heuristic (LEH).} The LEH constructs a 
task sequence by greedily appending the task with the least expiration time to the 
end of current task sequence. It first orders the tasks by their expiration time, 
then check each task on the ascending order of their expiration time. If one 
task can be conducted by the worker, which means the worker can arrive at the 
location of the task before its deadline, then the algorithm adds the task to the 
end of the current task sequence. Finally, the task sequence is sent to workers to conduct one by one.

\noindent \textbf{Nearest neighbor heuristic (NNH).} The NNH utilizes the 
spatial proximity between tasks through keeping selecting the nearest valid 
task to the last added task in the current task sequence, where the valid task 
means the worker can arrive at its location before its deadline. The heuristic 
greedily adds more tasks to the end to the task sequence until no more tasks 
can be selected, then it returns the task sequence to the worker who is querying 
the available tasks.

\noindent \textbf{Most promising heuristic (MPH).} The MPH is a heuristic for 
the branch-and-bound algorithm in Section \ref{sec:bb_mts} to choose the most 
promising branches when it is exploring the search tree, where the most promising 
branch for each level can be the branch having the nodes with the highest upper bound at 
that level. In addition, MPH just reports the first found candidate task sequence.

As the heuristic algorithms run fast, on real-world application, the system can run 
the three heuristic algorithms at the same time, noted as Heuristic Algorithm (\textsf{HA}), and just reports the best result to 
improve the utility of the final result but without harming the user experience of workers.

\subsection{Progressive Algorithms}
The idea of progressive algorithms (\textsf{PRS}) is to report a small number of spatial tasks 
to a worker quickly at the beginning, and then to keep incrementally building the 
rest of the task sequence off-line and report the newly added tasks to 
the worker before he/she finishes all the tasks already reported to them. 
Under this framework, one progressive algorithm can use approximation 
algorithms to response one worker very fast at the beginning, then utilizes one 
exact algorithm to progressively construct the rest task sequence.

The advantage of progressive algorithms is that they can response a worker faster 
than exact algorithms and report more accurate results than heuristic ones. 
On the other hand, the potential tasks for a worker may be promoted to other 
workers when they are conducting the initial tasks, and they cannot see 
the entire task sequence at the beginning which may lead to a worse user experience compared to that of the other online algorithms.

\section{Experimental Study}
\label{sec:exp}

\begin{table}[ht!]
	\parbox[b]{\linewidth}{
		\caption{\small Experiments Settings.} \label{tab:experiment}
		{\small\scriptsize
			\begin{tabular}{l|l}
				{\bf \qquad \qquad \quad Parameters} & {\bf \qquad \qquad \qquad Values} \\ \hline \hline
				number of tasks, $m$ & \textbf{7.5K}, 10K, 12.5K, 15K, 17.5K \\
				number of workers, $n$ & \textbf{7.5K}, 10K, 12.5K, 15K, 17.5K \\
				task duration range, $[rt^-, rt^+]$  & \textbf{[1, 2]}, [2, 3], [3, 4], [4, 5]\\
				required answers range, $[b^-, b^+]$  & [1, 3], \textbf{[3, 5]}, [5, 7], [7, 9]\\
				capacity range, $[c^-, c^+]$  & \textbf{[2, 3]}, [3, 4], [4, 5], [5, 6]\\
				required quality level range, $[q^-, q^+]$  & [0.65, 0.7], \textbf{[0.75, 0.8]}, [0.8, 0.85], [0.85, 0.9]\\
				reliability range, $[r^-, r^+]$  & [0.65, 0.7], \textbf{[0.75, 0.8]}, [0.8, 0.85], [0.85, 0.9]\\
				side length range, $[a^-, a^+]$  & \textbf{[0.05, 0.1]}, [0.1, 0.15], [0.15, 0.2], [0.2, 0.25]\\
				worker velocity, $v$  & 0.01, 0.05, 0.1, \textbf{0.15}\\	
				time slot length, $\phi$  & 30, 60, \textbf{120}, 180\\
				mean of Gaussian distribution, $\mu$  & 0.1, 0.3, \textbf{0.5}, 0.7, 0.9\\
				variance of Gaussian distribution, $\sigma^2$  & $0.01^2$, $0.03^2$, \textbf{$\mathbf{0.05^2}$}, $0.07^2$, $0.1^2$\\
				number of Gaussian distributed  & 1, \textbf{3}, 5, 7\\ clusters in the skewed distribution $\Lambda$ \\
				\hline
			\end{tabular}
		}
	}\vspace{-2ex}
\end{table}

\subsection{Experiments Setup}

\noindent \textbf{Data Sets.} We use both real and synthetic data to
test task assignment methods in batch-based mode and online mode. 

For real data set, we utilize the data provided by DiDi Chuxing \cite{didi, didi_gaia}. Specifically, the real data set includes the temporal locations of taxis and orders, which is retrieved from the time period between 7:30 am and 8:30 am in a normal day in the urban area of Beijing (with latitude from \ang{39.7558} to \ang{40.0229} 
and longitude from \ang{116.1996}  to \ang{116.5457}). There are 10,816 orders and 13,892 taxis in the data set.  For simplicity, we first linearly map check-in locations from DiDi Chuxing into a $[0, 1]^2$ data space. Then, we use the taxi records to initialize locations and timestamps of workers, and utilize the order records to set up the required locations and creation timestamps of spatial tasks. In the experiments, we treat every $\phi$ seconds as a time slot (i.e., the temporal unit  in the experiments).



For synthetic data, we generate locations of workers and tasks in a
2D data space $[0, 1]^2$ following Uniform (UNIF), Gaussian (GAUS), 
Skewed (SKEW), as different distributions may affect the validation relationships of worker-and-task pairs (i.e., satisfying working area constraint and deadline constraint). For Uniform distribution,
we uniformly generate the locations of tasks in the 2D data space.
For Gaussian distribution, we generate the locations of tasks/workers in a 
Gaussian cluster (with mean  $\mu$ and variance $\sigma^2$).
Similarly, we also generate the tasks/workers with the Skewed 
distribution through locating 90\% of them into
$\Lambda$ Gaussian clusters (with mean of $0.5$, variance of $0.05^2$ and randomly chosen centers),
and distributing the rest workers/tasks uniformly in the 2D data space. We present the illustrations of the distributions with different parameters in Appendix A. For each synthetic dataset, we generate 50 time slots.

To simulate the synthetic data and the other properties of real data, we use a toolbox, SCAWG \cite{to2016scawg}, to generate data records for each time slot.  
For both real and synthetic data sets, we simulate the working ranges
each worker as squares whose centers are at the locations of workers, and the length of sides of the squares are generated with Gaussian distribution within range $[a^-, a^+]$ \cite{kazemi2012geocrowd, to2015server}. In addition, we set the velocity of each worker as $v$. When the workers are idle, they move randomly within their working areas. Each worker will locate at the position of his/her latest task after finishing it. For the count of required answers to each task and the capacity of each worker, we generate them following the Gaussian distributions within the range $[b^-, b^+]$ and the range $[c^-, c^+]$, respectively \cite{kazemi2012geocrowd, to2015server}. Meanwhile, for the required confidence of each task and the reliability of each worker, we produce them following the Gaussian distributions within the range $[q^-, q^+]$ and the range $[r^-, r^+]$, respectively \cite{kazemi2013geotrucrowd}.
For temporal constraints on tasks, we also generate the
deadlines for tasks according to the range $[rt^-, rt^+]$ of the duration of tasks with
Gaussian distribution \cite{cheng2016task, cheng2015reliable, deng2013maximizing}.
Here, for Gaussian distributions, we linearly map data
samples within $[-1, 1]$ of a Gaussian distribution $\mathcal{N}(0, 0.2^2)$
to the target ranges.

\noindent \textbf{Evaluation Metrics.}
To evaluate the efficiency and effectiveness of the tested approaches, we report the most important metrics for spatial crowdsourcing systems as follows:
\begin{itemize}[leftmargin=*]
	\item \textit{Average moving distance of each worker (AvgMD).} For workers, they want to accomplish maximum number of tasks with minimum moving distance. Then, higher average moving distance of each worker may harm the benefit of workers, which should be avoided. Thus, algorithms achieving results with lower AvgMDs are better.\vspace{-1ex}
	\item \textit{Number of fully assigned tasks (NFT).} For the spatial crowdsourcing platforms, they want to fully assign as many tasks as possible, which can reflect their effectiveness. Here, one fully assigned task means it is assigned with the required number of workers. Higher NFT is better.\vspace{-1ex}
	\item \textit{Number of confidently assigned tasks (NCT).} Only when the expected accuracy (calculated with Equation \ref{eq:task_accuracy}) of the assigned workers $W_j$ of task $t_j$ is higher than the required quality level $q_j$, task $t_j$ is considered as a confidently  assigned task. When the total number of tasks is fixed, algorithms achieving higher NCTs are better.\vspace{-1ex}
	\item \textit{Running time (RT).} The running time represents the total execution time of the tested algorithm for resolving a given \textsf{TA-GSC} problem. Lower RT is better.
\end{itemize}

\begin{figure*}[t!]\centering
	\subfigure{
		\scalebox{0.4}[0.4]{\includegraphics{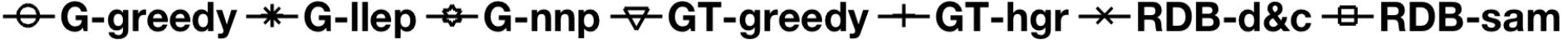}}}\hfill\\\vspace{-2ex}
	\addtocounter{subfigure}{-1}
	\subfigure[][{\small Moving Distances}]{
		\scalebox{0.2}[0.2]{\includegraphics{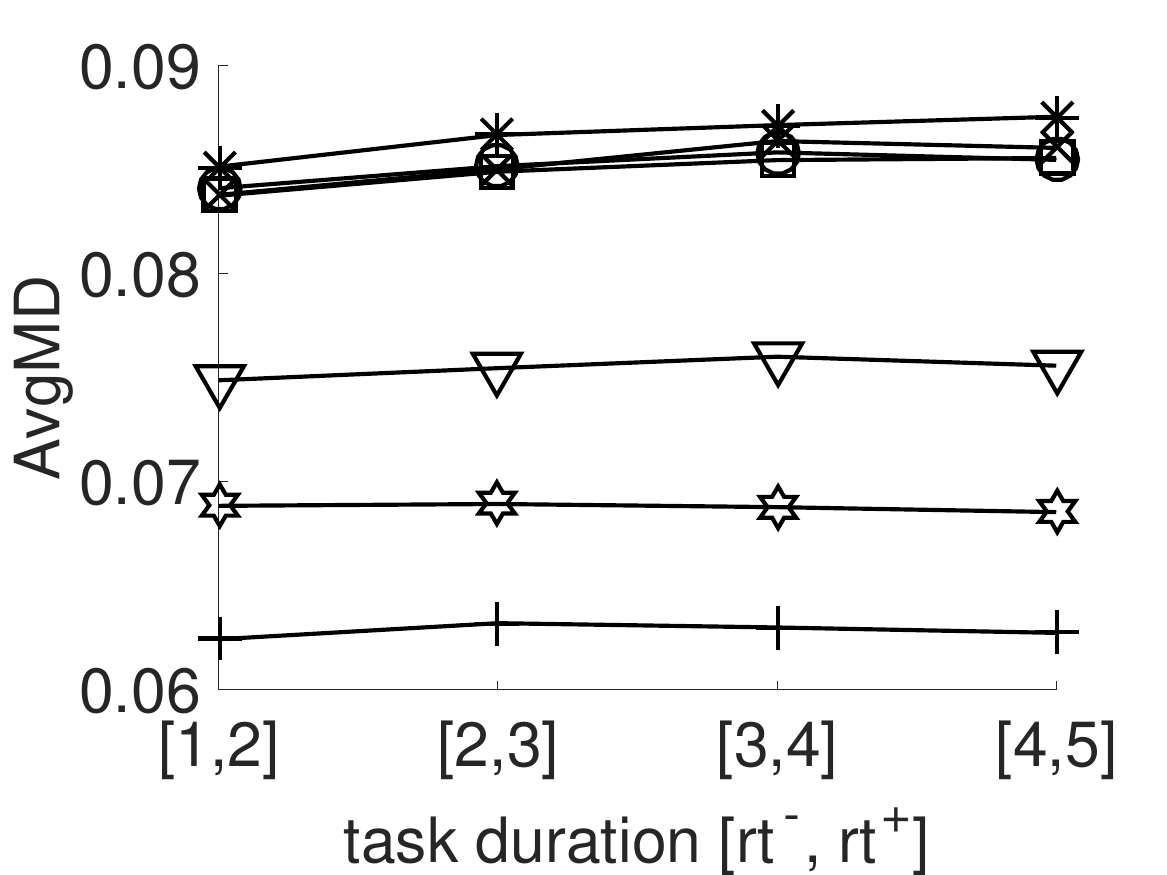}}
		\label{subfig:rt_avg_moving_distance}}\hfill
	\subfigure[][{\small Fully Assigned Tasks}]{
		\scalebox{0.2}[0.2]{\includegraphics{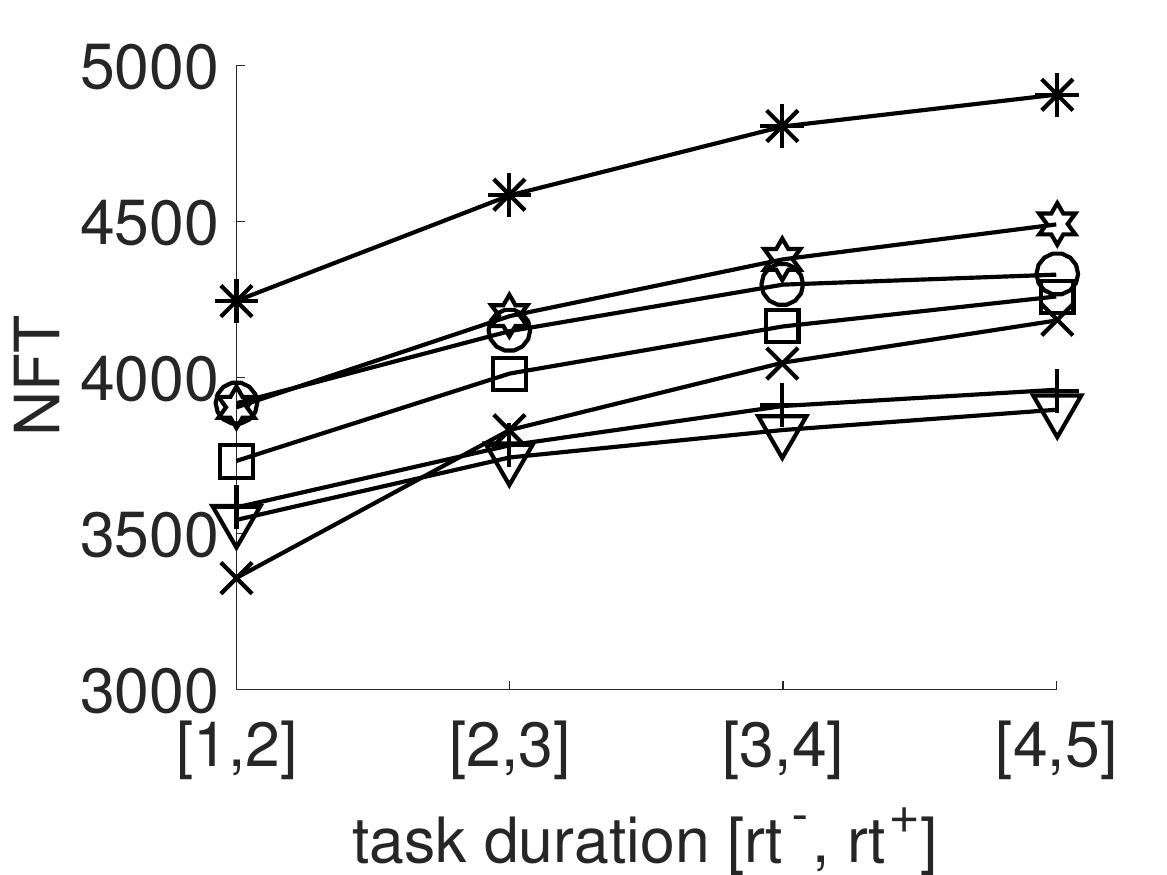}}
		\label{subfig:rt_finished_task_number}}\hfill
	\subfigure[][{\small Confidently Assigned Tasks}]{
		\scalebox{0.2}[0.2]{\includegraphics{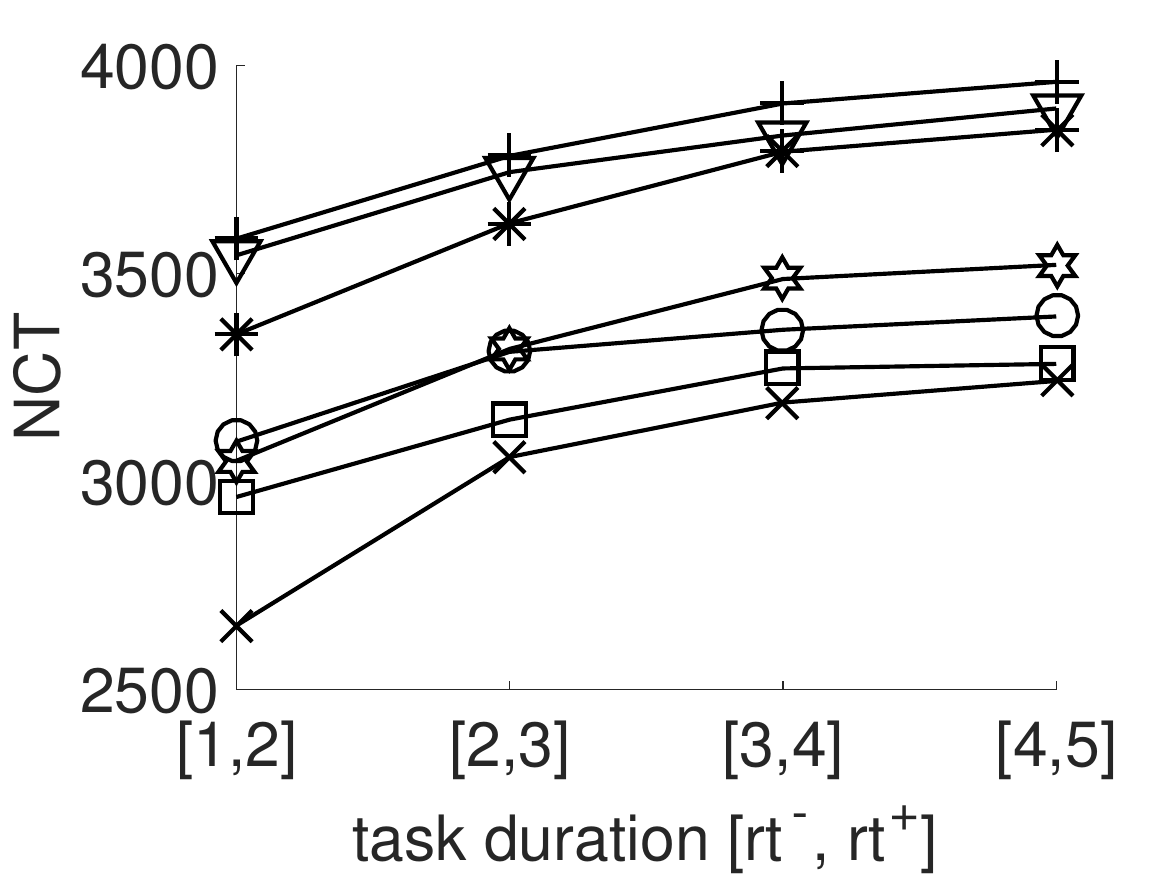}}
		\label{subfig:rt_finished_task_number_conf}}
	\subfigure[][{\small Running Times}]{
		\scalebox{0.2}[0.2]{\includegraphics{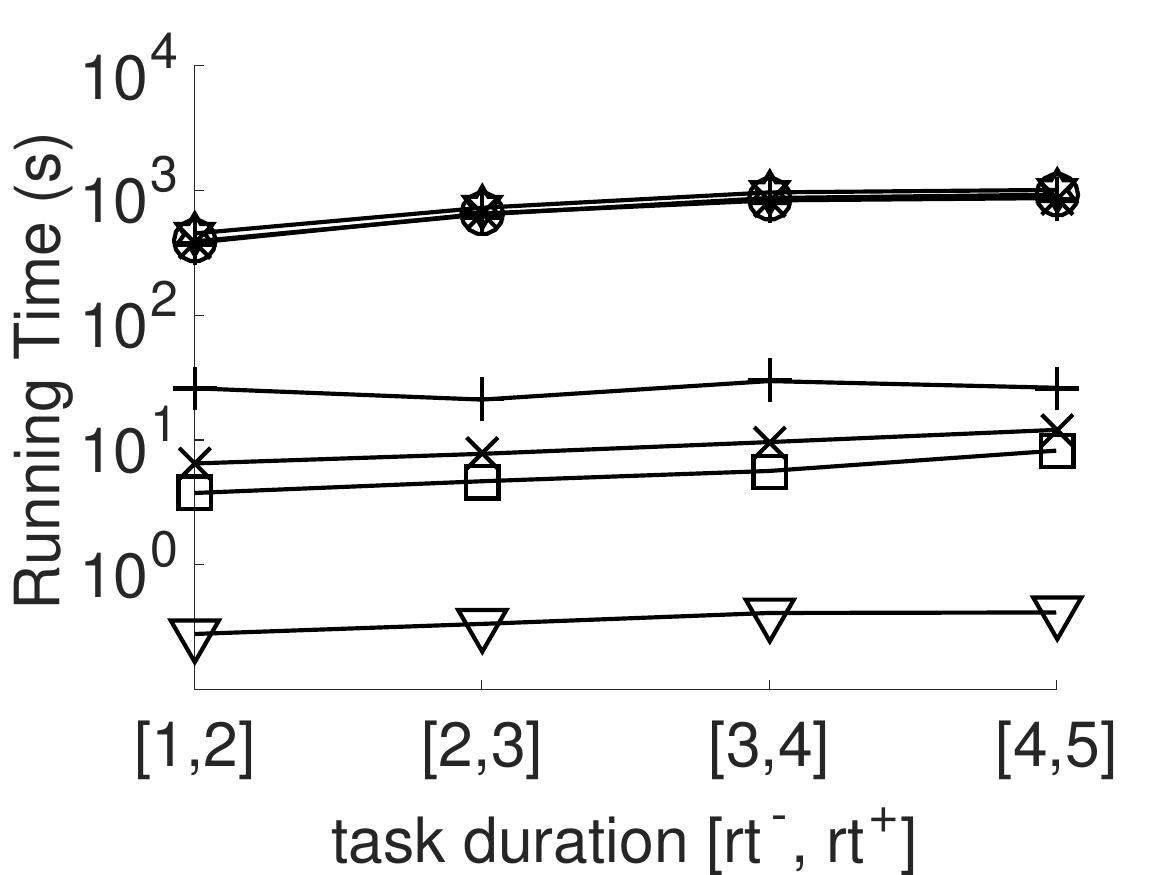}}
		\label{subfig:rt_running_time}}\hfill\vspace{-2ex}
	\caption{\small Effects of Task Duration $rt$ (Batch-based Mode, Real).}
	\label{fig:batch_effect_expiration}\vspace{-2ex}
\end{figure*}

\begin{figure*}[t!]\centering
	\subfigure{
		\scalebox{0.2}[0.2]{\includegraphics{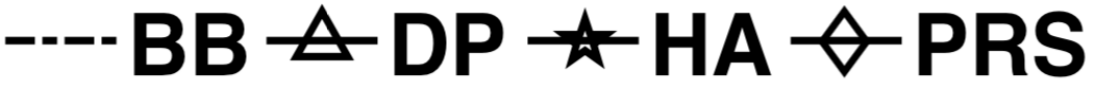}}}\hfill\\\vspace{-2ex}
	\addtocounter{subfigure}{-1}
	\subfigure[][{\small Moving Distances}]{
		\scalebox{0.2}[0.2]{\includegraphics{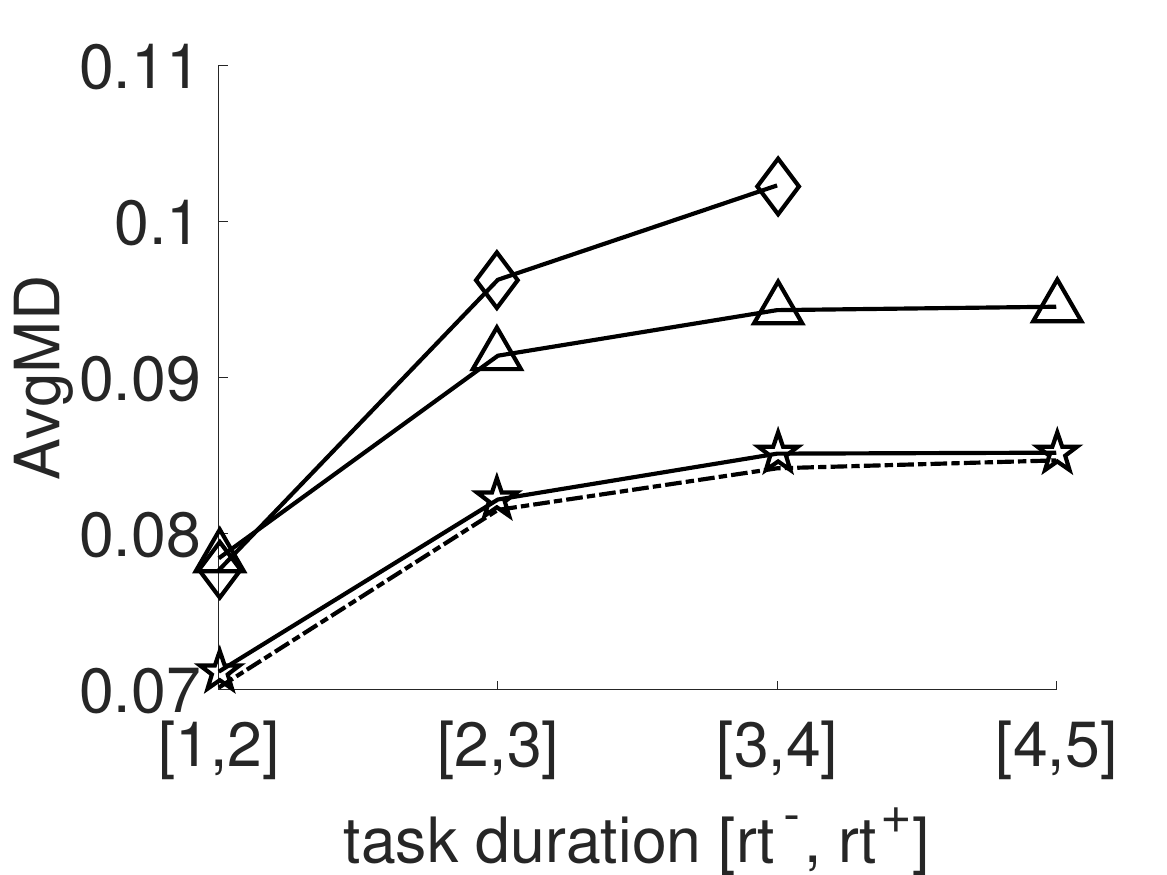}}
		\label{subfig:online_rt_avg_moving_distance}}\hfill
	\subfigure[][{\small Fully Assigned Tasks}]{
		\scalebox{0.2}[0.2]{\includegraphics{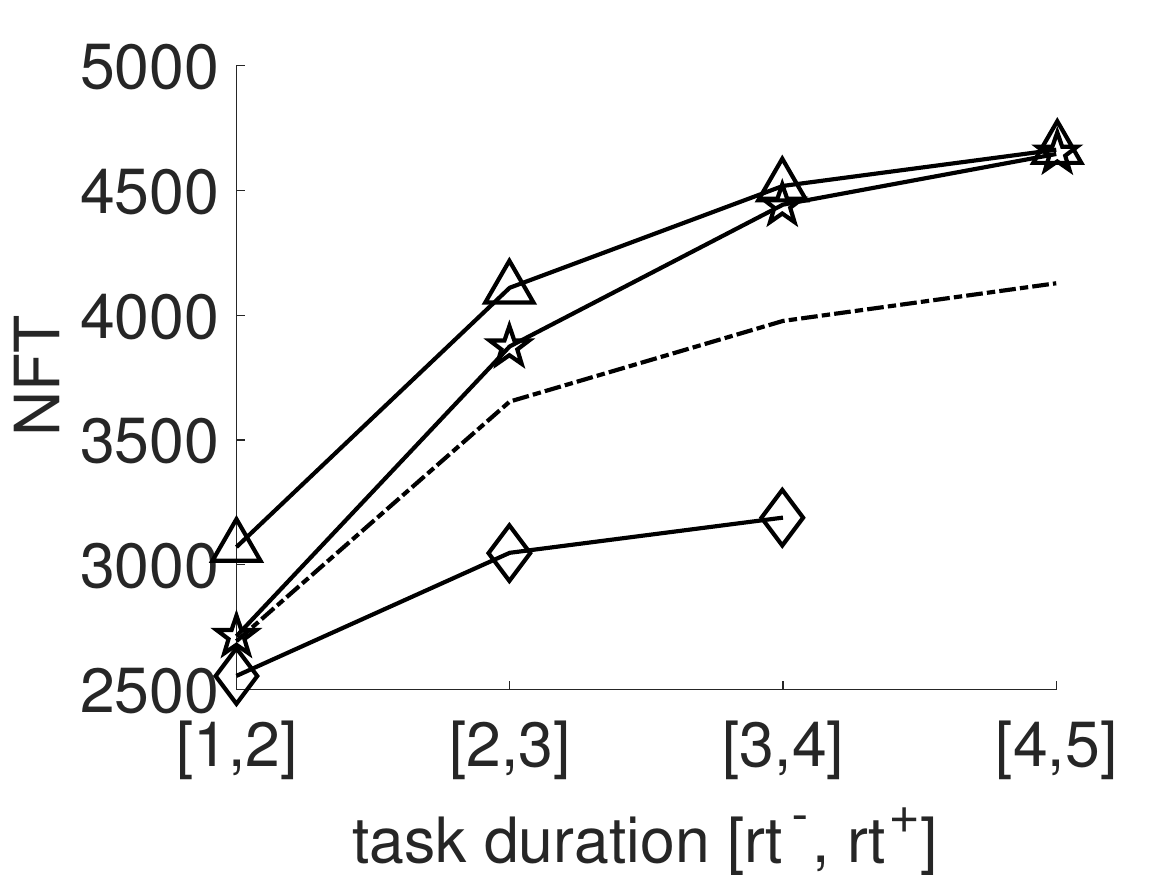}}
		\label{subfig:online_rt_finished_task_number}}\hfill
	\subfigure[][{\small Confidently Assigned Tasks}]{
		\scalebox{0.2}[0.2]{\includegraphics{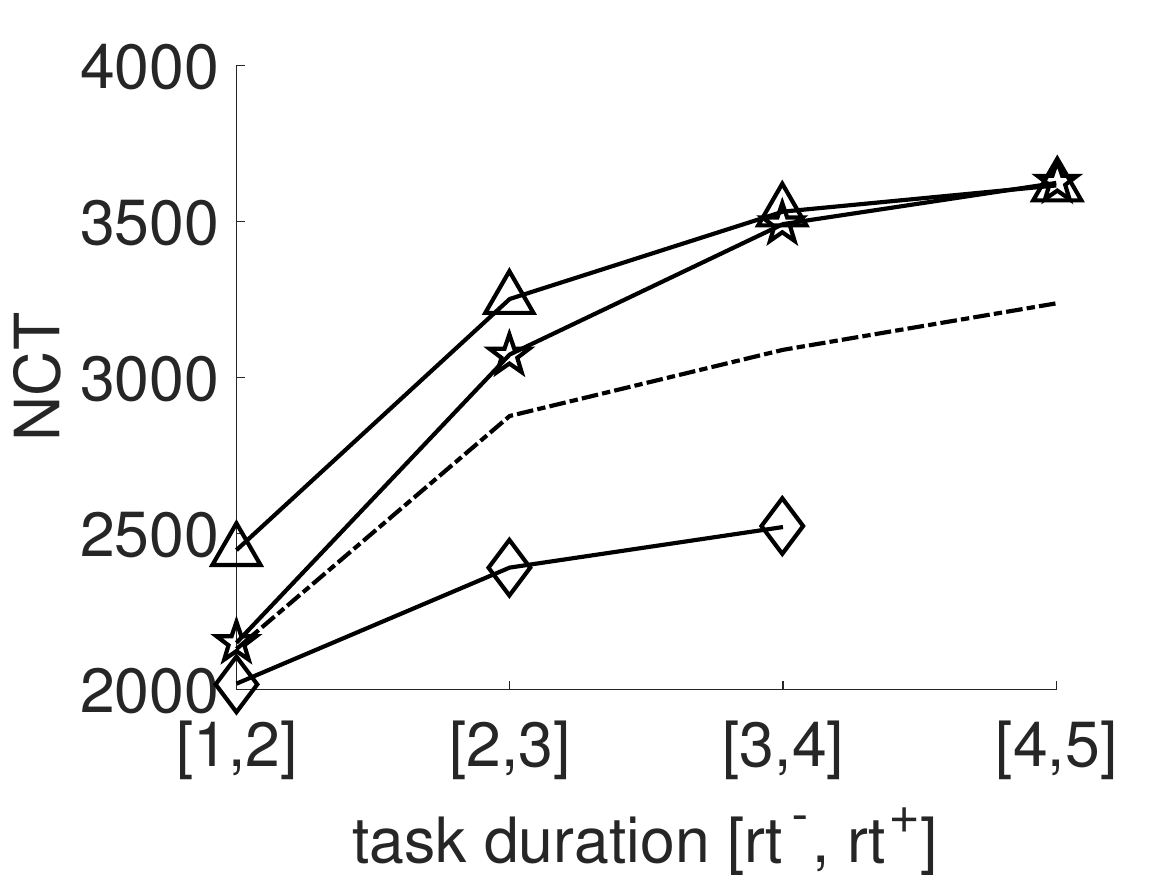}}
		\label{subfig:online_rt_finished_task_number_conf}}
	\subfigure[][{\small Running Times}]{
		\scalebox{0.2}[0.2]{\includegraphics{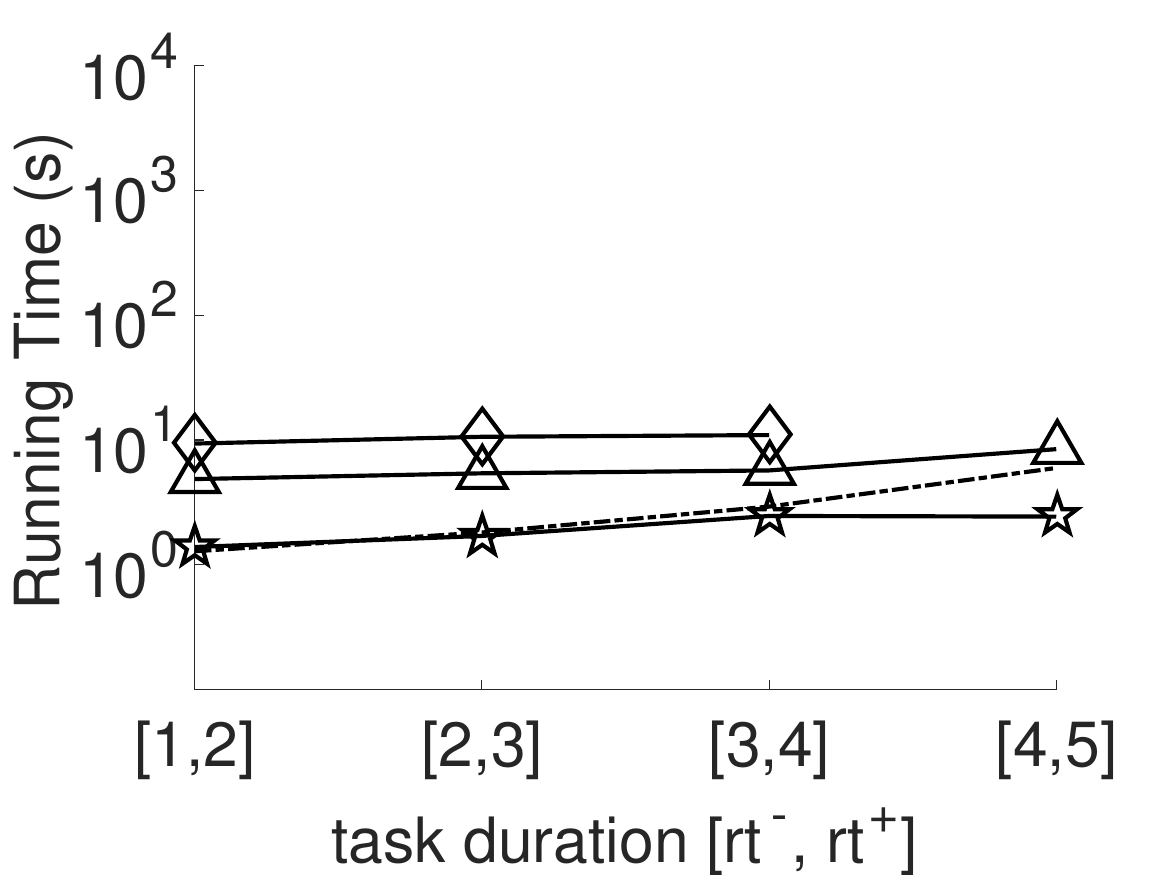}}
		\label{subfig:online_rt_running_time}}\hfill\vspace{-2ex}
	\caption{\small Effects of Task Duration $rt$ (Online Mode, Real).}
	\label{fig:online_effect_expiration}\vspace{-2ex}
\end{figure*}

\noindent\textbf{Tested Approaches.}
Table \ref{tab:experiment} depicts our experimental settings, where the
default values of parameters are in bold font. In each set of
experiments, we vary one parameter, while setting other parameters
to their default values. For each experiment, we report the measured metrics of all tested approaches, which 
includes the algorithms for batch-based mode: 
maximum flow based greedy algorithm (\textsf{G-greedy}), 
maximum flow with least location entropy priority heuristic algorithm (\textsf{G-llep}), 
maximum flow with nearest neighbor priority heuristic algorithm (\textsf{G-nnp}), 
greedy algorithm for trustworthy query (\textsf{GT-greedy}), 
heuristic-en-hanced greedy algorithm for trustworthy query (\textsf{GT-hgr}), 
sampling-based algorithm (\textsf{RDB-sam}) 
and divide-and-conquer-based algorithm (\textsf{RDB-d\&c}), 
and the algorithms in online mode: 
dynamic programming algorithm (\textsf{DP}), 
branch-and-bound algorithm (\textsf{BB}), 
heuristic ensemble algorithm (\textsf{HA}, here we run three heuristic 
algorithms, LEH, NNH and MPH, introduced in Section \ref{sec:mts_heuristic} 
and report the best result of the results of them) and progress algorithm (\textsf{PRS}). Table \ref{tab:algorithms_comp} summarizes all the tested algorithms, where $E$ is the number of valid worker-and-task pairs, $\max|f|$ is the size of the maximum flow, $m$ is the number of tasks and $n$ is the number of workers.

All our experiments were run on an Intel Xeon X5675 CPU @3.07 GHZ with 32 GB
RAM in Python. The source code to generate the testing data sets and implementations of tested algorithms can be found on our GitHub repositories \cite{datasetGenerator, algorithms}.

\begin{figure*}[t!]\centering
	\subfigure{
		\scalebox{0.4}[0.4]{\includegraphics{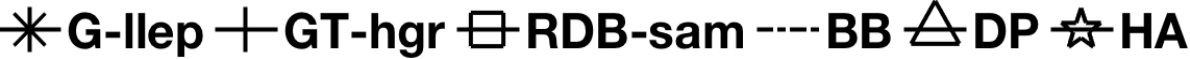}}}\hfill\\\vspace{-2.5ex}
	\addtocounter{subfigure}{-1}
	\subfigure[][{\small Moving Distance }]{
		\scalebox{0.2}[0.2]{\includegraphics{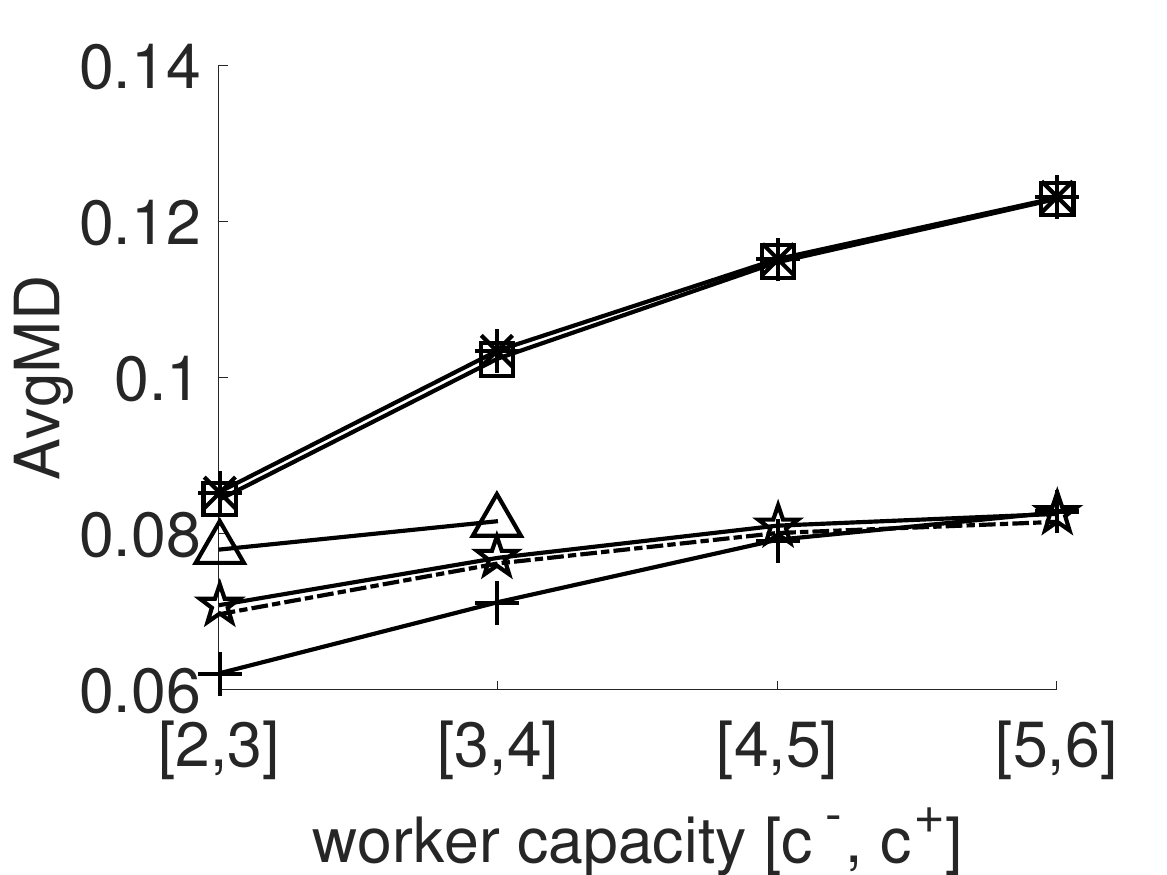}}\vspace{-2ex}
		\label{subfig:c_avg_moving_distance_real_all}}\hfill\vspace{-2ex}
	\subfigure[][{\small Fully Assigned Tasks }]{
		\scalebox{0.2}[0.2]{\includegraphics{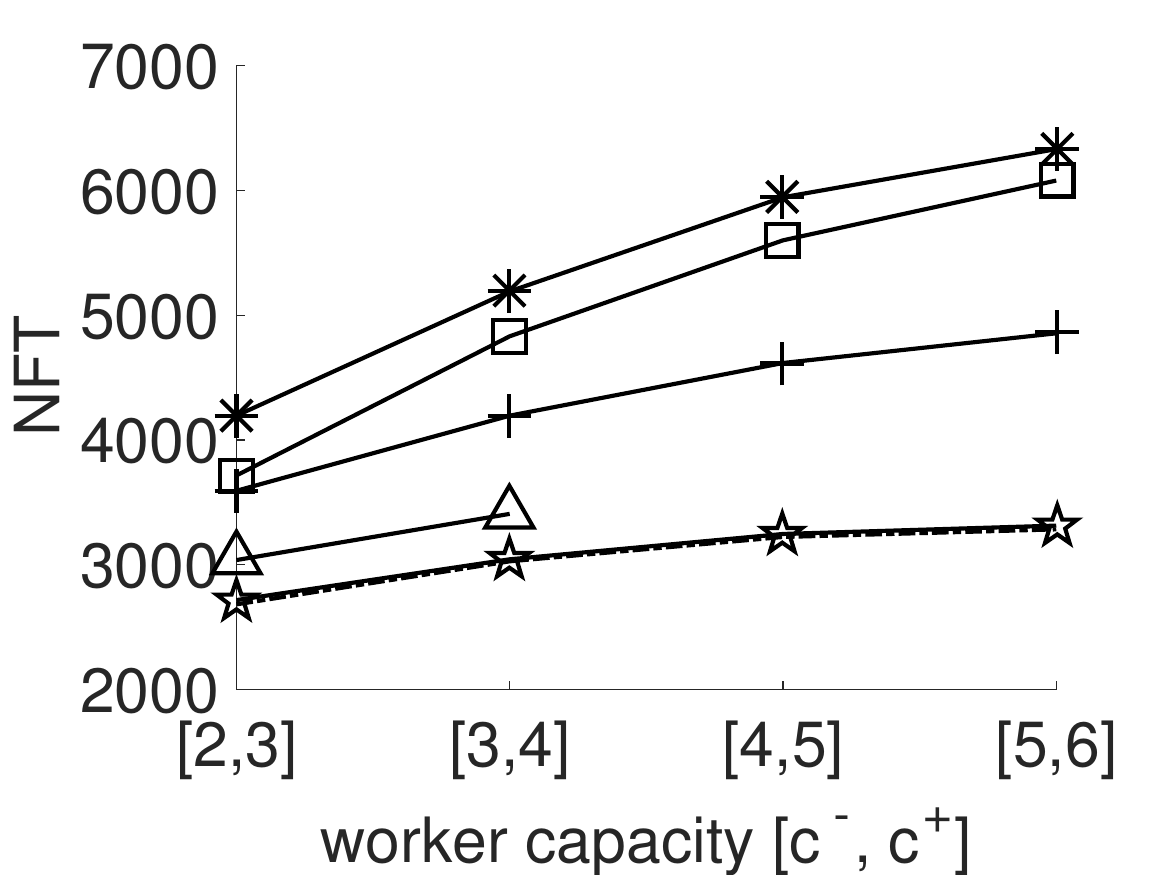}}
		\label{subfig:c_finished_task_number_real_all}}\hfill
	\subfigure[][{\small Confidently Assigned Tasks }]{
		\scalebox{0.2}[0.2]{\includegraphics{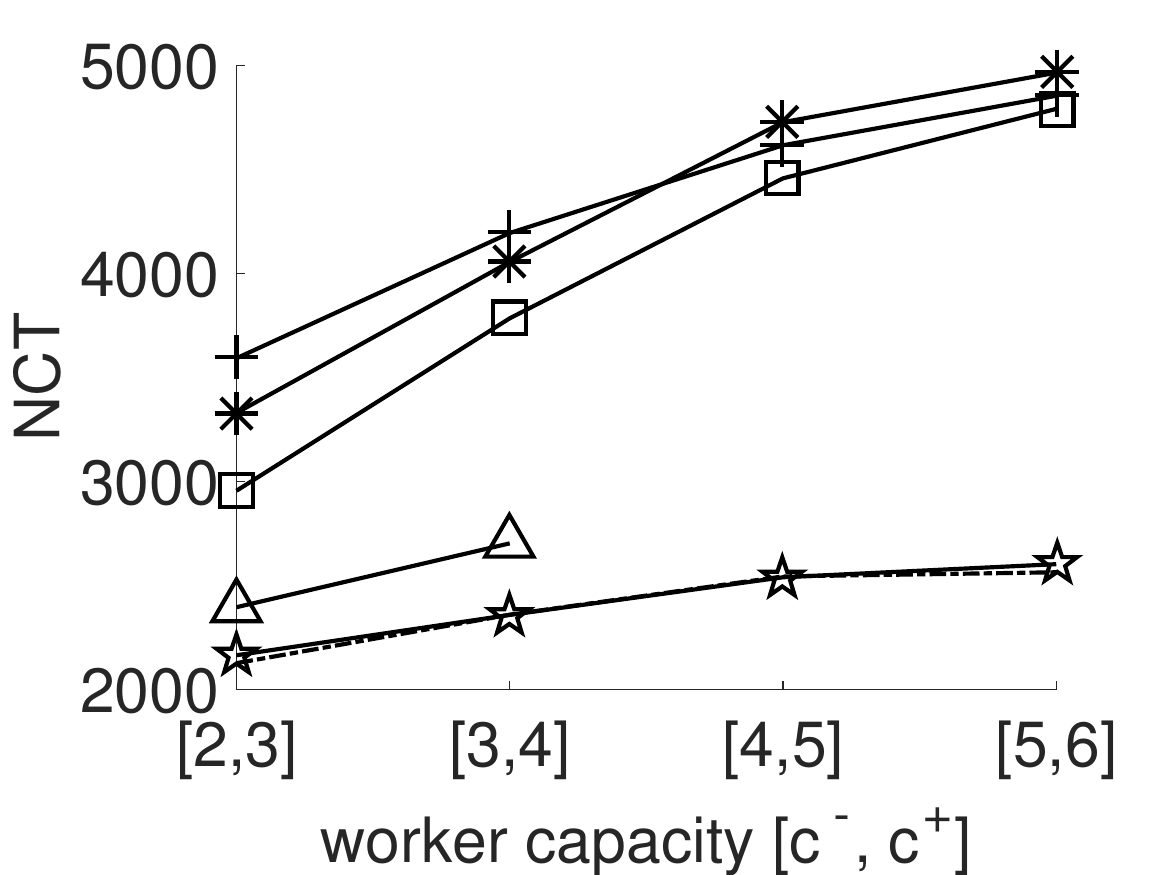}}
		\label{subfig:c_finished_task_number_conf_real_all}}\hfill\vspace{-1ex}
	\subfigure[][{\small Running Times }]{
		\scalebox{0.2}[0.2]{\includegraphics{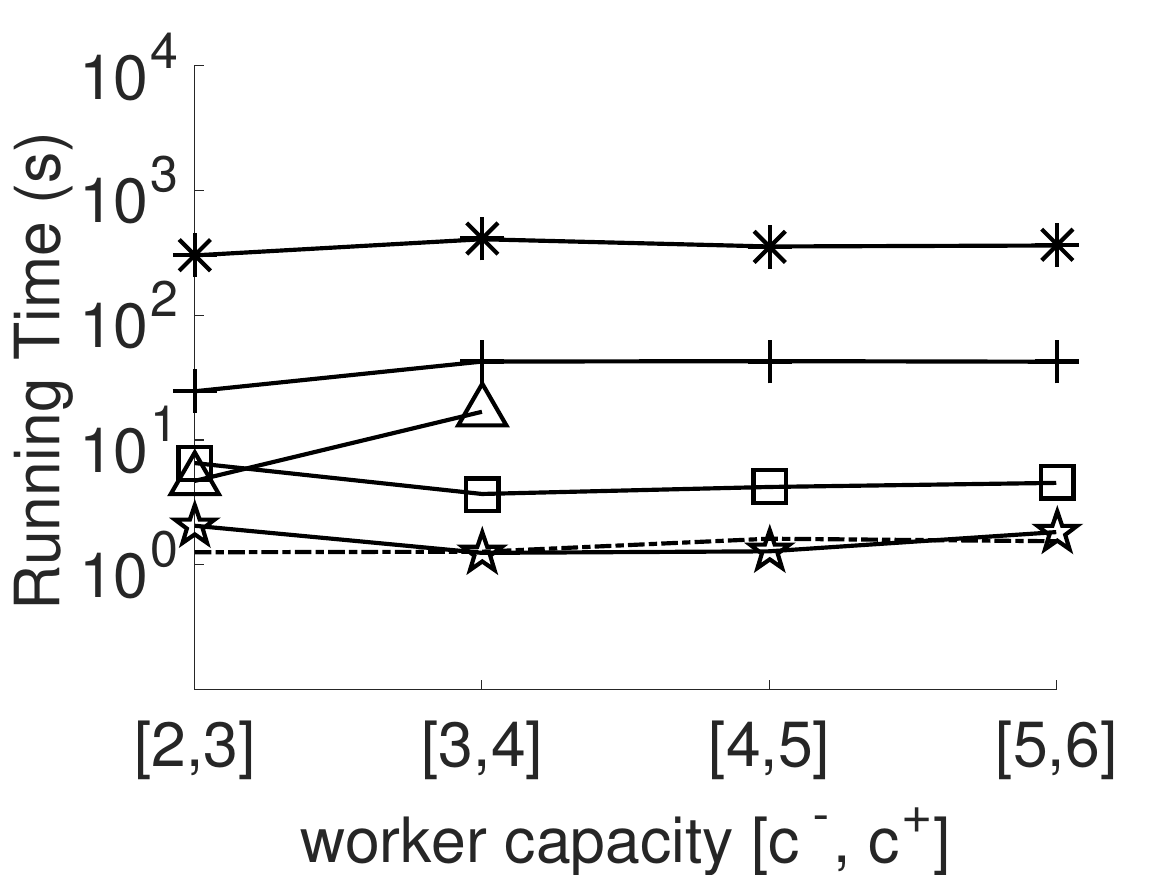}}
		\label{subfig:c_running_time_real_all}}\hfill
	\caption{\small Effects of Worker Capacity $c$ (Real).}
	\label{fig:effect_capacity_real_all}\vspace{-2ex}
\end{figure*}

\begin{figure*}[t!]\centering
	\subfigure{
		\scalebox{0.4}[0.4]{\includegraphics{bar_mix-eps-converted-to.pdf}}}\hfill\\\vspace{-2.5ex}
	\addtocounter{subfigure}{-1}
	\subfigure[][{\small Moving Distance}]{
		\scalebox{0.2}[0.2]{\includegraphics{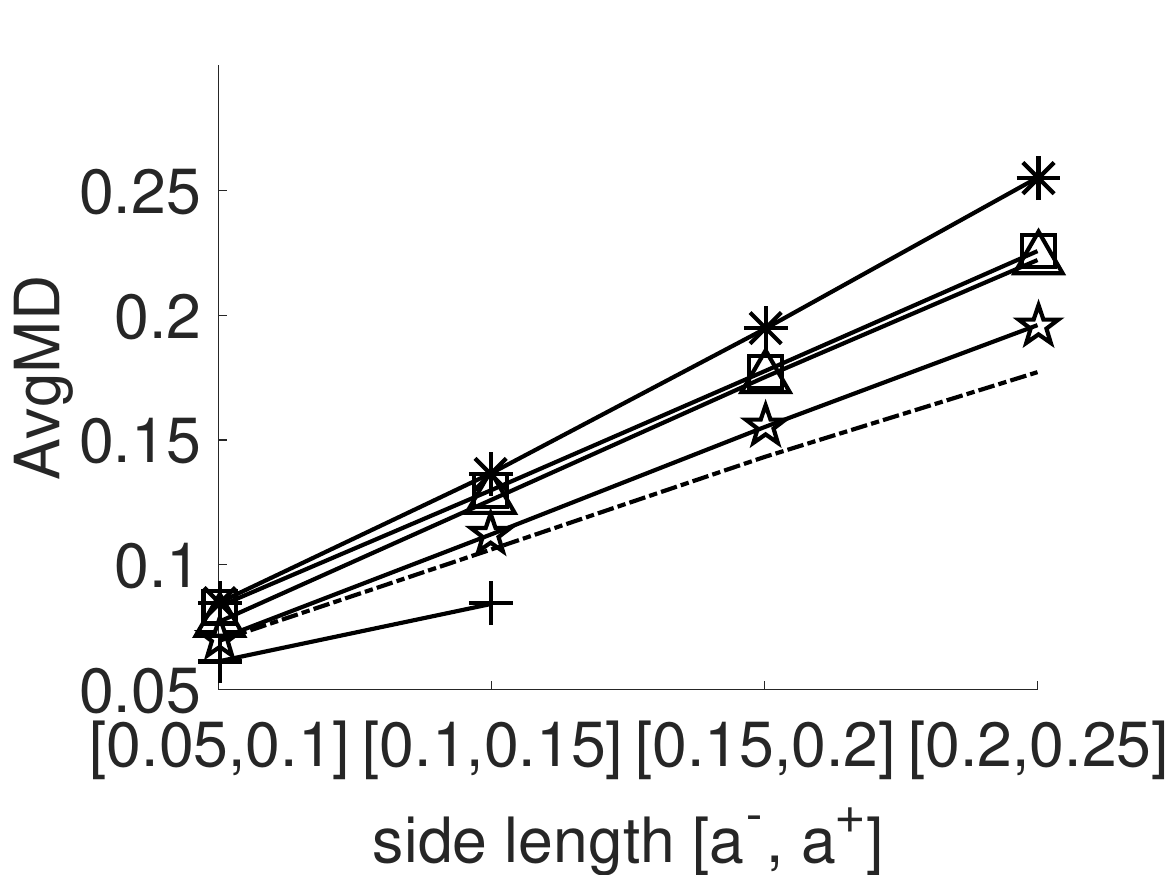}}
		\label{subfig:a_avg_moving_distance_real}}\hfill
	\subfigure[][{\small Fully Assigned Tasks}]{
		\scalebox{0.2}[0.2]{\includegraphics{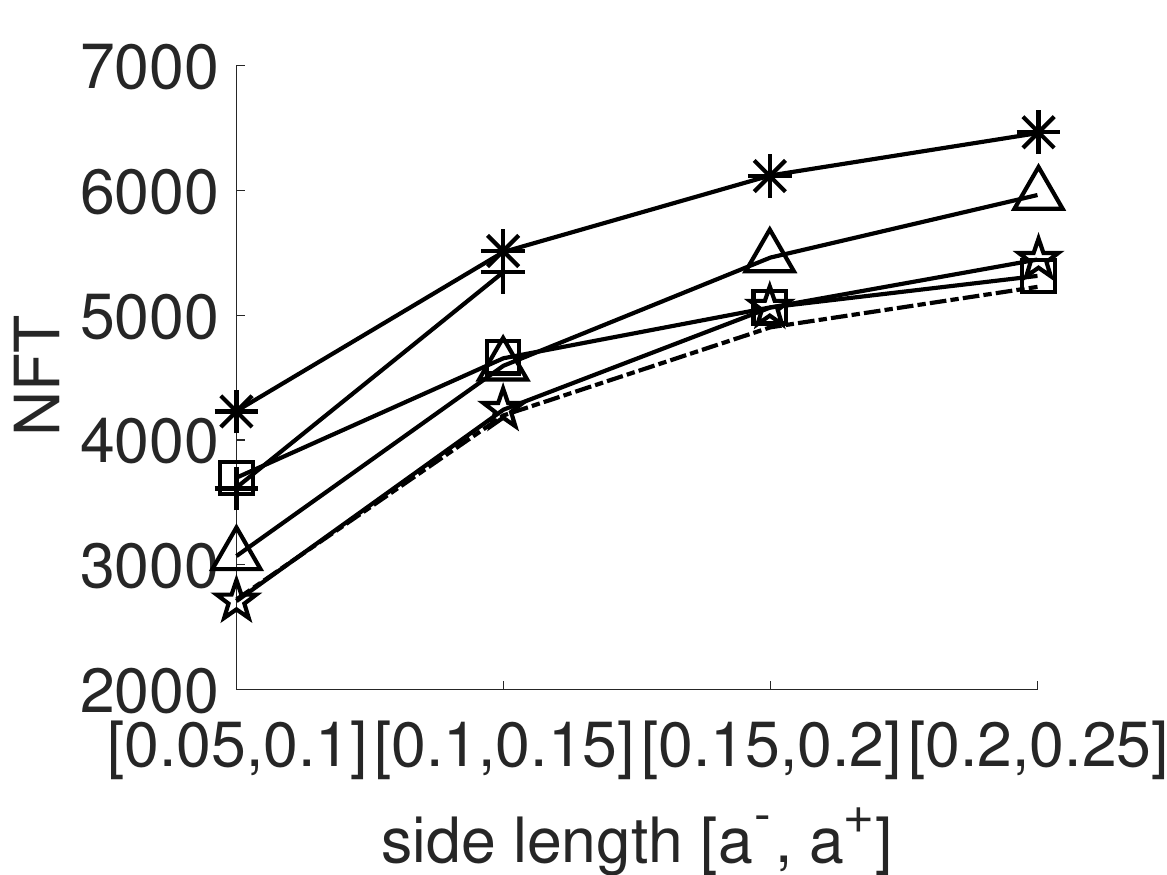}}
		\label{subfig:a_finished_task_number_real}}\hfill
	\subfigure[][{\small Confidently Assigned Tasks}]{
		\scalebox{0.2}[0.2]{\includegraphics{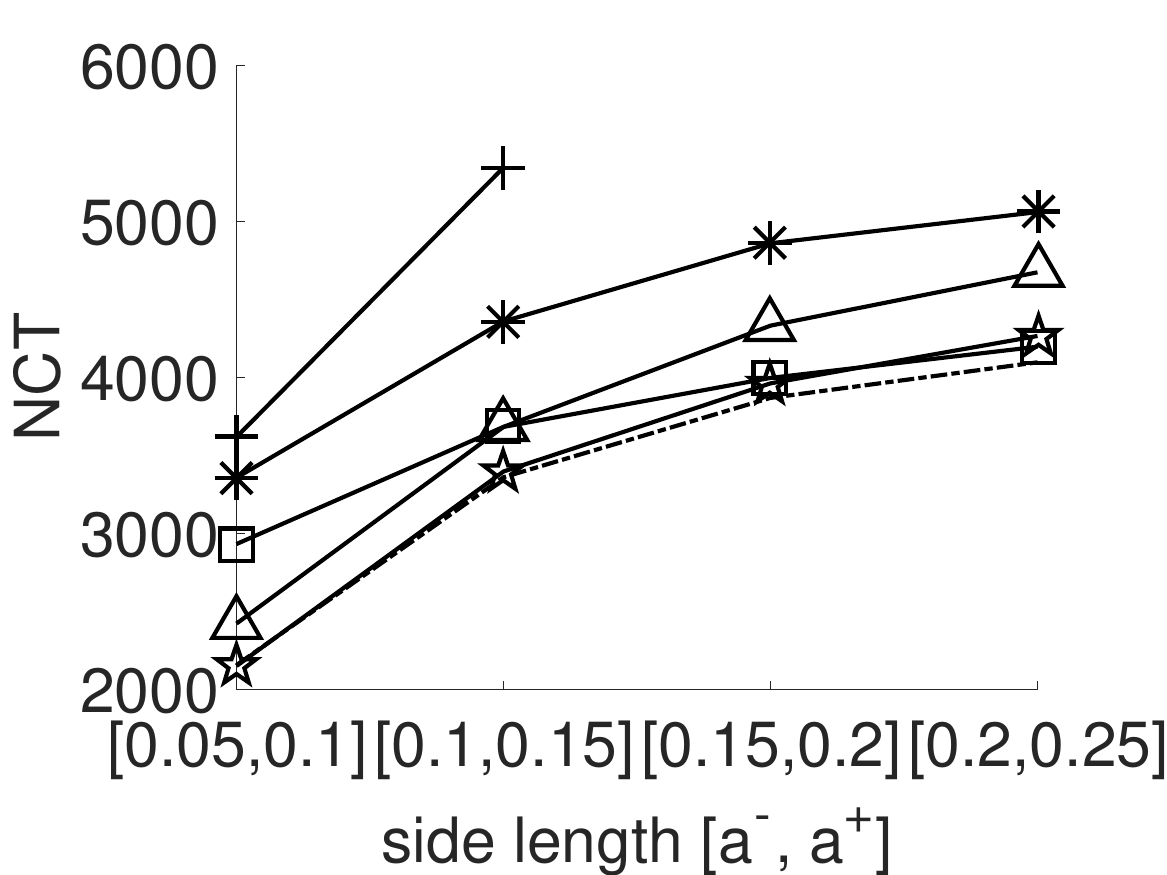}}
		\label{subfig:a_finished_task_number_conf_real}}\vspace{-1ex}
	\subfigure[][{\small Running Times}]{
		\scalebox{0.2}[0.2]{\includegraphics{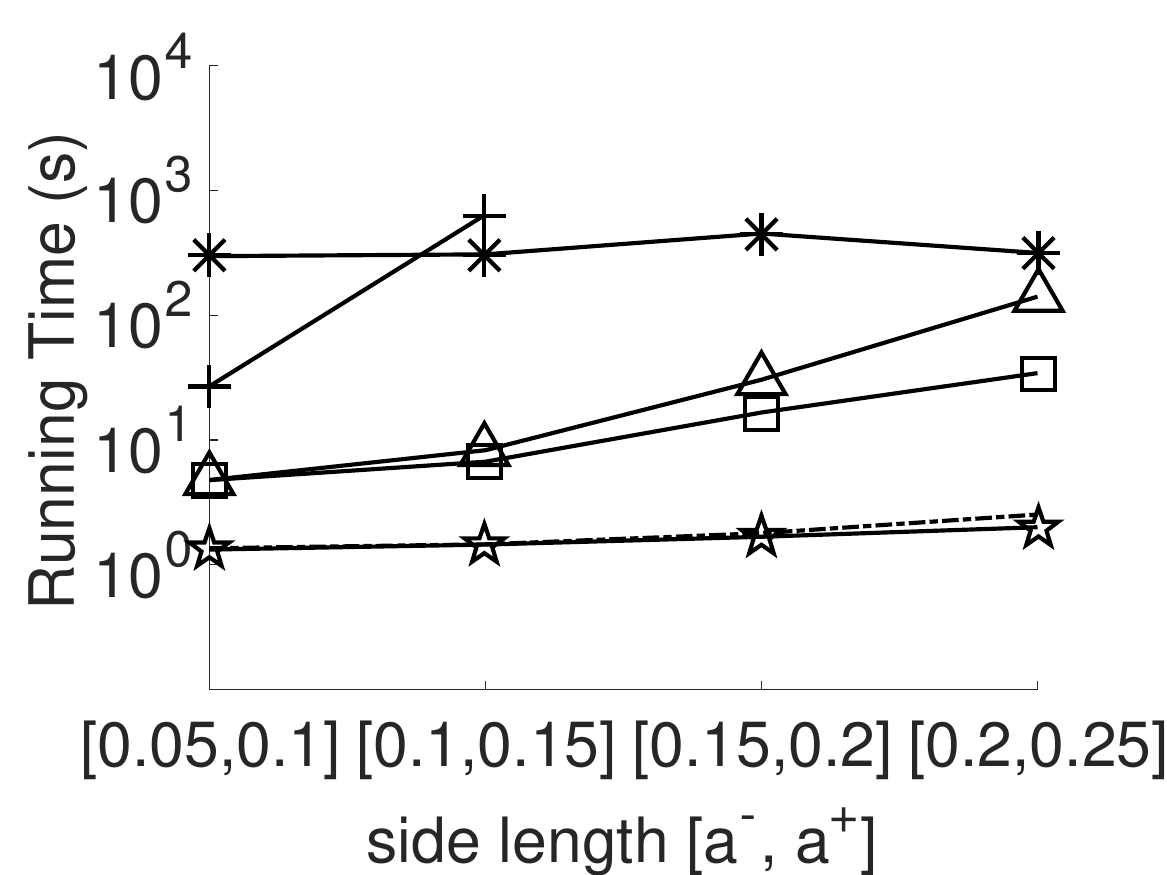}}
		\label{subfig:a_running_time_real}}\hfill\vspace{-2ex}
	\caption{\small Effects of Range of Side Length of Worker Working Area $a$ (Real).}
	\label{fig:effect_working_range_real}\vspace{-2ex}
\end{figure*}

\subsection{Experimental Results}
\subsubsection{Experiments on Real Data}
In this subsection, we show the results on the real data set and vary the range of task durations $rt$, the range $[a^-, a^+]$ of the side length workers' working areas, the range $[q^-, q^+]$ of tasks' required quality levels, the range  $[r^-, r^+]$ of workers' reliabilities, the range $[b^-, b^+]$ of tasks' required answers, the range $[c^-, c^+]$ of workers' capacities, the velocity $v$ of workers and the length of the time slot $\phi$.

\noindent {\bf Effect of the range, $[rt^-, rt^+]$, of
	tasks' durations.}
We show the effect of the range, $[rt^-, rt^+]$, of tasks' durations on the performances of tested approaches through varying $[rt^-, rt^+]$ from [1, 2] to [4, 5] (the unit is time slot). 

Figure \ref{fig:batch_effect_expiration} illustrates the results of batch-based algorithms.
In Figure \ref{subfig:rt_avg_moving_distance}, AvgMDs of the results of batch-based algorithms almost do not change when tasks' durations increase, because in each batch workers are assigned with tasks whose numbers reach their capacities in the real data set (tasks and workers are well mixed as shown in Figure \ref{fig:beijing_location_distribution}). \textsf{G-llep} causes workers to move the longest average distances, as it prefers to assign workers to tasks positioned at farther locations but with lower location entropies.  \textsf{GT-greedy} and \textsf{GT-hgr} only assign correct matches (one correct match is a task-and-workers 
pair $\langle t_j, W_j\rangle$ whose expected accuracy $Pr(W_j)$ is not less than the required quality level $q_j$ of task $t_j$), thus the two algorithms will use limited workers to confidently finish fewer tasks. As a result, the AvgMDs of results of \textsf{GT-greedy} and \textsf{GT-hgr} are small. As for \textsf{G-nnp}, it assigns workers to their nearest tasks, thus AvgMDs of its results are smaller than \textsf{GT-greedy} but higher than \textsf{GT-hgr}. \textsf{G-greedy}, \textsf{RDB-d\&c} and \textsf{RDB-sam} tend to assign as many worker-and-task pairs as possible thus have high AvgMDs.
In Figure \ref{subfig:rt_finished_task_number}, when the tasks' durations increase, all the batch-based approaches can fully assign more tasks, as each task will last for more batches such that more workers will be available for it. \textsf{G-llep} can fully assign the most number of tasks, which shows the effectiveness of its least location entropy priority strategy. \textsf{G-greedy} and \textsf{G-nnp} can fully assign fewer tasks than \textsf{G-llep} but more tasks than other batch-based algorithms. 
\textsf{GT-greedy} and \textsf{GT-hgr} can fully assign fewer tasks than other algorithms, as they just assign correct matches.  In addition, \textsf{RDB-d\&c} and \textsf{RDB-sam} can complete more tasks than \textsf{GT-greedy} and \textsf{GT-hgr} but fewer than other batch-based algorithms.
When we consider the quality of assigned workers to tasks as shown in 
Figure \ref{subfig:rt_finished_task_number_conf}, \textsf{GT-greedy} and \textsf{GT-hgr} can complete most tasks than other algorithms, as they are proposed to handle the quality issues of tasks. The other batch-based algorithms keep their ranks in Figure \ref{subfig:rt_finished_task_number}. Note that many fully assigned tasks are in fact not confidently assigned.
In Figure \ref{subfig:rt_running_time}, when the tasks' durations increase, all the batch-based algorithms need more time to resolve the problems, because the average number of tasks in each batch increases, which leads to the problem space increases. \textsf{G-greedy}, \textsf{G-llep} and \textsf{G-nnp} use more time than other batch-based algorithms, because they all need to invoke the time-consuming Ford-Fulkerson algorithm or its variants. \textsf{GT-greedy} runs fastest among the batch-based algorithms. As \textsf{RDB-sam} just quickly sample worker-and-task pairs, it runs fast but still slower than \textsf{GT-greedy}. \textsf{RDB-d\&c} is slower than \textsf{RDB-sam} but faster than \textsf{GT-hgr}.

Figure \ref{fig:online_effect_expiration} shows the results of online algorithms. In Figure \ref{subfig:online_rt_avg_moving_distance}, when tasks' durations increase, AvgMDs of results of online algorithms will increase, because workers can arrive at tasks located at farther locations leading to that online algorithms schedule workers to farther tasks. \textsf{PRS} achieves results with the highest AvgMDs.  \textsf{PRS} first assigns one task to each worker then use \textsf{BB} to plan other valid tasks. In the first step of \textsf{PRS}, some tasks may already be fully assigned with workers. Then in the second step of \textsf{PRS}, workers may be scheduled with farther tasks compared with using \textsf{DP} or \textsf{BB} directly. \textsf{HA} will result in larger AvgMDs than \textsf{BB} but smaller AvgMDs than other tested online algorithms. In Figure \ref{subfig:online_rt_finished_task_number}, when the tasks' durations increase, similarly, all the tested online approaches also can fully assign more tasks. \textsf{PRS} fully assigns the least tasks among online algorithms while \textsf{DP} fully assigns the most tasks. Similar ranking of results achieved by the tested online approaches can be observed when the quality levels of tasks are considered, as shown in Figure \ref{subfig:online_rt_finished_task_number_conf}. However, the number of confidently assigned tasks is less than the number of fully assigned tasks for all the results achieved by the tested online algorithms. In Figure \ref{subfig:online_rt_running_time}, when the tasks durations increase, all the tested online algorithms consume more time to achieve results. \textsf{HA} and \textsf{PRS} are the fastest  and slowest approaches among the tested online algorithms, respectively. In addition, \textsf{BB} is faster than \textsf{DP}.

To compare the algorithms in batch-based mode and online mode together, we select three algorithms performing well from each category and place the results of them in the same figures to compare clearly. Specifically, we select \textsf{G-llep}, \textsf{GT-hgr} and \textsf{RDB-sam} from algorithms in batch-based mode, and select \textsf{BB}, \textsf{DP} and \textsf{HA} from algorithms in online mode. In the following discussion, we just show the results of the six selected algorithms.

\noindent {\bf Effect of the range, $[c^-, c^+]$, of
	workers' capacities.}
Figure \ref{fig:effect_capacity_real_all} shows the effect of the range of workers' capacities on the performances of tested approaches through varying $[c^-, c^+]$ from [2, 3] to [5, 6]. As the running time of \textsf{DP} increases dramatically, we do not  report the results of \textsf{DP} when $[c^-, c^+]$ is  [4, 5] and [5,  6].

When the capacities
of workers increase, each worker may need to move longer to finish more tasks as shown in Figure \ref{subfig:c_avg_moving_distance_real_all}. However, we find \textsf{G-llep} in fact sacrifices the efficiency of moving distances to fully assign more tasks. When some tasks are located in far positions with low location entropies, \textsf{G-llep} will assign these tasks with higher priorities such that AvgMD will increase. In addition, we find AvgMDs of the results of batch-based algorithms, except for \textsf{GT-hgr}, are higher than that of online algorithms. The reason is that online algorithms schedule the assigned tasks for each worker with the minimum total travel cost.
NFTs of the tested algorithms are shown in Figure \ref{subfig:c_finished_task_number_real_all}. When the capacities of workers increase, NFTs of the tested algorithm  increase. Moreover, batch-based algorithms can fully assign more tasks than online algorithms. For NCTs shown in Figure \ref{subfig:c_finished_task_number_conf_real_all}, batch-based algorithms can also confidently assign more tasks than online algorithms. NCTs of \textsf{GT-hgr} are higher than that of \textsf{G-llep} when the worker capacities are lower than 4. However, when the worker capacities are higher than 4, \textsf{G-llep} can confidently assign more tasks than \textsf{GT-hgr}. The reason is that although \textsf{G-llep} does not consider the expected quality of the fully assigned tasks, when NFT of \textsf{G-llep} is high enough, NCT of \textsf{G-llep} can beat that of \textsf{GT-hgr}, the one particularly designed to focus on the quality of tasks.
For the running times of the tested approaches as shown in Figure \ref{subfig:c_running_time_real_all}, \textsf{DP} is the slowest when worker capacities are higher than 4. \textsf{BB} and \textsf{HA} are faster than batch-based algorithms. \textsf{G-llep} is slower than \textsf{GT-hgr}, as \textsf{G-llep} needs to keep updating the entropies of many positions.

\begin{figure*}[t!]\centering
	\subfigure{
		\scalebox{0.4}[0.4]{\includegraphics{bar_mix-eps-converted-to.pdf}}}\hfill\\\vspace{-2ex}
	\addtocounter{subfigure}{-1}
	\subfigure[][{\small Moving Distance}]{
		\scalebox{0.2}[0.2]{\includegraphics{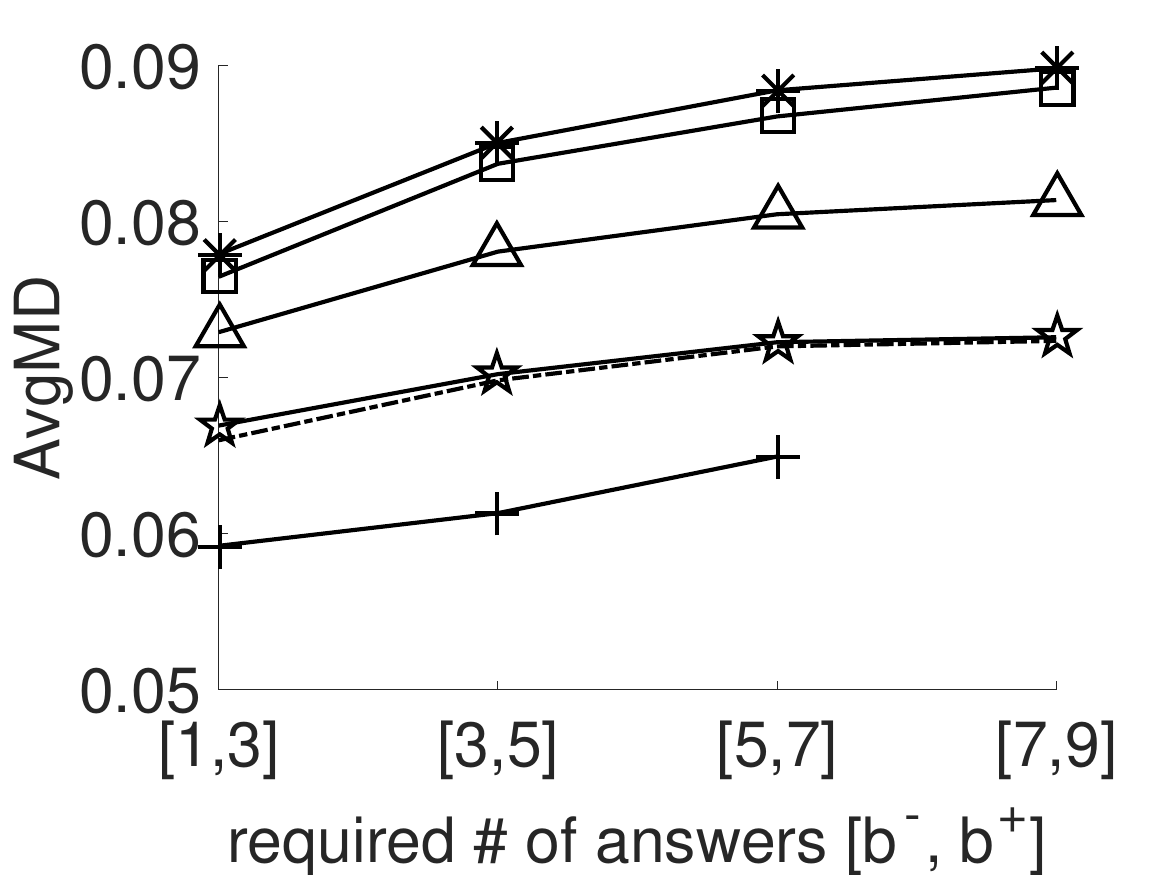}}
		\label{subfig:b_avg_moving_distance_real_all}}\hfill\vspace{-2ex}
	\subfigure[][{\small Fully Assigned Tasks}]{
		\scalebox{0.2}[0.2]{\includegraphics{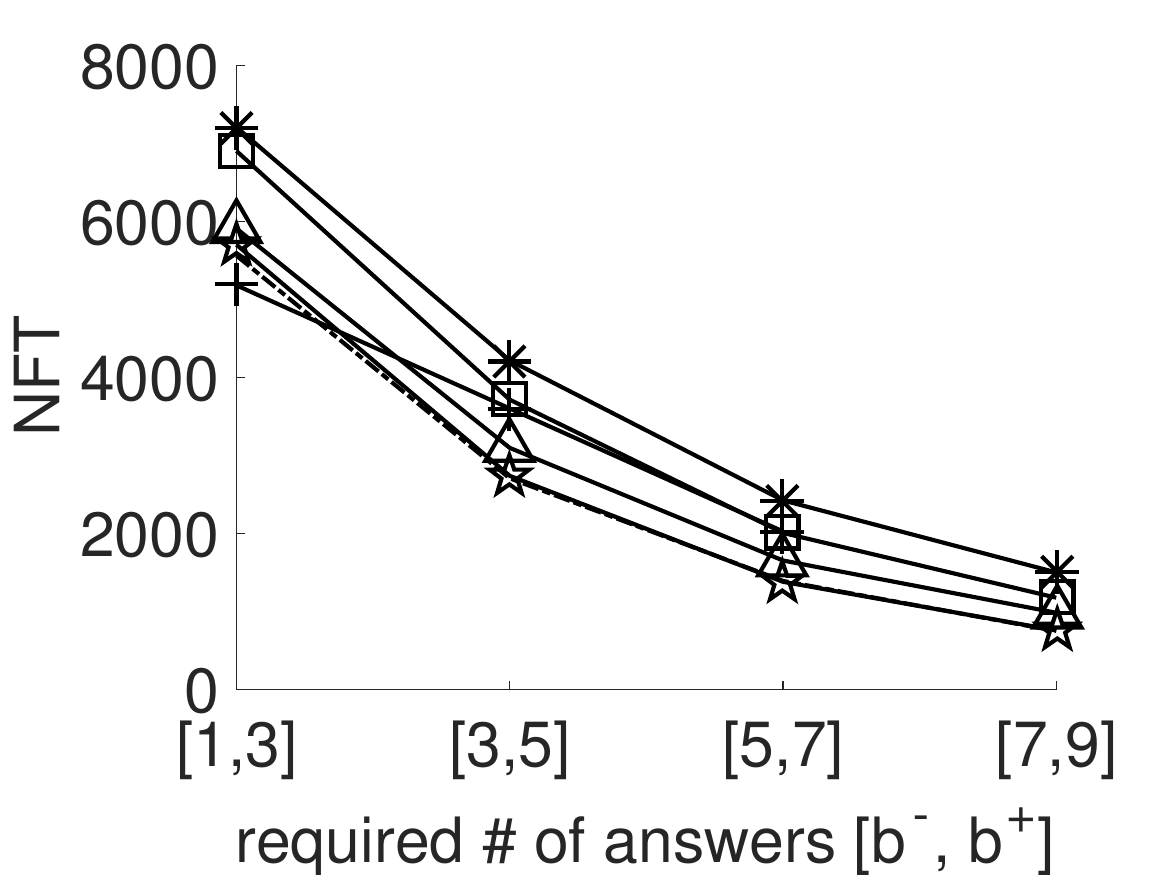}}
		\label{subfig:b_finished_task_number_real_all}}\hfill
	\subfigure[][{\small Confidently Assigned Tasks}]{
		\scalebox{0.2}[0.2]{\includegraphics{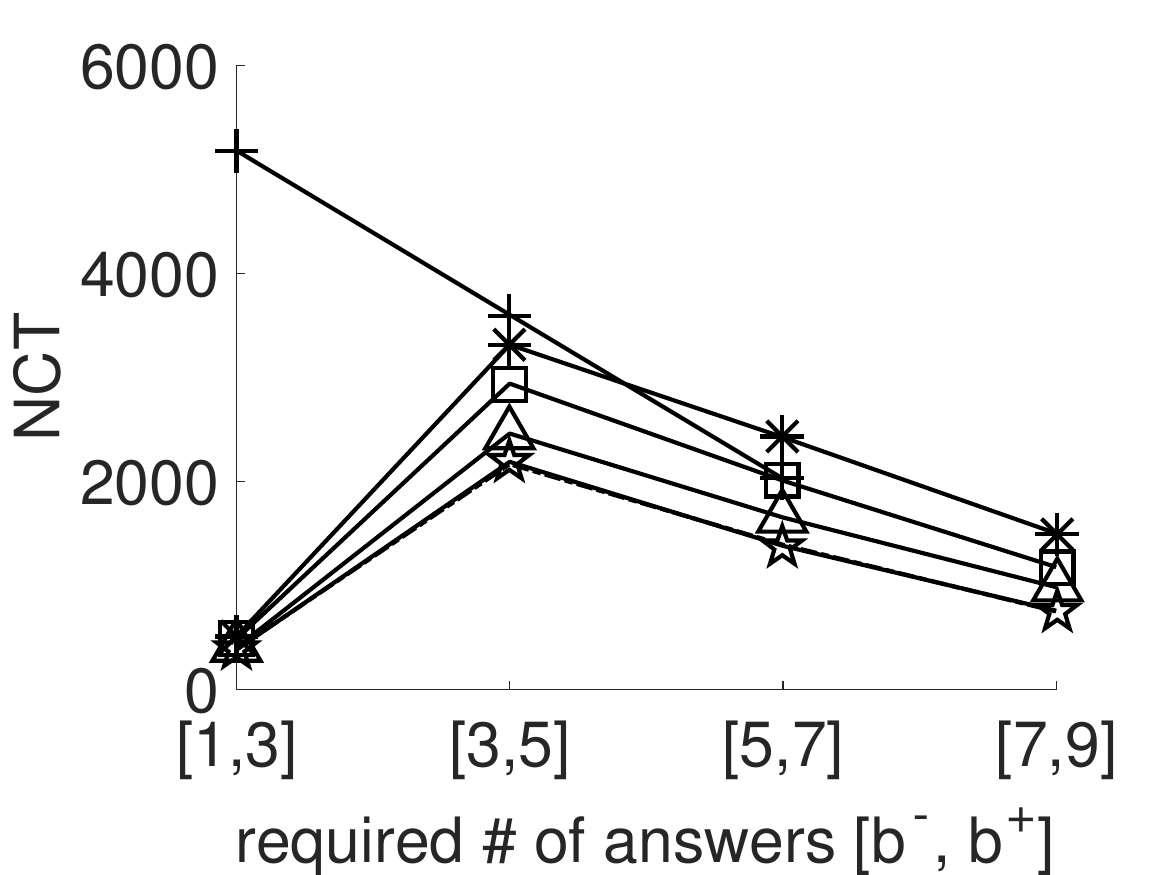}}
		\label{subfig:b_finished_task_number_conf_real_all}}\hfill\vspace{-1ex}
	\subfigure[][{\small Running Time}]{
		\scalebox{0.2}[0.2]{\includegraphics{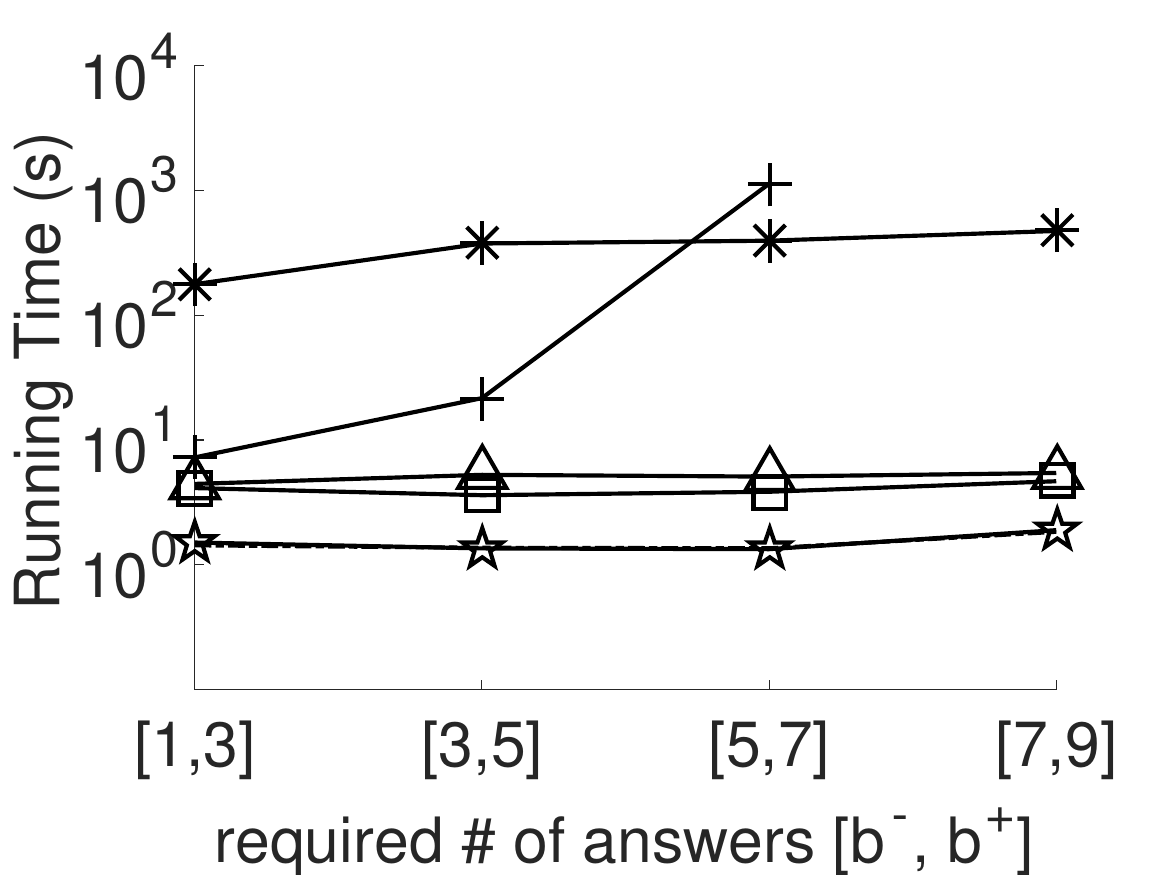}}
		\label{subfig:b_running_time_real_all}}\hfill
	\caption{\small Effects of Required Answer Count $b$ (Real).}
	\label{fig:effect_task_require_answer_count_real}\vspace{-2ex}
\end{figure*}

\begin{figure*}[ht!]\centering
	\subfigure{
		\scalebox{0.4}[0.4]{\includegraphics{bar_mix-eps-converted-to.pdf}}}\hfill\\\vspace{-2ex}
	\addtocounter{subfigure}{-1}
	\subfigure[][{\small Moving Distance}]{
		\scalebox{0.2}[0.2]{\includegraphics{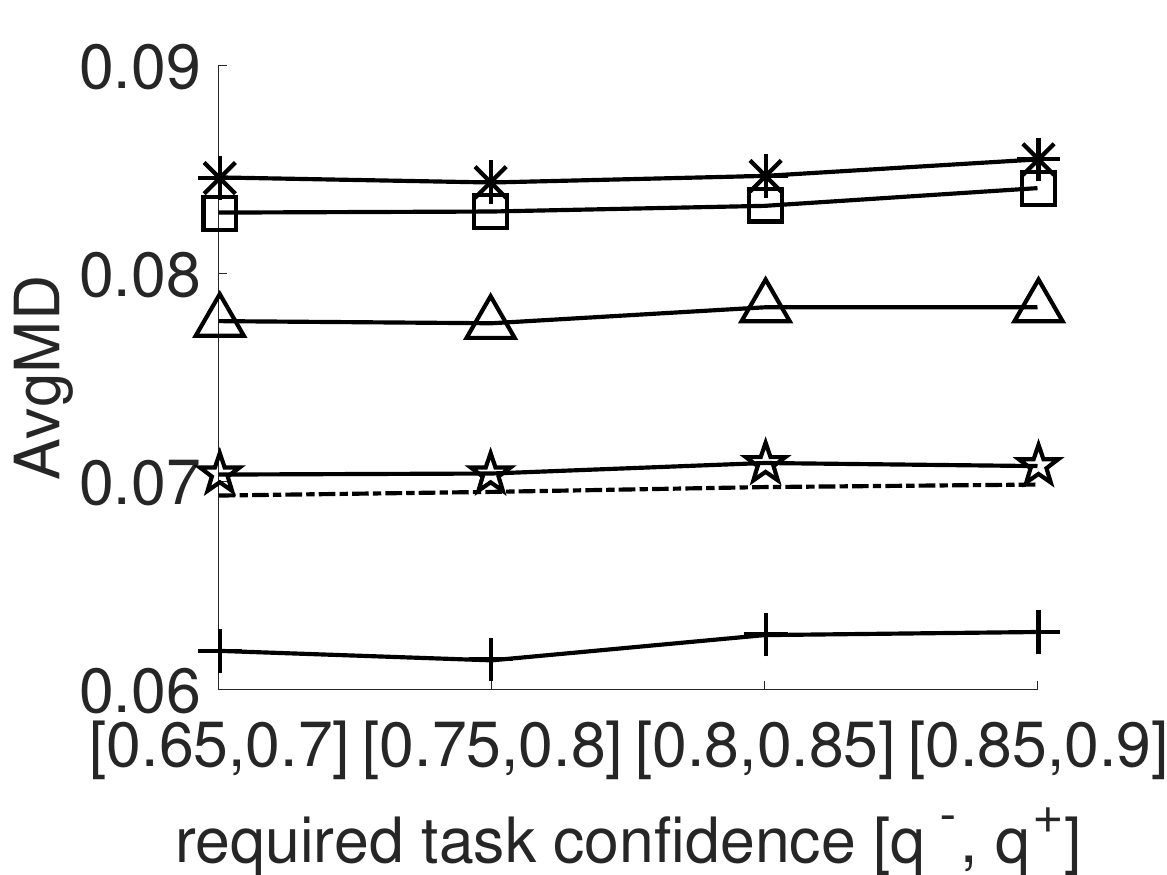}}\vspace{-2ex}
		\label{subfig:q_avg_moving_distance_real}}\hfill
	\subfigure[][{\small Fully Assigned Tasks }]{
		\scalebox{0.2}[0.2]{\includegraphics{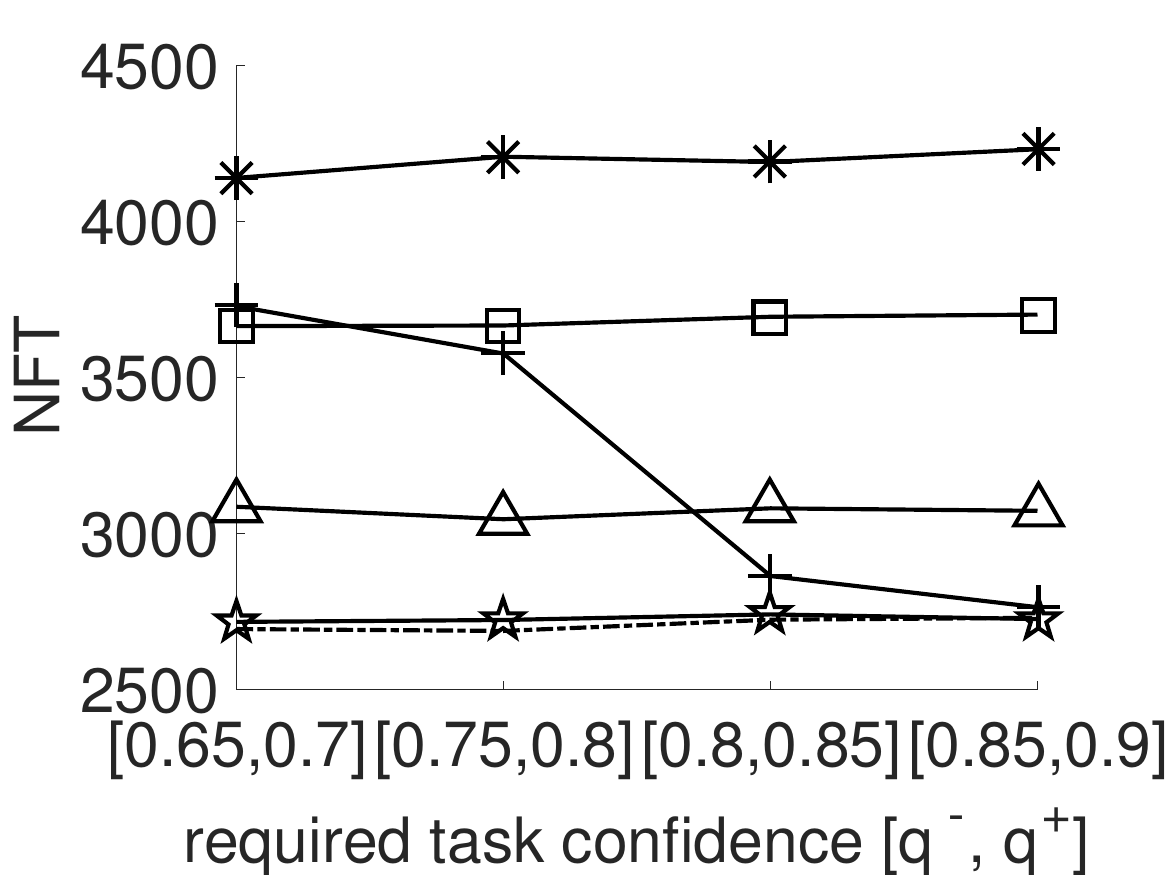}}
		\label{subfig:q_finished_task_number_real}}\hfill
	\subfigure[][{\small Confidently Assigned Tasks}]{
		\scalebox{0.2}[0.2]{\includegraphics{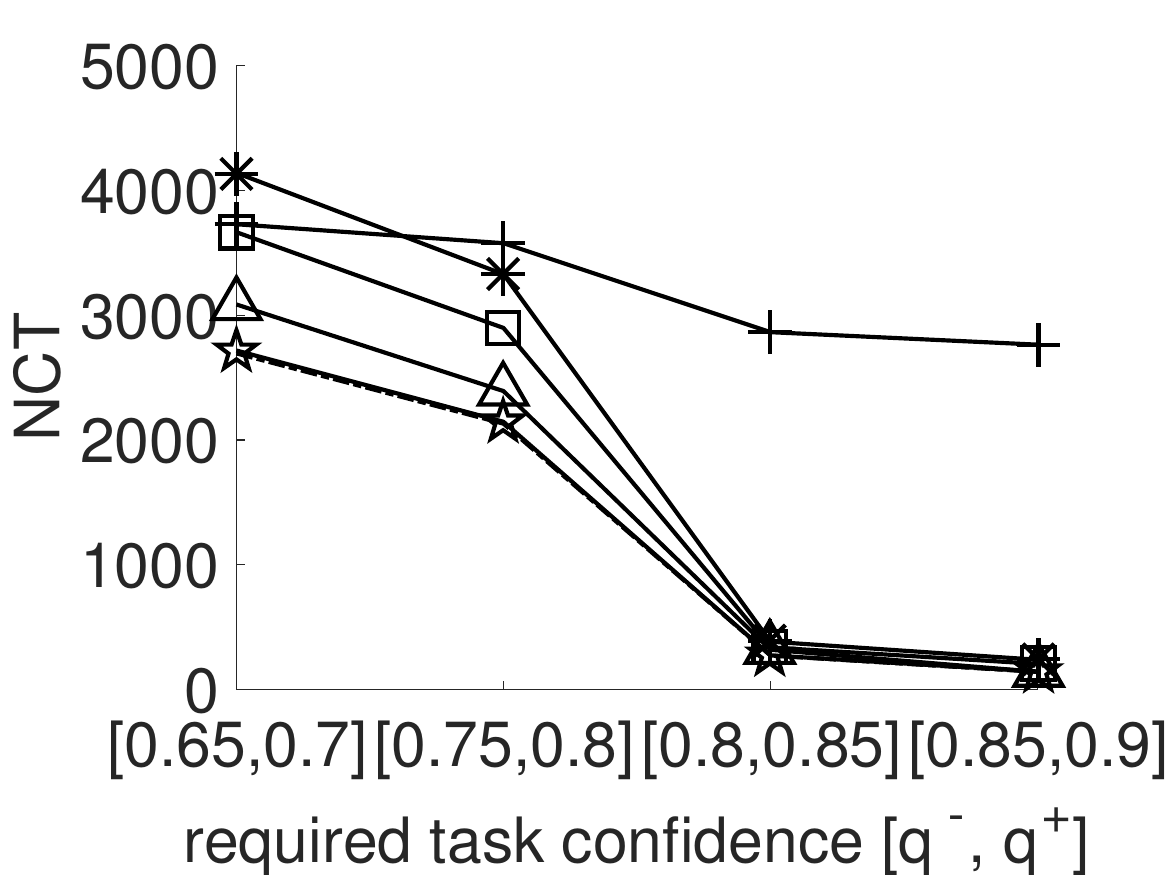}}
		\label{subfig:q_finished_task_number_conf_real}}
	\subfigure[][{\small Running Time}]{
		\scalebox{0.2}[0.2]{\includegraphics{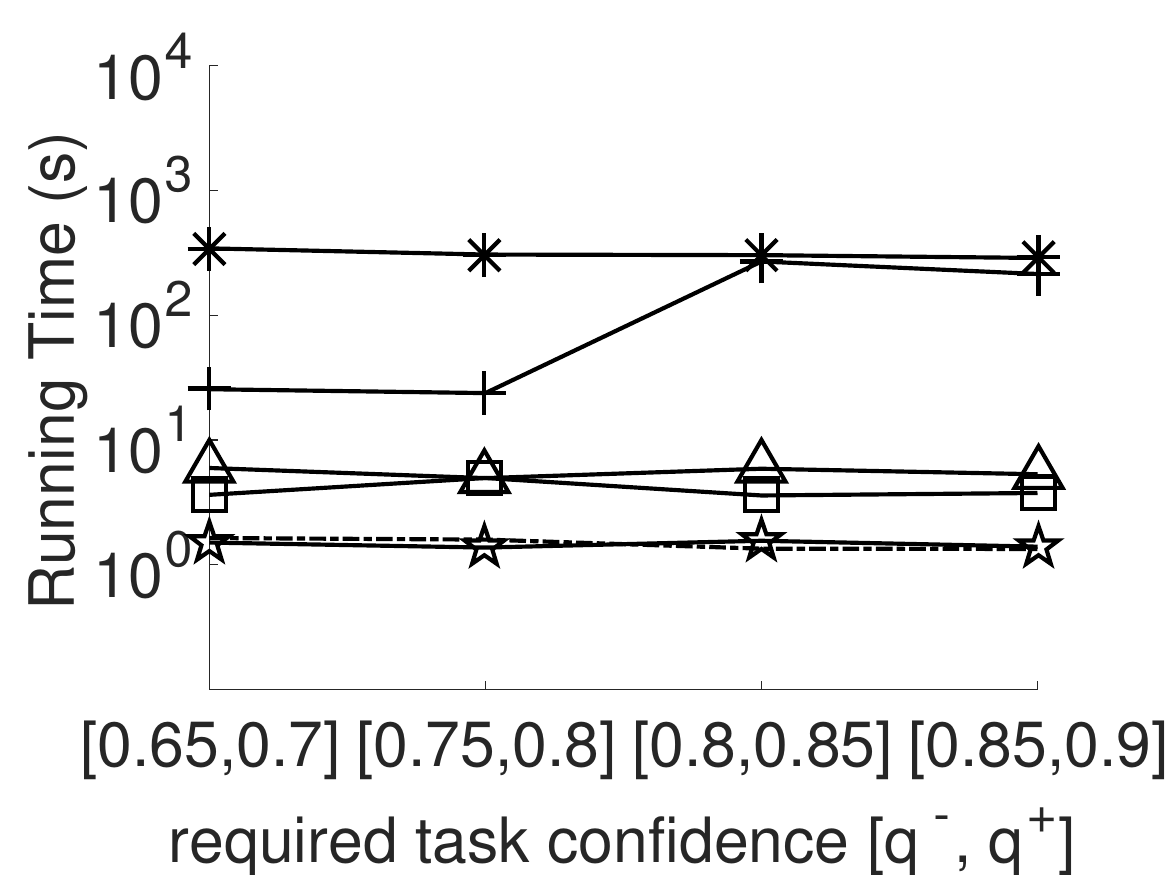}}
		\label{subfig:q_running_time_real}}\hfill\vspace{-2ex}
	\caption{\small Effects of Required Quality Level $q$ (Real).}
	\label{fig:effect_task_require_confidence_real}\vspace{-2ex}
\end{figure*}

\noindent {\bf Effect of the range, $[a^-, a^+]$, of 
 the side length of	workers' working areas.}
When workers' working areas get larger, there will be more available tasks located in the working area of each worker leading to the number of valid worker-and-task pairs increases. As the running time of \textsf{GT-hgr} increases dramatically when workers' working areas get larger, we do not  report the results of \textsf{GT-hgr} when $[a^-, a^+]$ is  [0.15, 0.2] and [0.2,  0.25].

In Figure \ref{subfig:a_avg_moving_distance_real}, as the working areas get larger, AvgMDs of the results achieved by all the tested approaches increase obviously, because the worker can reach tasks located further. 
In Figure \ref{subfig:a_finished_task_number_real}, all the tested approaches can fully assign more tasks when the range of side length of working areas $a$ increases, as each task can be reached by more workers and can be fully assigned with a higher probability. Specifically, the increasing speed of NFT of the tested online algorithms is higher than that of \textsf{G-llep} and \textsf{RDB-sam}.
In Figure \ref{subfig:a_finished_task_number_conf_real}, \textsf{GT-hgr} still can achieve the highest NCT than other tested algorithms. When the range of side length of working areas reaches [0.15, 0.2], online algorithms can achieve similar or even higher NCT than \textsf{RDB-sam}, as far workers be scheduled to farther tasks by online algorithms. 
In Figure \ref{subfig:a_running_time_real}, the running time of all the tested approaches increases when the range of $a_i$ increases, as more valid worker-and-task pairs need to process. When $[a^-, a^+]$ is higher than [0.1, 0.15], \textsf{GT-hgr} needs much more time than other tested approaches.
Running time of DP increases faster than other approaches except for \textsf{GT-hgr}, as for each worker the computation complexity of DP is $O(m^2\cdot 2^m)$.

\noindent {\bf Effect of the range, $[b^-, b^+]$, of the number of tasks' required answers.}
When the range $[b^-, b^+]$ increases, AvgMDs achieved by  the tested approaches increase simultaneously shown in Figure \ref{subfig:b_avg_moving_distance_real_all}. The reason is the worker labor does not increase, when a task $t_j$ needs more workers, the platform will to schedule farther workers to join. 
For the number of finished tasks as shown in Figure \ref{subfig:b_finished_task_number_real_all}, when the range $[b^-, b^+]$ increases, NFTs of all approaches decrease as the worker labor does not increase. For NCTs shown in Figure \ref{subfig:b_finished_task_number_conf_real_all}, different approaches performed quite different. When the range $[b^-, b^+]$ increases, NCTs of \textsf{GT-hgr} decrease monotonously, because workers are just enough for \textsf{GT-hgr} to fully assign fewer tasks. For other approaches not caring the correctness of the assignment, when the range $[b^-, b^+]$ is too small, like (1, 3), each tasks' assigned workers will rarely satisfy its required quality level. When the range $[b^-, b^+]$ increases a little, more fully assigned tasks will become confidently assigned tasks. But when the range $[b^-, b^+]$ becomes larger, as NFTs decrease, NCTs also decrease.
When each task requires more workers, all the tested algorithms need more time to achieve results, as shown in Figure \ref{subfig:b_running_time_real_all}. Specifically, the running time of \textsf{GT-hgr} increases dramatically, as for a task $t_j$, when more workers can be assigned to it, the number of correct matches for $t_j$ will increase quickly.

\noindent {\bf Effect of  the range, $[q^-, q^+]$, of tasks' required quality levels.} When the range of tasks' required quality levels changes, only \textsf{GT-hgr} will be affected in all the metrics and other algorithms will only be affected in NCTs. 
In Figure \ref{subfig:q_avg_moving_distance_real}, the required quality levels does not affect the average moving distance of the results achieved by the tested algorithms.
In Figure \ref{subfig:q_finished_task_number_real}, \textsf{GT-hgr} will assign fewer workers when the range $[q^-, q^+]$ gets higher, as the number of correct matches will decrease leading to NFT of \textsf{GT-hgr} decreasing. 
In Figure \ref{subfig:q_finished_task_number_conf_real}, when the range of $q_j$ increases, NCTs of all the tested approaches will decrease. We  notice that although NCT of \textsf{G-llep} is higher than that of \textsf{GT-hgr} when $[q^-, q^+]$ is [0.65, 0.7], \textsf{GT-hgr} can confidently assign more tasks when $[q^-, q^+]$ becomes larger (e.g., 0.75 to 0.9), which shows the effectiveness of the trustworthy query. 
In Figure \ref{subfig:q_running_time_real}, when $[q^-, q^+]$ increases, only the running time of \textsf{GT-hgr} increases, as it is harder to select a correct match for each task from fewer correct matches.

\begin{figure*}[t!]\centering
	\subfigure{
		\scalebox{0.4}[0.4]{\includegraphics{bar_mix-eps-converted-to.pdf}}}\hfill\\\vspace{-2ex}
	\addtocounter{subfigure}{-1}
	\subfigure[][{\small Moving Distance}]{
		\scalebox{0.2}[0.2]{\includegraphics{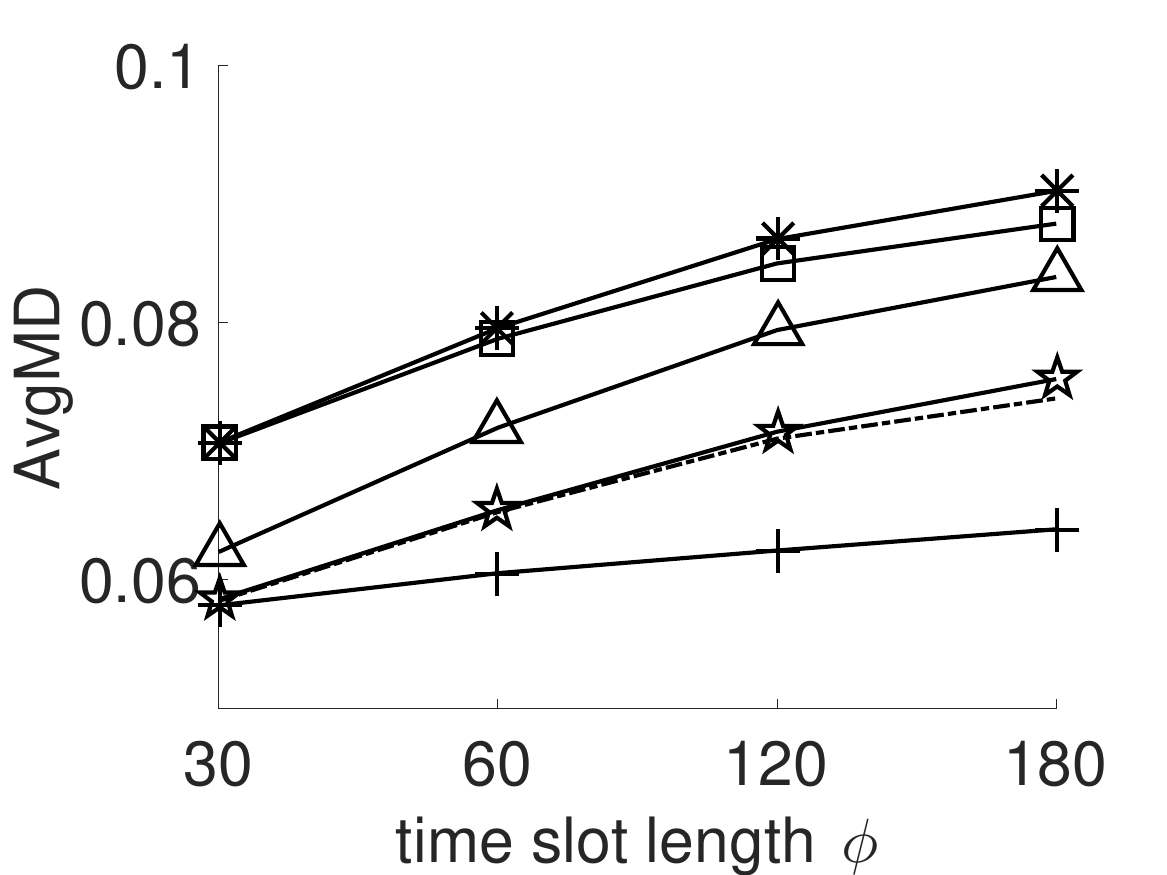}}
		\label{subfig:phi_avg_moving_distance_real}}\hfill\vspace{-2ex}
	\subfigure[][{\small Fully Assigned Tasks}]{
		\scalebox{0.2}[0.2]{\includegraphics{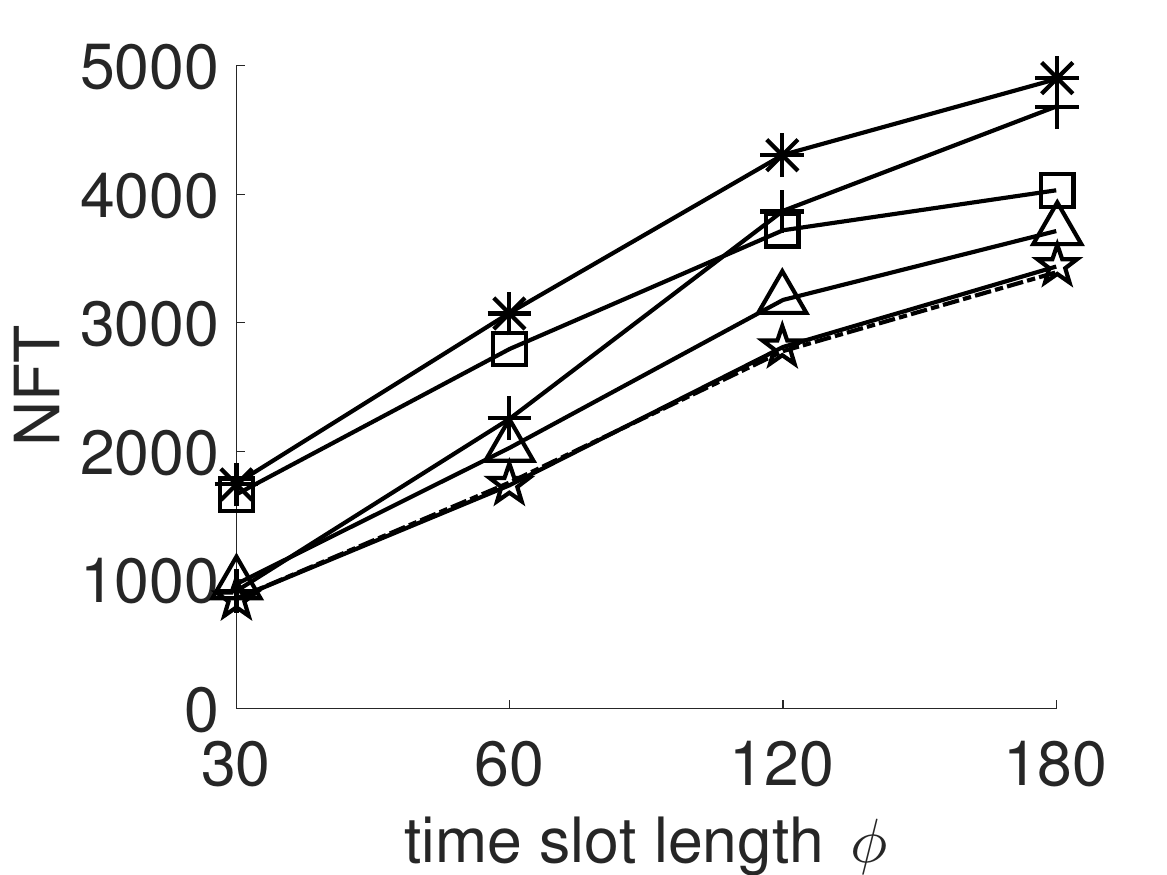}}
		\label{subfig:phi_finished_task_number_real}}\hfill
	\subfigure[][{\small Confidently Assigned Tasks}]{
		\scalebox{0.2}[0.2]{\includegraphics{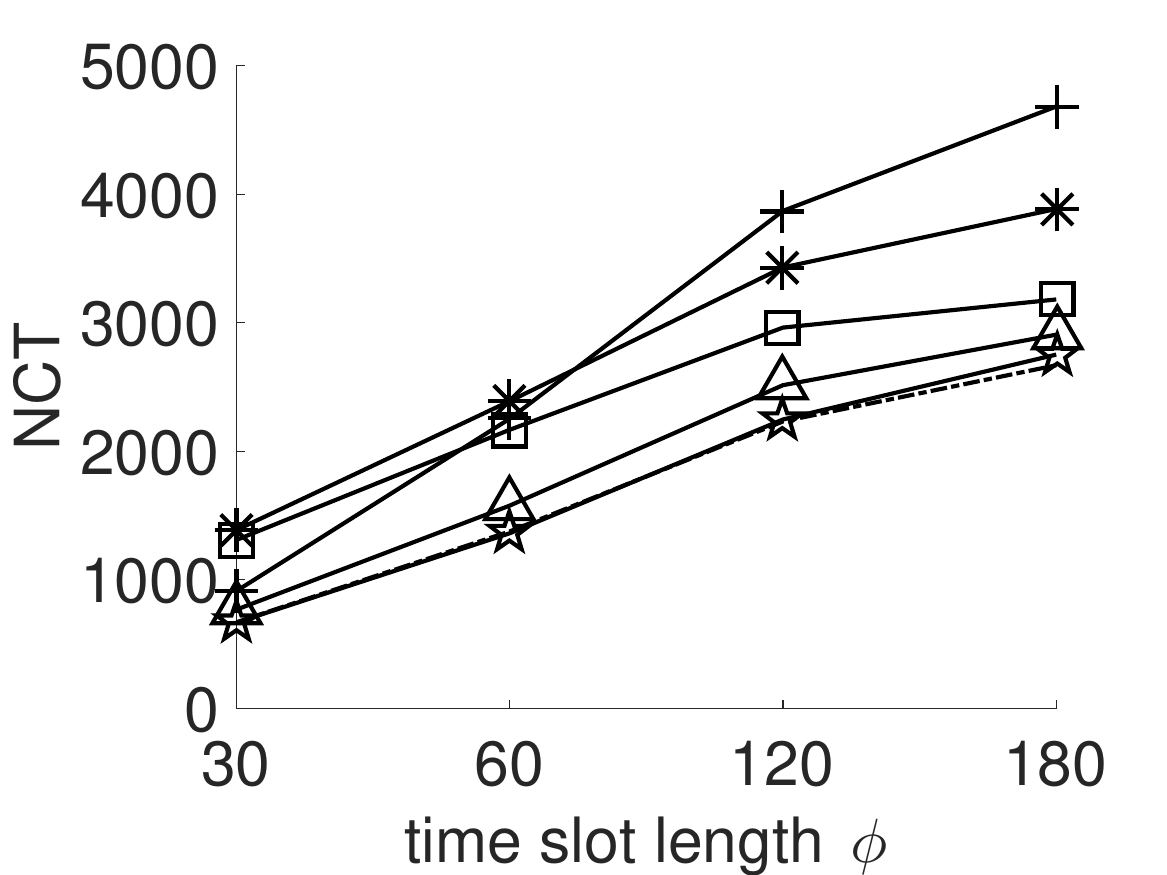}}
		\label{subfig:phi_finished_task_number_conf_real}}\hfill\vspace{-1ex}
	\subfigure[][{\small Running Time}]{
		\scalebox{0.2}[0.2]{\includegraphics{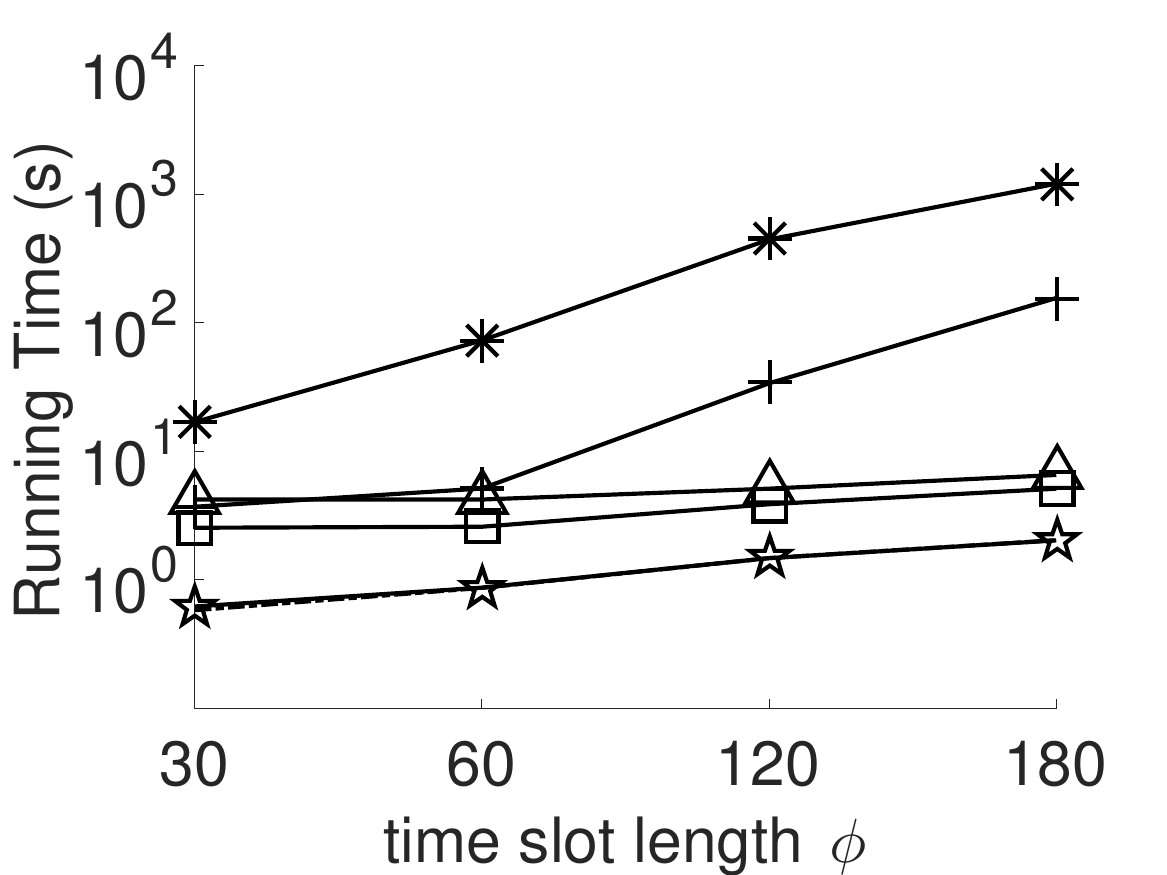}}
		\label{subfig:phi_running_time_real}}\hfill
	\caption{\small Effects of Time Slot Length $\phi$ (Real).}
	\label{fig:effect_task_time_slot_length_real}\vspace{-2ex}
\end{figure*}

\begin{figure*}[t!]\centering
	\subfigure{
		\scalebox{0.4}[0.4]{\includegraphics{bar_mix-eps-converted-to.pdf}}}\hfill\\\vspace{-2.5ex}
	\addtocounter{subfigure}{-1}
	\subfigure[][{\small Moving Distance}]{
		\scalebox{0.2}[0.2]{\includegraphics{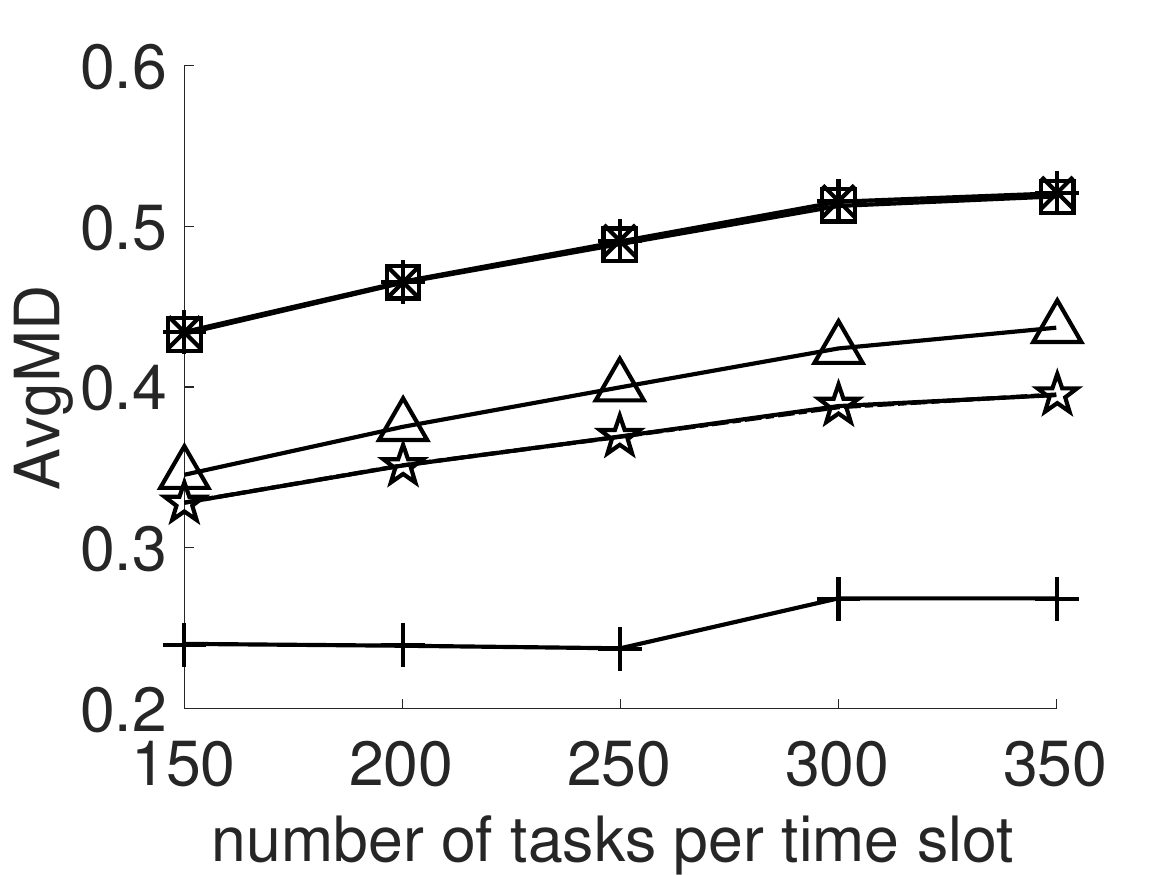}}
		\label{subfig:task_number_avg_moving_distance_unif}}\hfill\vspace{-2ex}
	\subfigure[][{\small Fully Assigned Tasks}]{
		\scalebox{0.2}[0.2]{\includegraphics{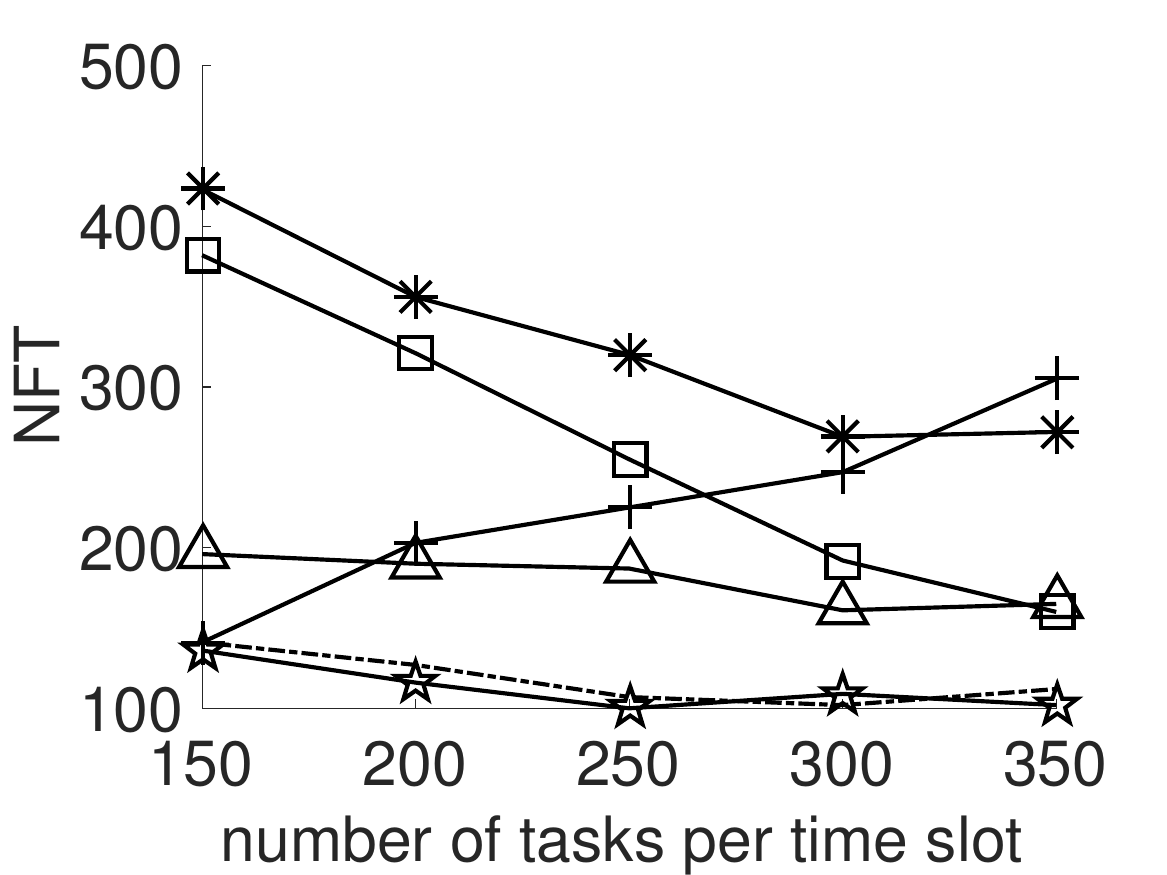}}
		\label{subfig:task_number_finished_task_number_unif}}\hfill
	\subfigure[][{\small Confidently Assigned Tasks}]{
		\scalebox{0.2}[0.2]{\includegraphics{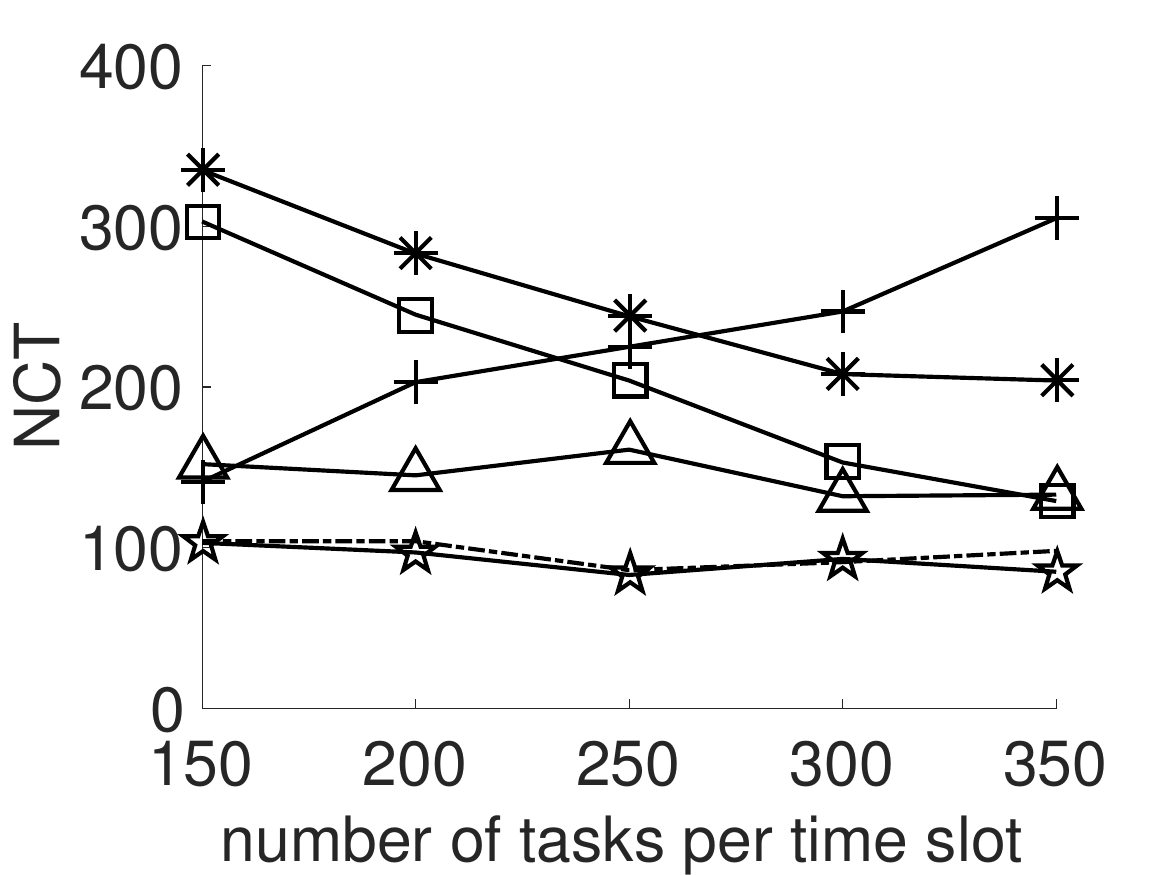}}
		\label{subfig:task_number_finished_task_number_conf_unif}}\hfill\vspace{-1ex}
	\subfigure[][{\small Running Time}]{
		\scalebox{0.2}[0.2]{\includegraphics{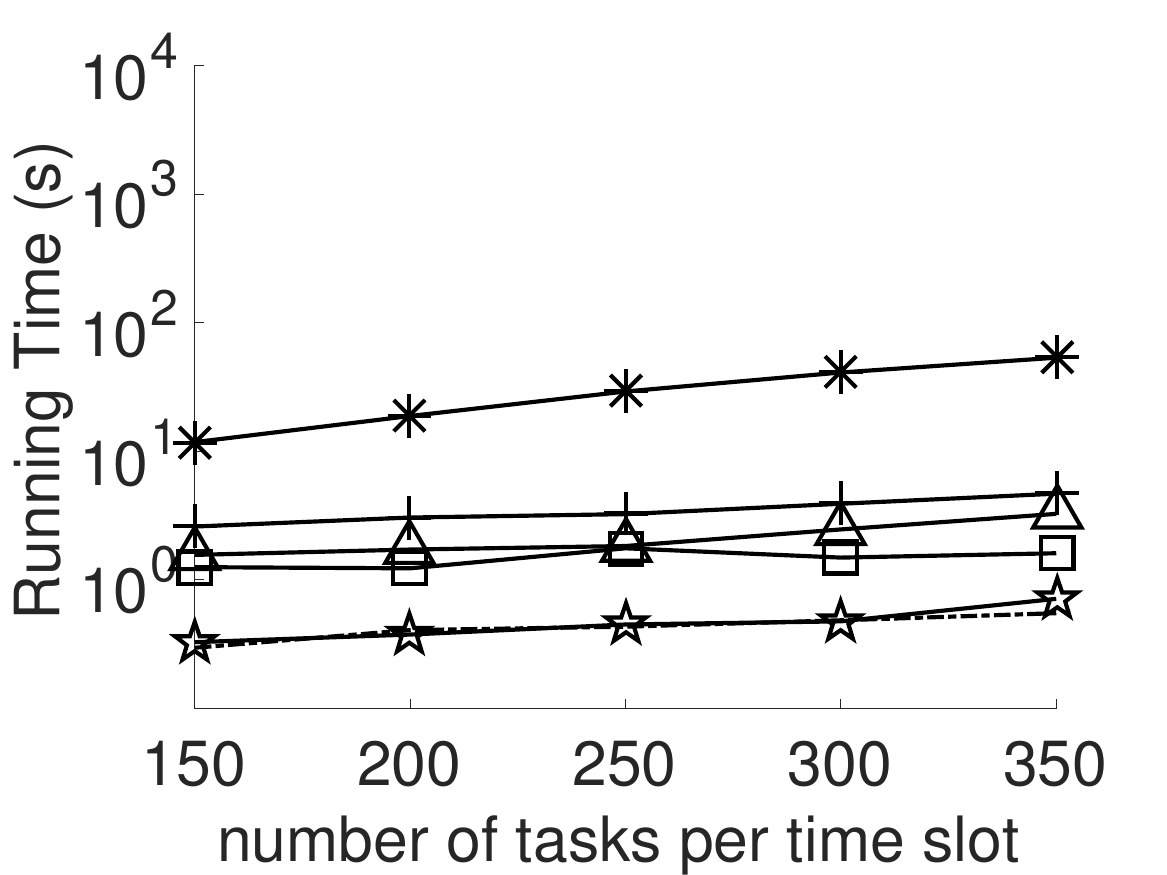}}
		\label{subfig:task_number_running_time_unif}}\hfill
	\caption{\small Effects of Number of Tasks Per Time Slot (Synthetic).}
	\label{fig:effect_task_number_synthetic}\vspace{-2ex}
\end{figure*}

\begin{figure*}[t!]\centering
	\subfigure{
		\scalebox{0.4}[0.4]{\includegraphics{bar_mix-eps-converted-to.pdf}}}\hfill\\\vspace{-2ex}
	\addtocounter{subfigure}{-1}
	\subfigure[][{\small Moving Distance}]{
		\scalebox{0.2}[0.2]{\includegraphics{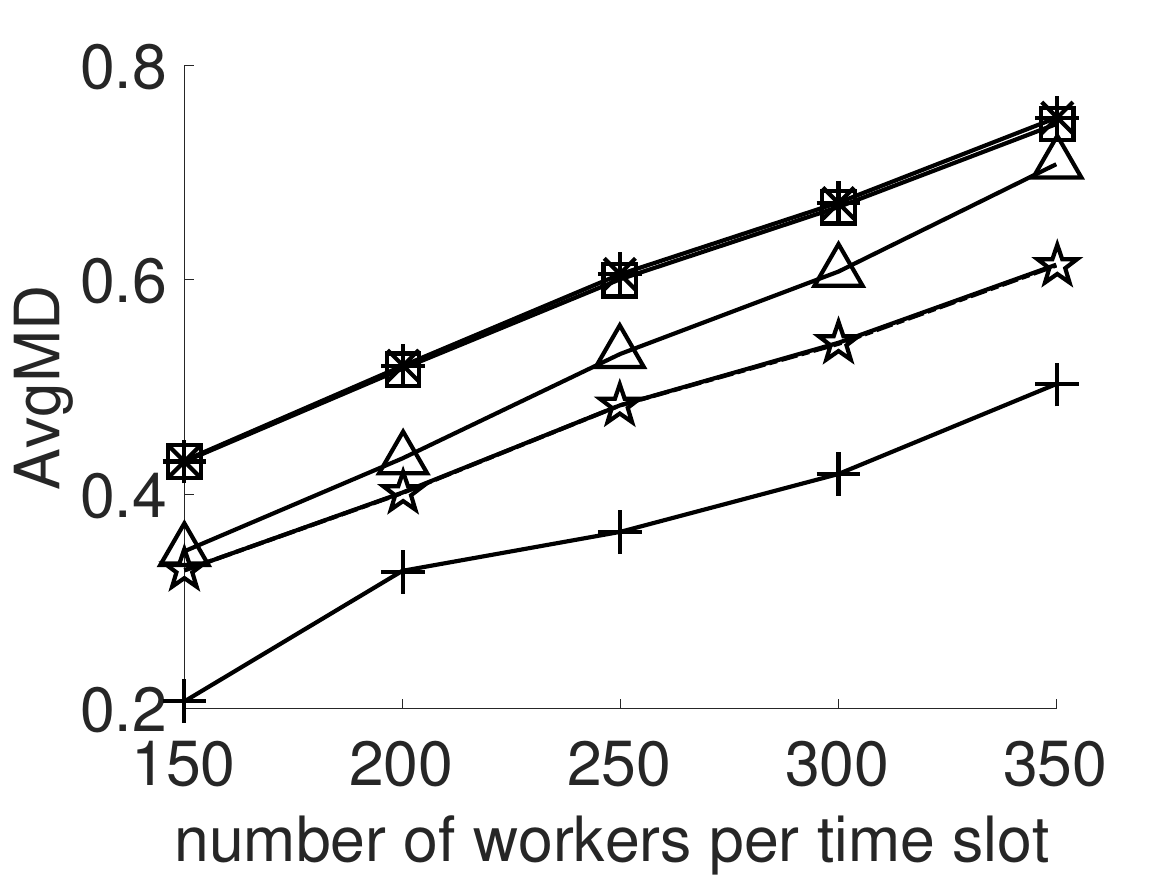}}
		\label{subfig:worker_number_avg_moving_distance_unif}}\hfill\vspace{-2ex}
	\subfigure[][{\small Fully Assigned Tasks}]{
		\scalebox{0.2}[0.2]{\includegraphics{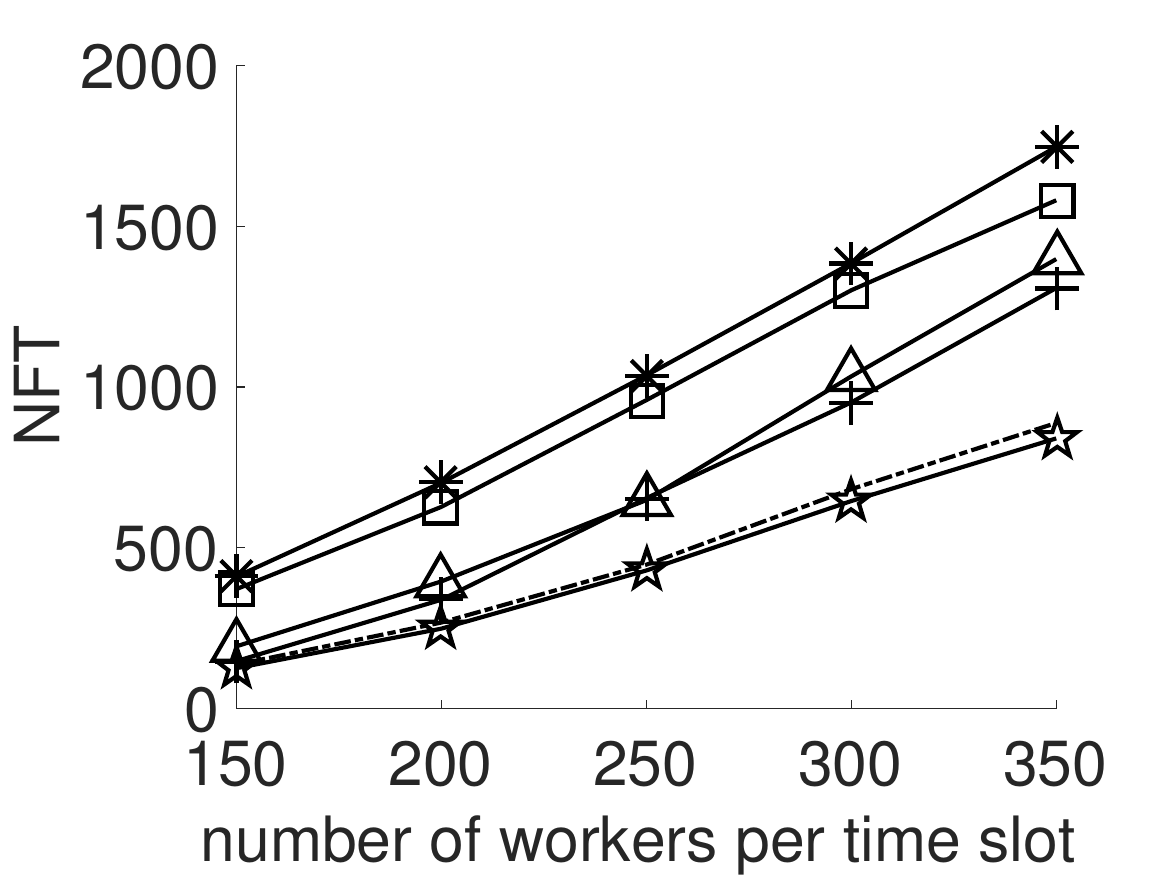}}
		\label{subfig:worker_number_finished_task_number_unif}}\hfill
	\subfigure[][{\small Confidently Assigned Tasks}]{
		\scalebox{0.2}[0.2]{\includegraphics{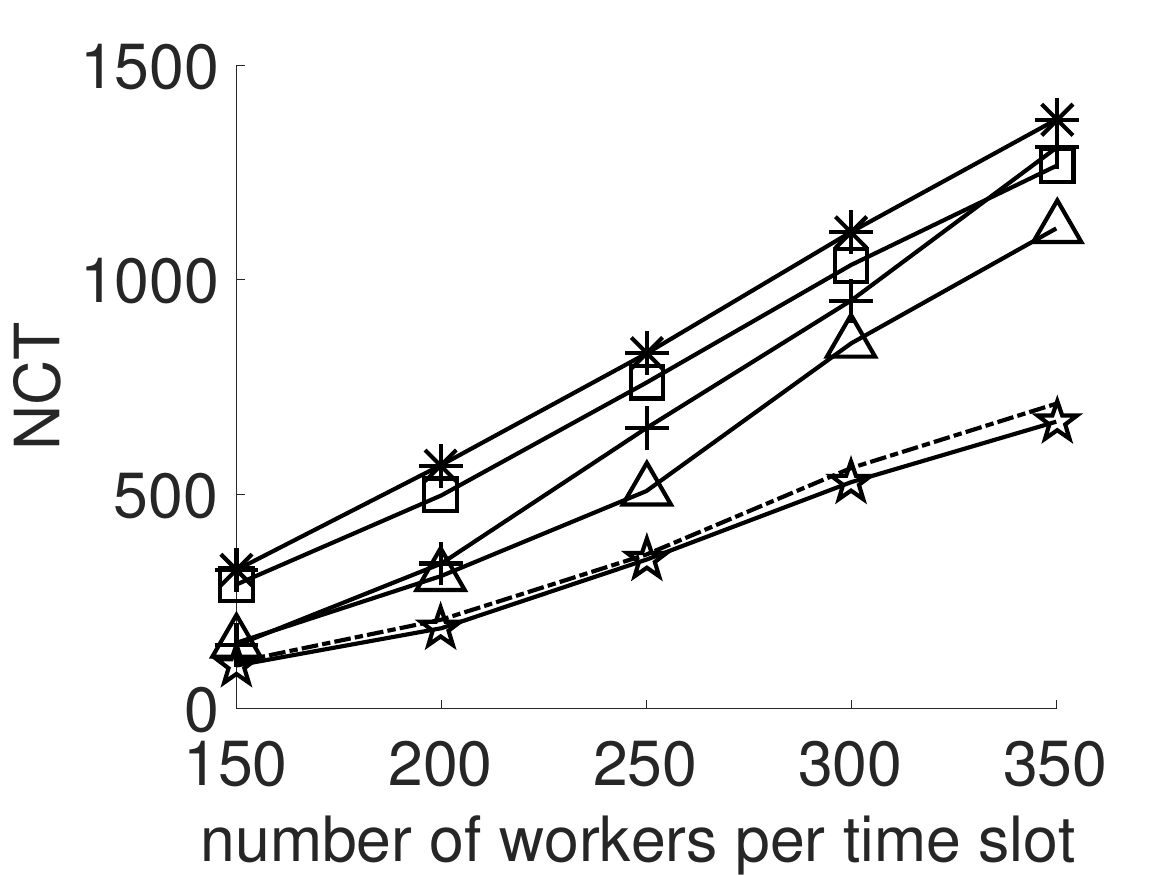}}
		\label{subfig:worker_number_finished_task_number_conf_unif}}\hfill\vspace{-1ex}
	\subfigure[][{\small Running Time}]{
		\scalebox{0.2}[0.2]{\includegraphics{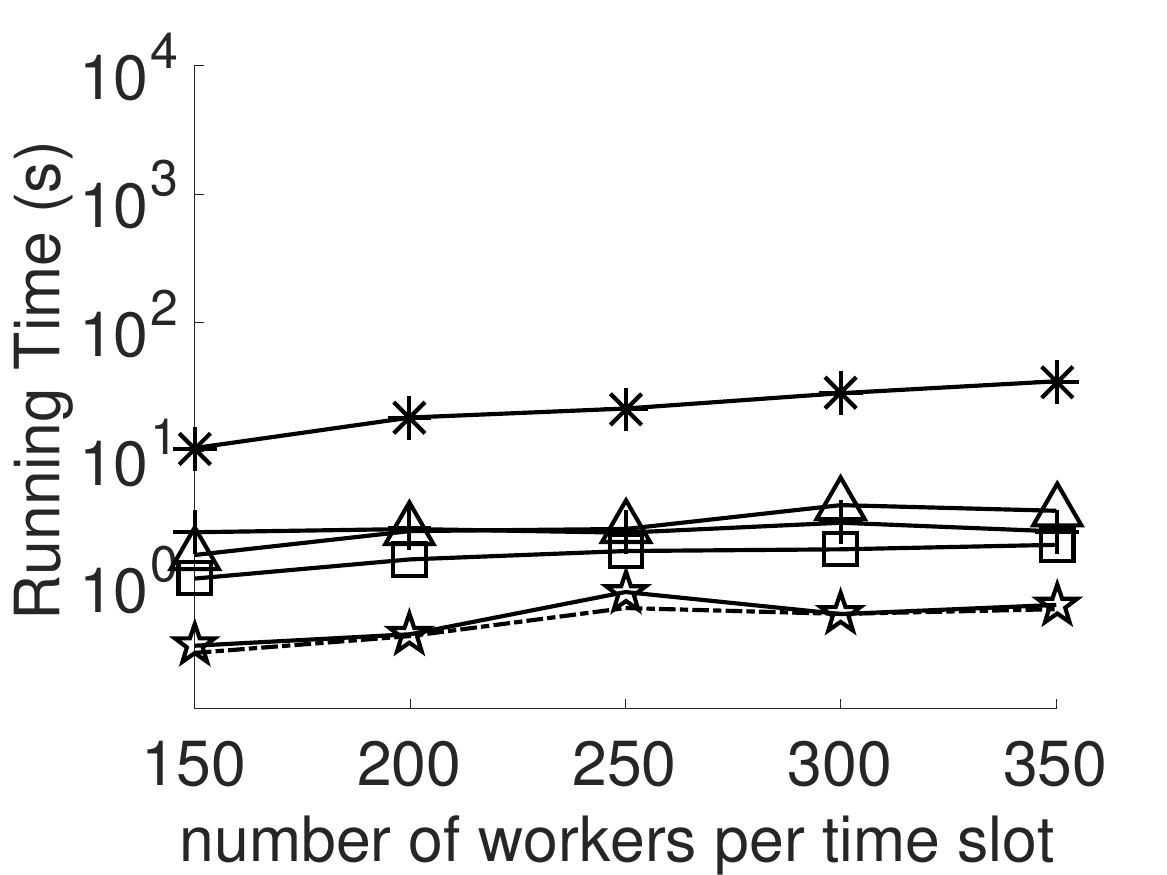}}
		\label{subfig:worker_number_running_time_unif}}\hfill
	\caption{\small Effects of Number of Workers Per Time Slot (Synthetic).}
	\label{fig:effect_worker_number_synthetic}\vspace{-3ex}
\end{figure*}

\noindent{\bf Effect of  the length of time slot $\phi$.} Figure \ref{fig:effect_task_time_slot_length_real} presents the effects of the length of time slot $\phi$ on the performances of the tested approaches by varying $\phi$ from 30 seconds to 180 seconds.

When the length of the time slot increases, AvgMDs of all the tested approaches increase as shown in Figure \ref{subfig:phi_avg_moving_distance_real} and NFTs of them also increase as shown in Figure \ref{subfig:phi_finished_task_number_real}. The reason is that when the time slot length increases, during each time slot the number of workers will increase leading to more tasks can be fully assigned. As more tasks are fully assigned, AvgMDs of all the tested algorithms will increase. We can still observe that \textsf{GT-hgr} has the lowest AvgMDs as it only selects correct matches. In addition, as \textsf{G-llep} may give higher priorities to far tasks located in positions with low location entropies, workers may move farther to conduct tasks in the results of \textsf{G-llep}. For NCTs  shown in Figure \ref{subfig:phi_finished_task_number_conf_real}, all the tested algorithms can confidently assign more tasks when the time slot length increases, as more tasks are fully assigned. Specifically, when the time slot length is short (e.g., $\phi=30$), \textsf{G-llep} and \textsf{RDB-sam} can confidently assign more tasks than \textsf{GT-hgr}. When the time slot length is longer than 60 seconds, \textsf{GT-hgr} has the highest NCTs.  As GT-hgr algorithms only assign correct matches, when more workers are available, the number of correct matches will increase exponentially leading to that NCT of \textsf{GT-hgr} increases similarly. For the running times shown in Figure \ref{subfig:phi_running_time_real}, when $\phi$ increases, the running times of \textsf{G-llep} and \textsf{GT-hgr} increase dramatically. Because the numbers of valid worker-and-task pairs and the correct matches increase quickly when the numbers of workers and tasks increase due to the increase of length of time slots.

We also conducted experiments on the real dataset with varied  workers' reliabilities and workers' velocities.  In addition, we also tested the algorithms when the working areas of workers are circles whose diameters are configured with a Gaussian distribution within range $[a^-, a^+]$. For details, please refer to Appendix B.

\subsubsection{Experimental Results on Synthetic Data}
\label{subsec:exp_synthetic}

In this subsection, we show the performances of tested approaches on synthetic dataset by varying the number of tasks  $m$, and the number of workers  $n$ when the locations of workers/tasks both follow Uniform (UNIF) distribution.  Due to the space limitation, we put the results about the effects of the location distributions of workers/tasks in Appendix C.

\noindent {\bf Effect of the number of tasks, $m$.}
Figure \ref{fig:effect_task_number_synthetic} shows the effect of the number $m$ of spatial 
tasks on the performances of tested approaches, where we vary $m$ from 7.5K to 17.5K.

In Figure \ref{subfig:task_number_avg_moving_distance_unif}, the assigned workers of all the tested approaches will have higher average moving distances for larger $m$. The reason is that the approaches select tasks in perspectives different from proximity of tasks. When the number of tasks per time slot increases, they may select the most suitable tasks located further. In addition, as the \textsf{GT-hgr} assigns much fewer workers, the average moving distance of the results achieved by it is small.
In Figure \ref{subfig:task_number_finished_task_number_unif}, except for \textsf{GT-hgr}, NFTs of the results achieved by all the tested algorithms will decrease when the number of tasks per time slot increases. Except for \textsf{GT-hgr,} other tested algorithms do not concern the minimum required number of answers by each task. When the number of tasks increases, the workers are distributed to more tasks and the average number of workers for tasks will decrease, which leads to NCFs decrease. However, \textsf{GT-hgr} only assigns correct matches, which can guarantee that each assigned task will have a set of workers to satisfy the required number of workers. Meanwhile, when the number of tasks increases, \textsf{GT-hgr} will produce more correct matches as more suitable tasks are available to be selected such that NFT of \textsf{GT-hgr} increases.
 In addition, we notice that although some tasks in the results of \textsf{G-llep} and \textsf{RDB-sam} algorithms are assigned with required number of workers, their expected accuracy values may be not satisfied (their aggregation reputation scores may be smaller than their required quality levels). The reason is that \textsf{G-llep} and \textsf{RDB-sam} algorithms only assign as many worker-and-task pairs as possible without considering the required quality level.
Similar results can be observed in Figure \ref{subfig:task_number_finished_task_number_conf_unif} due to the same reason. In Figure \ref{subfig:task_number_running_time_unif}, when each time slot has more tasks, the running time of all the tested algorithms increases slightly, as more tasks need to be checked and maintained. \textsf{DP} runs much slower than other online algorithms.

\noindent {\bf Effect of the Number of Workers, $n$.}
Figure \ref{fig:effect_worker_number_synthetic} shows the effect of the number $n$ of spatial 
workers on the performances of tested approaches, where we vary $n$ from 7.5K to 17.5K. 

In Figure \ref{subfig:worker_number_avg_moving_distance_unif}, when the number of workers in each time slot increases, the average moving distance of workers in the results achieved by all the tested algorithms also increases. The reason is that when there are more workers in each time slot,
 the working areas of workers can cover more tasks, to conduct more tasks the AvgMDs will increase. In addition, we find that \textsf{G-llep} requires workers to move more to conduct tasks. The reason is \textsf{G-llep} gives higher priorities to the tasks created at locations with fewer workers, then more far tasks are assigned to workers.
As shown in Figure \ref{subfig:worker_number_finished_task_number_unif}, when $m$ increases, all the tested approaches can complete more tasks. Online algorithms can complete fewer tasks than batch-based algorithms. \textsf{G-llep} algorithms can complete more tasks than other algorithms. The reason is compared to other algorithms, \textsf{G-llep} can finish more far tasks as explained above. \textsf{RDB-d\&c} can finish many tasks but still slightly fewer than that of {\sf G-llep}. 
In Figure \ref{subfig:worker_number_finished_task_number_conf_unif}, all the approaches can achieve results with higher NCTs when there are more workers available in each time slot. Moreover, the increasing rate of NCT of \textsf{GT-hgr} is faster than other approaches. As \textsf{GT-hgr} only assigns correct matches, when more workers are available, the number of correct matches will increase exponentially leading to that NCT of \textsf{GT-hgr} increases similarly.
In Figure \ref{subfig:worker_number_running_time_unif},  the running time of \textsf{G-llep} algorithms increases obviously when the number of workers per time slot increases, as the complexity of maximum flow algorithm increases linearly with respect to the number of edges of the graph, which increases super-linearly w.r.t $n$. \textsf{BB} and \textsf{HA} are faster than other algorithms, as BB can quickly assign enough tasks for each worker and \textsf{HA} just assigns tasks based on very simple heuristics (e.g., selecting next nearest neighbor). \textsf{G-llep} is slower than other five algorithms.

\subsection{Summary}
\label{sec:summary}

With the experimental studies, we summarized one grade table that describes the pros and cons of each algorithms under different metrics. Specifically, for a set of experimental results, we grade the performance of algorithm $\Psi_j$ on metric $M_i$ with  Equation \ref{eq:grade} as follows:\vspace{-3ex}
{\small \begin{equation}
	G(M_i, \Psi_j)=\left\{
	\begin{array}{ll}
	5 \cdot \frac{V_{ij} - L_i}{U_i - L_i} , & M_i \in \{NFT, NCT\} \\
	5\cdot (1 - \frac{V_{ij} - L_i}{U_i - L_i}), & M_i \in \{AvgMD, RT\}
	\end{array}
	\right. \label{eq:grade}
	\end{equation}
}\vspace{-2ex}

\noindent where $V_{ij}$ is the result value of algorithm $\Psi_j$ on metric $M_i$, and $U_i$ and $L_i$ are the upper and lower values among all the tested algorithms on metric $M_i$, respectively. For example, in Figure \ref{subfig:c_avg_moving_distance_real_all}, when $[c^-, c^+] =[2,3]$, AvgMD of \textsf{GT-hgr} is the lowest, then the grade of \textsf{GT-hgr} on AvgMD is 5 in this set of experiments. Then, in Table \ref{tab:algorithms_summary_synthetic}, we report the average grades of each tested algorithms for the four metrics.
Therefore, users can find a good option given a \textsf{TA-GSC} application. 

Another important issue is about location privacy of workers. In batch-based mode, spatial crowdsourcing systems need to trace the location of workers, which may scare away some potential workers. However, in online mode, workers only need to reveal their locations when they are requesting the available tasks, which is much more acceptable for most workers.

Online algorithms usually have good efficiency, which means the spatial crowdsourcing systems can response to the worker requests quickly, which leads to a better user experience than batch-based mode. However, systems in batch-based mode also can reduce the time interval between two adjacent batches such that they can also response to the worker requests quickly.

\begin{table}[t!]\centering\vspace{-3ex}
	\caption{\small Grades of algorithms for different metrics on our data sets. The grade varies from zero to five, and a higher grade indicates that the algorithm is better at the corresponding metric. “B” and “O” stand for batch-based mode and online mode, respectively. }\label{tab:algorithms_summary_synthetic}\vspace{1ex}
		\begin{tabular}{l|c|c|c|c|c}
			{\bf Algorithm} &{\bf Mode}& {\bf  AvgMD} & \textbf{NFT} & \textbf{NCT} &\textbf{RT}\\
			\hline \hline
			\textsf{G-greedy} & B& 1.6 & 3.6 & 3.1 & 0.5  \\
			\textsf{G-llep} & B& 1.4 & 5.0 & 4.5 & 0.7 \\
			\textsf{G-nnp} & B& 4.1 & 3.8 & 3.3 & 0.0 \\
			\textsf{GT-greedy} & B&3.1 & 2.3 & 4.8 & 5.0 \\
			\textsf{GT-hgr} & B& 5.0 & 2.5 & 5.0 & 4.8 \\
			\textsf{RDB-d\&c} & B& 1.6 & 2.6 & 2.2 & 4.9 \\
			\textsf{RDB-sam} & B& 1.7 & 3.2 & 2.7 & 5.0 \\
			\textsf{DP} & O & 1.2 & 3.3 & 2.9 & 5.0 \\
			\textsf{BB} & O & 2.6 & 1.9 & 1.6 & 5.0 \\
			\textsf{HA} & O & 2.5 & 2.8 & 2.5 & 5.0 \\
			\textsf{PRS} & O &0.4 & 0.0 & 0.0 & 4.9\\
			\hline
		\end{tabular}
	\end{table}

We provide the following high-level suggestions for choosing algorithms for \textsf{TA-GSC} applications.
\begin{enumerate}[leftmargin=*]
	\item When the expected accuracy of tasks is important for the platforms (e.g., mobile audit services, such as Field Agent \cite{fieldagent}), \textsf{GT-greedy} or \textsf{GT-hgr} should be selected, as they can guarantee the quality of tasks, especially for  applications for tasks with high required quality levels and workers with low reliabilities. On the contrast, when the required quality levels of tasks are low, the reliabilities of workers are high or the required numbers of answers are high, the expected accuracy of tasks will be high, then  there is no need to particularly care the expected quality of tasks. As a result, \textsf{G-llep}, \textsf{RDB-sam} and  \textsf{DP} are good choices.  \vspace{-1ex}	
	\item When the travel costs of workers is the key measure for the platforms (e.g., Uber \cite{uber} and DiDi Chuxing \cite{didi}), \textsf{GT-hgr} and \textsf{G-nnp} should be chosen. Moreover, \textsf{GT-hgr} can also guarantee the quality of tasks.\vspace{-1ex}	
	\item When the responding speed for the workers is the key issue for the platforms (e.g., car-hailing platforms, such as Uber \cite{uber}), the maximum flow based algorithms, such as \textsf{G-greedy}, \textsf{G-llep} and \textsf{G-nnp}, should be avoided due to their high running time.
	
\end{enumerate}

\section{Conclusion}
\label{sec:conclusion}

In this paper, we present a comprehensive experimental comparison 
of most existing algorithms on task assignment in spatial crowdsourcing. 
Specifically, we first give some general definitions about spatial workers 
and spatial tasks based on definitions in the existing works studying task 
assignment problems in spatial crowdsourcing such that the existing 
algorithms can be applied on the same synthetic and real data sets. We 
uniformly implement tested algorithms in both batch-based and online modes. 
With the experimental results of the tested algorithms on synthetic 
and real datasets, we show the effectiveness and efficiencies of the 
algorithms through their performances on five important metrics.  According to the experimental results, we summarize the performance of tested algorithms on synthetic and real data sets through grading, which can guide users on selecting algorithms for real applications under different situations.

\bibliographystyle{abbrv}
\bibliography{add}

\appendix
\begin{figure*}[t!]\centering
	\subfigure[][{\tiny GAUS ($\mu=0.1, \sigma^2=0.05$)}]{
		\scalebox{0.165}[0.165]{\includegraphics{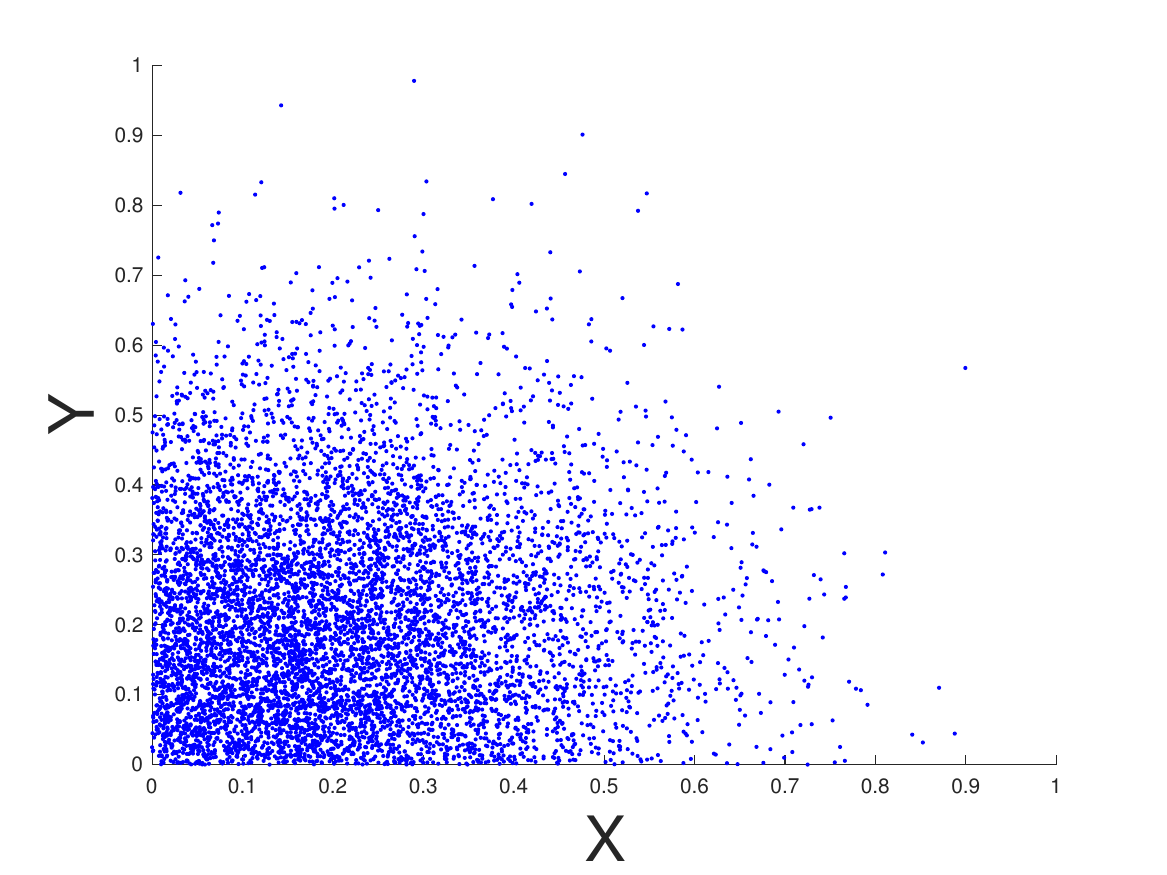}}
		\label{subfig:gaus_mean_0.1}}\hfill
	\subfigure[][{\tiny GAUS ($\mu=0.3, \sigma^2=0.05$)}]{
		\scalebox{0.165}[0.165]{\includegraphics{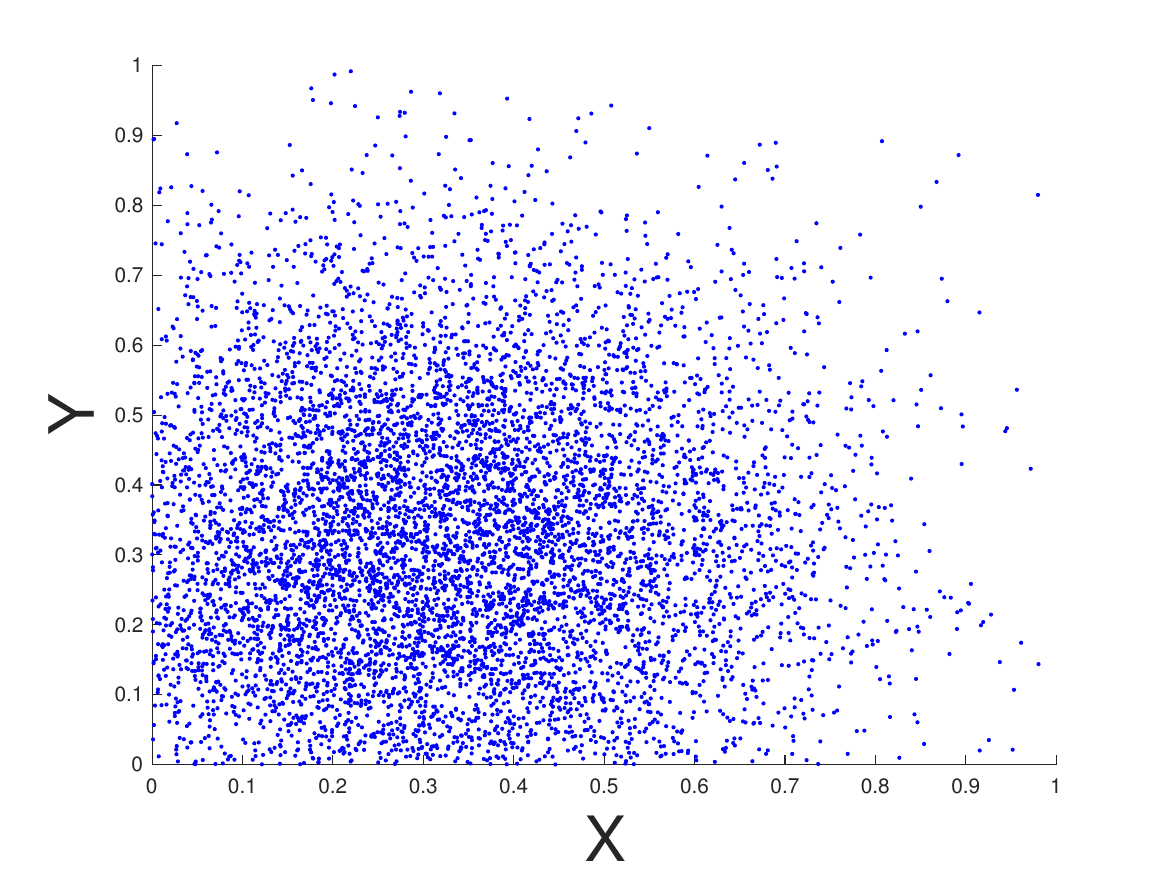}}
		\label{subfig:gaus_mean_0.3}}\hfill
	\subfigure[][{\tiny GAUS ($\mu=0.5, \sigma^2=0.05$)}]{
		\scalebox{0.165}[0.165]{\includegraphics{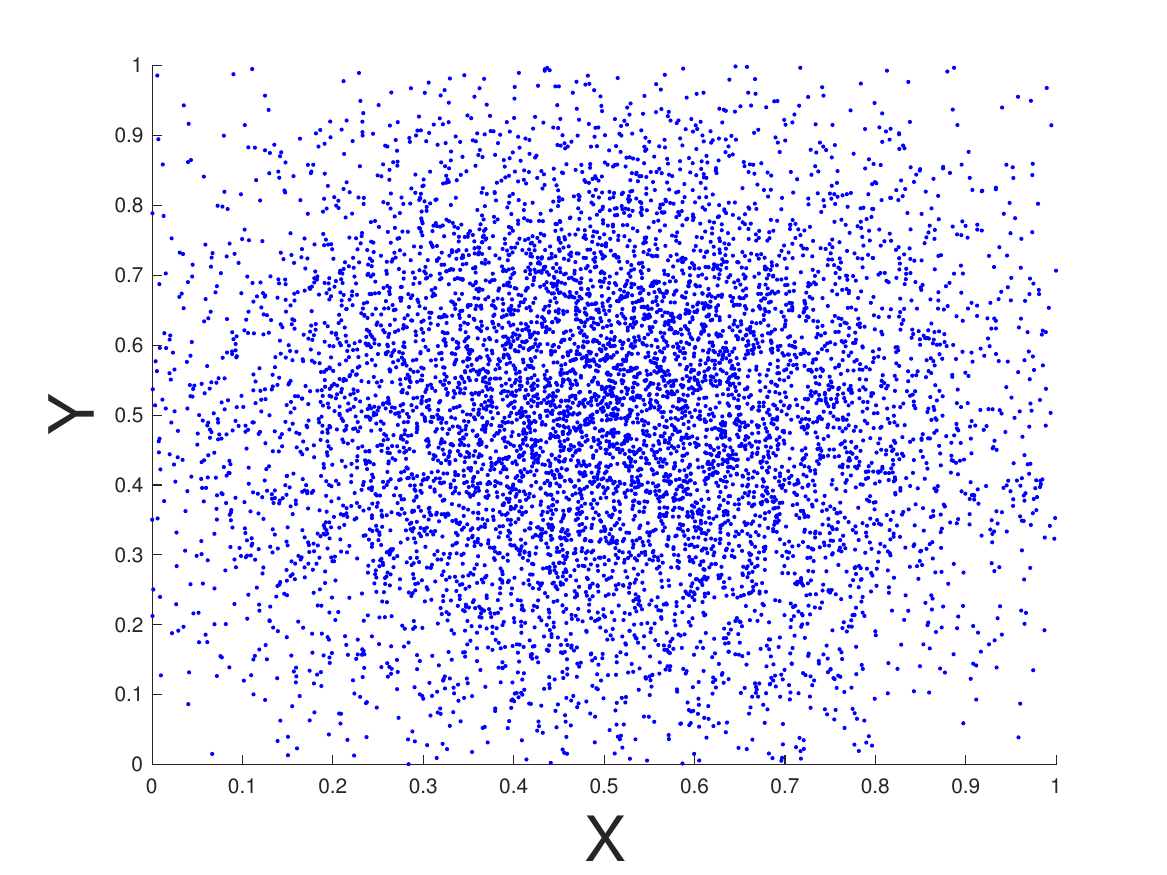}}
		\label{subfig:gaus_mean_0.5}}\hfill
	\subfigure[][{\tiny GAUS ($\mu=0.7, \sigma^2=0.05$)}]{
		\scalebox{0.165}[0.165]{\includegraphics{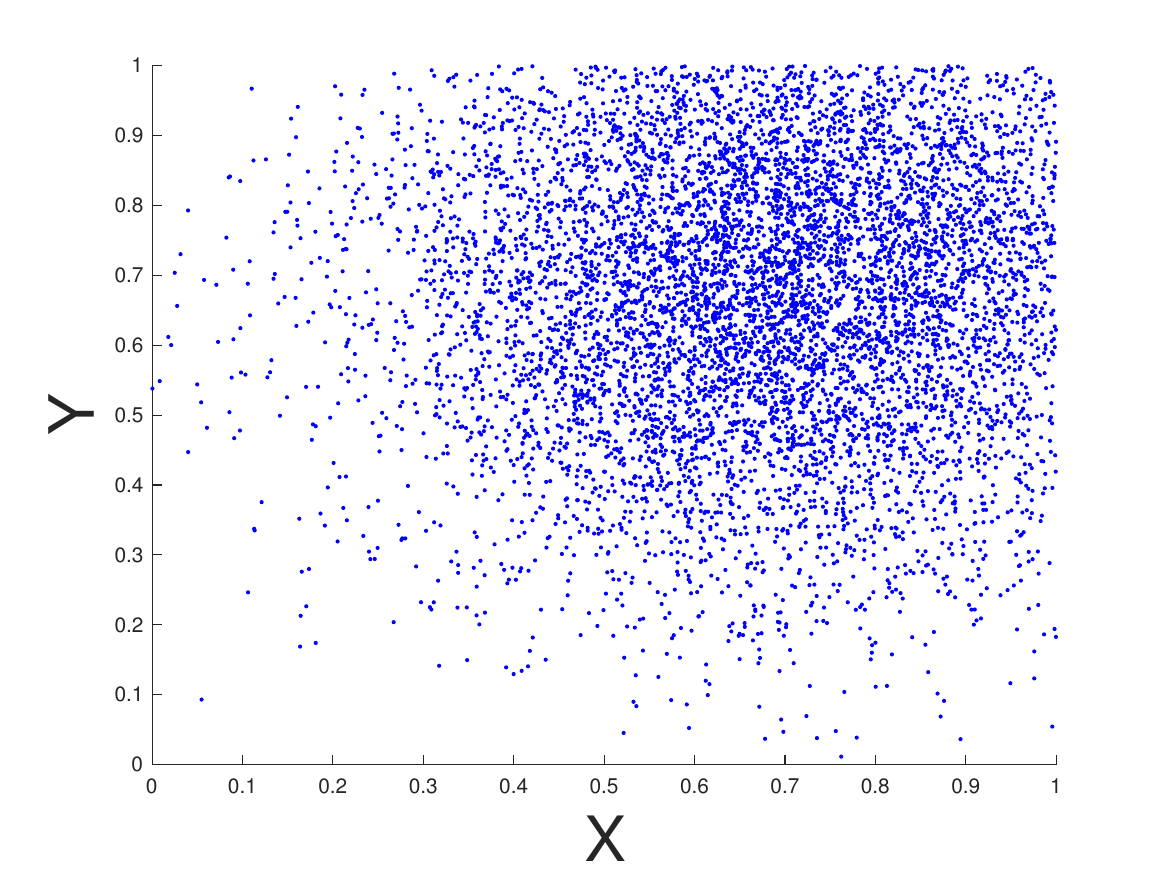}}
		\label{subfig:gaus_mean_0.7}}\hfill
	\subfigure[][{\tiny GAUS ($\mu=0.9, \sigma^2=0.05$)}]{
		\scalebox{0.165}[0.165]{\includegraphics{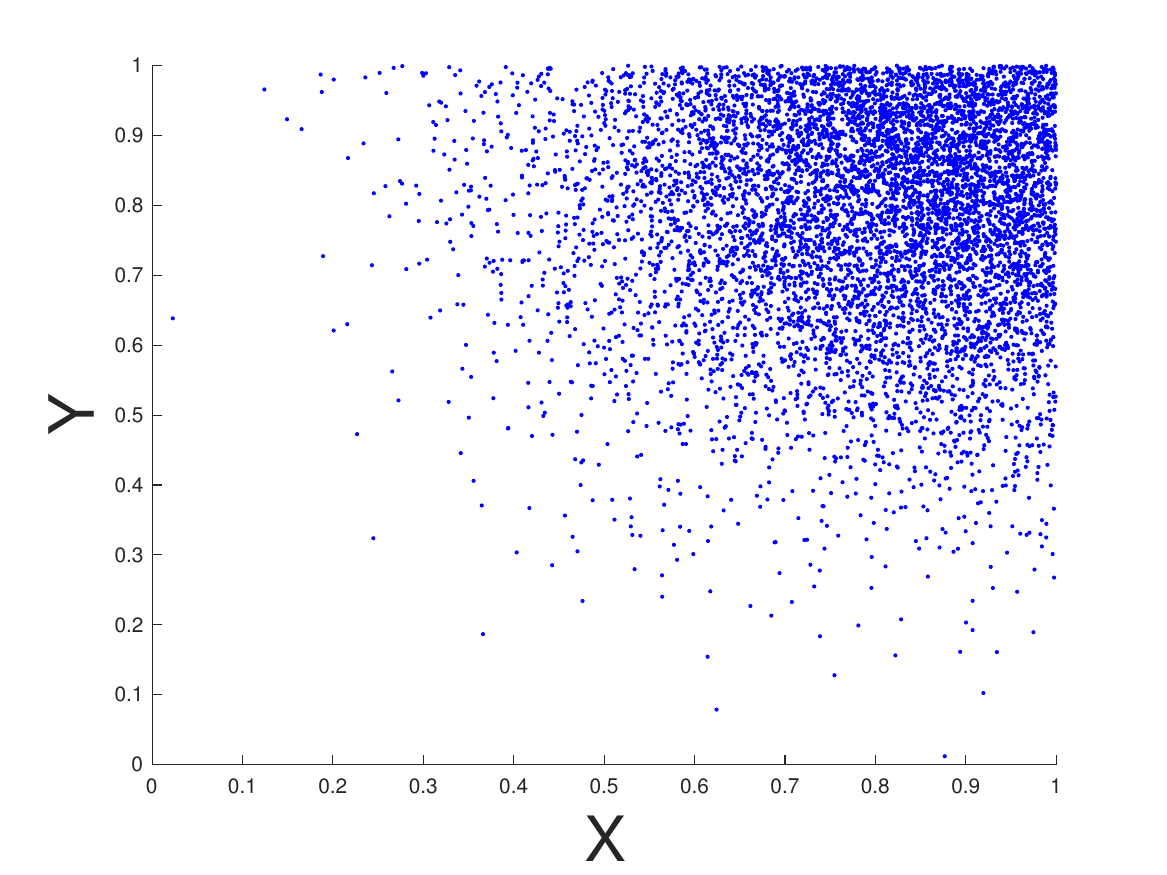}}
		\label{subfig:gaus_mean_0.9}}\hfill
	\subfigure[][{\tiny GAUS ($\mu=0.5, \sigma^2=0.01$)}]{
		\scalebox{0.165}[0.165]{\includegraphics{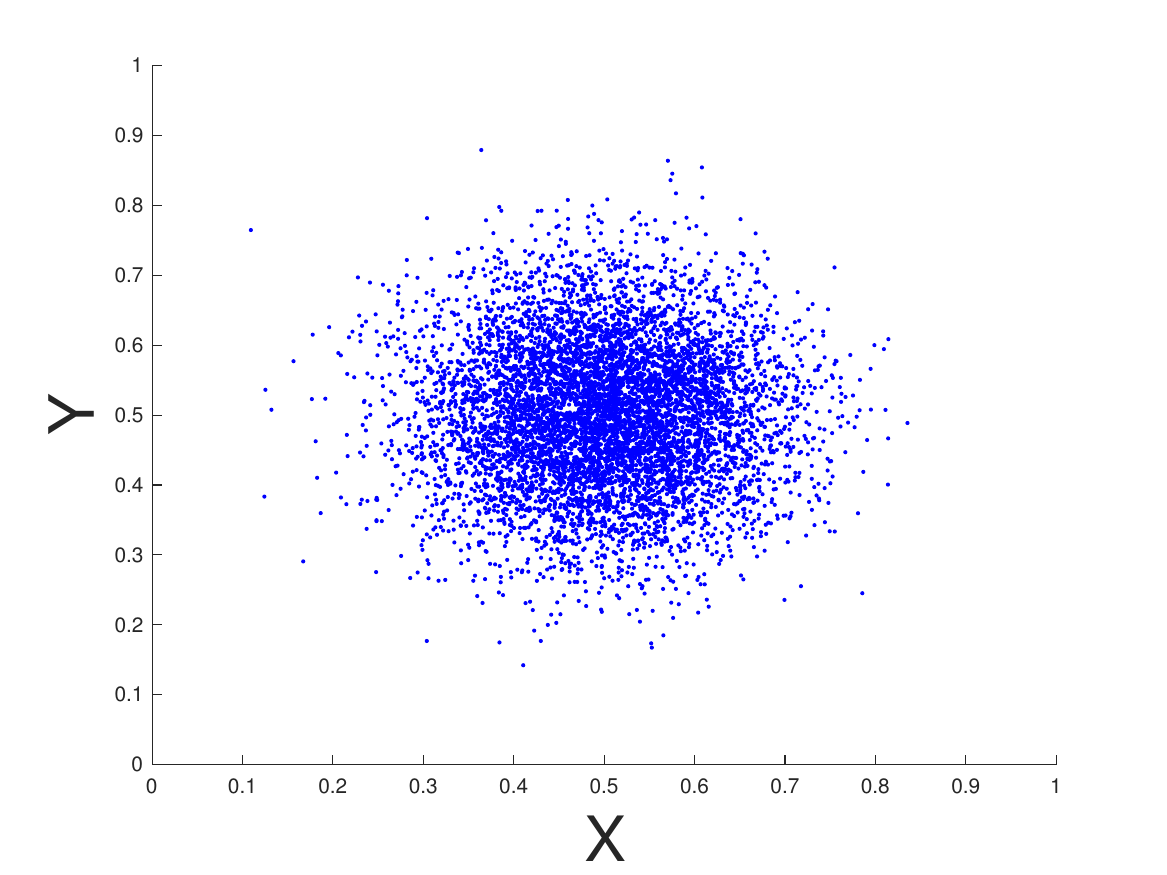}}
		\label{subfig:gaus_variance_0.01}}\hfill
	\subfigure[][{\tiny GAUS ($\mu=0.5, \sigma^2=0.03$)}]{
		\scalebox{0.165}[0.165]{\includegraphics{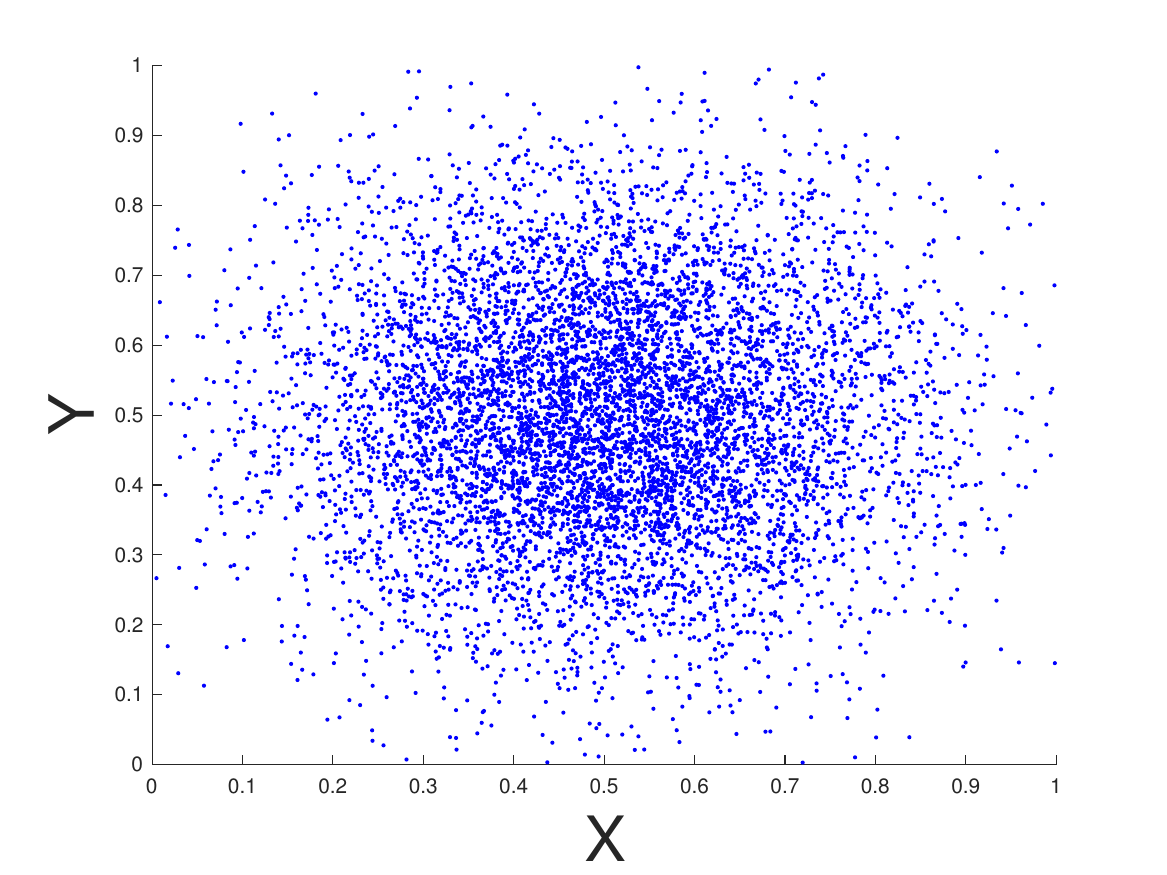}}
		\label{subfig:gaus_variance_0.03}}\hfill
	\subfigure[][{\tiny GAUS ($\mu=0.5, \sigma^2=0.05$)}]{
		\scalebox{0.165}[0.165]{\includegraphics{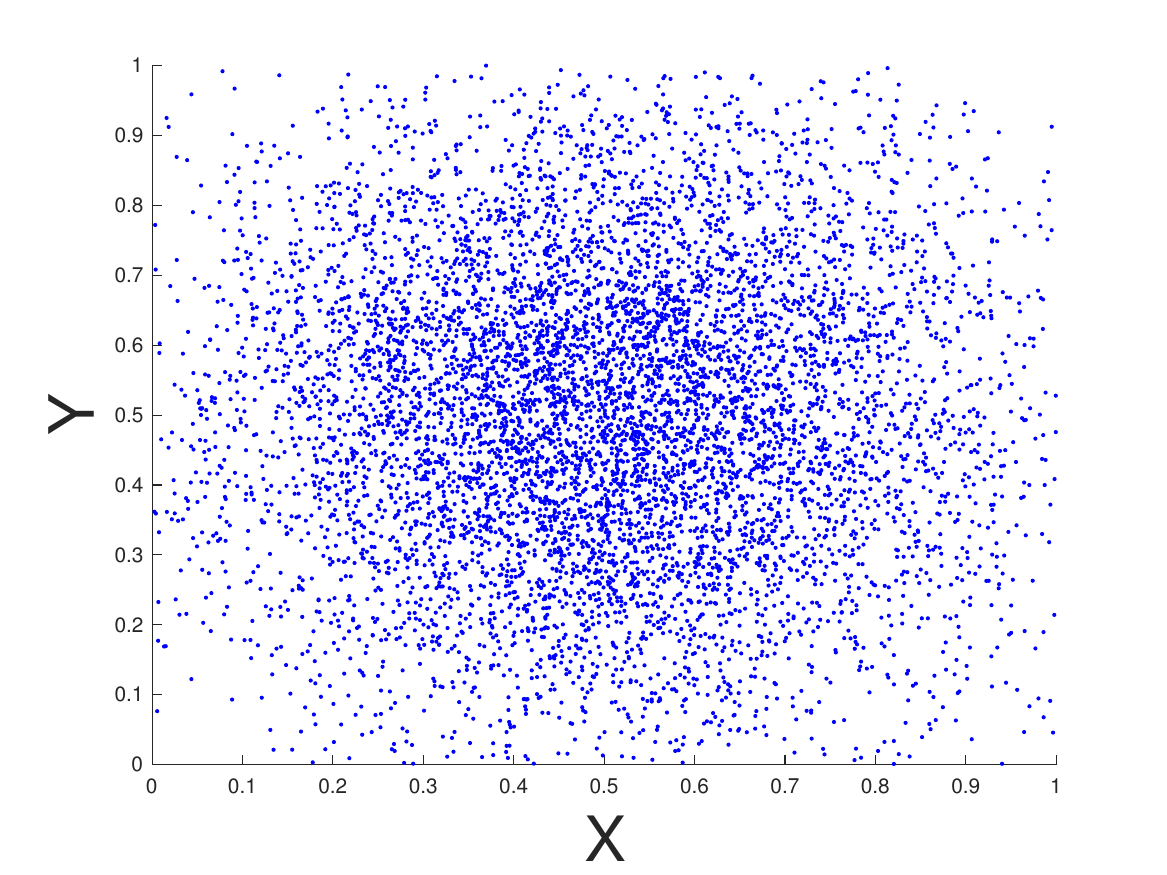}}
		\label{subfig:gaus_variance_0.05}}\hfill
	\subfigure[][{\tiny GAUS ($\mu=0.5, \sigma^2=0.07$)}]{
		\scalebox{0.165}[0.165]{\includegraphics{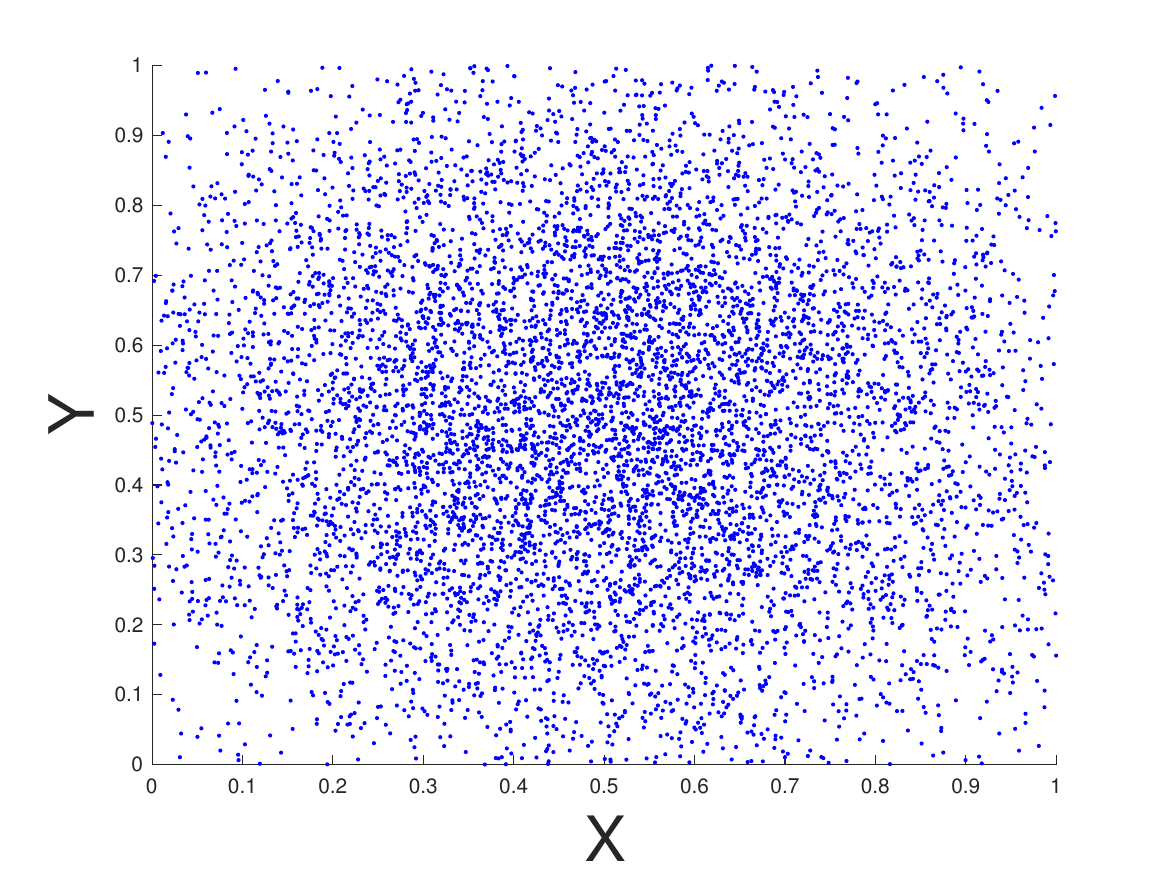}}
		\label{subfig:gaus_variance_0.07}}\hfill
	\subfigure[][{\tiny GAUS ($\mu=0.5, \sigma^2=0.1$)}]{
		\scalebox{0.165}[0.165]{\includegraphics{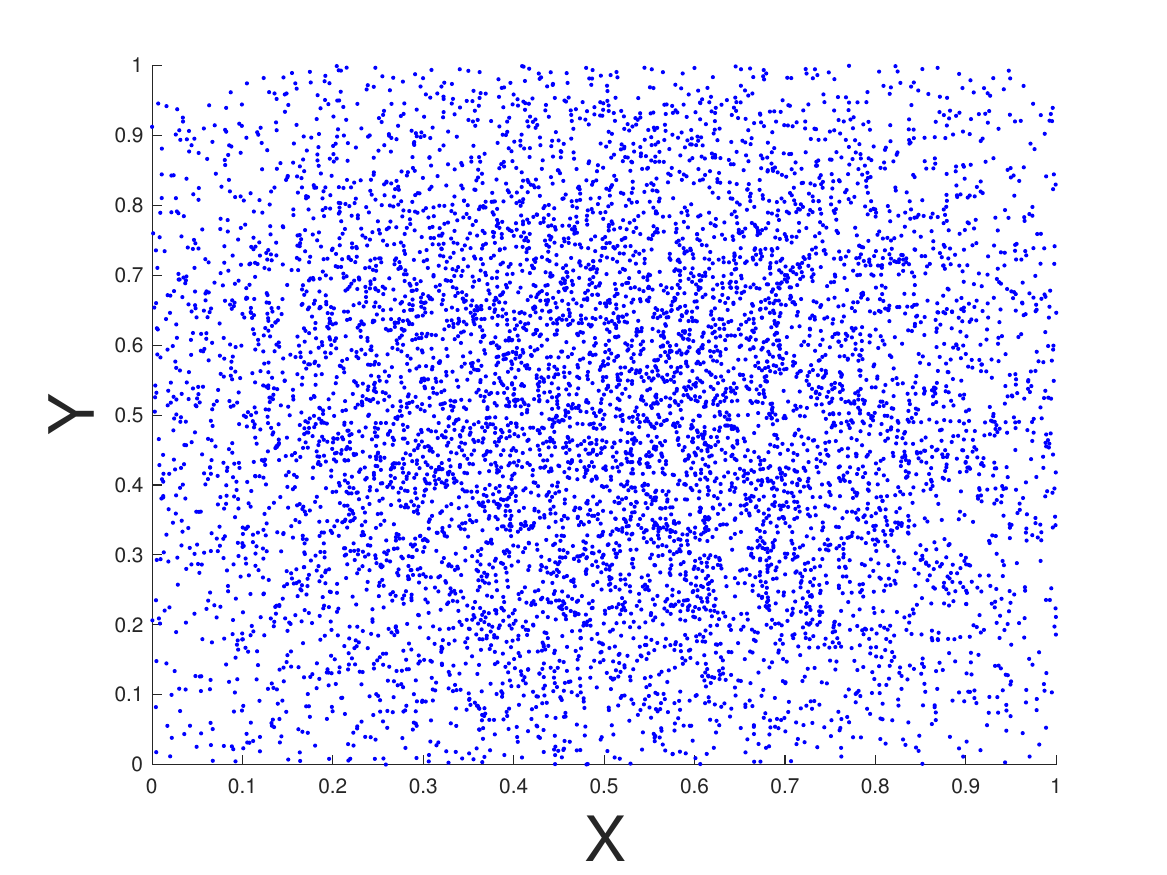}}
		\label{subfig:gaus_variance_0.1}}\hfill
	\subfigure[][{\tiny SKEW ($\Lambda=1$)}]{
		\scalebox{0.165}[0.165]{\includegraphics{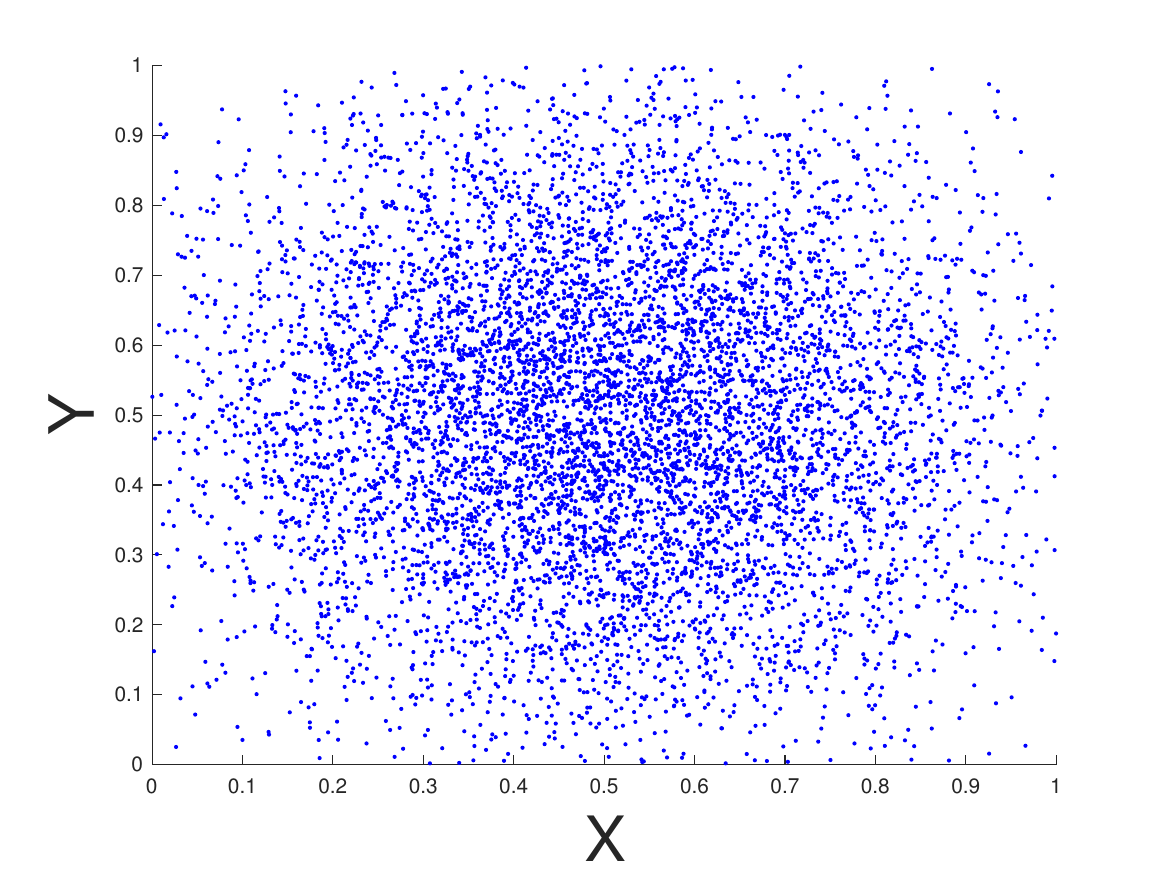}}
		\label{subfig:skew_center_1}}\hfill
	\subfigure[][{\tiny SKEW ($\Lambda=3$)}]{
		\scalebox{0.165}[0.165]{\includegraphics{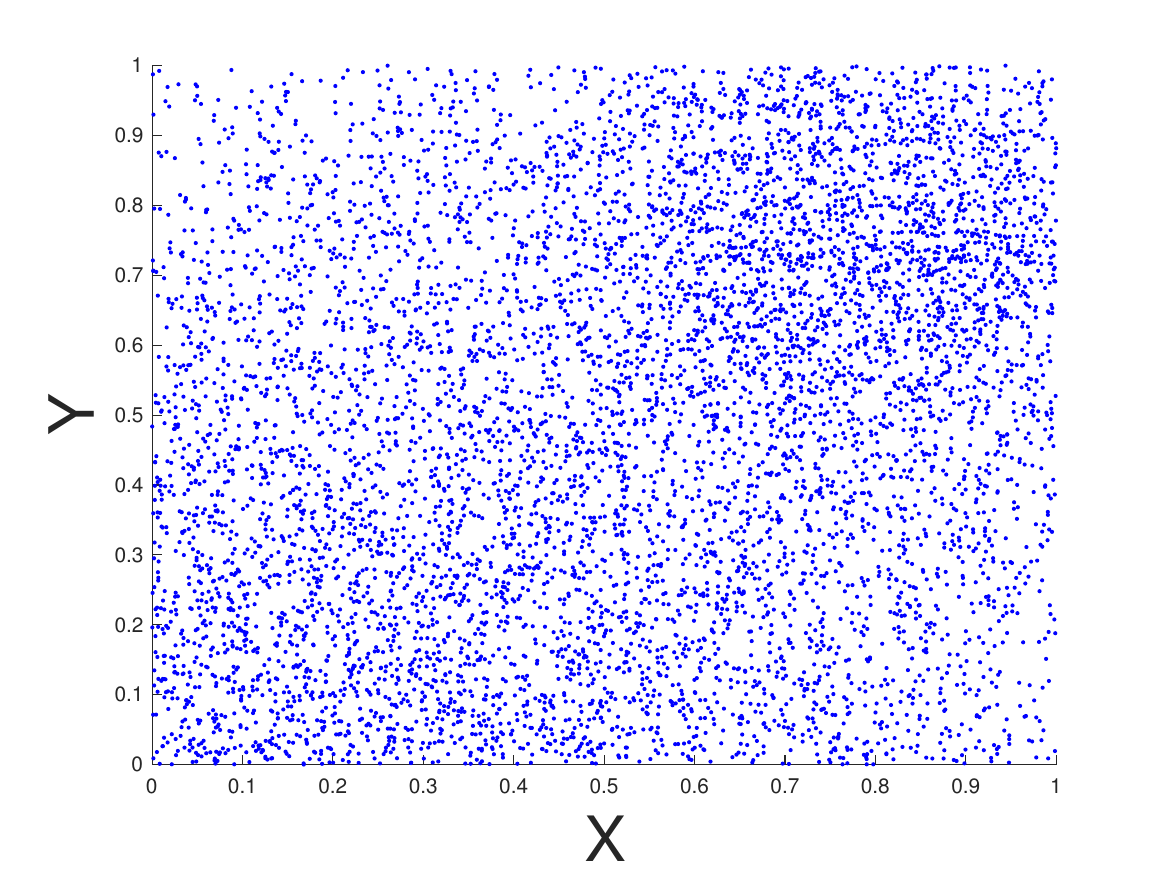}}
		\label{subfig:skew_center_3}}\hfill
	\subfigure[][{\tiny SKEW ($\Lambda=5$)}]{
		\scalebox{0.165}[0.165]{\includegraphics{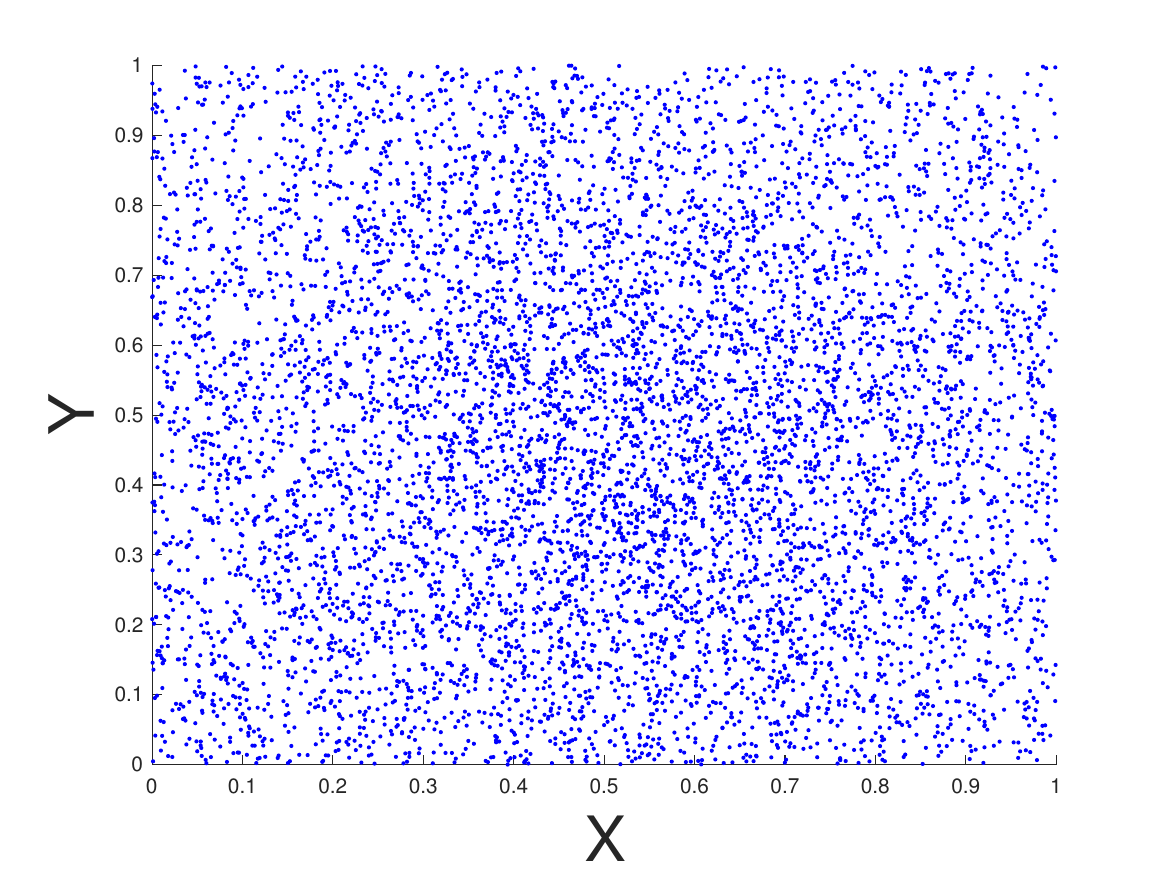}}
		\label{subfig:skew_center_5}}\hfill
	\subfigure[][{\tiny SKEW ($\Lambda=7$)}]{
		\scalebox{0.165}[0.165]{\includegraphics{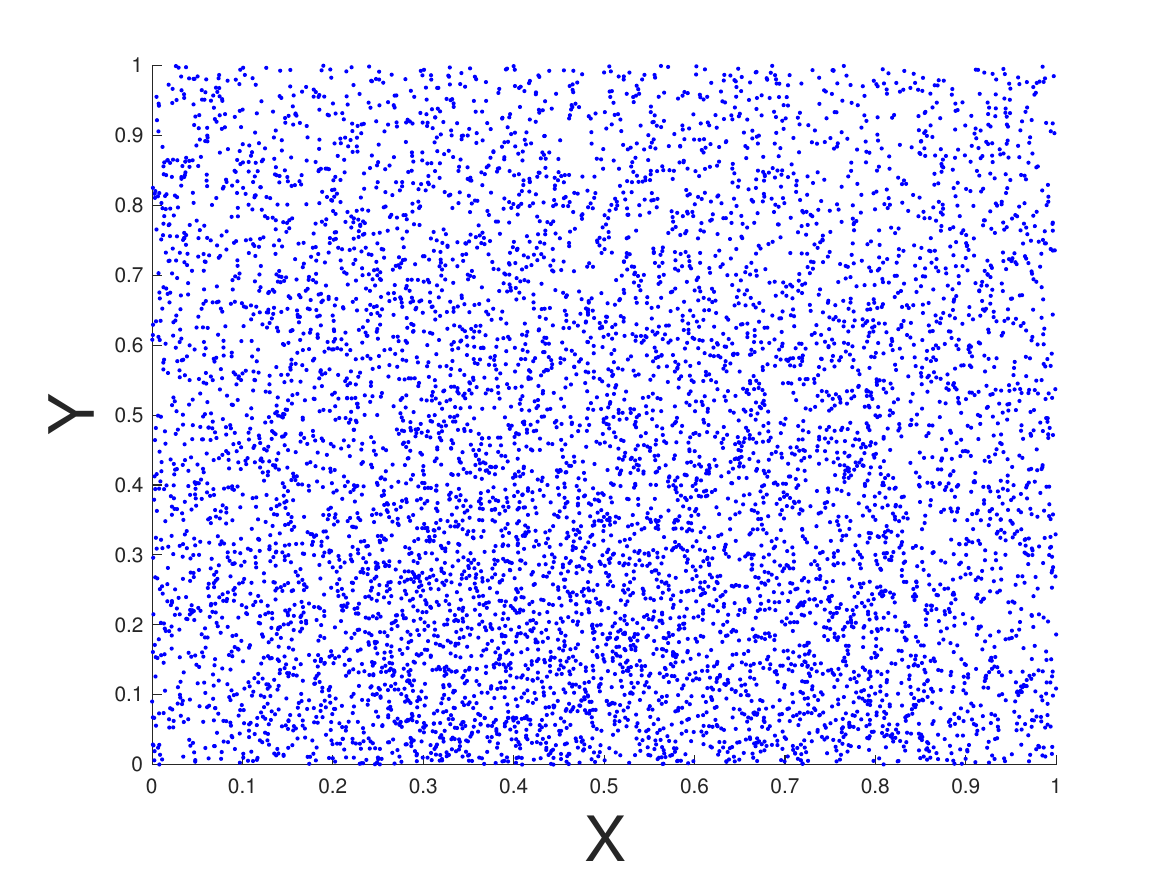}}
		\label{subfig:skew_center_7}}\hfill
	\subfigure[][{\tiny UNIF}]{
		\scalebox{0.165}[0.165]{\includegraphics{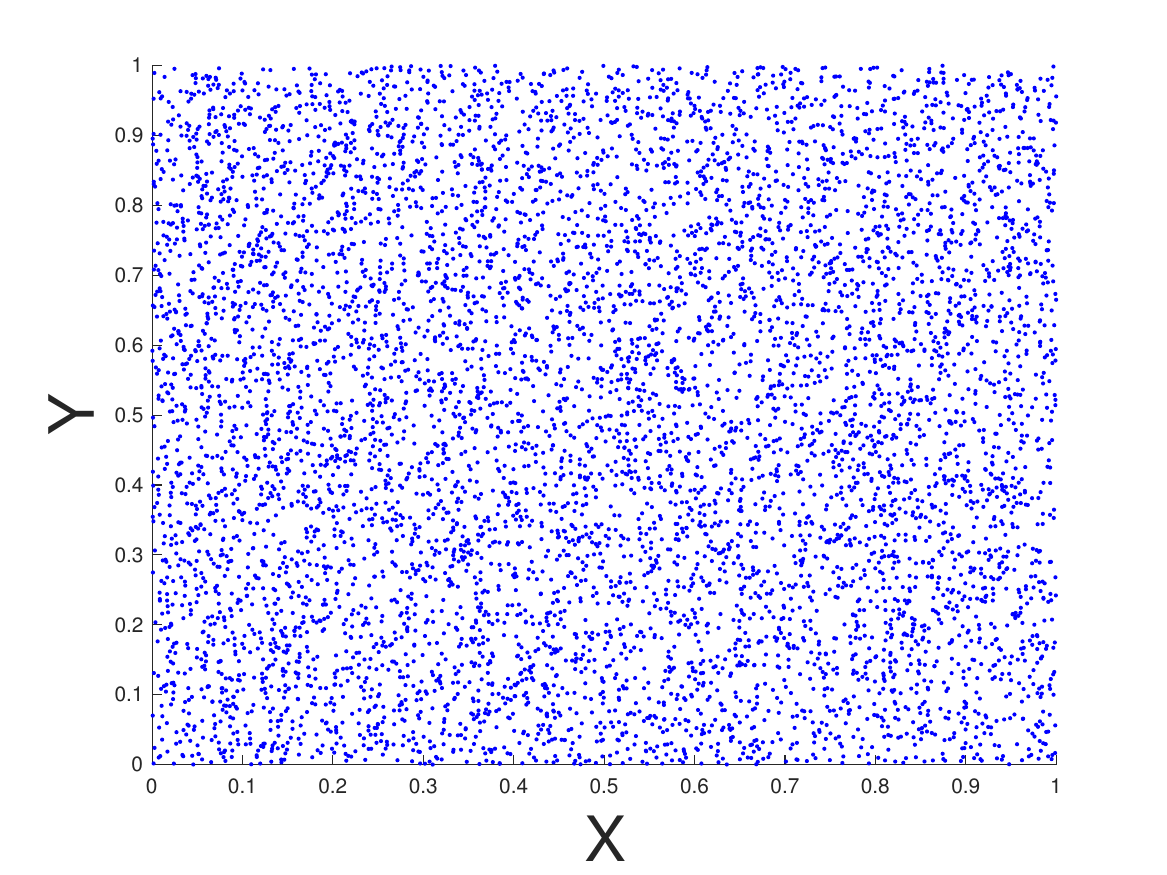}}
		\label{subfig:beijing_distributions}}\hfill
	\caption{\small Illustrations of GAUS, SKEW and UNIF location distributions.}
	\label{fig:distributon_illustration}\vspace{-2ex}
\end{figure*}

\section{Illustrations of Distributions}

We present the illustrations of different distributions used in generating the synthetic data as shown in Figure \ref{fig:distributon_illustration} to show the visual and high level patterns. For the first row, when the center of the GAUS cluster is more close to point ($0.5$, $0.5$), the location points are more evenly distributed as the effect of constraint of the spatial space is smaller. For the second row, when the variance of GAUS increases, the location points are more evenly distributed. For the third row, when the number, $\Lambda$, of Gaussian distributed clusters in the skewed distribution increases, the location points are more uniformly distributed.

\section{Effect of workers' reliabilities, velocities and circle working areas}

\begin{figure*}[t!]\centering
	\subfigure{
		\scalebox{0.4}[0.4]{\includegraphics{bar_mix-eps-converted-to.pdf}}}\hfill\\\vspace{-2ex}
	\addtocounter{subfigure}{-1}
	\subfigure[][{\small Moving Distance }]{
		\scalebox{0.2}[0.2]{\includegraphics{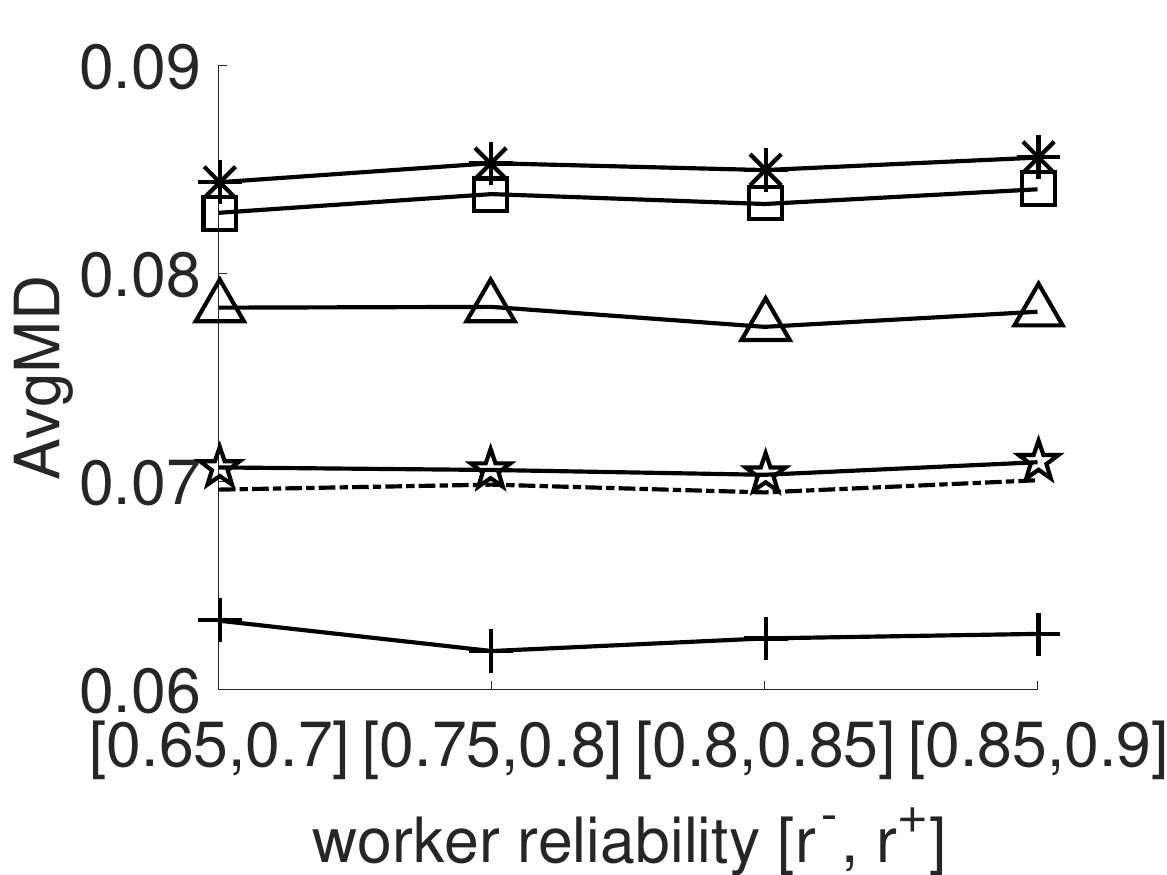}}\vspace{-2ex}
		\label{subfig:r_avg_moving_distance_real}}\hfill
	\subfigure[][{\small Fully Assigned Tasks}]{
		\scalebox{0.2}[0.2]{\includegraphics{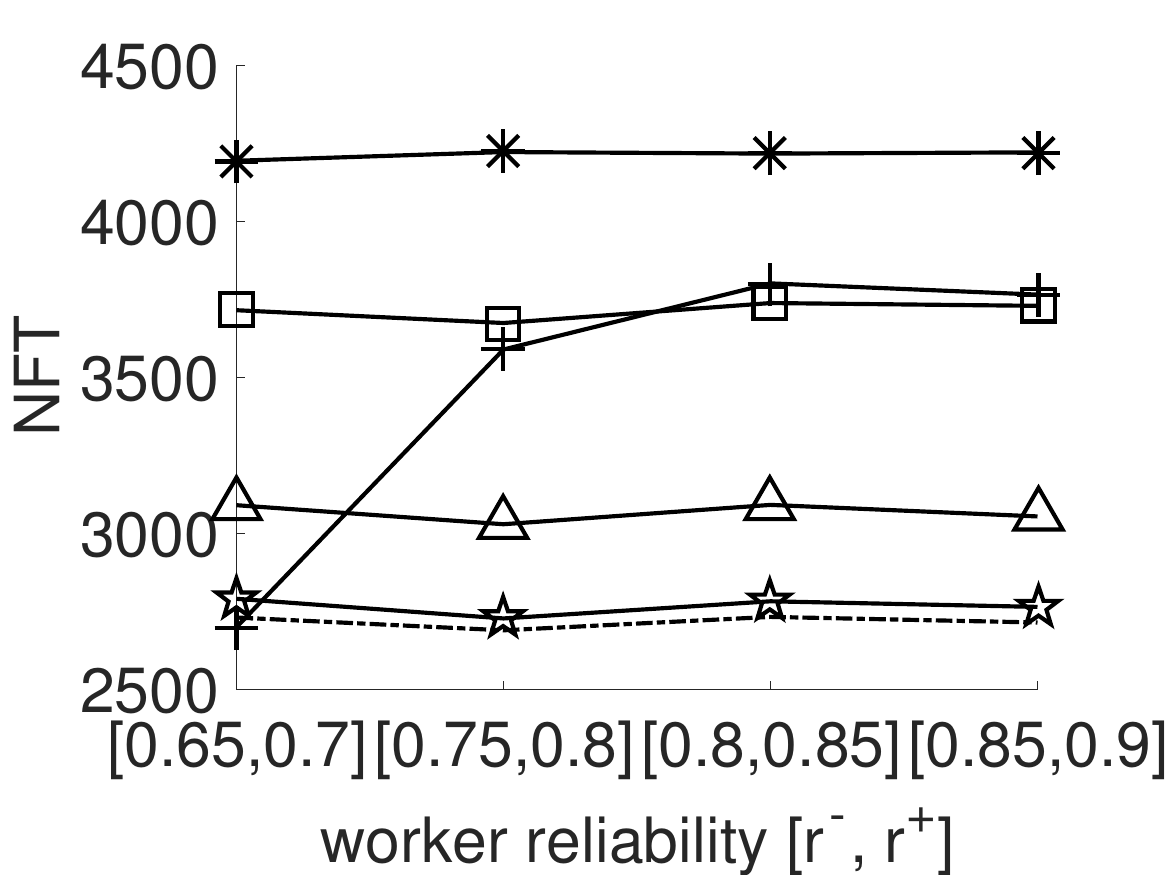}}
		\label{subfig:r_finished_task_number_real}}\hfill
	\subfigure[][{\small Confidently Assigned Tasks}]{
		\scalebox{0.2}[0.2]{\includegraphics{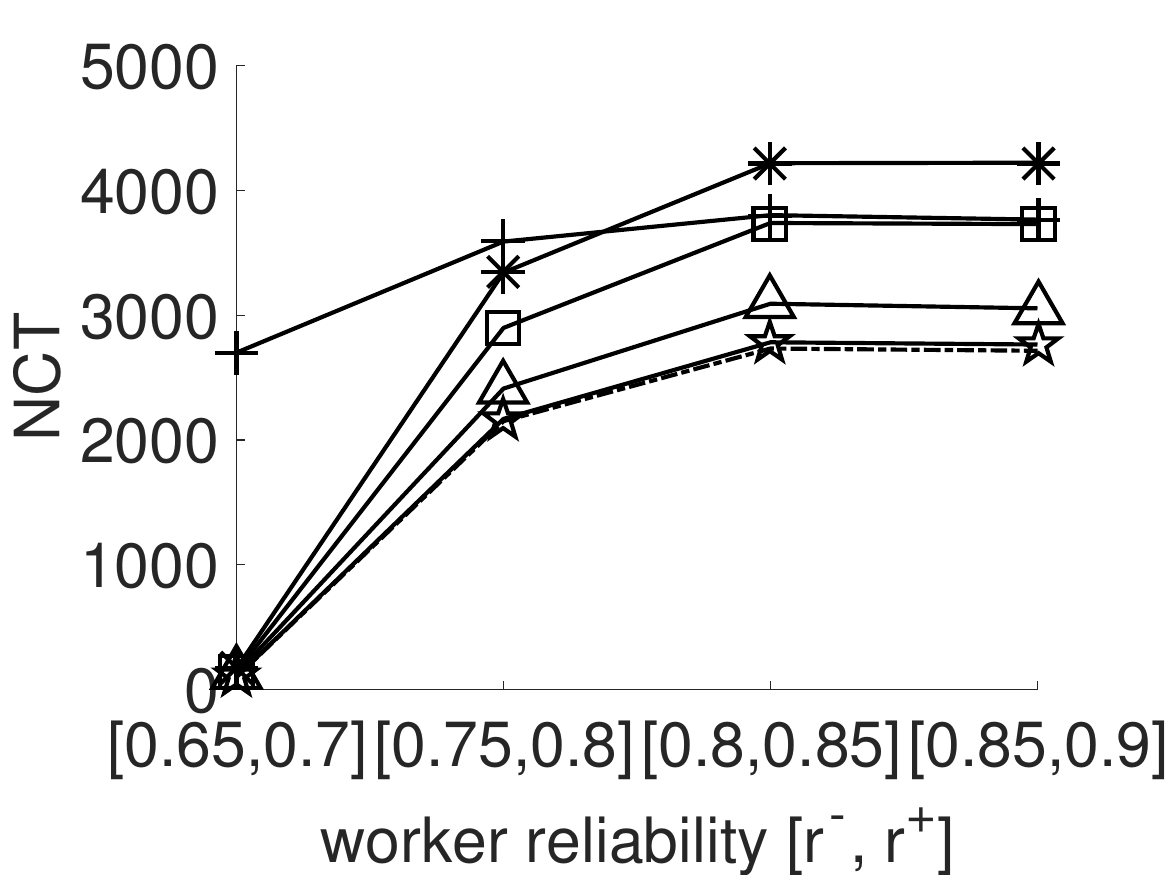}}
		\label{subfig:r_finished_task_number_conf_real}}\hfill\vspace{-2ex}
	\subfigure[][{\small Running Times}]{
		\scalebox{0.2}[0.2]{\includegraphics{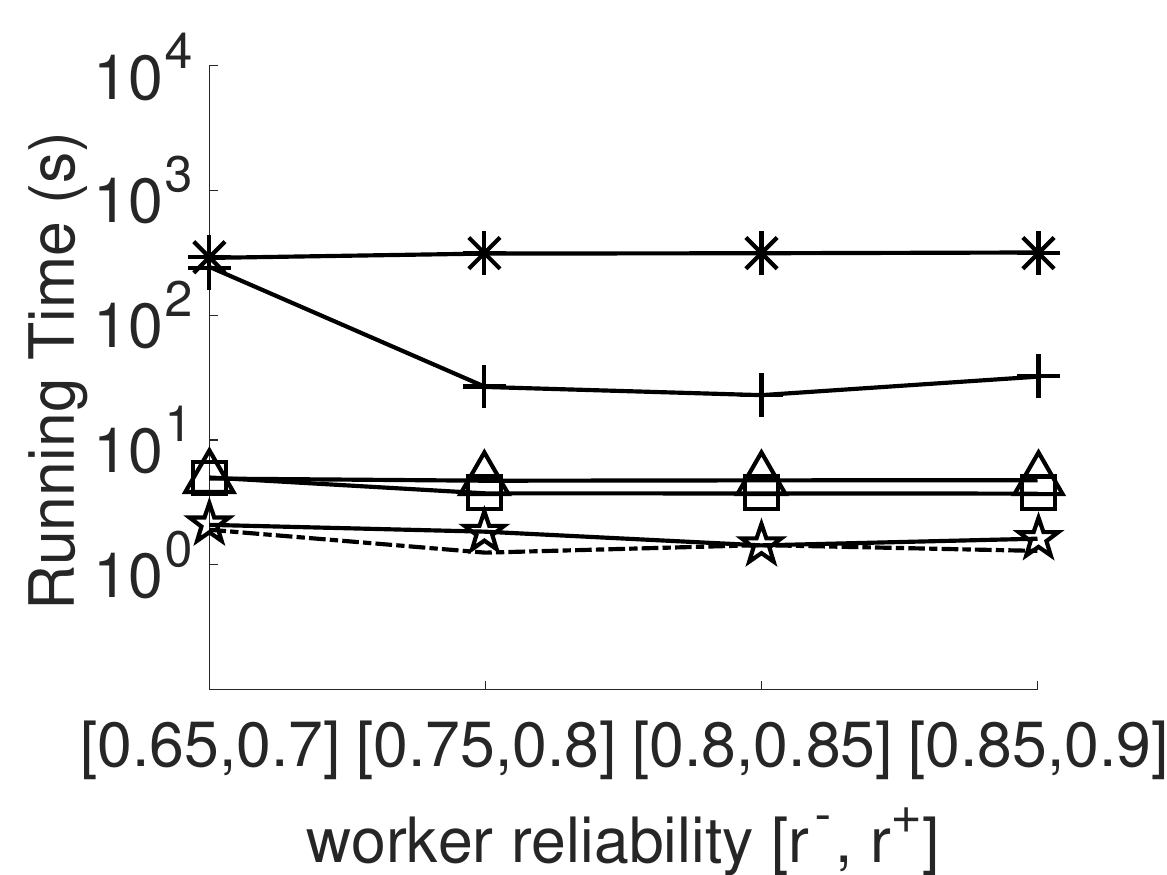}}
		\label{subfig:r_running_time_real}}\hfill\vspace{-2ex}
	\caption{\small Effects of Worker Reliability $r$ (Real).}
	\label{fig:effect_reliability_real}\vspace{-2ex}
\end{figure*}

\begin{figure*}[t!]\centering
	\subfigure{
		\scalebox{0.4}[0.4]{\includegraphics{bar_mix-eps-converted-to.pdf}}}\hfill\\\vspace{-2ex}
	\addtocounter{subfigure}{-1}
	\subfigure[][{\small Moving Distance }]{
		\scalebox{0.2}[0.2]{\includegraphics{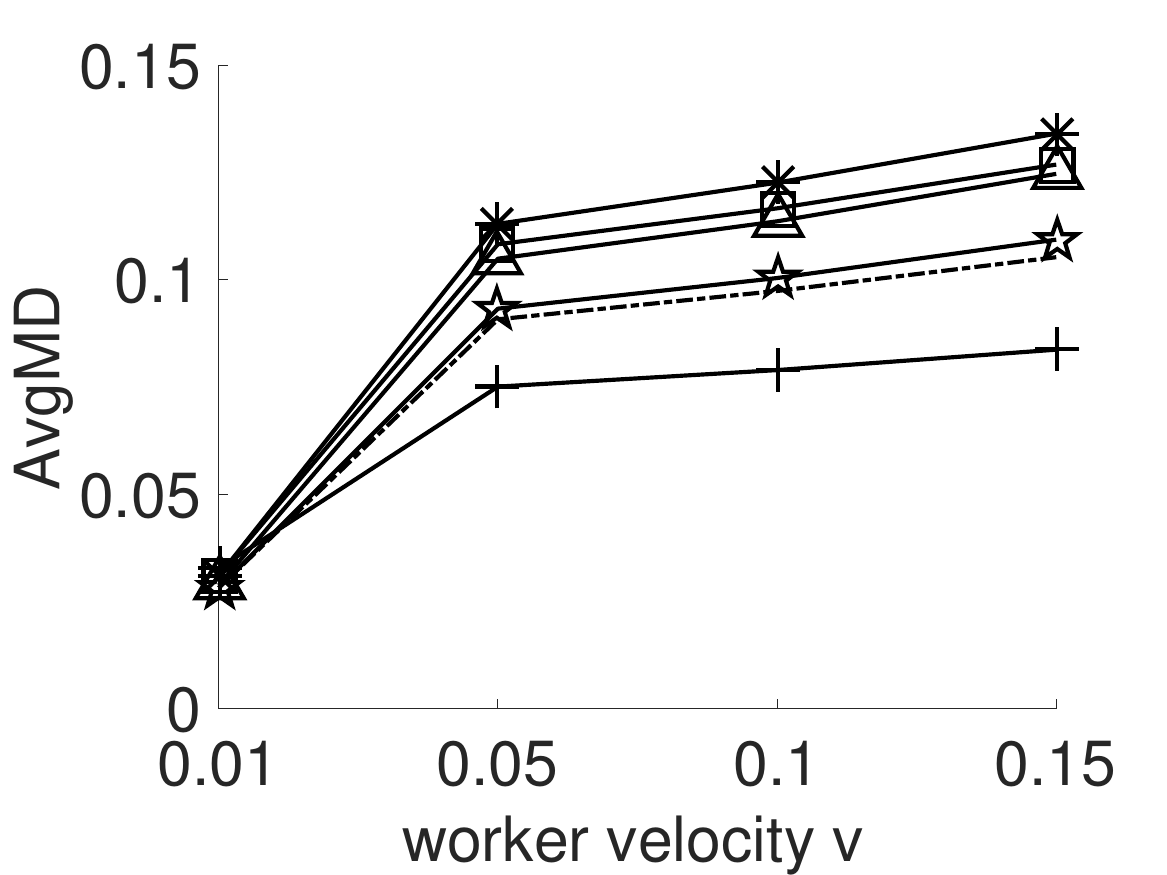}}\vspace{-2ex}
		\label{subfig:v_avg_moving_distance_real_all}}\hfill\vspace{-2ex}
	\subfigure[][{\small Fully Assigned Tasks }]{
		\scalebox{0.2}[0.2]{\includegraphics{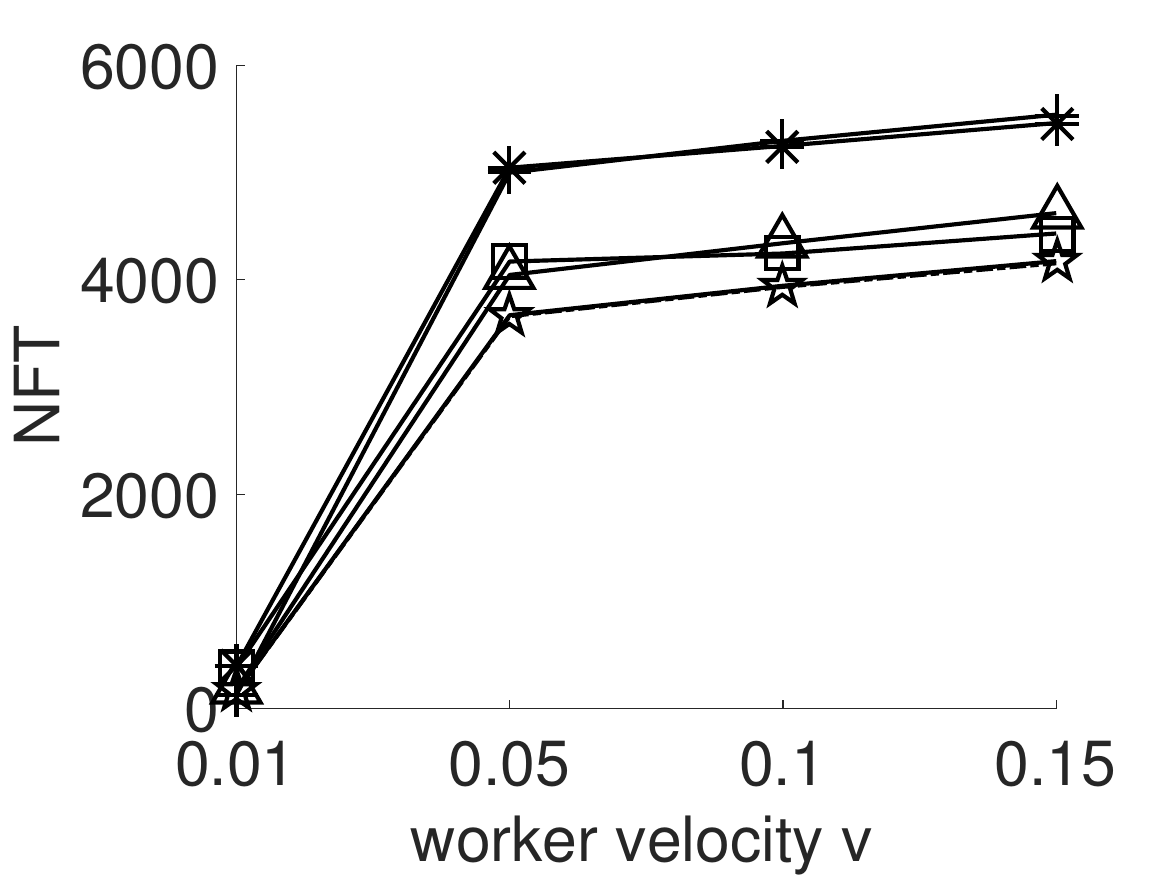}}
		\label{subfig:v_finished_task_number_real_all}}\hfill
	\subfigure[][{\small Confidently Assigned Tasks }]{
		\scalebox{0.2}[0.2]{\includegraphics{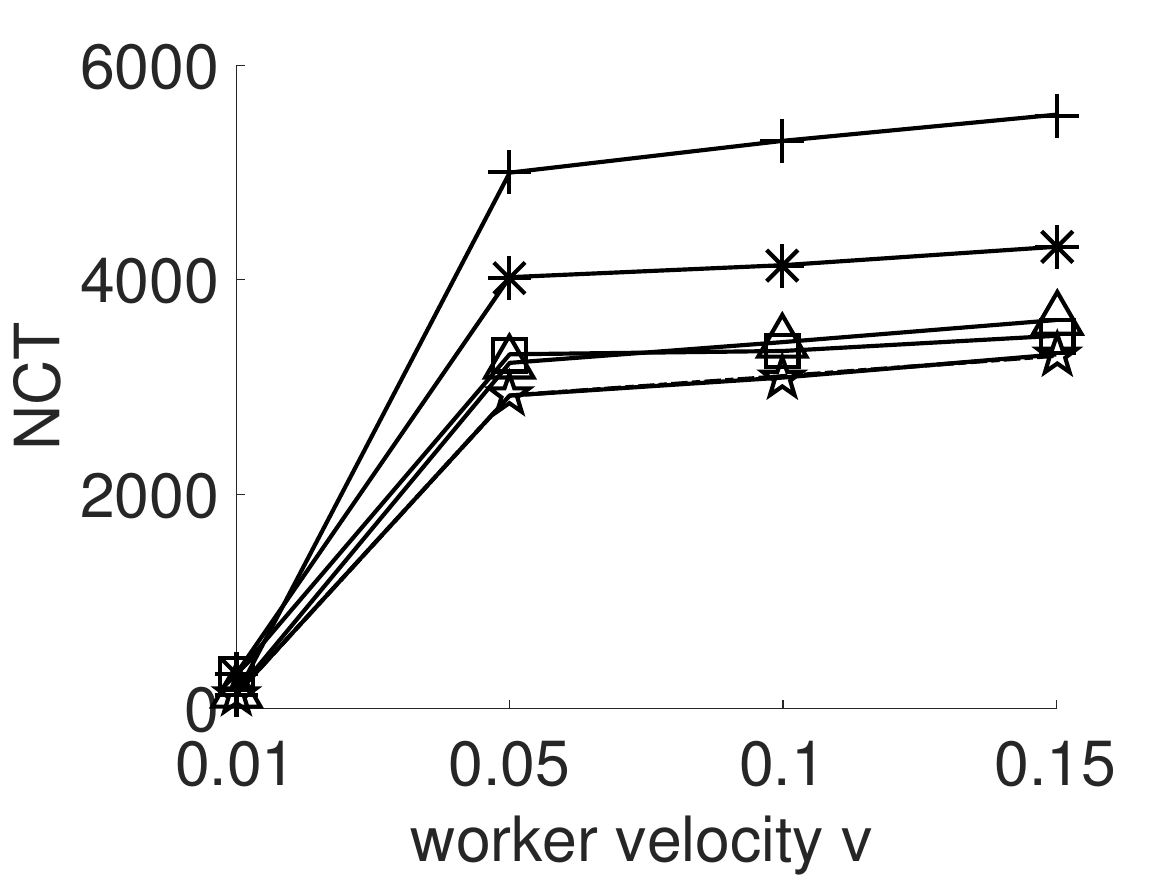}}
		\label{subfig:v_finished_task_number_conf_real_all}}\hfill\vspace{-2ex}
	\subfigure[][{\small Running Times }]{
		\scalebox{0.2}[0.2]{\includegraphics{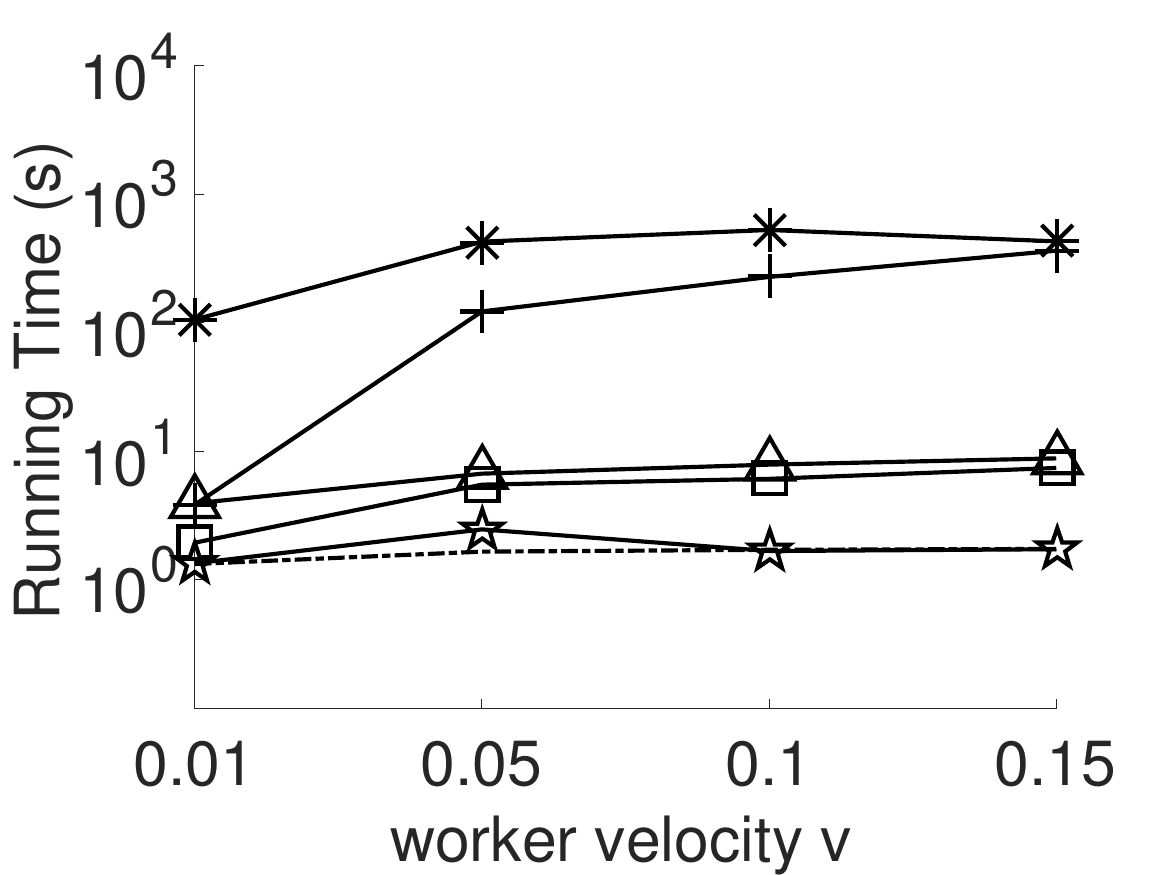}}
		\label{subfig:v_running_time_real_all}}\hfill
	\caption{\small Effects of worker velocity $v$ (Real).}
	\label{fig:effect_speed_real_all}\vspace{-2ex}
\end{figure*}

\begin{figure*}[t!]\centering
	\subfigure{
		\scalebox{0.4}[0.4]{\includegraphics{bar_mix-eps-converted-to.pdf}}}\hfill\\\vspace{-2ex}
	\addtocounter{subfigure}{-1}
	\subfigure[][{\small Moving Distance }]{
		\scalebox{0.2}[0.2]{\includegraphics{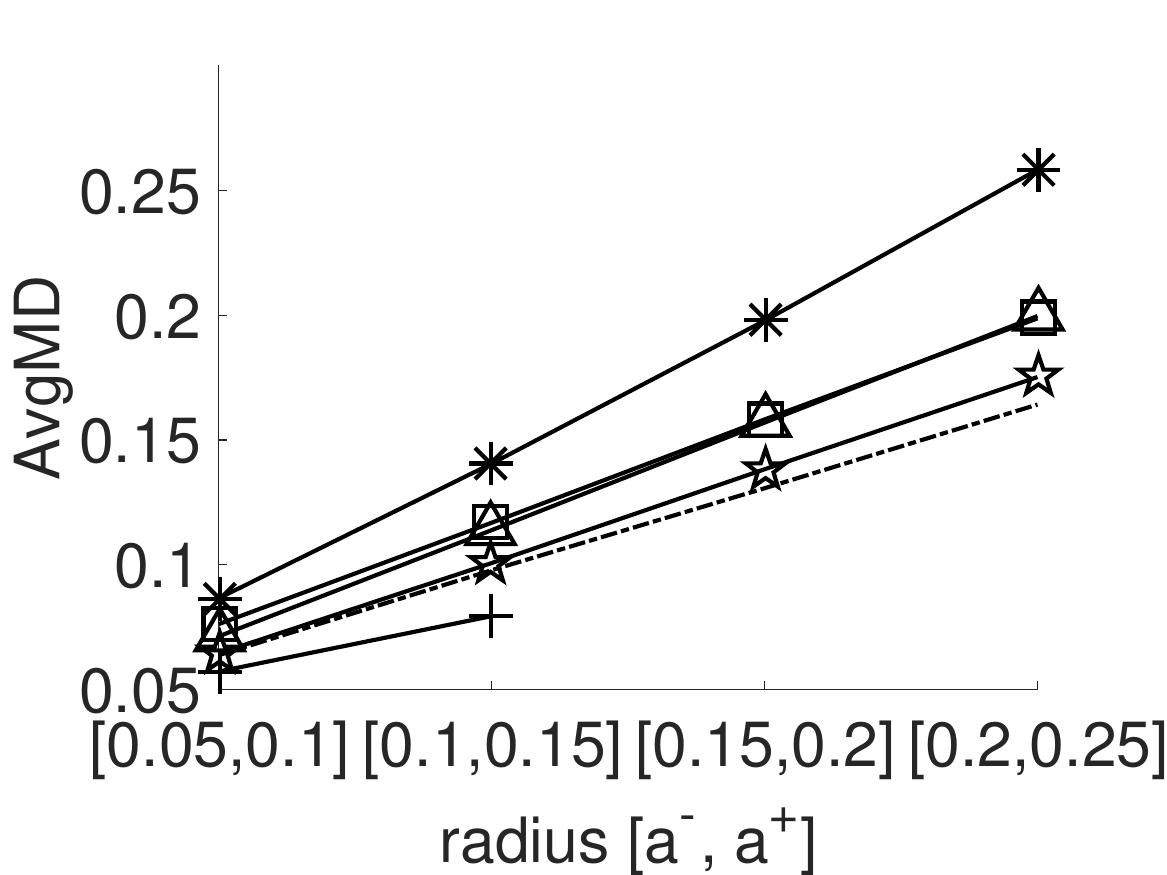}}\vspace{-2ex}
		\label{subfig:ac_avg_moving_distance_real}}\hfill
	\subfigure[][{\small Fully Assigned Tasks}]{
		\scalebox{0.2}[0.2]{\includegraphics{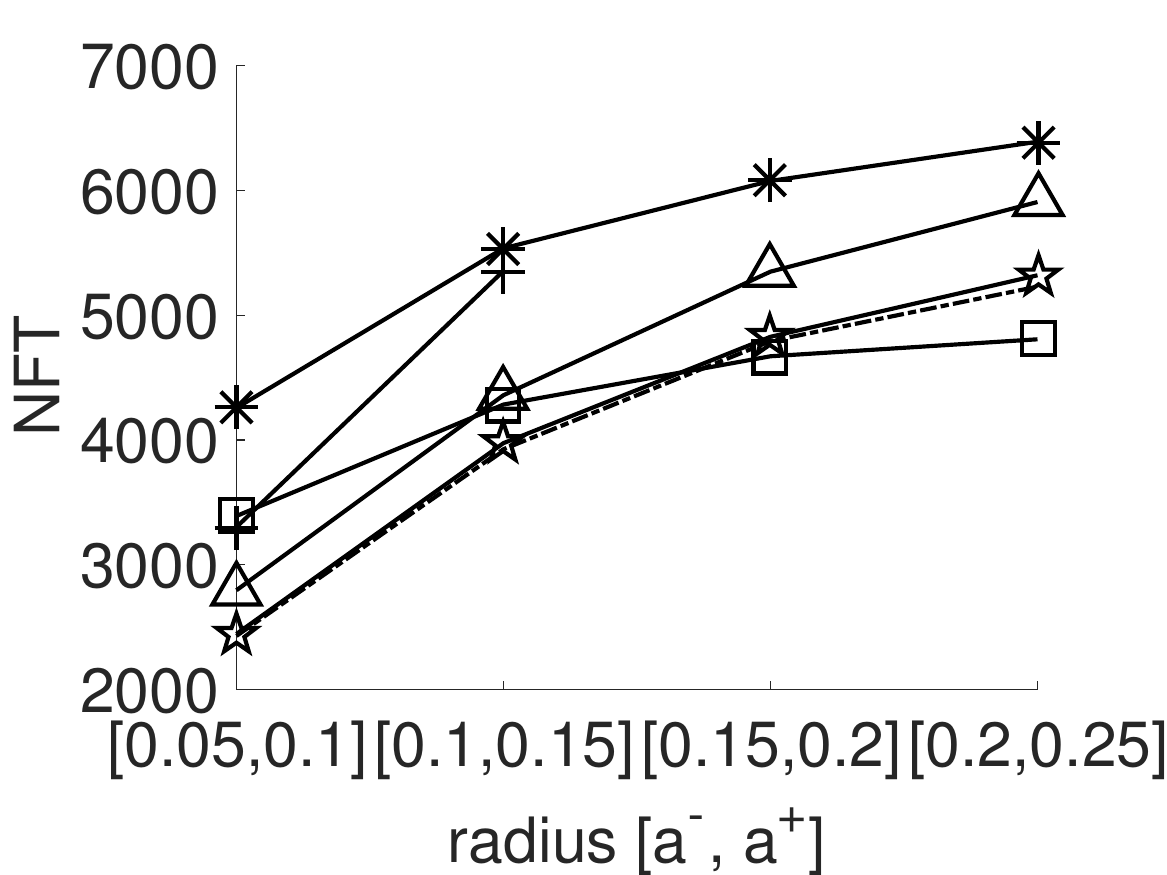}}
		\label{subfig:ac_finished_task_number_real}}\hfill
	\subfigure[][{\small Confidently Assigned Tasks}]{
		\scalebox{0.2}[0.2]{\includegraphics{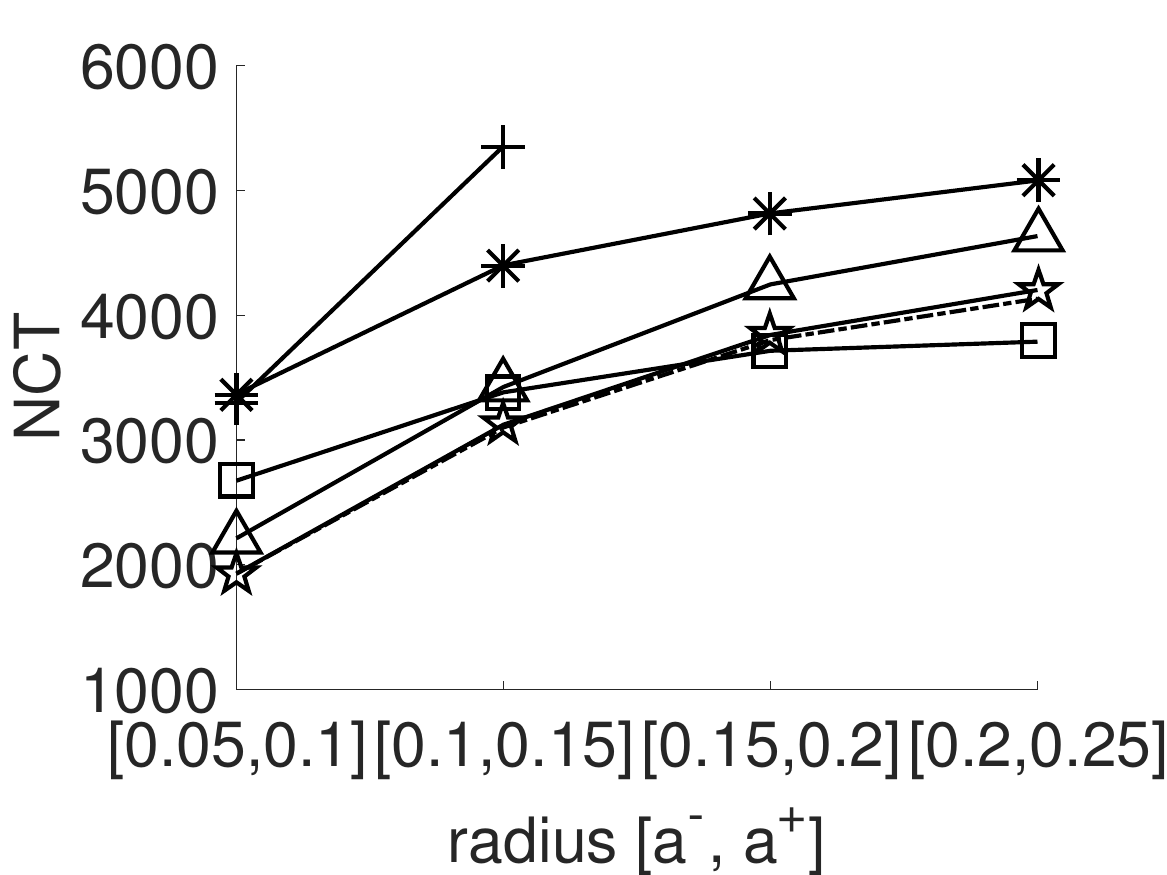}}
		\label{subfig:ac_finished_task_number_conf_real}}\hfill\vspace{-2ex}
	\subfigure[][{\small Running Times}]{
		\scalebox{0.2}[0.2]{\includegraphics{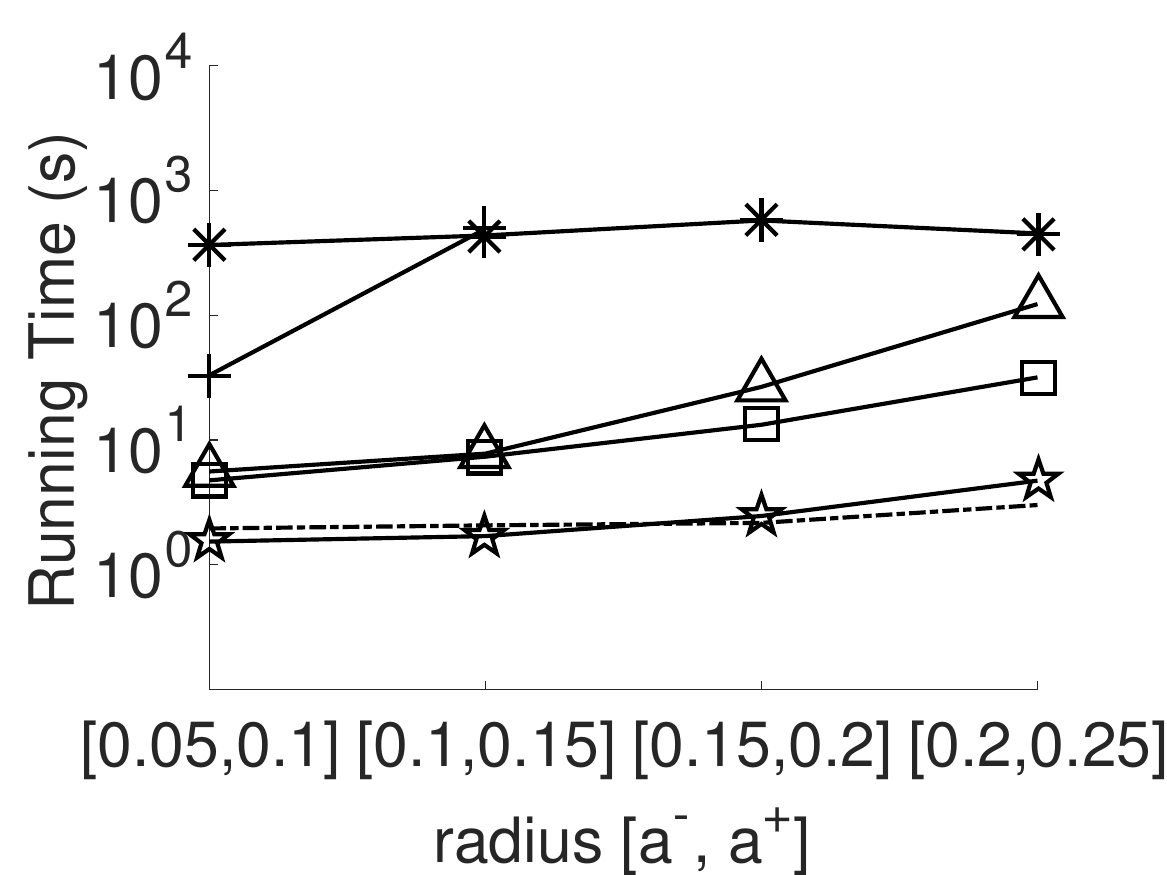}}
		\label{subfig:ac_running_time_real}}\hfill\vspace{-2ex}
	\caption{\small  Effects of Radii of  Working Areas $a$ (Real).}
	\label{fig:effect_circle_real}\vspace{-3ex}
\end{figure*}

\noindent {\bf Effect of the range, $[r^-, r^+]$, of
	workers' reliabilities.}  
Figure \ref{fig:effect_reliability_real} shows the effect of the range, $[r^-, r^+]$, of workers' reliabilities by varying it from [$0.65$, $0.7$] to [$0.85$, $0.9$]. We find that the range of workers' reliabilities does not affect AvgMDs of the results achieved by the tested algorithms, as shown in Figure \ref{subfig:r_avg_moving_distance_real}.
In Figure \ref{subfig:r_finished_task_number_real}, GT-hgr can achieve higher NFTs when the range $[r^-, r^+]$ gets larger, as the number of correct matches will increase leading to more tasks will be fully assigned by GT-hgr. 
In Figure \ref{subfig:r_finished_task_number_conf_real}, when the range $[r^-, r^+]$ increases, all the tested approaches can achieve results with higher NCTs. We notice that NCTs of G-llep and RDB-sam is higher than that of GT-hgr when $[r^-, r^+]$ is large, but NCTs of GT-hgr is the highest when $[r^-, r^+]$ is small (e.g., $0.65$ to $0.7$), which shows the effectiveness of the trustworthy query of GT-hgr. When the quality of workers is low, GT-hgr can guarantee the correctness/quality of the fully assigned tasks.
In Figure \ref{subfig:r_running_time_real}, only the running time of GT-hgr decreases when the range $[r^-, r^+]$ gets larger, as GT-hgr can easily use more correct matches to satisfy the required quality levels of tasks.

\noindent \textbf{Effect of the velocities, $v$, of workers.} Figure \ref{fig:effect_speed_real_all} shows the effect of the velocities, $v$, of workers by varying it from $0.01$ to $0.15$ per time slot. As shown in Figure \ref{subfig:v_avg_moving_distance_real_all}, when the velocities of workers increase from $0.01$ to $0.05$, each worker can reach more (far) tasks before their deadlines. Then more tasks can be fully assigned. As a result, AvgMDs of the results achieved by all the tested algorithms increase. However, when $v$ increases from $0.05$ to $0.15$, AvgMDs will stop increasing because the constraint of the working areas prevents workers from moving too far. Similar phenomena can be found in Figures \ref{subfig:v_finished_task_number_real_all} and \ref{subfig:v_finished_task_number_conf_real_all},  NFTs and NCTs increase first when $v$ increases from $0.01$ to $0.05$, then stop increasing when $v$ increases from $0.05$ to $0.15$. The running times of the tested algorithms are shown in Figure \ref{subfig:v_running_time_real_all}, they are also first increase when when $v$ increases from $0.01$ to $0.05$, then stop increasing when $v$ increases from $0.05$ to $0.15$. The reason is that when $v$ increases from  $0.01$ to $0.05$, more tasks can be reached by workers, thus the running times increase.

 \noindent {\bf Effect of the range, $[a^-, a^+]$, of 
 	the diameters of workers' circle working areas.}
 When workers' circle working areas get larger, there will be more available tasks located in the working area of each worker leading to the number of valid worker-and-task pairs increases. 
 Similar to the results when the working areas are squares,
 in Figure \ref{subfig:ac_avg_moving_distance_real}, as the working areas get larger, AvgMDs of the tested approaches increase obviously, because the worker can reach tasks located further. 
 In Figure \ref{subfig:ac_finished_task_number_real}, all the tested approaches can fully assign more tasks when the range of diameters of working areas $a$ increases, as each task can be reached by more workers and can be fully assigned with a higher probability. Specifically, the increasing speed of NFT of the tested online algorithms is higher than that of G-llep and RDB-sam.
 In Figure \ref{subfig:ac_finished_task_number_conf_real}, GT-hgr still can achieve the highest NCT than other tested algorithms.
 In Figure \ref{subfig:ac_running_time_real}, the running time of all the tested approaches increases when the range of $a_i$ increases, as more valid worker-and-task pairs need to process. When $[a^-, a^+]$ is higher than [0.1, 0.15], GT-hgr needs much more time than other tested approaches.

\section{Results on Other Distributions}

Before discussing the effects of different distributions of locations of workers/tasks, we introduce three notions: 1) covered task (CT) denoting the task can be reached by any workers; 2) confidently covered task (CCT) referring to the task that there is at least one correct match for it; 3) number of reachable workers per covered task (W/CT). Intuitively, when the number of CT is large, AvgMD will be large, since for each worker he/she can be assigned with more tasks. What is more, when CCT increases, GT-hgr can confidently assign more tasks. In addition, when W/CT increases, all algorithms, except for GT-hgr, will fully assign more tasks.

\begin{figure*}[th!]\centering
	\subfigure{
		\scalebox{0.4}[0.4]{\includegraphics{bar_mix-eps-converted-to.pdf}}}\hfill\\\vspace{-2ex}
	\addtocounter{subfigure}{-1}
	\subfigure[][{\small Moving Distance}]{
		\scalebox{0.2}[0.2]{\includegraphics{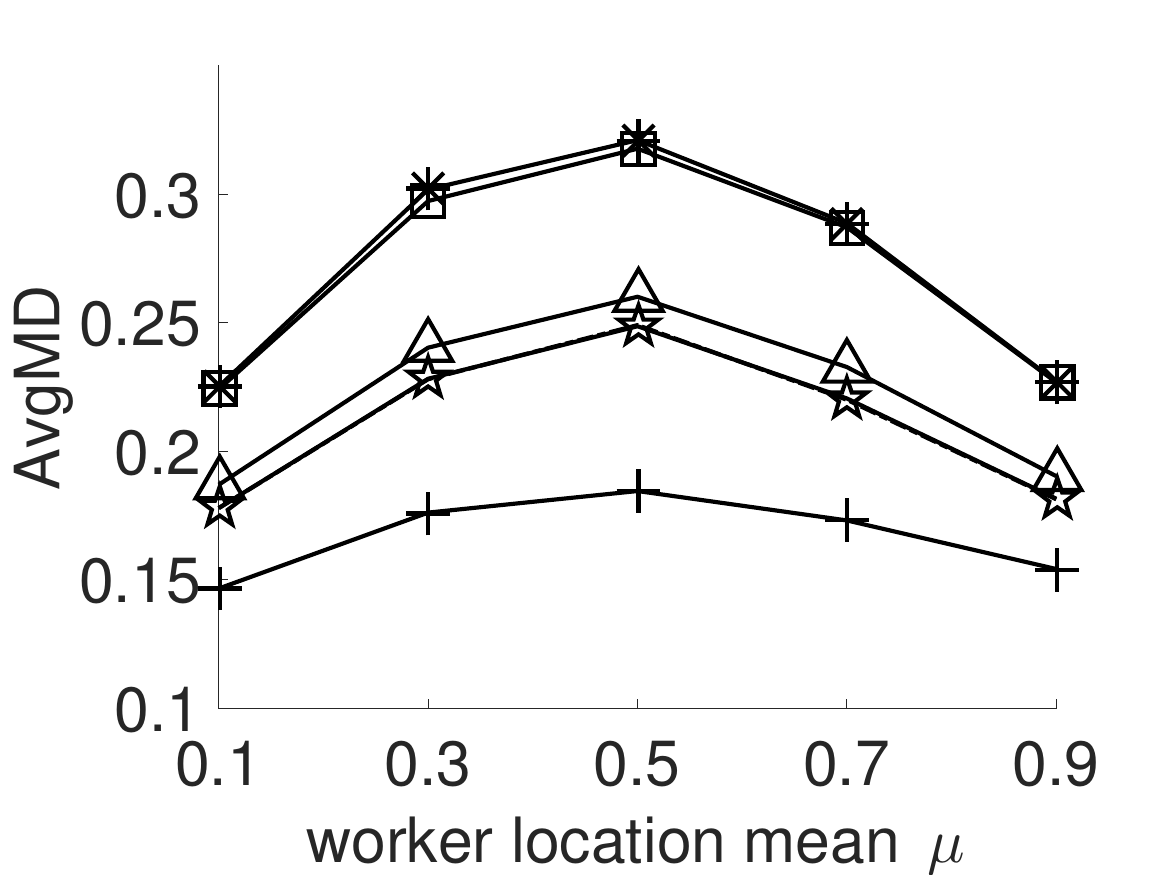}}
		\label{subfig:mean_avg_moving_distance_unif}}\hfill\vspace{-2ex}
	\subfigure[][{\small Fully Assigned Tasks}]{
		\scalebox{0.2}[0.2]{\includegraphics{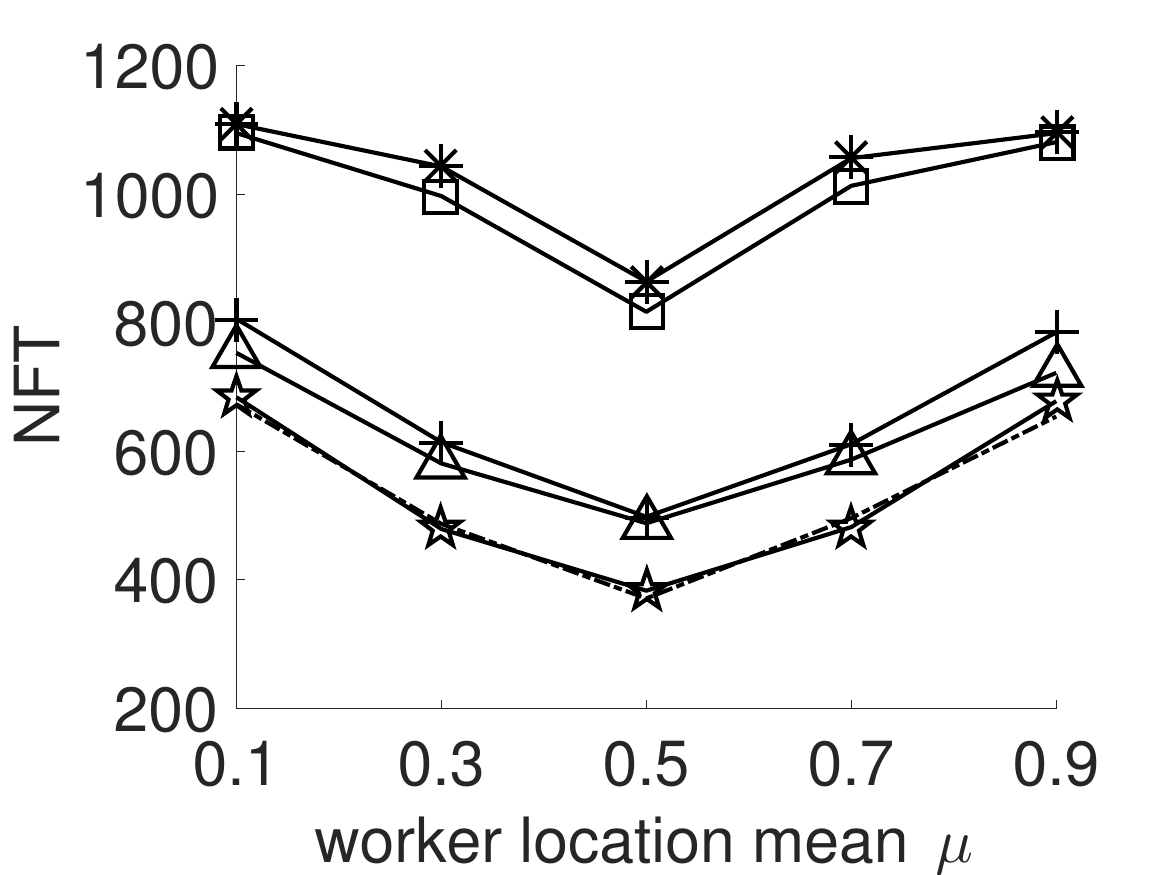}}
		\label{subfig:mean_finished_task_number_unif}}\hfill
	\subfigure[][{\small Confidently Assigned Tasks}]{
		\scalebox{0.2}[0.2]{\includegraphics{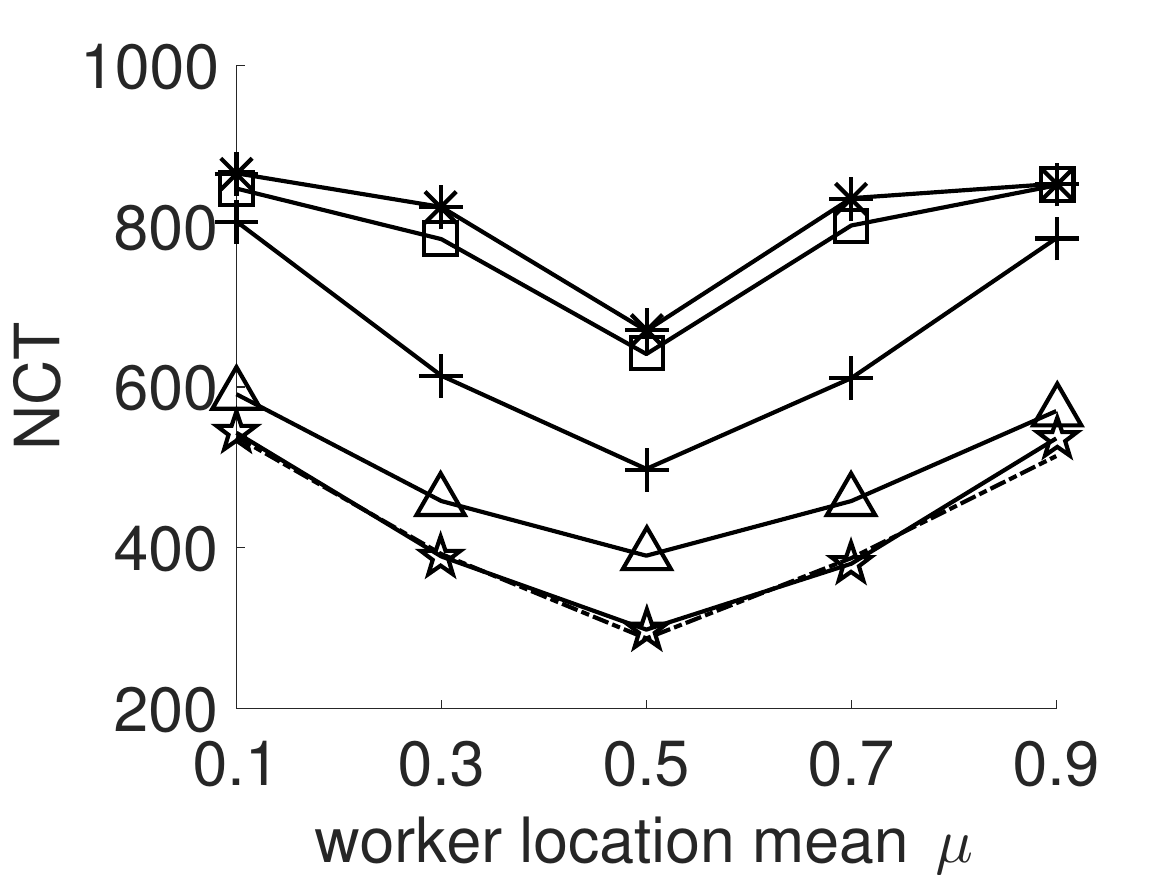}}
		\label{subfig:mean_finished_task_number_conf_unif}}\hfill
	\subfigure[][{\small Running Time}]{
		\scalebox{0.2}[0.2]{\includegraphics{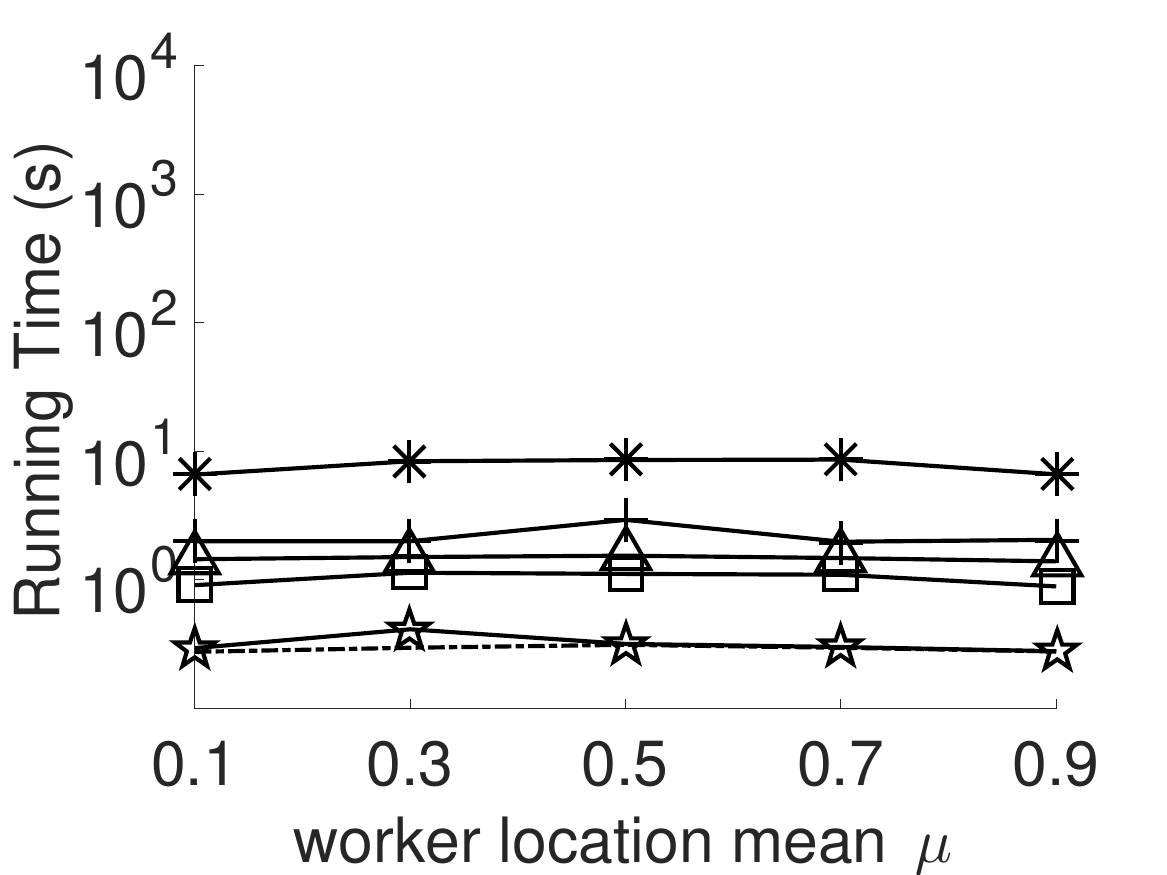}}
		\label{subfig:mean_running_time_unif}}\hfill
	
	\subfigure[][{\small Moving Distance}]{
		\scalebox{0.2}[0.2]{\includegraphics{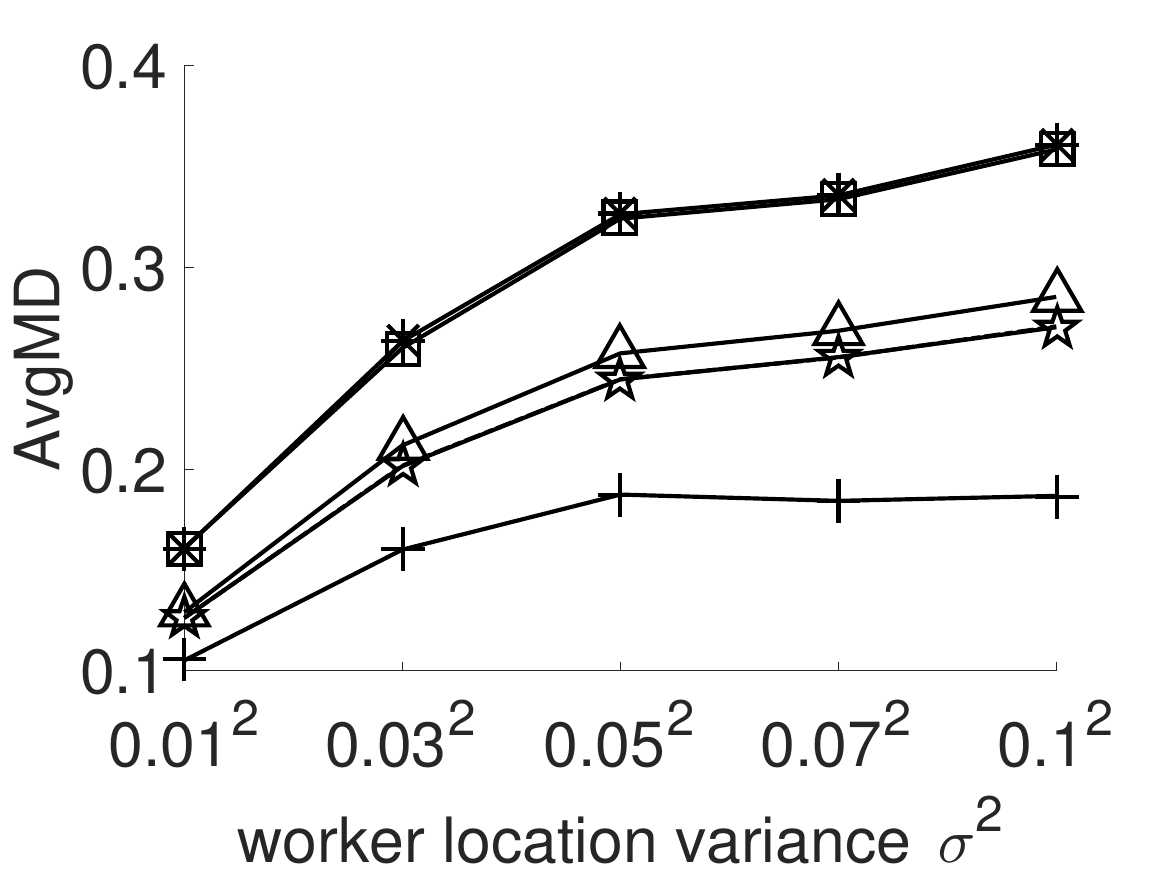}}
		\label{subfig:variance_avg_moving_distance_unif}}\hfill\vspace{-2ex}
	\subfigure[][{\small Fully Assigned Tasks}]{
		\scalebox{0.2}[0.2]{\includegraphics{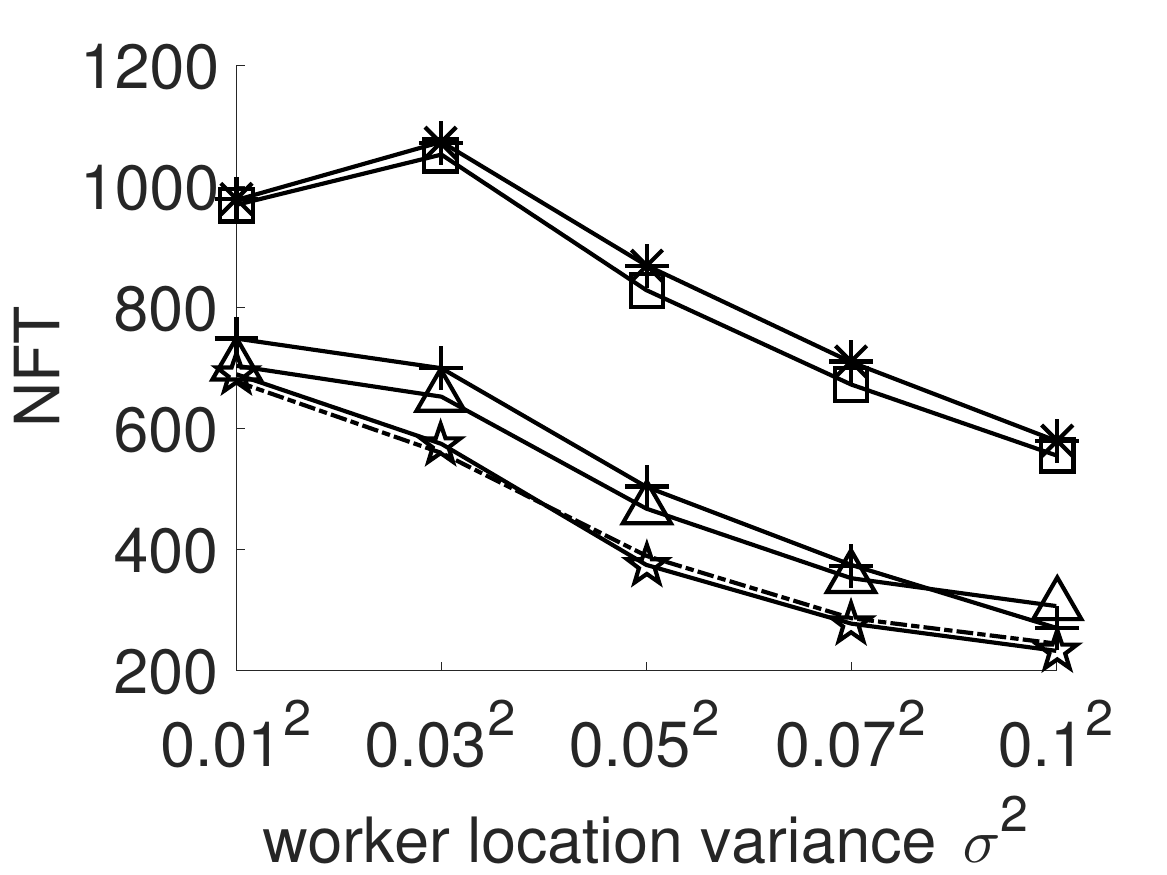}}
		\label{subfig:variance_finished_task_number_unif}}\hfill
	\subfigure[][{\small Confidently Assigned Tasks}]{
		\scalebox{0.2}[0.2]{\includegraphics{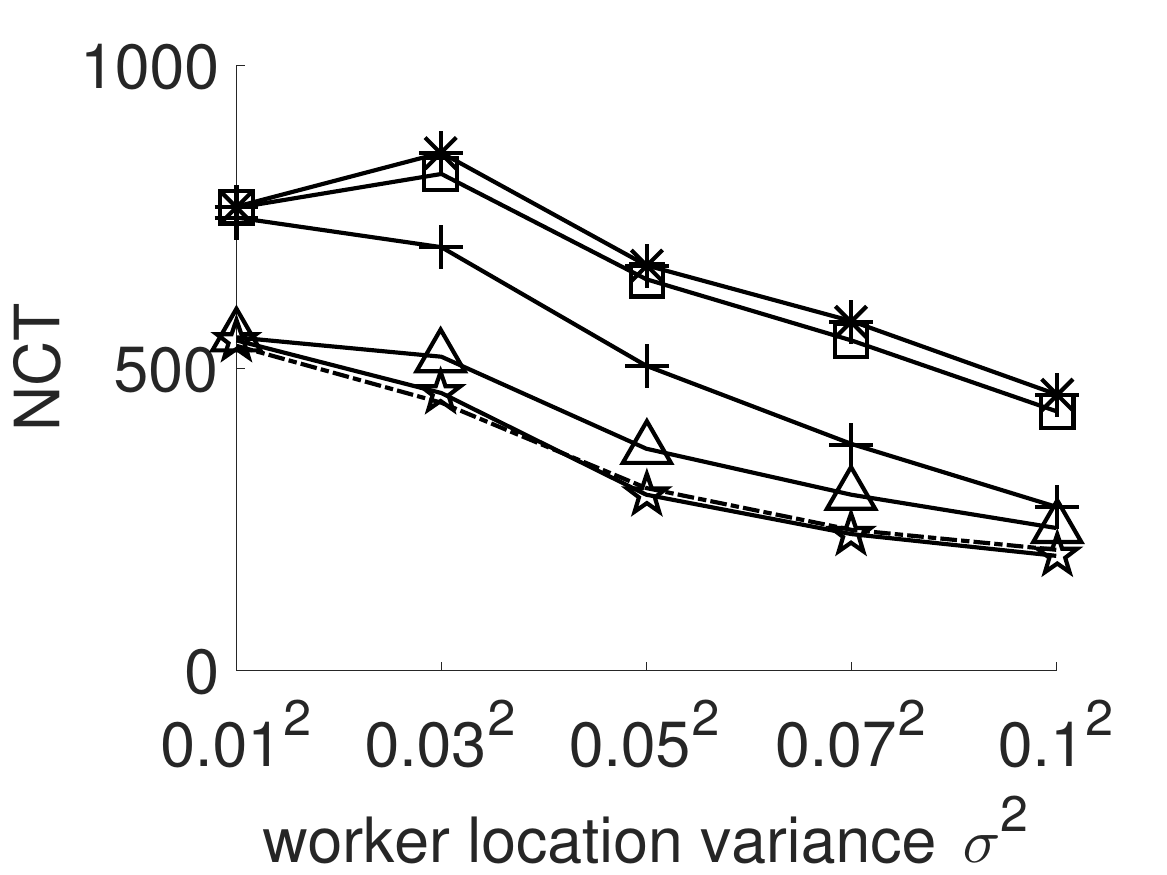}}
		\label{subfig:variance_finished_task_number_conf_unif}}
	\subfigure[][{\small Running Time}]{
		\scalebox{0.2}[0.2]{\includegraphics{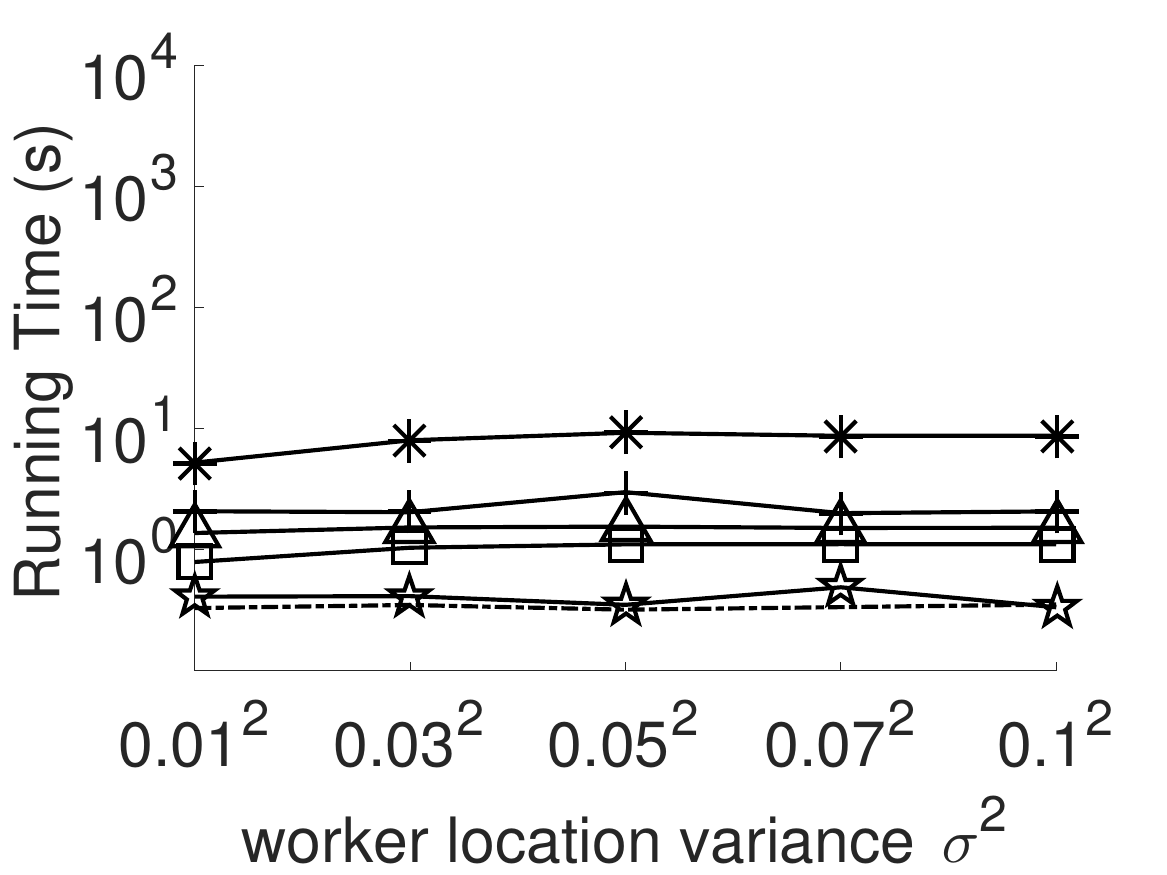}}
		\label{subfig:variance_running_time_unif}}\hfill
	
		\subfigure[][{\small Moving Distance}]{
		\scalebox{0.2}[0.2]{\includegraphics{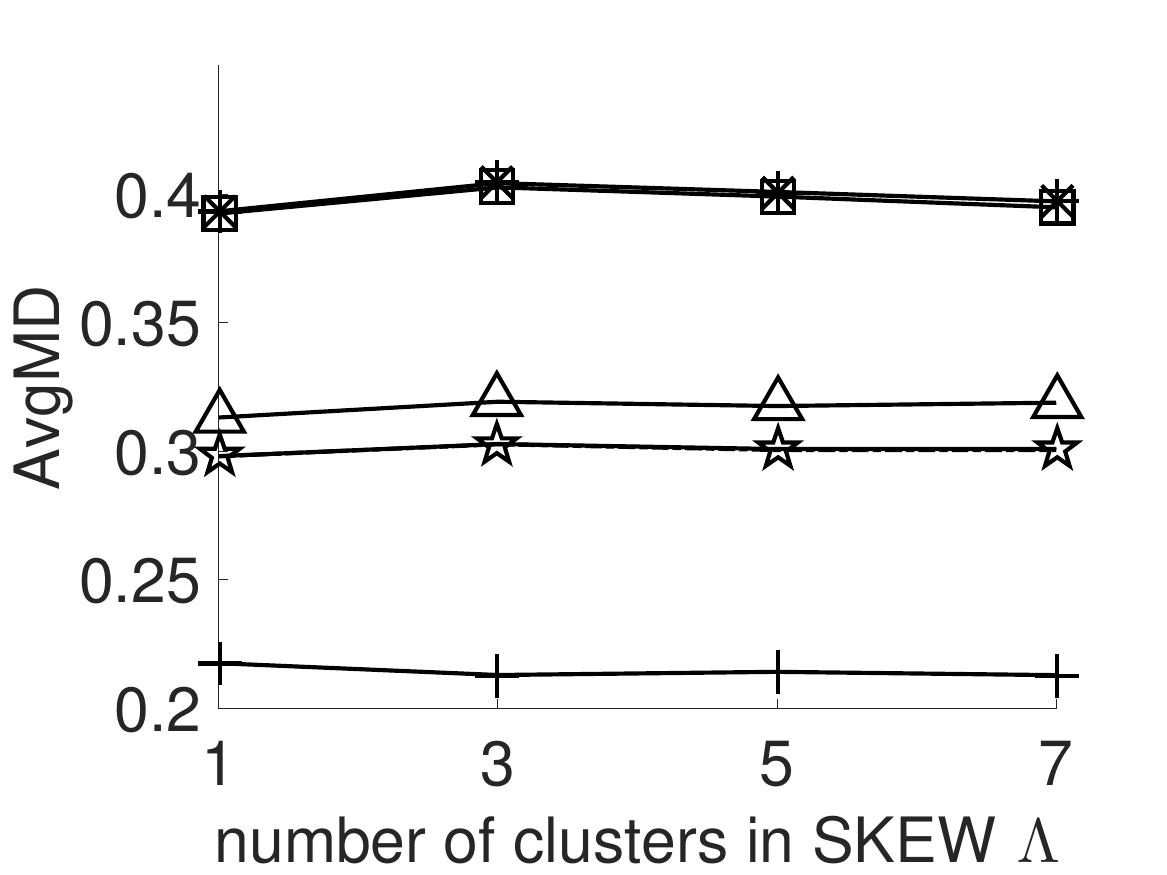}}
		\label{subfig:cluster_avg_moving_distance_unif}}\hfill\vspace{-2ex}
	\subfigure[][{\small Fully Assigned Tasks}]{
		\scalebox{0.2}[0.2]{\includegraphics{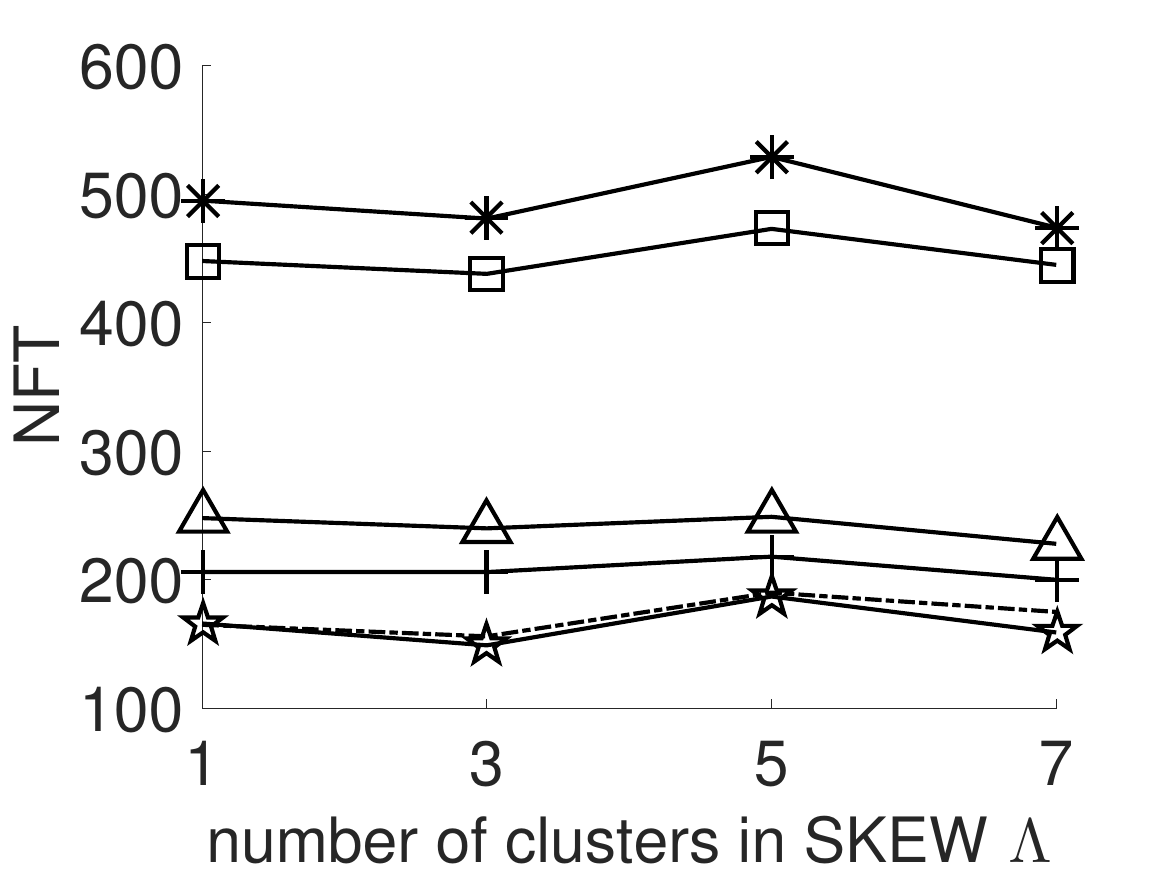}}
		\label{subfig:cluster_finished_task_number_unif}}\hfill
	\subfigure[][{\small Confidently Assigned Tasks}]{
		\scalebox{0.2}[0.2]{\includegraphics{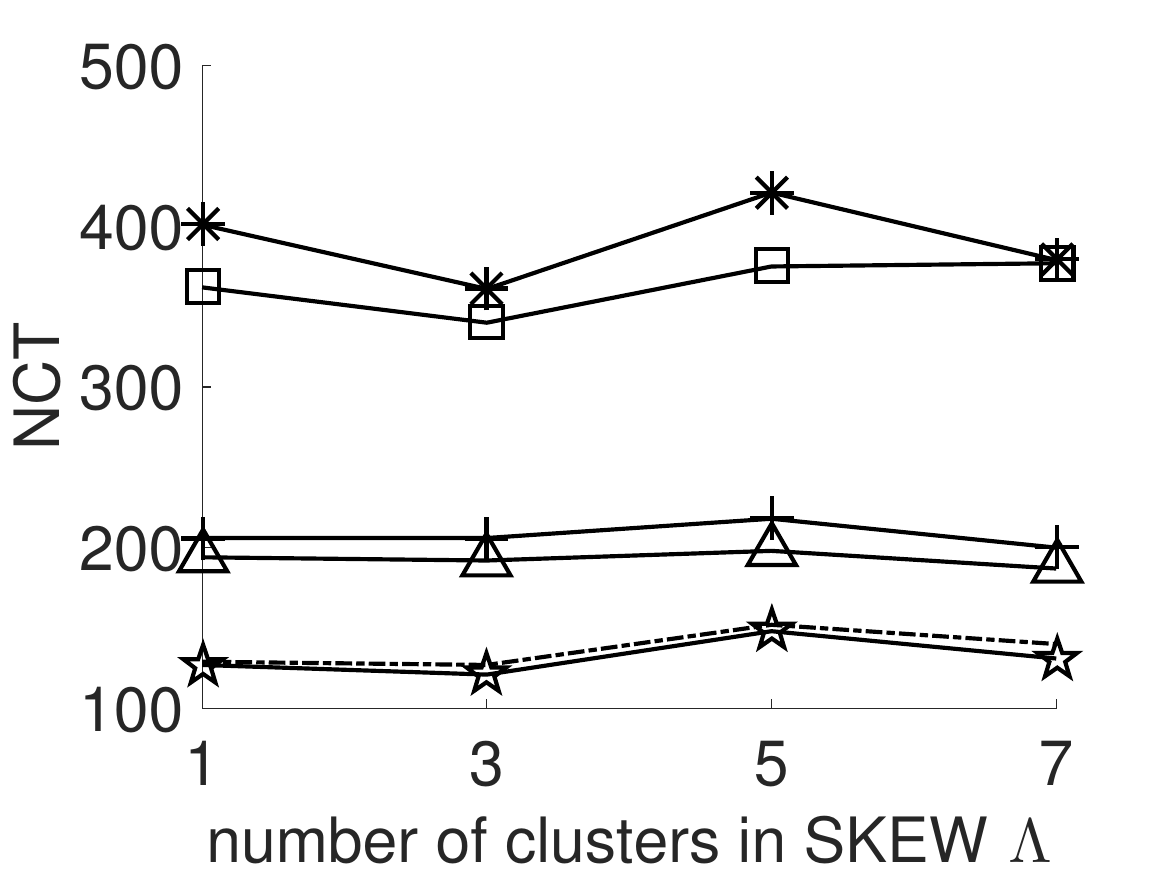}}
		\label{subfig:cluster_finished_task_number_conf_unif}}\hfill\vspace{-1ex}
	\subfigure[][{\small Running Time}]{
		\scalebox{0.2}[0.2]{\includegraphics{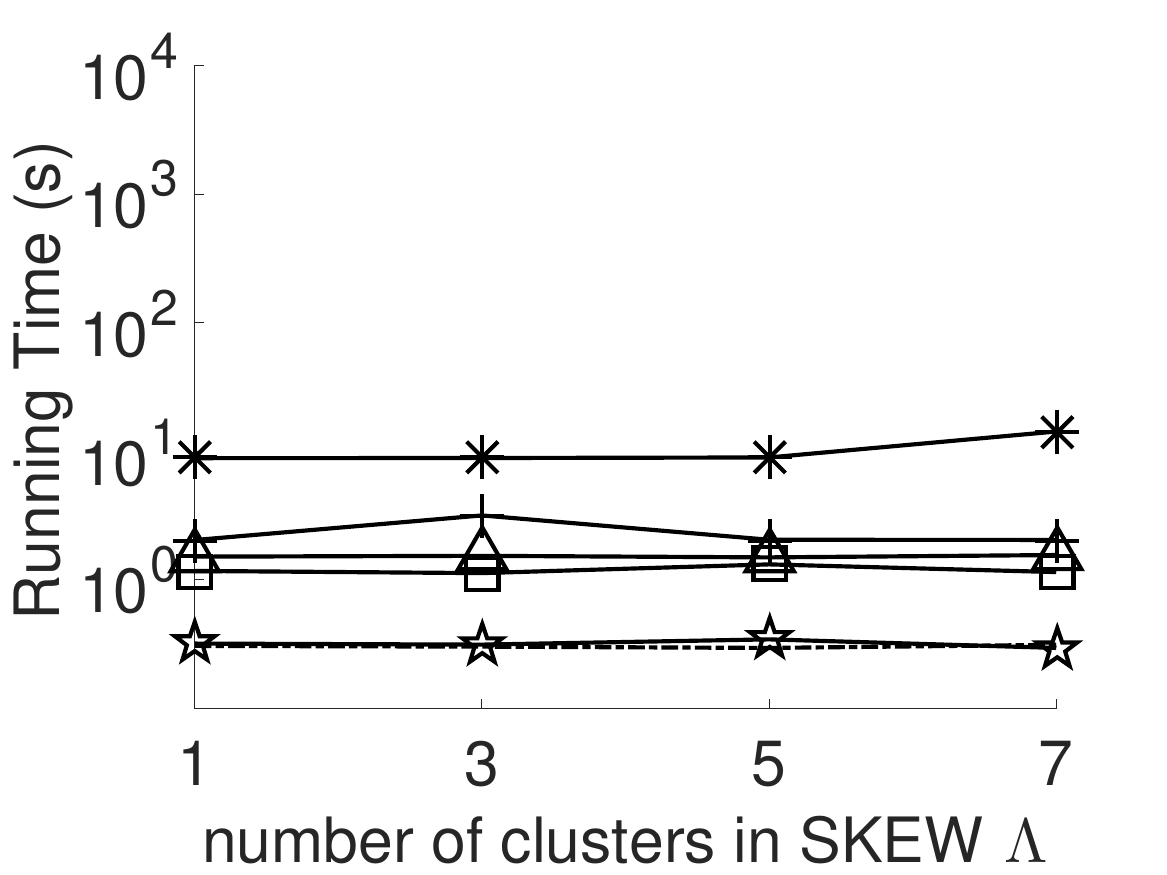}}
		\label{subfig:cluster_running_time_unif}}\hfill
	\caption{\small Results that the locations of workers follow GAUS and SKEW  while the locations of tasks follow UNIF. (Synthetic).}
	\label{fig:effect_worker_distribution_task_unif}\vspace{-2ex}
\end{figure*}

\noindent \textbf{Effect of locations of tasks following UNIF distribution.} Figure \ref{fig:effect_worker_distribution_task_unif} shows the results when the locations of tasks follow UNIF distribution and the locations of workers follow GAUS and SKEW. In Figure \ref{subfig:mean_avg_moving_distance_unif}, when the center of the GAUS distributed workers moves from the left bottom corner to the right top corner, CT increases when $\mu$ changes from $0.1$ to $0.5$ and decreases when $\mu$
 changes from $0.5$ to $0.9$ as shown in the first row of Figure \ref{fig:distributon_illustration}. Thus, AvgMDs of the results achieved by all the tested algorithms first increase then decrease. In addition, when locations of tasks follow UNIF, the number of CCT and W/CT will decrease  when $\mu$ changes from $0.1$ to $0.5$ and increase when $\mu$
 changes from $0.5$ to $0.9$. For CCT, since tasks are uniformly distributed, when the center of Gaussian distributed locations of workers is close to the location point ($0.5$, $0.5$), the workers are distributed more sparsely leading to CCT decrease. Thus, NFTs and NCTs of all the tested algorithms increase when $\mu$ changes from $0.1$ to $0.5$ and decrease when $\mu$
 changes from $0.5$ to $0.9$ as shown in Figures \ref{subfig:mean_finished_task_number_unif} and  \ref{subfig:mean_finished_task_number_conf_unif}, respectively. The running times of the tested approaches do not change obviously when $\mu$ changes as shown in Figure \ref{subfig:mean_running_time_unif}.
 
 The second row of Figure \ref{fig:effect_worker_distribution_task_unif} shows the results of the tested algorithms when the variance $\sigma^2$ increases from $0.01^2$ to $0.1^2$ when the locations of tasks follow the UNIF distribution. Specifically, the number of CT increases because the workers are distributed deviating more from the center point ($0.5$, $0.5$). Thus, AvgMDs of all the tested algorithms increase as shown in Figure \ref{subfig:variance_avg_moving_distance_unif}. For W/CT, when $\sigma^2$ increases, it decreases since the total number of workers does not change and CT increases. Under the effects of increase of CT and decrease of W/CT, NFTs of G-llep and RDB-sam increase first then drop, and NFTs of online algorithms keep decreasing as shown in Figure \ref{subfig:variance_finished_task_number_unif}.
 For the number of CCT, it decreases slightly when $\sigma^2$ increases from $0.01^2$ to $0.03^2$ then dramatically when $\sigma^2$ increases from $0.03^2$ to $0.1^2$. Although when $\sigma^2$ increases from $0.01^2$ to $0.03^2$ CT increases, the density of workers decreases, which leads to that CCT still decreases. Thus,  NCTs of GT-hgr decrease slightly first then drop quickly as shown in Figure \ref{subfig:variance_finished_task_number_conf_unif}. The running times of the tested algorithms do not change obviously as shown in Figure \ref{subfig:variance_running_time_unif}.
 
 The third row of Figure \ref{fig:effect_worker_distribution_task_unif}  shows the results of the tested algorithms when when the number, $\Lambda$, of Gaussian distributed clusters in the skewed distribution of workers' locations increases from 1 to 7 and the locations of workers follow UNIF. The changes of AvgMDs and the running times are not obvious as shown in Figures \ref{subfig:cluster_avg_moving_distance_unif} and \ref{subfig:cluster_running_time_unif}, respectively. In addition, the changes of NFTs and NCTs of the tested algorithms are small and randomly as shown in Figures \ref{subfig:cluster_finished_task_number_unif} and \ref{subfig:cluster_finished_task_number_conf_unif}, respectively. The reason is that the centers of Gaussian clusters in SKEW are randomly selected.
 
\begin{figure*}[th!]\centering
	\subfigure{
		\scalebox{0.4}[0.4]{\includegraphics{bar_mix-eps-converted-to.pdf}}}\hfill\\\vspace{-2ex}
	\addtocounter{subfigure}{-1}
	\subfigure[][{\small Moving Distance}]{
		\scalebox{0.2}[0.2]{\includegraphics{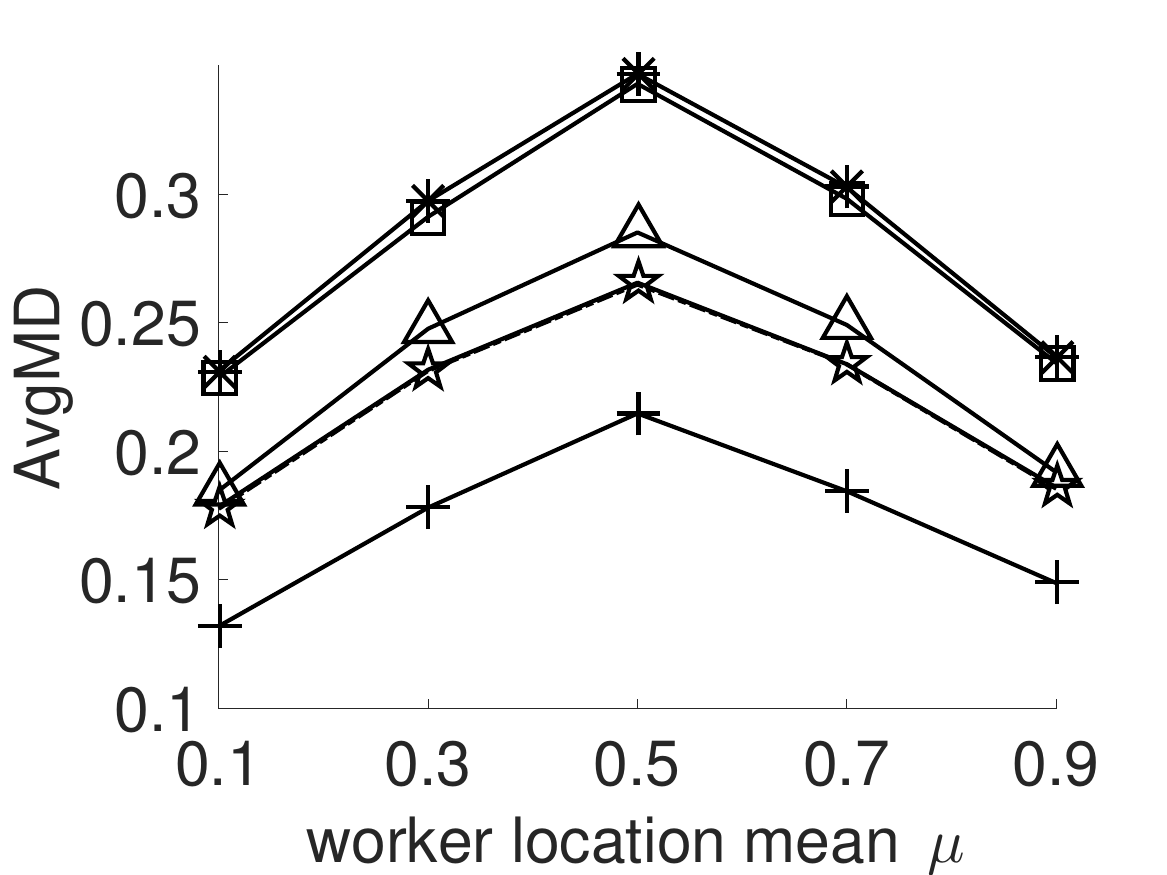}}
		\label{subfig:mean_avg_moving_distance_gaus}}\hfill\vspace{-2ex}
	\subfigure[][{\small Fully Assigned Tasks}]{
		\scalebox{0.2}[0.2]{\includegraphics{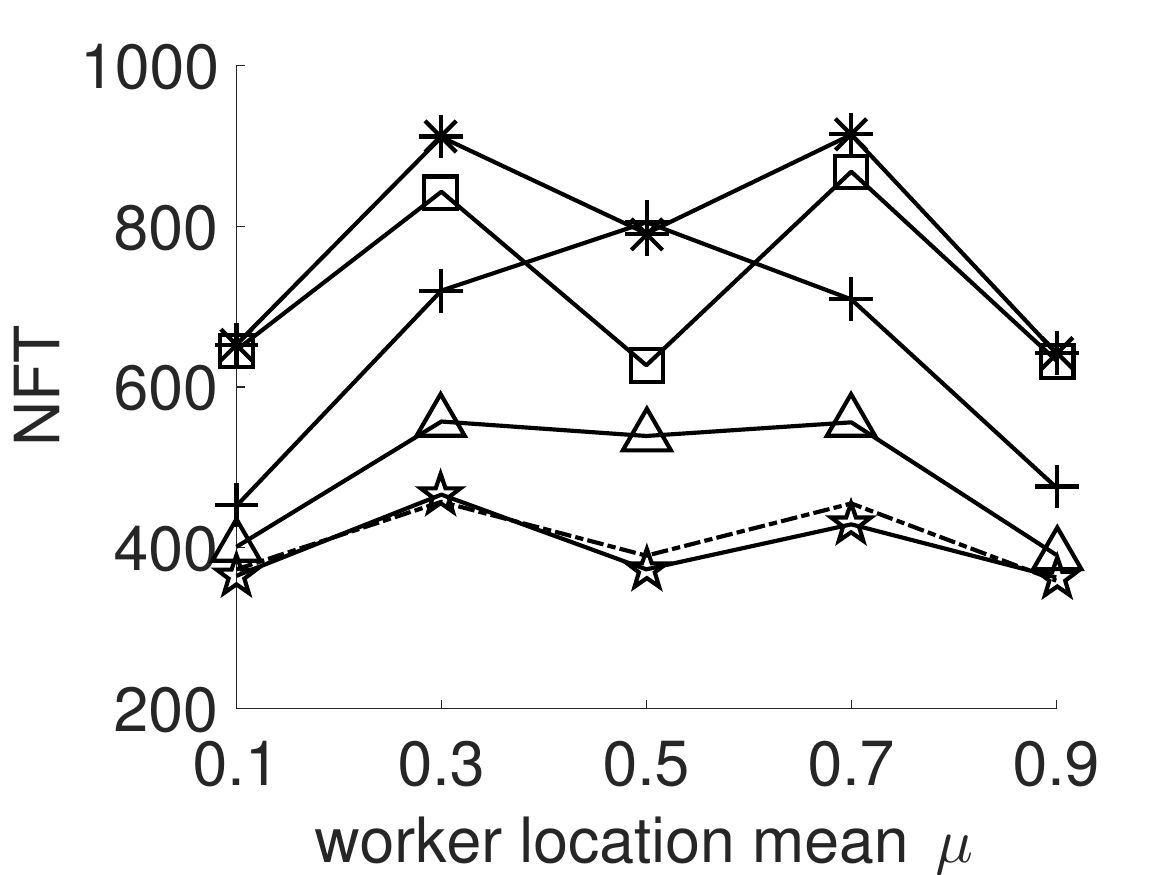}}
		\label{subfig:mean_finished_task_number_gaus}}\hfill
	\subfigure[][{\small Confidently Assigned Tasks}]{
		\scalebox{0.2}[0.2]{\includegraphics{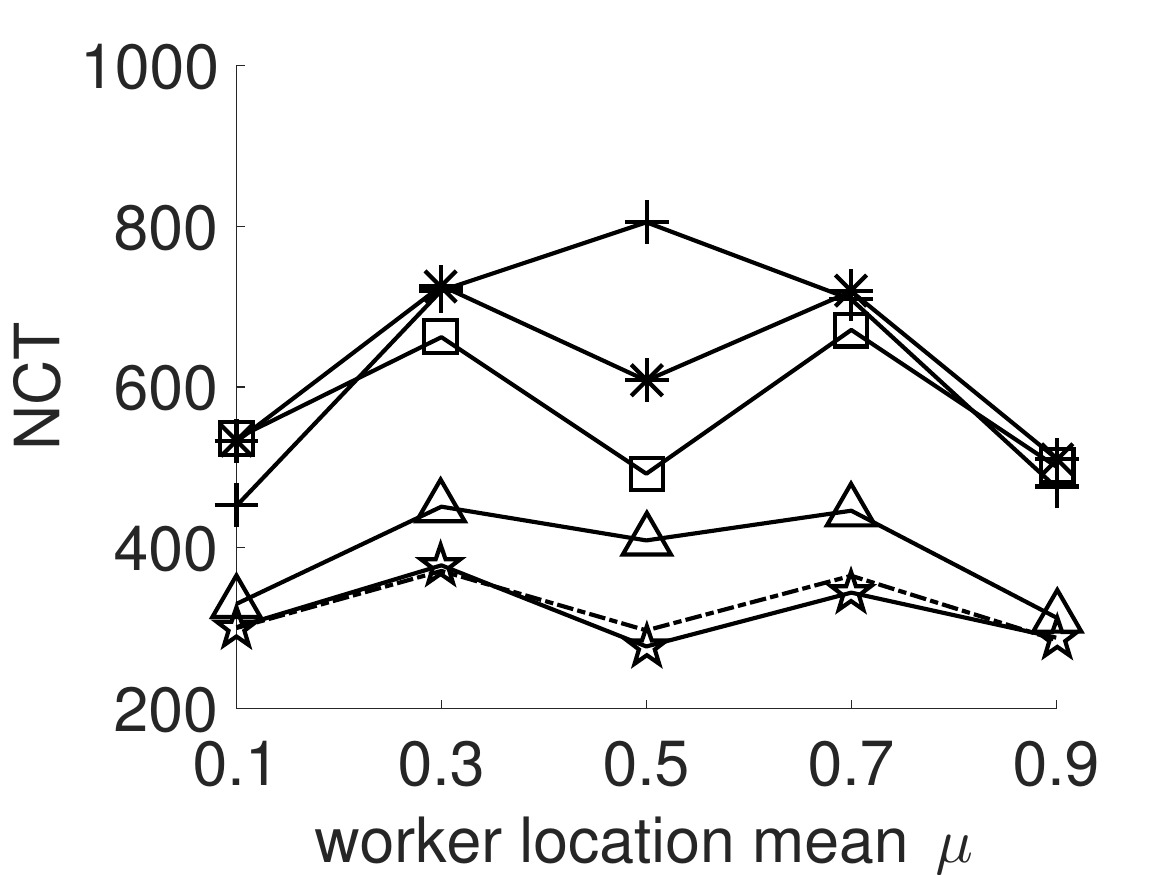}}
		\label{subfig:mean_finished_task_number_conf_gaus}}\hfill
	\subfigure[][{\small Running Time}]{
		\scalebox{0.2}[0.2]{\includegraphics{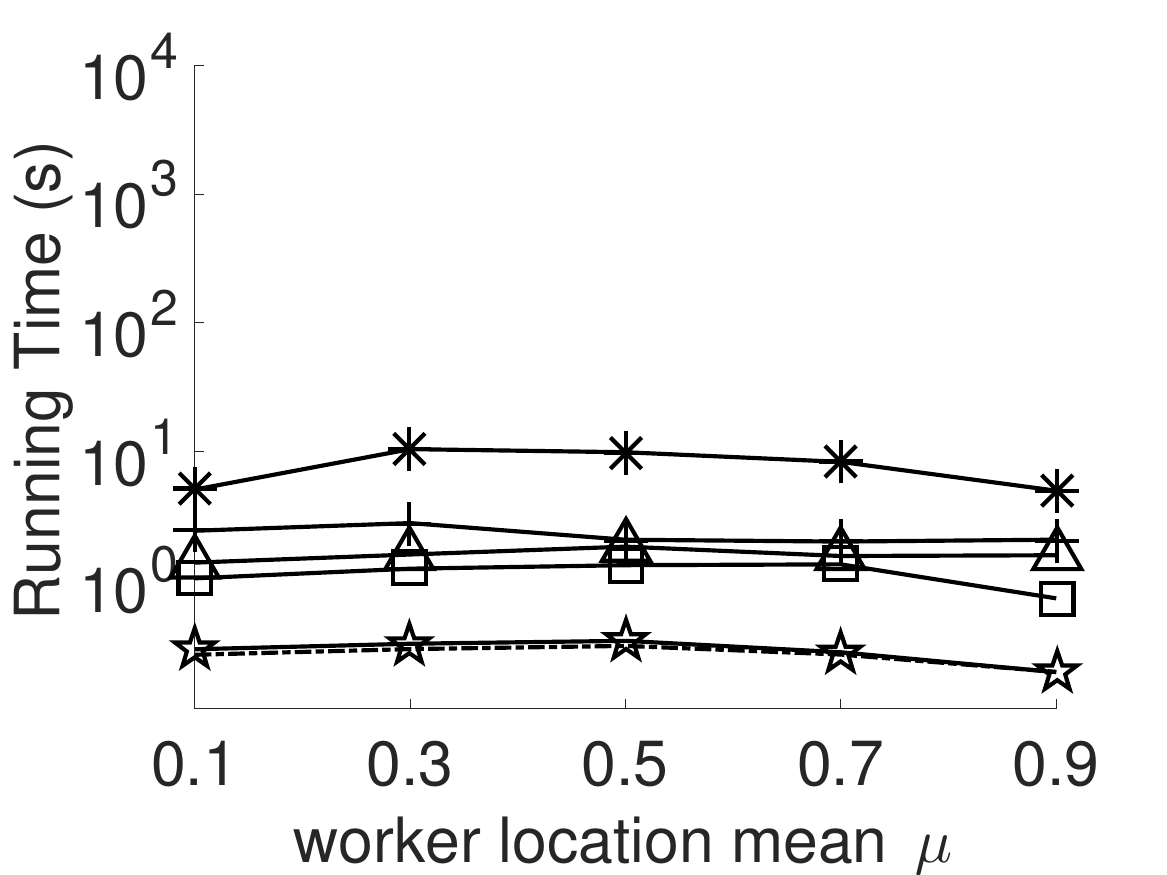}}
		\label{subfig:mean_running_time_gaus}}\hfill
	
	\subfigure[][{\small Moving Distance}]{
		\scalebox{0.2}[0.2]{\includegraphics{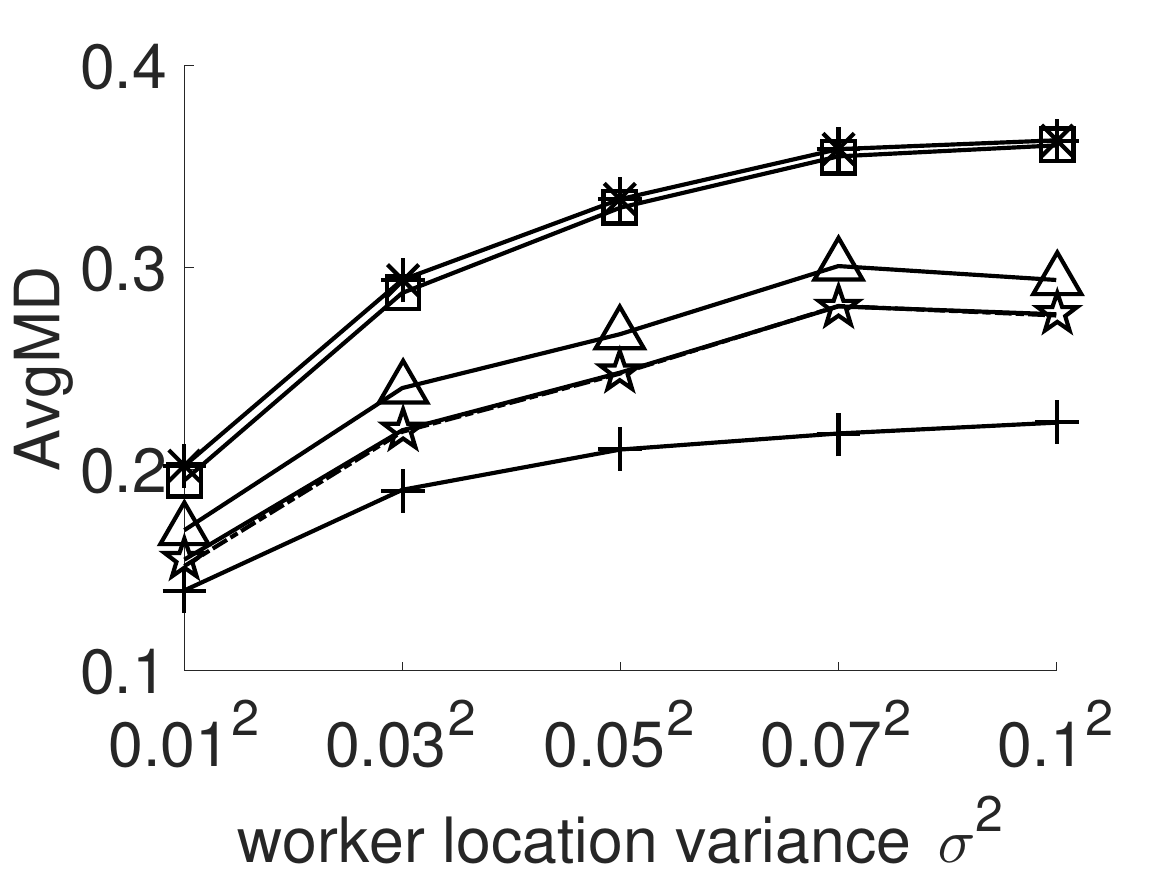}}
		\label{subfig:variance_avg_moving_distance_gaus}}\hfill\vspace{-2ex}
	\subfigure[][{\small Fully Assigned Tasks}]{
		\scalebox{0.2}[0.2]{\includegraphics{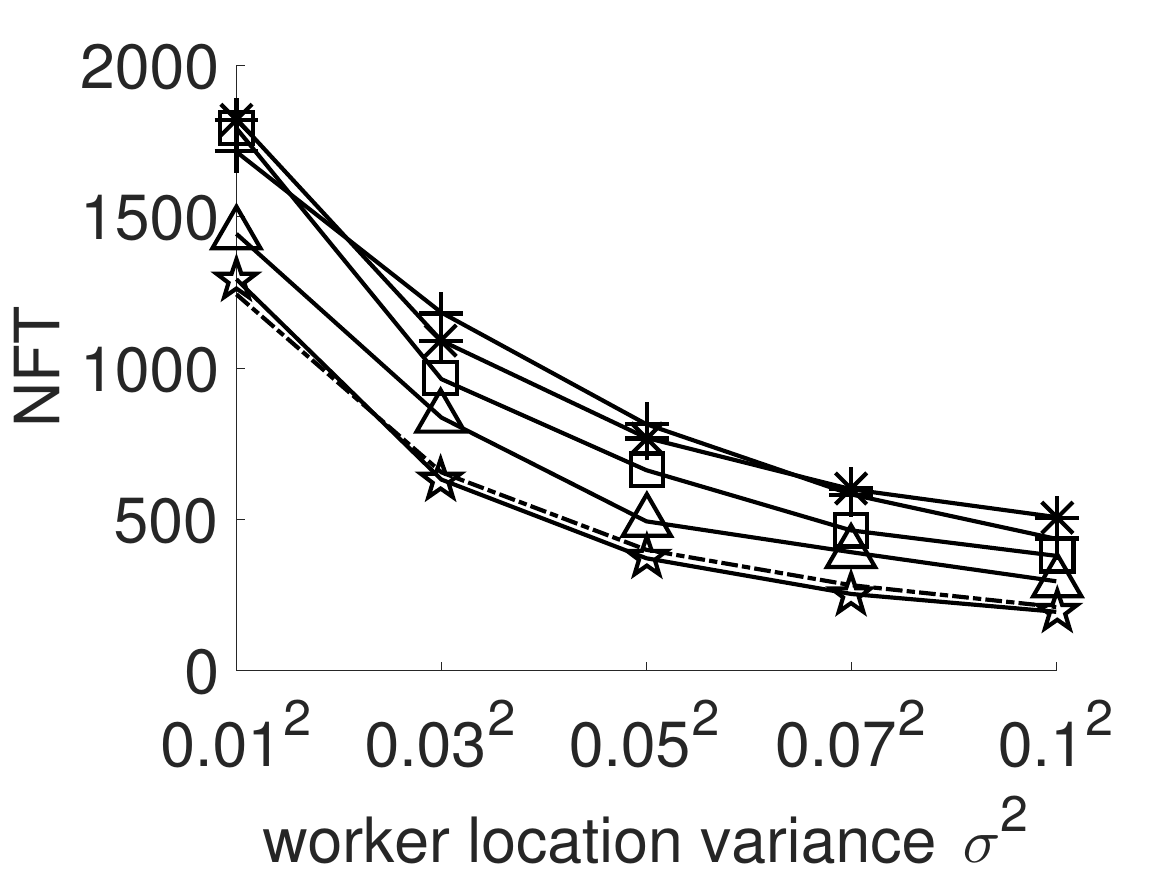}}
		\label{subfig:variance_finished_task_number_gaus}}\hfill
	\subfigure[][{\small Confidently Assigned Tasks}]{
		\scalebox{0.2}[0.2]{\includegraphics{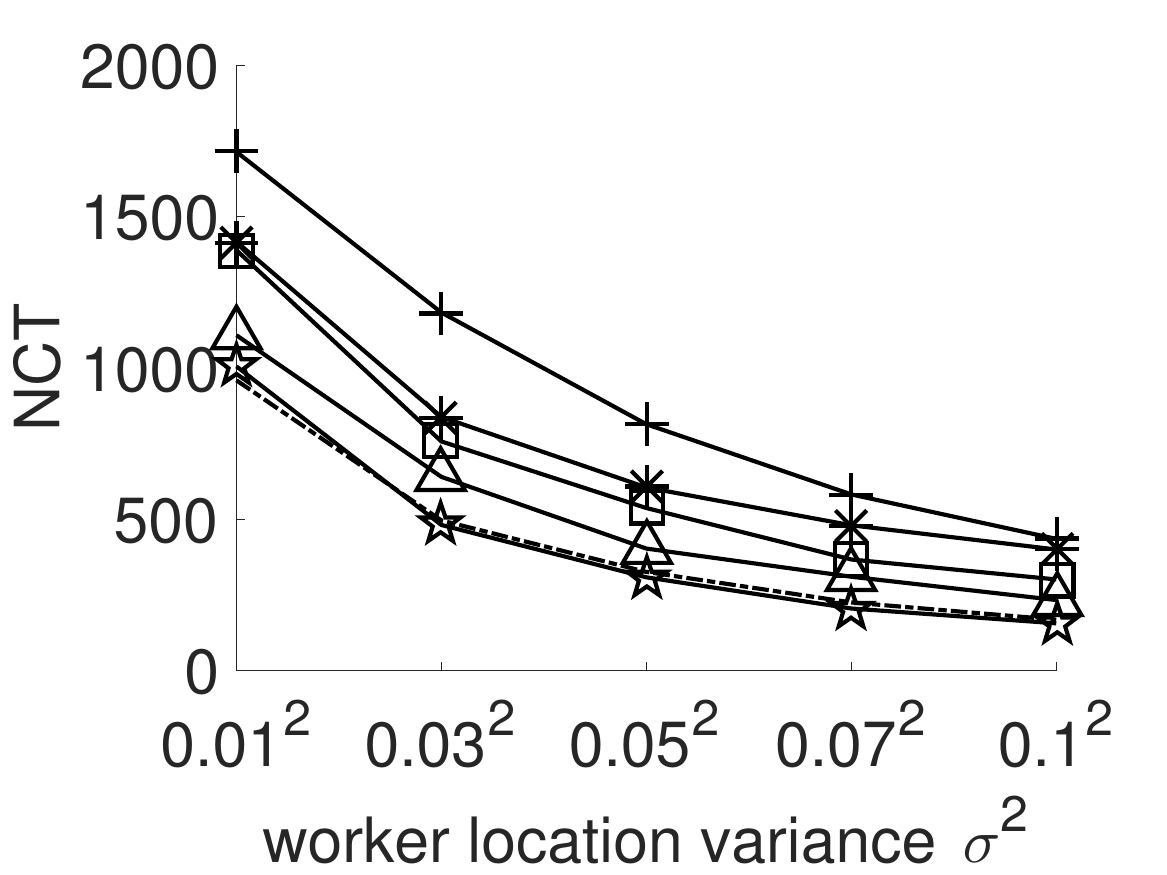}}
		\label{subfig:variance_finished_task_number_conf_gaus}}
	\subfigure[][{\small Running Time}]{
		\scalebox{0.2}[0.2]{\includegraphics{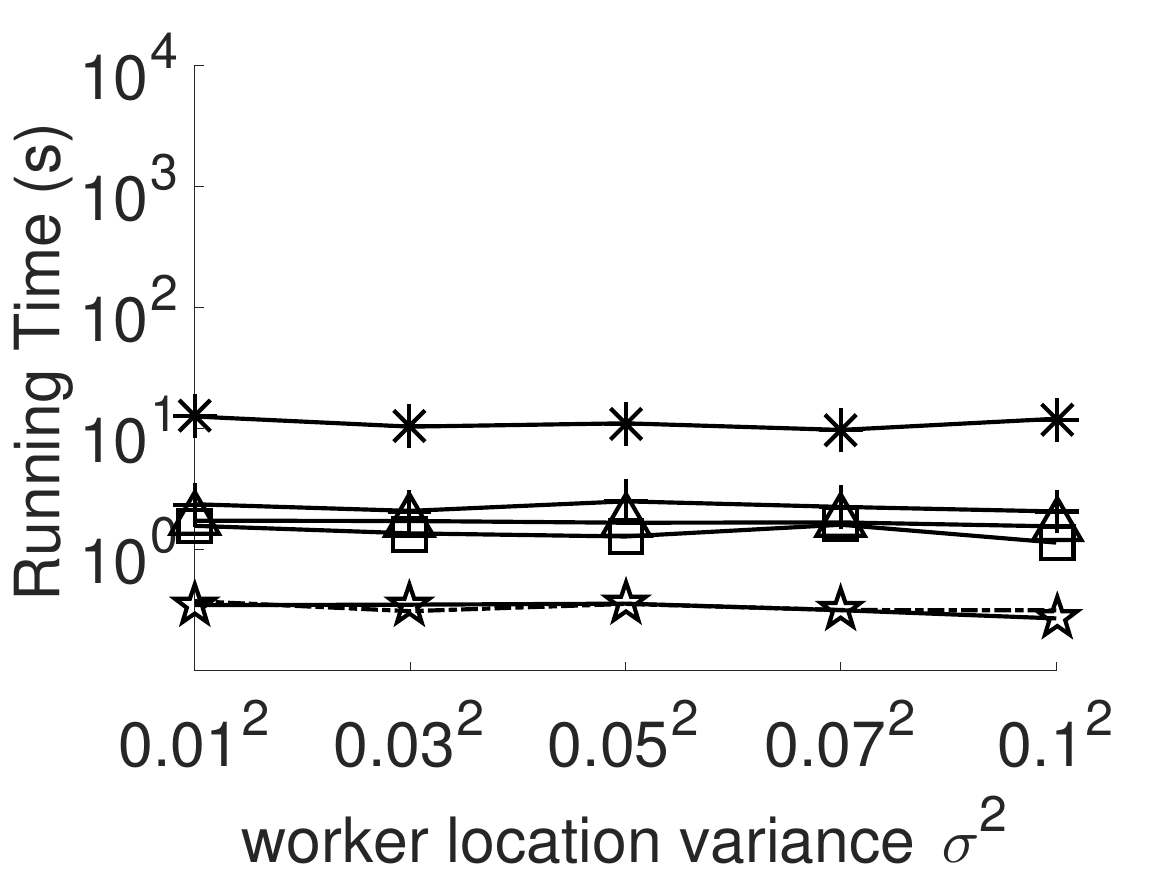}}
		\label{subfig:variance_running_time_gaus}}\hfill
	
	\subfigure[][{\small Moving Distance}]{
		\scalebox{0.2}[0.2]{\includegraphics{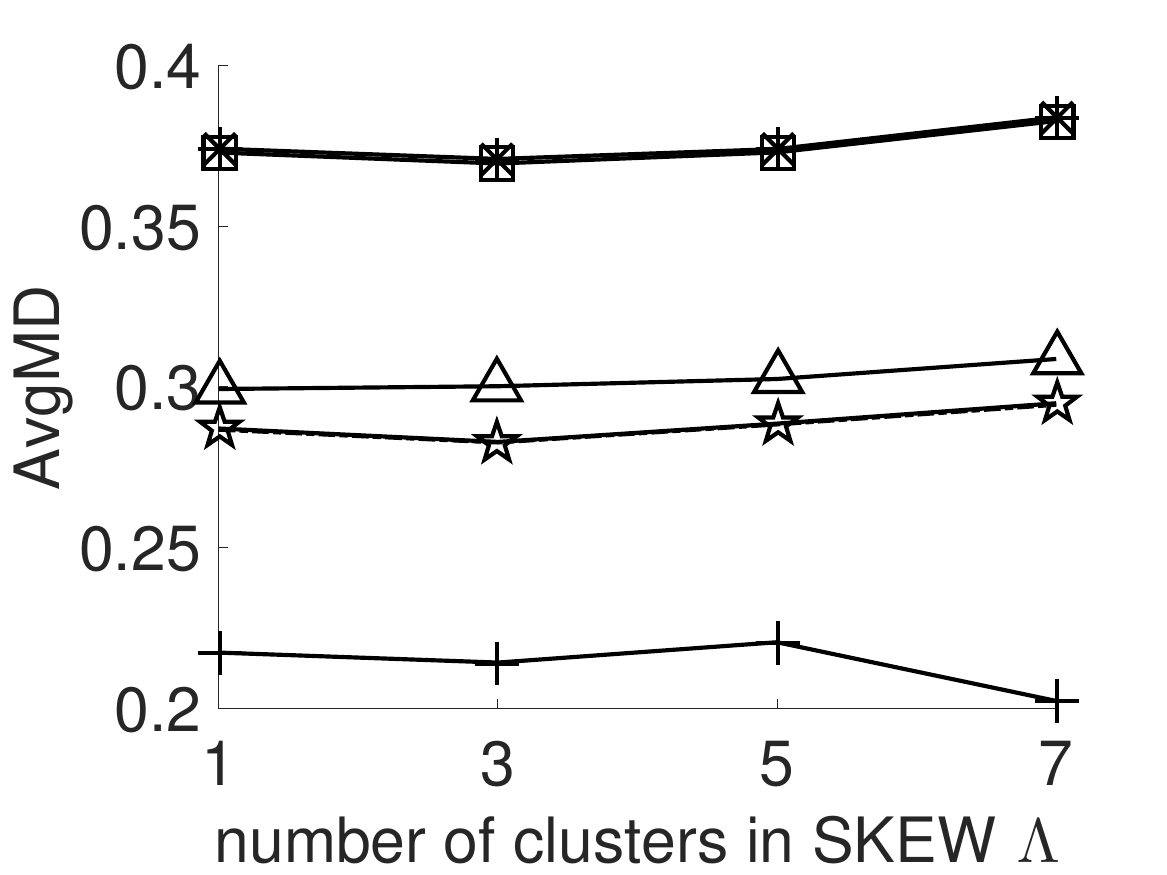}}
		\label{subfig:cluster_avg_moving_distance_gaus}}\hfill\vspace{-2ex}
	\subfigure[][{\small Fully Assigned Tasks}]{
		\scalebox{0.2}[0.2]{\includegraphics{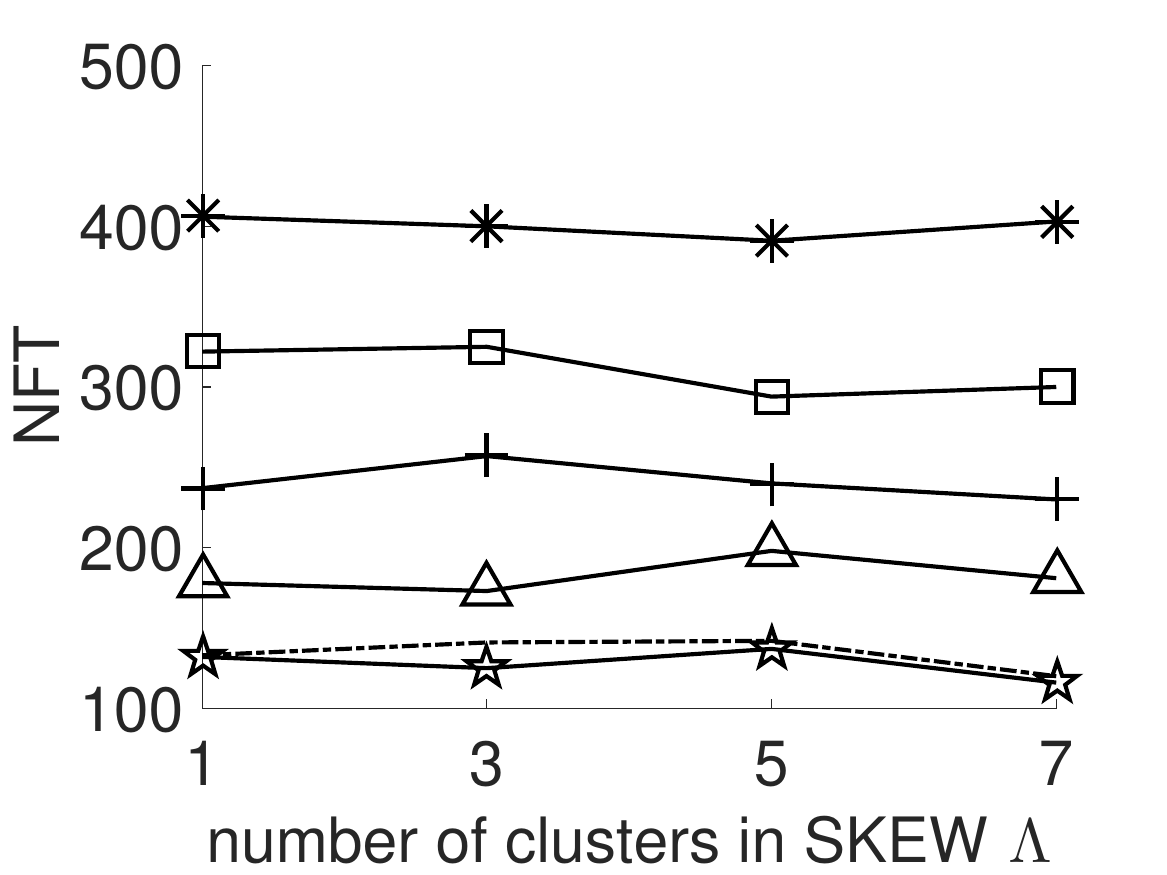}}
		\label{subfig:cluster_finished_task_number_gaus}}\hfill
	\subfigure[][{\small Confidently Assigned Tasks}]{
		\scalebox{0.2}[0.2]{\includegraphics{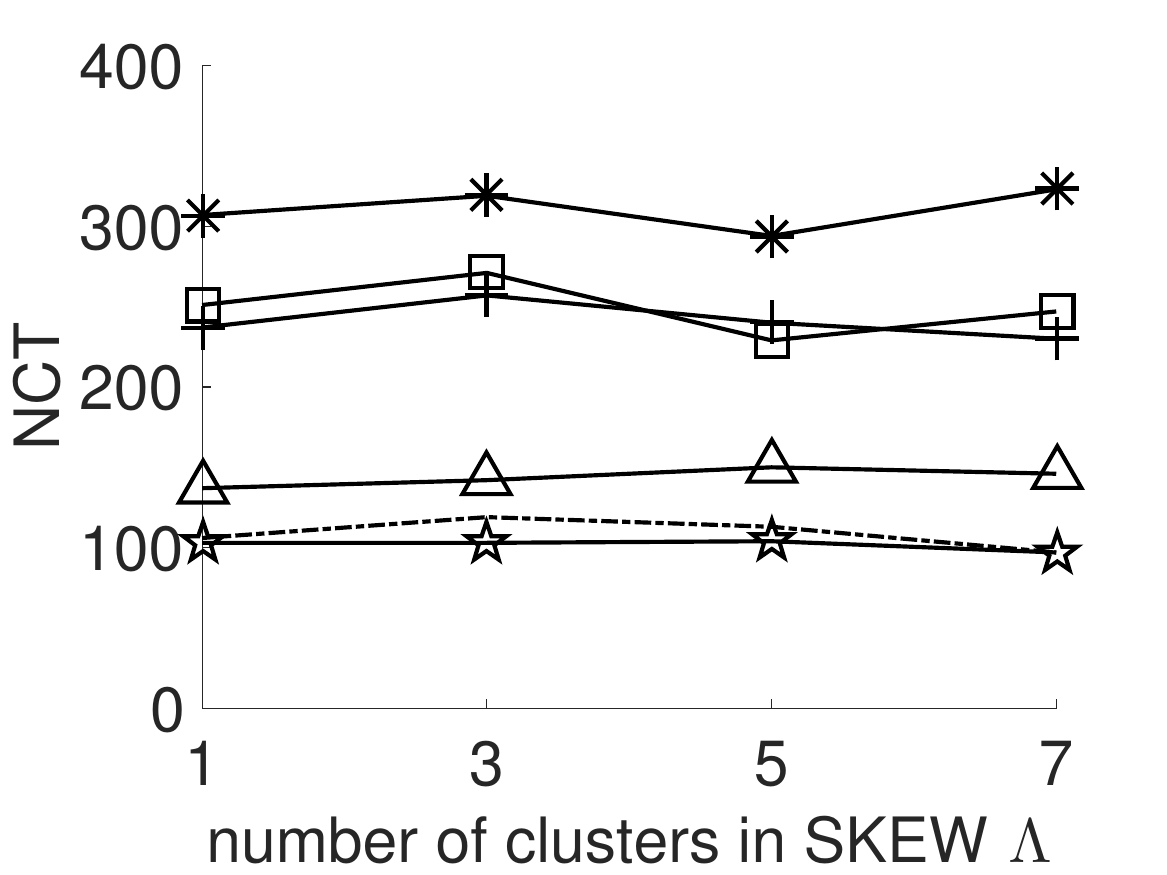}}
		\label{subfig:cluster_finished_task_number_conf_gaus}}\hfill\vspace{-1ex}
	\subfigure[][{\small Running Time}]{
		\scalebox{0.2}[0.2]{\includegraphics{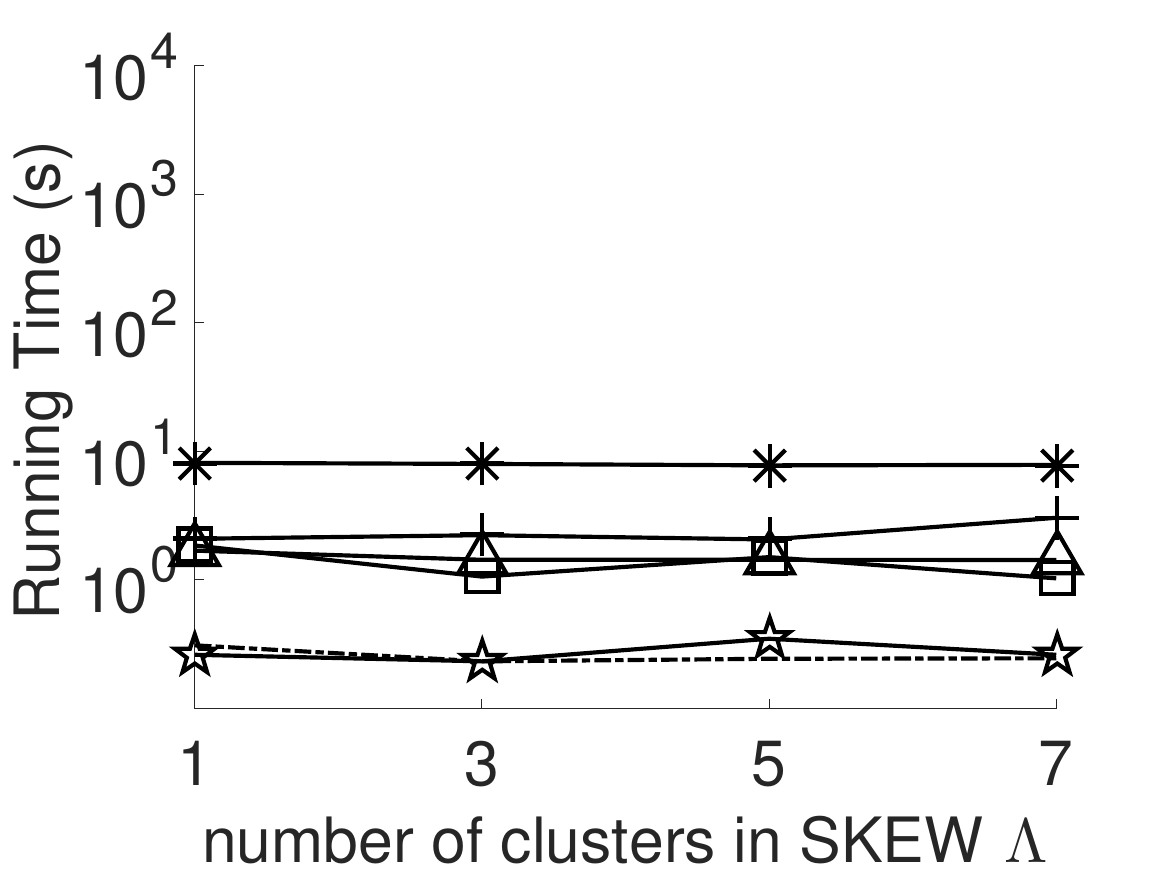}}
		\label{subfig:cluster_running_time_gaus}}\hfill
	\caption{\small Results that the locations of workers follow GAUS and SKEW  while the locations of tasks follow GAUS. (Synthetic).}
	\label{fig:effect_worker_distribution_task_gaus}\vspace{-2ex}
\end{figure*}

\begin{figure*}[th!]\centering
	\subfigure{
		\scalebox{0.4}[0.4]{\includegraphics{bar_mix-eps-converted-to.pdf}}}\hfill\\\vspace{-2ex}
	\addtocounter{subfigure}{-1}
	\subfigure[][{\small Moving Distance}]{
		\scalebox{0.2}[0.2]{\includegraphics{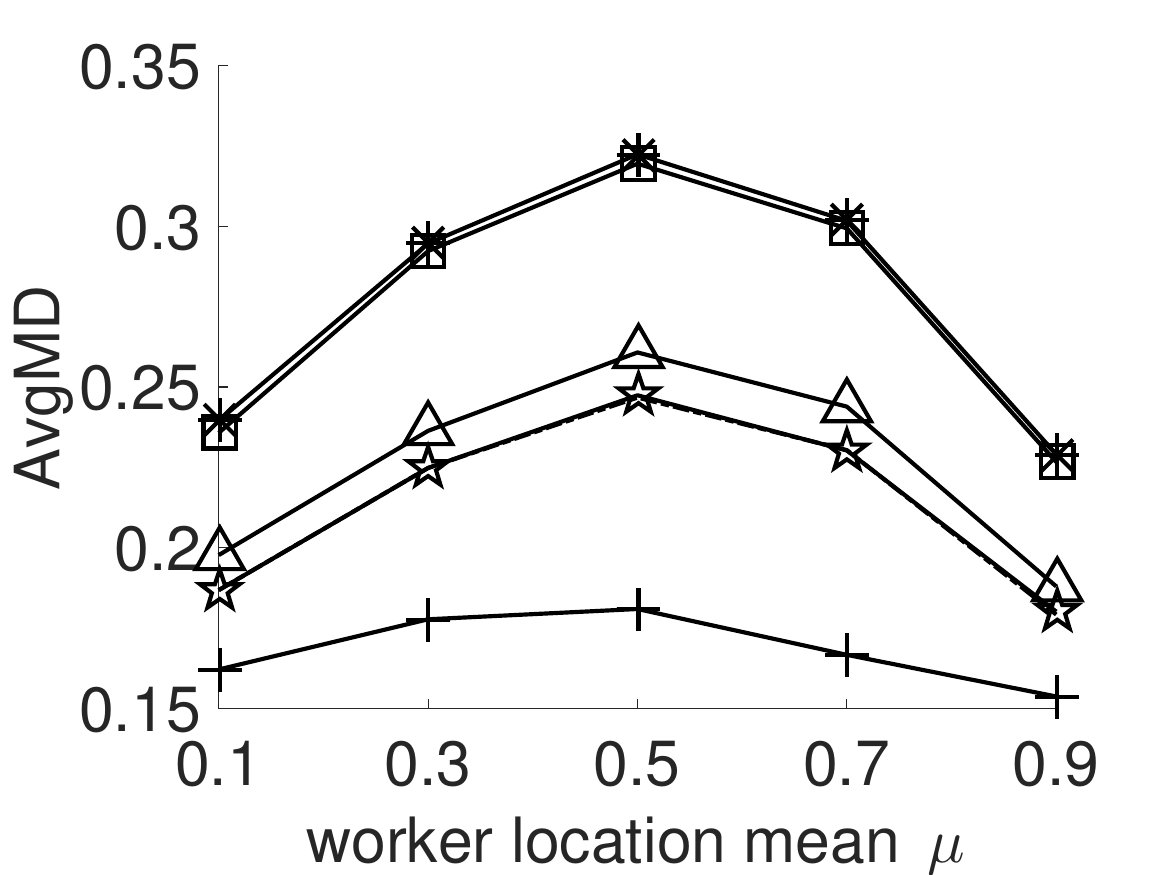}}
		\label{subfig:mean_avg_moving_distance_skew}}\hfill\vspace{-2ex}
	\subfigure[][{\small Fully Assigned Tasks}]{
		\scalebox{0.2}[0.2]{\includegraphics{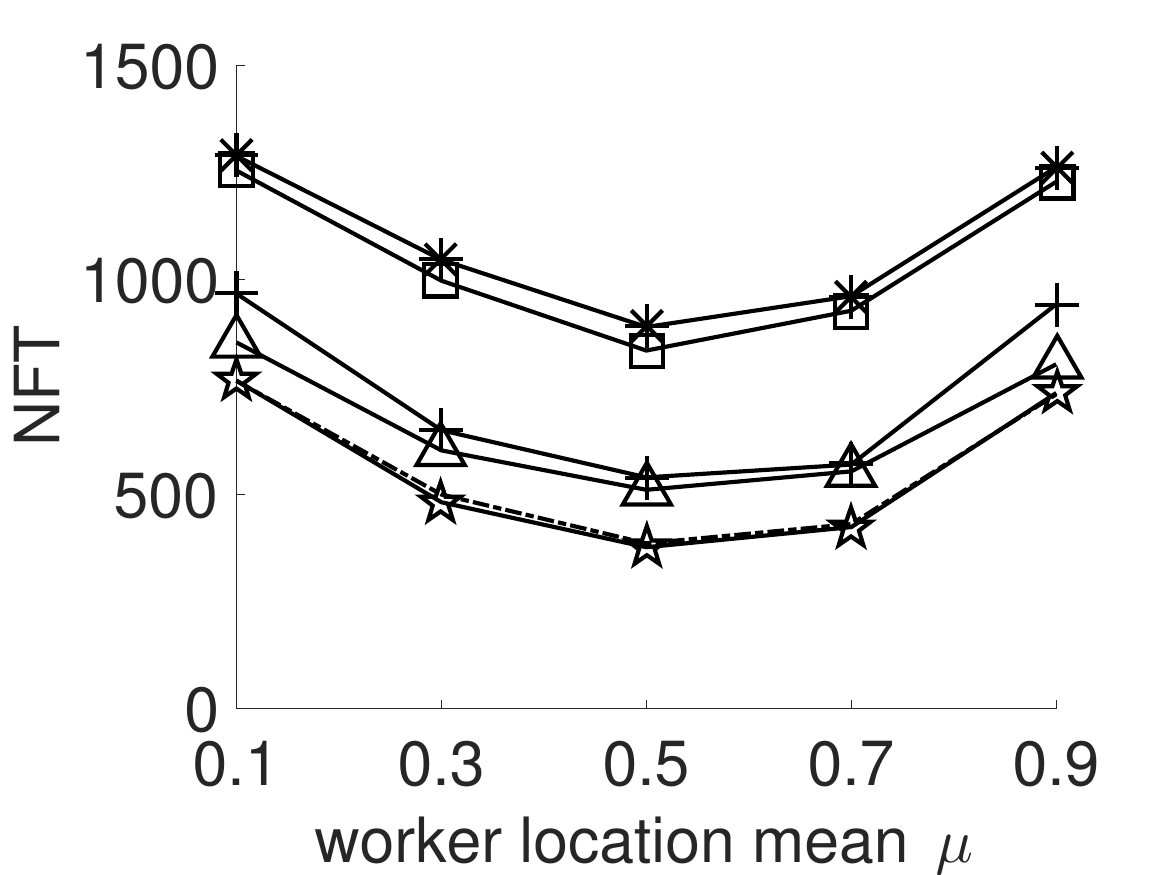}}
		\label{subfig:mean_finished_task_number_skew}}\hfill
	\subfigure[][{\small Confidently Assigned Tasks}]{
		\scalebox{0.2}[0.2]{\includegraphics{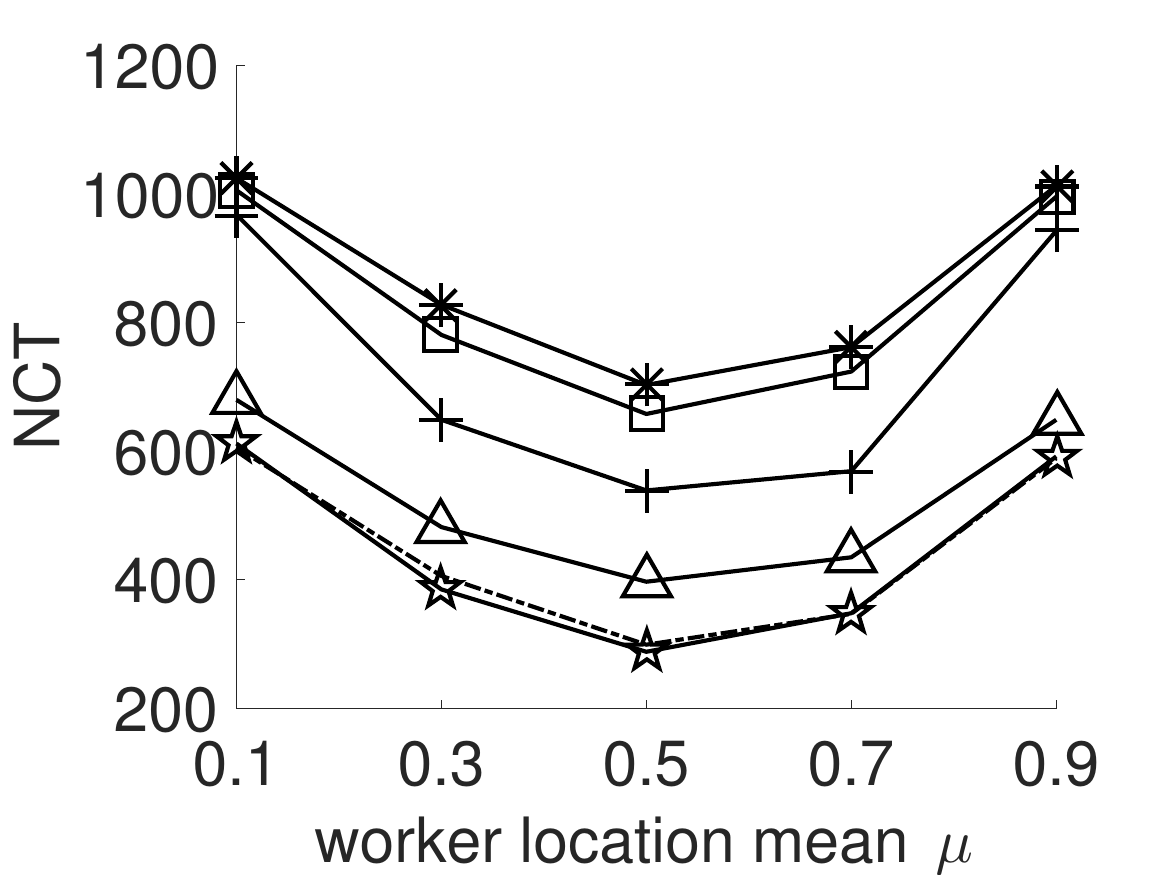}}
		\label{subfig:mean_finished_task_number_conf_skew}}\hfill
	\subfigure[][{\small Running Time}]{
		\scalebox{0.2}[0.2]{\includegraphics{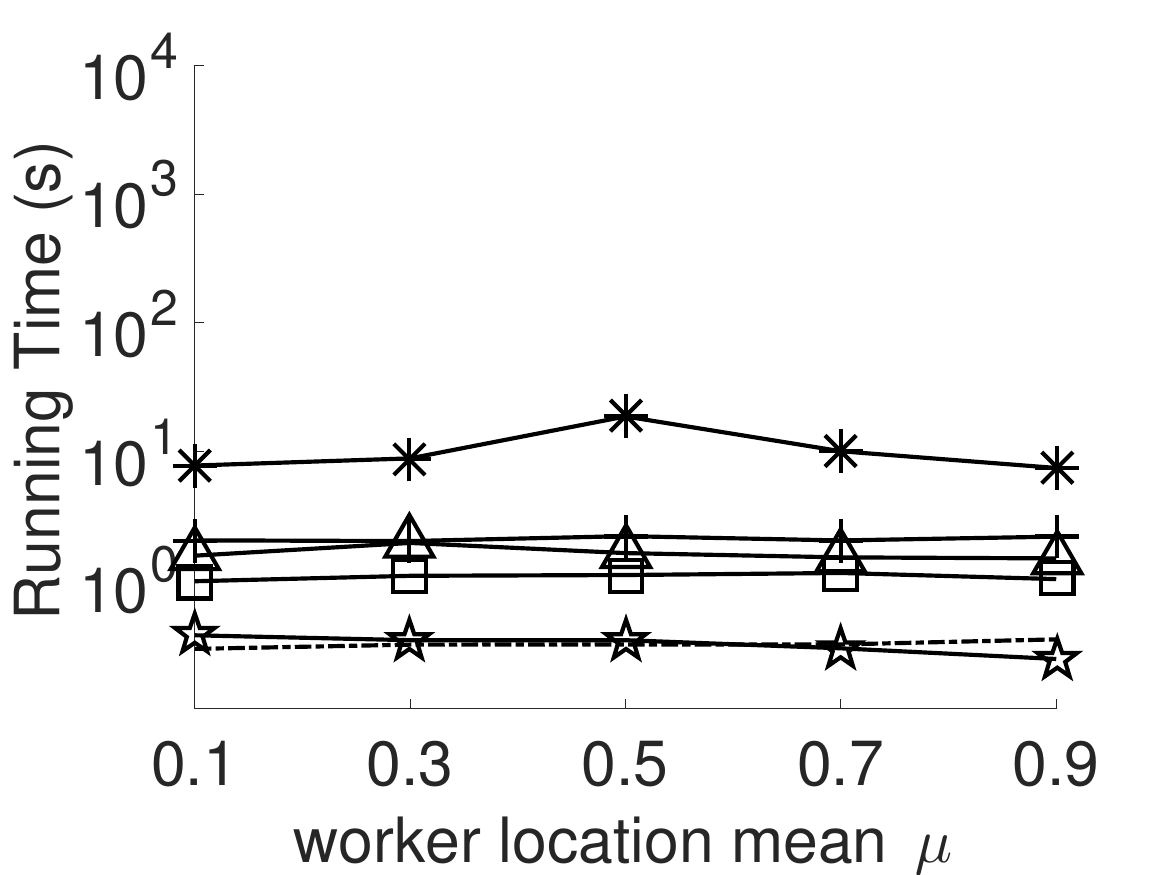}}
		\label{subfig:mean_running_time_skew}}\hfill
	
	\subfigure[][{\small Moving Distance}]{
		\scalebox{0.2}[0.2]{\includegraphics{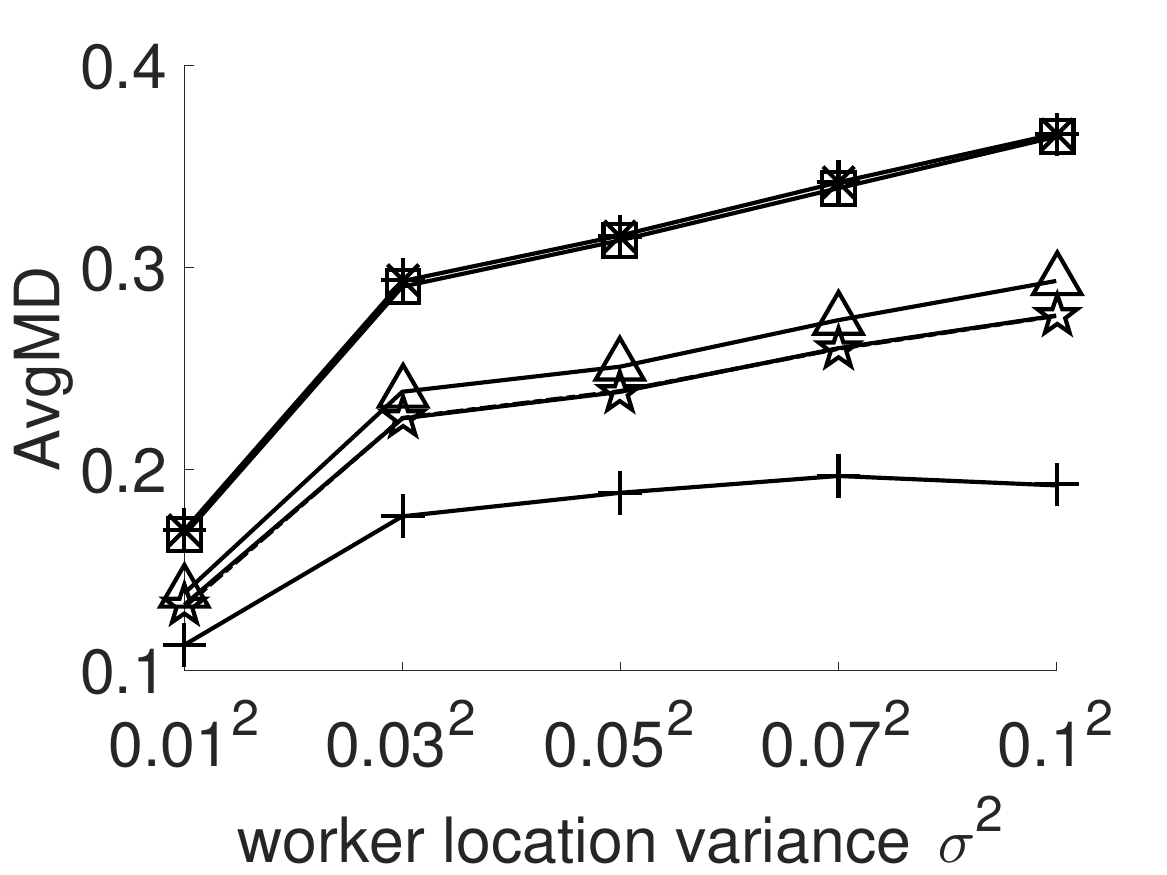}}
		\label{subfig:variance_avg_moving_distance_skew}}\hfill\vspace{-2ex}
	\subfigure[][{\small Fully Assigned Tasks}]{
		\scalebox{0.2}[0.2]{\includegraphics{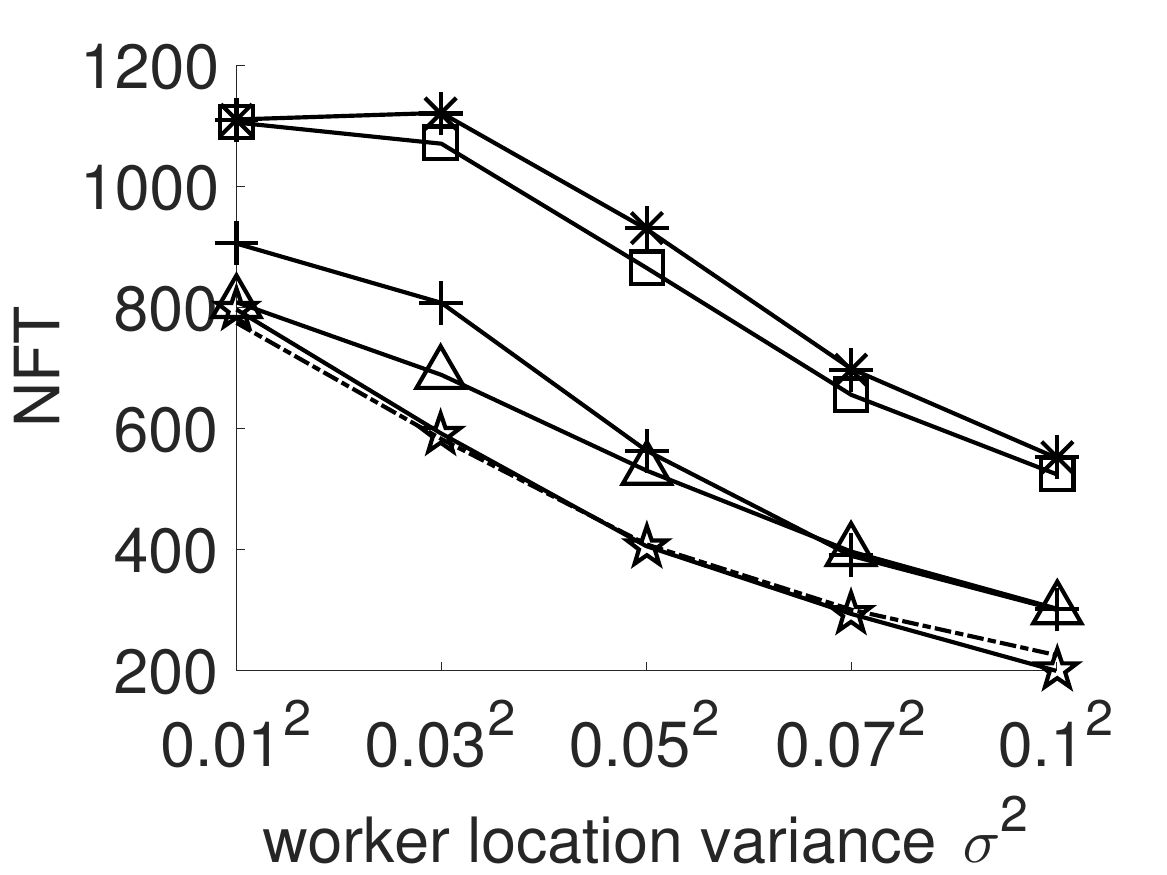}}
		\label{subfig:variance_finished_task_number_skew}}\hfill
	\subfigure[][{\small Confidently Assigned Tasks}]{
		\scalebox{0.2}[0.2]{\includegraphics{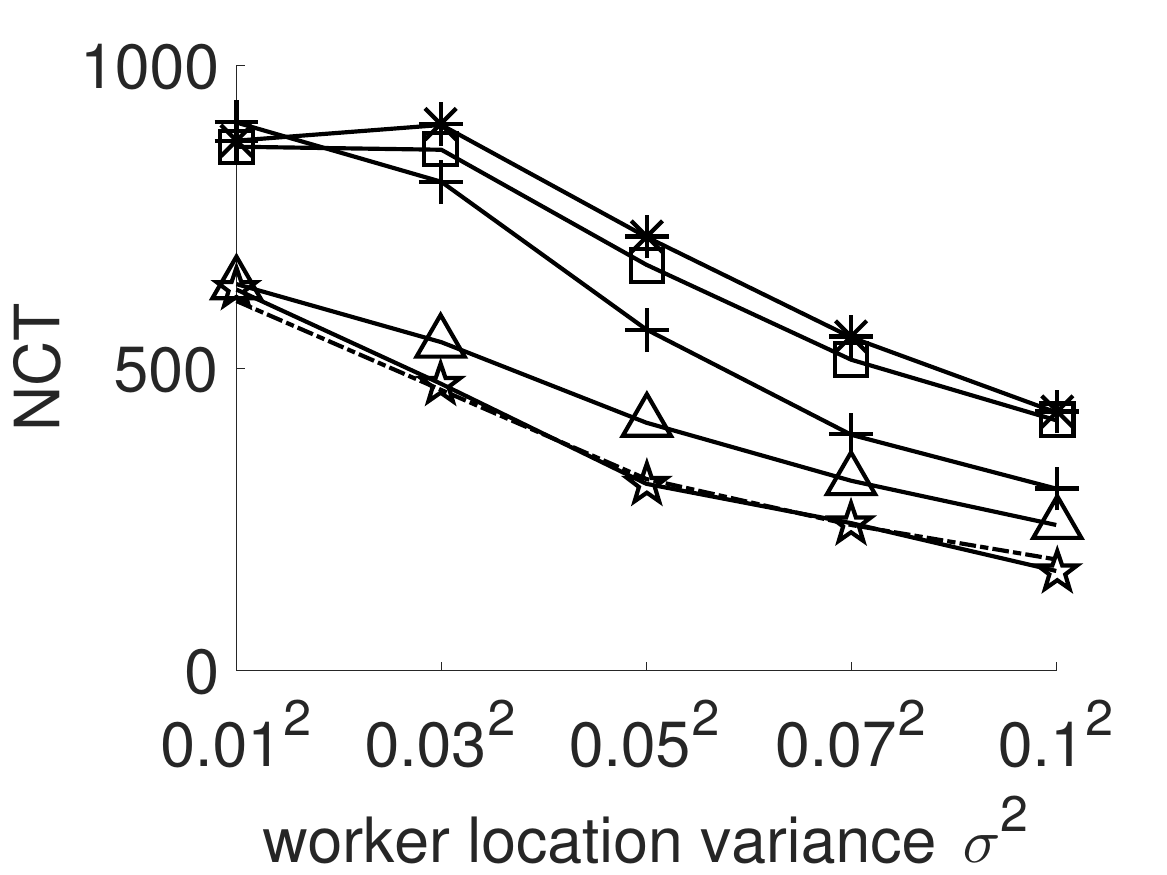}}
		\label{subfig:variance_finished_task_number_conf_skew}}
	\subfigure[][{\small Running Time}]{
		\scalebox{0.2}[0.2]{\includegraphics{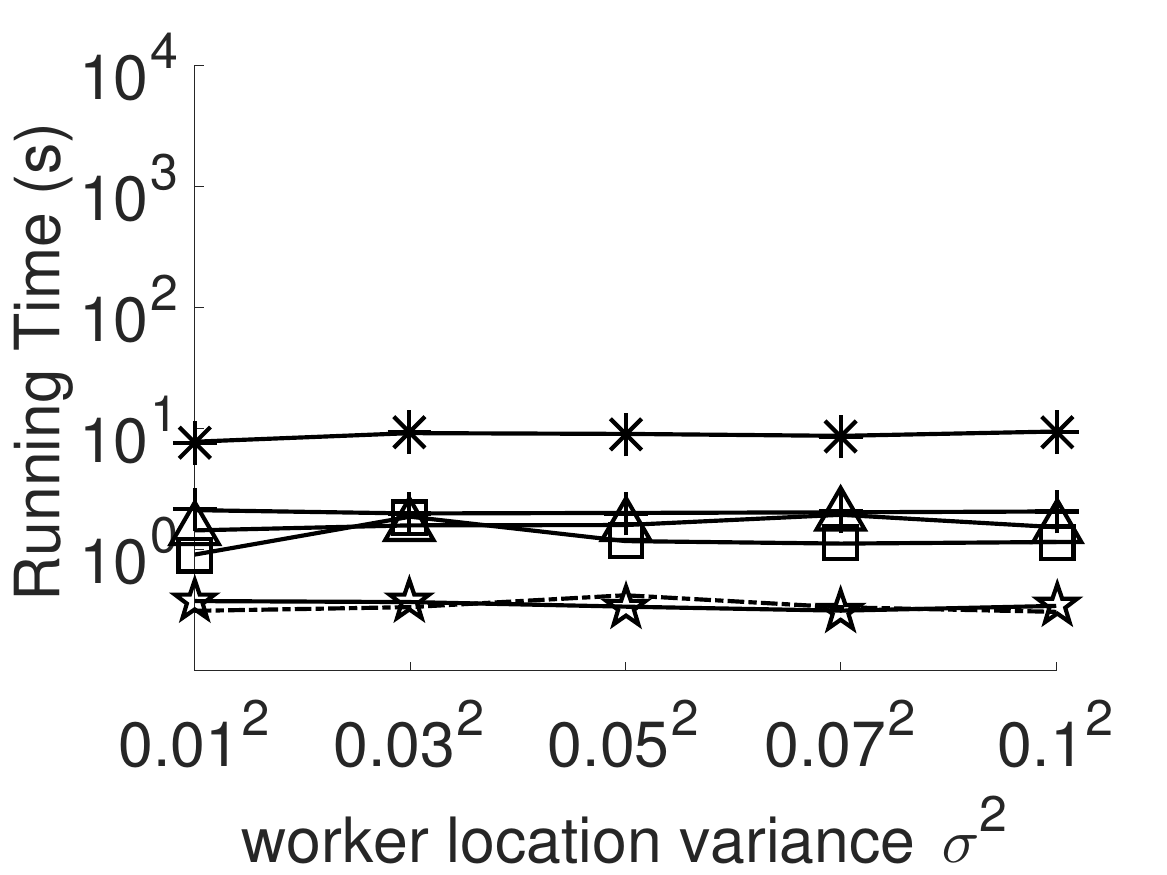}}
		\label{subfig:variance_running_time_skew}}\hfill
	
	\subfigure[][{\small Moving Distance}]{
		\scalebox{0.2}[0.2]{\includegraphics{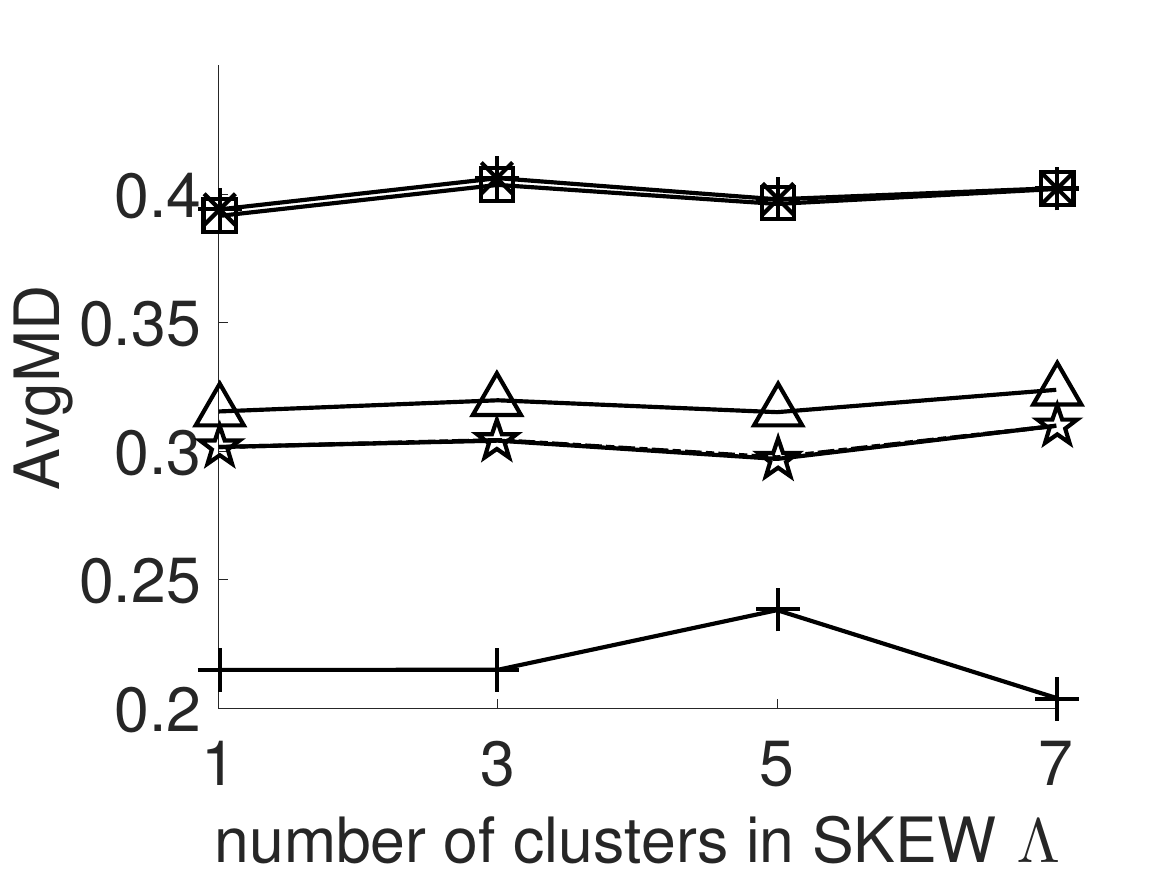}}
		\label{subfig:cluster_avg_moving_distance_skew}}\hfill\vspace{-2ex}
	\subfigure[][{\small Fully Assigned Tasks}]{
		\scalebox{0.2}[0.2]{\includegraphics{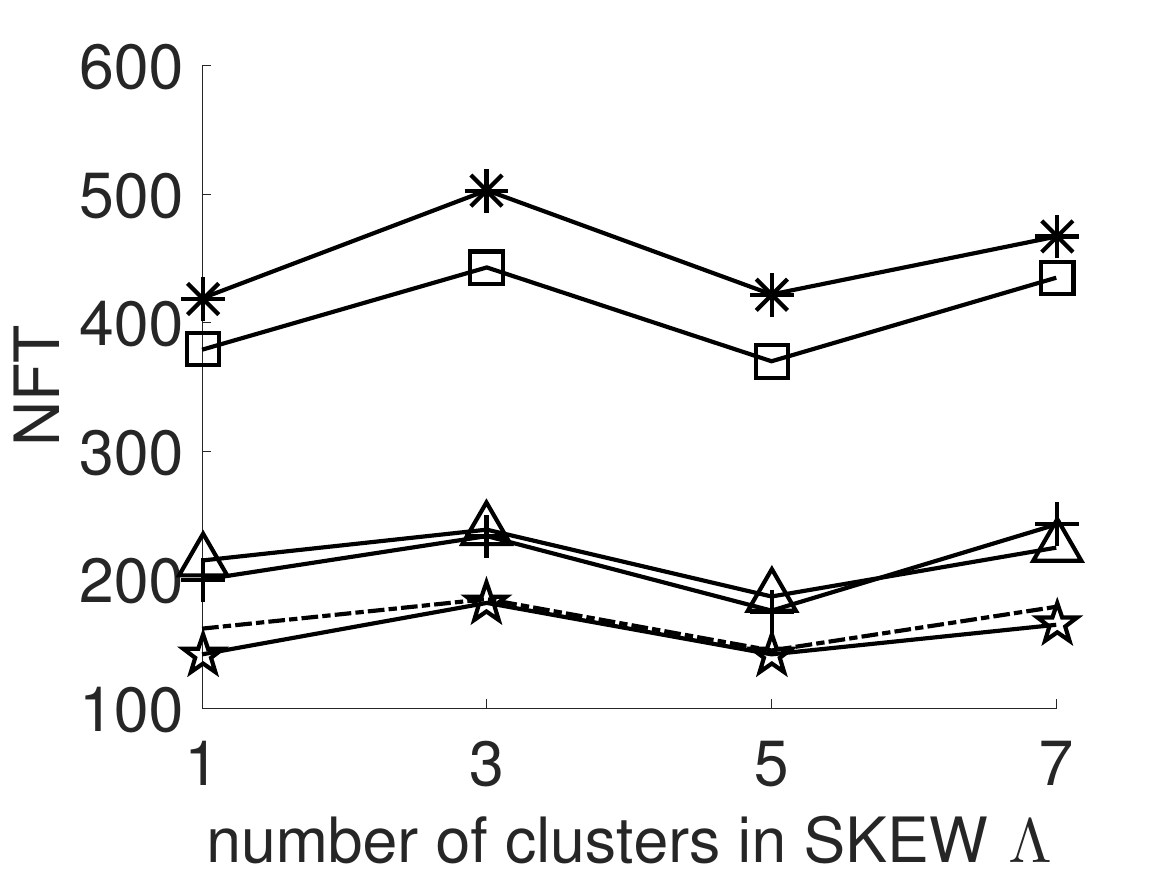}}
		\label{subfig:cluster_finished_task_number_skew}}\hfill
	\subfigure[][{\small Confidently Assigned Tasks}]{
		\scalebox{0.2}[0.2]{\includegraphics{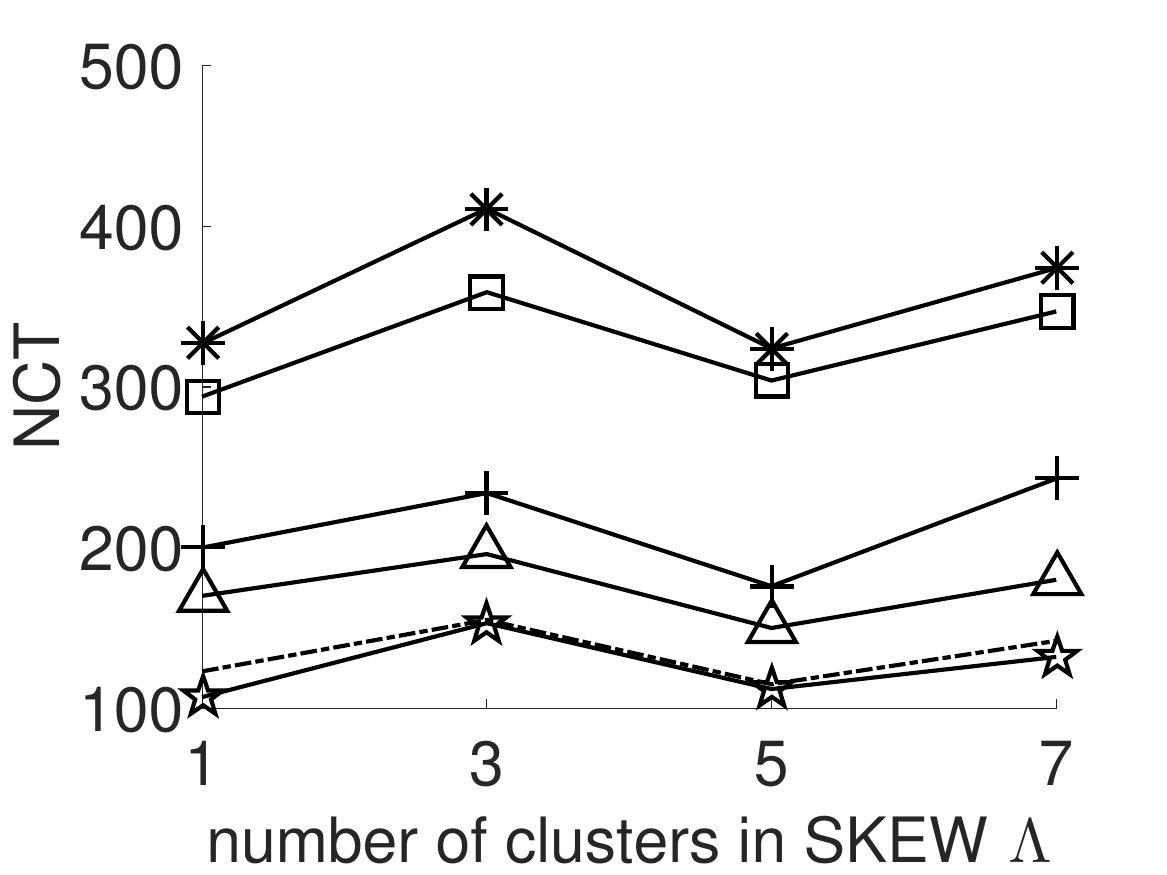}}
		\label{subfig:cluster_finished_task_number_conf_skew}}\hfill\vspace{-1ex}
	\subfigure[][{\small Running Time}]{
		\scalebox{0.2}[0.2]{\includegraphics{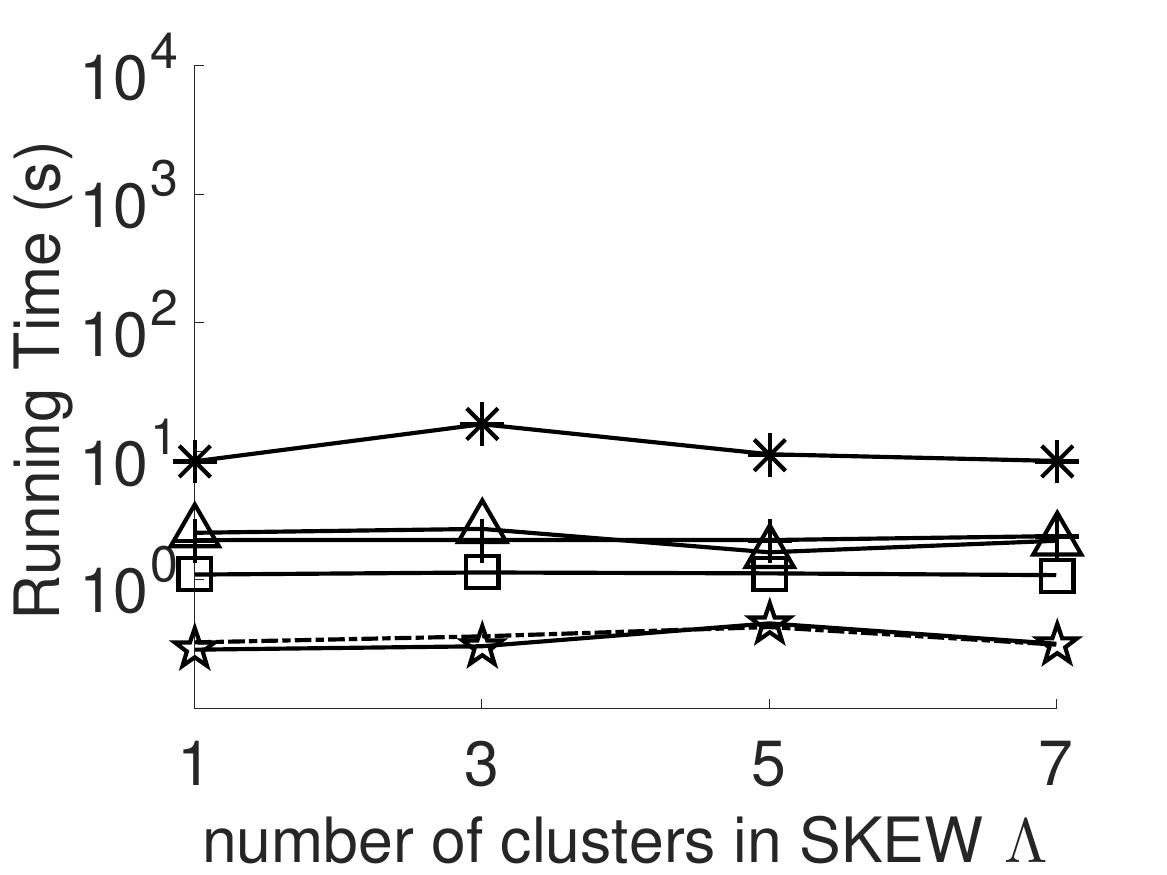}}
		\label{subfig:cluster_running_time_skew}}\hfill
	\caption{\small Results that the locations of workers follow GAUS and SKEW  while the locations of tasks follow SKEW. (Synthetic).}
	\label{fig:effect_worker_distribution_task_skew}\vspace{-2ex}
\end{figure*}

 For locations of tasks following GAUS and SKEW as shown in Figures \ref{fig:effect_worker_distribution_task_gaus} and \ref{fig:effect_worker_distribution_task_skew} respectively, the results are similar to that when the locations of tasks follow UNIF. We will just discuss the different situations in Figures \ref{subfig:mean_finished_task_number_gaus} and \ref{subfig:mean_finished_task_number_conf_gaus}.
 The difference between Figures \ref{subfig:mean_finished_task_number_unif} and \ref{subfig:mean_finished_task_number_gaus} is that when the mean, $\mu$, of workers' locations' GAUS  increases from $0.1$ to $0.3$, NFTs of all the tested algorithms except for GT-hgr increase when the locations of tasks follow GAUS and decrease when the locations of tasks follow UNIF.  When the locations of tasks follow GAUS, the tasks are crowded close to the center point ($0.5$, $0.5$). Then when  the mean, $\mu$, of workers' locations' GAUS  increases from $0.1$ to $0.3$, CT increases, but W/CT does not drop since more workers can cover tasks. However, When the locations of tasks follow UNIF, when  the mean, $\mu$, of workers' locations' GAUS  increases from $0.1$ to $0.3$, CT increases and W/CT drops.
 
\end{document}